%% file: docnet.tex
\documentstyle[12pt,twoside,a4wide,theading,epsf,rotate,equations]{article}
\newcommand{\nwc}{\newcommand}
\nwc{\cl}  {$\clubsuit$}

\nwc{\hyp} {\hyphenation} 
\nwc{\be}  {\begin{equation}}
\nwc{\ee}  {\end{equation}}
\nwc{\ba}  {\begin{array}}
\nwc{\ea}  {\end{array}}
\nwc{\bdm} {\begin{displaymath}}
\nwc{\edm} {\end{displaymath}}
\nwc{\bea} {\be\ba{rcl}}
\nwc{\eea} {\ea\ee}
\nwc{\ben} {\begin{eqnarray}}
\nwc{\een} {\end{eqnarray}}
\nwc{\bda} {\bdm\ba{lcl}}
\nwc{\eda} {\ea\edm}
\nwc{\bc}  {\begin{center}}
\nwc{\ec}  {\end{center}}
\nwc{\ds}  {\displaystyle}
\nwc{\bmat}{\left(\ba}
\nwc{\emat}{\ea\right)}
\nwc{\non} {\nonumber}
\nwc{\bib} {\bibitem}
\nwc{\lra} {\longrightarrow}
\nwc{\Llra}{\Longleftrightarrow}
\nwc{\ra}  {\rightarrow}
\nwc{\Ra}  {\Rightarrow}
\nwc{\lmt} {\longmapsto}
\nwc{\prl} {\partial}
\nwc{\iy}  {\infty}
\nwc{\ol}  {\overline}
\nwc{\hm}  {\hspace{3mm}}
\nwc{\lf}  {\left}
\nwc{\ri}  {\right}
\nwc{\lm}  {\limits}
\nwc{\lb}  {\lbrack}
\nwc{\rb}  {\rbrack}
\nwc{\ov}  {\over}
\nwc{\pri}  {\prime}
\nwc{\nnn} {\nonumber \vspace{.2cm} \\ }
\nwc{\Sc}  {{\cal S}}
\nwc{\Lc}  {{\cal L}}
\nwc{\Rc}  {{\cal R}}
\nwc{\Dc}  {{\cal D}}
\nwc{\Oc}  {{\cal O}}
\nwc{\Cc}  {{\cal C}}
\nwc{\Pc}  {{\cal P}}
\nwc{\Mc}  {{\cal M}}
\nwc{\Ec}  {{\cal E}}
\nwc{\Fc}  {{\cal F}}
\nwc{\Hc}  {{\cal H}}
\nwc{\Kc}  {{\cal K}}
\nwc{\Xc}  {{\cal X}}
\nwc{\Gc}  {{\cal G}}
\nwc{\Zc}  {{\cal Z}}
\nwc{\Nc}  {{\cal N}}
\nwc{\fca} {{\cal f}}
\nwc{\xc}  {{\cal x}}
\nwc{\Ac}  {{\cal A}}
\nwc{\Bc}  {{\cal B}}
\nwc{\Uc}  {{\cal U}}
\nwc{\Vc}  {{\cal V}}
\nwc{\Th} {\Theta}
\nwc{\th} {\theta}
\nwc{\vth} {\vartheta}
\nwc{\eps}{\epsilon}
\nwc{\si} {\sigma}
\nwc{\Gm} {\Gamma}
\nwc{\gm} {\gamma}
\nwc{\bt} {\beta}
\nwc{\La} {\Lambda}
\nwc{\la} {\lambda}
\nwc{\om} {\omega}
\nwc{\Om} {\Omega}
\nwc{\dt} {\delta}
\nwc{\Si} {\Sigma}
\nwc{\Dt} {\Delta}
\nwc{\al} {\alpha}
\nwc{\vp} {\varphi}
\nwc{\vph}{\varphi}
\nwc{\kp} {\kappa}
\def\tr{\mathop{\rm tr}}
\def\Tr{\mathop{\rm Tr}}

\def\VEV#1{\left\langle #1\right\rangle}

\def\abs#1{\left| #1\right|}
\def\pr#1{#1^\prime}
\def\ltap{\raisebox{-.4ex}{\rlap{$\sim$}} \raisebox{.4ex}{$<$}}
\def\gtap{\raisebox{-.4ex}{\rlap{$\sim$}} \raisebox{.4ex}{$>$}}
\nwc{\Id}  {{\bf 1}}
\nwc{\diag} {{\rm diag}}
\nwc{\inv}  {{\rm inv}}
\nwc{\mod}  {{\rm mod}}
\nwc{\hal} {\frac{1}{2}}
\nwc{\tpi}  {2\pi i}

\def\slash#1{#1\!\!\!/\!\,\,}

\def\PRL{\em Phys. Rev. Lett.}

\def\ijmpc#1{Int.\ J.\ Mod.\ Phys.\ {\bf C#1}}
\def\jpa#1{J.\ Phys.\ {\bf A#1}}

\def\npb#1{Nucl.\ Phys.\ {\bf B#1}}

\def\plb#1{Phys.\ Lett.\ {\bf B#1}}
\def\pr#1{Phys.\ Rev.\ {\bf #1}}
\def\pra#1{Phys.\ Rev.\ {\bf A#1}}
\def\prb#1{Phys.\ Rev.\ {\bf B#1}}

\def\prd#1{Phys.\ Rev.\ {\bf D#1}}
\def\prle#1{Phys.\ Rev.\ Lett.\ {\bf #1}}

\def\zpb#1{Z.\ Phys.\ {\bf B#1}}
\def\zpc#1{Z.\ Phys.\ {\bf C#1}}

\def\APP#1{Acta Phys.~Pol.~{\bf #1}}

\def\CNPP#1{Comm. Nucl. Part. Phys.~{\bf #1}}

\def\IJMP#1{Int. J. Mod. Phys.~{\bf #1}}

\def\NP#1{Nucl. Phys.~{\bf #1}}
\def\NPPS#1{Nucl. Phys. Proc. Suppl.~{\bf #1}}
\def\NC#1{Nuovo Cim.~{\bf #1}}

\def\PL#1{Phys. Lett.~{\bf #1}}
\def\PR#1{Phys. Rev.~{\bf #1}}
\def\PRP#1{Phys. Rep.~{\bf #1}}
\def\PRL#1{Phys. Rev. Lett.~{\bf #1}}

\def\PTP#1{Progr. Theor. Phys.~{\bf #1}}
\def\RMP#1{Rev. Mod. Phys.~{\bf #1}}

\def\ZP#1{Z. Phys.~{\bf #1}}

\def\MeV {\,{\rm  MeV}}
\def\GeV {\,{\rm  GeV}}

\def \lta {\mathrel{\vcenter
     {\hbox{$<$}\nointerlineskip\hbox{$\sim$}}}}
\def \gta {\mathrel{\vcenter
     {\hbox{$>$}\nointerlineskip\hbox{$\sim$}}}}
\newsavebox{\nnin} \sbox{\nnin}{$\hspace{1mm}\in\kern -.8em /
                   \hspace{1mm}$}

\newcommand{\sub}{\subset}
\newsavebox{\nnsub} \sbox{\nnsub}{$\hspace{1mm}\sub\kern -.9em /
            \hspace{1mm}$}

\def\KK{{\rm I\kern -.2em  K}}
\def\NN{{\rm I\kern -.16em N}}
\def\RR{{\rm I\kern -.2em  R}}
\def\ZZ{Z \kern -.43em Z}
\def\QQ{{\rm \kern .25em
             \vrule height1.4ex depth-.12ex width.06em\kern-.31em Q}}
\def\CC{{\rm \kern .25em
             \vrule height1.4ex depth-.12ex width.06em\kern-.31em C}}
\def\ZZZ{Z\kern -0.31em Z}

\nwc{\olnu}  {\ol{\nu}}
\nwc{\olla}  {\ol{\la}}
\nwc{\olm}   {\ol{m}}
\nwc{\olmu}  {\ol{\mu}}
\nwc{\olh}   {\ol{h}}
\nwc{\olpsi} {\ol{\psi}}
\nwc{\olsi}  {\ol{\sigma}}
\nwc{\olgm}  {\ol{\gm}}
\nwc{\prlt}  {\frac{\prl}{\prl t}}
\nwc{\ttau}  {\tilde{\tau}}
\nwc{\trho}  {\tilde{\rho}}
\nwc{\tP}    {\tilde{P}}
\nwc{\tU}    {\tilde{U}}
\nwc{\teps}  {\tilde{\eps}}
\nwc{\tla}   {\tilde{\la}}
\nwc{\tit}    {\tilde{t}}
\nwc{\iddq}  {\int\frac{d^dq}{(2\pi)^d}}
\nwc{\prpr}  {\prime\prime}
\nwc{\rN}    {\left(\frac{\rho}{N}\right)}
\nwc{\rNt}    {\left(\frac{\rho}{N}\right)^{\frac{N-2}{2}}}
\nwc{\rnN}   {\left(\frac{\rho_0}{N}\right)}
\nwc{\rnNt}    {\left(\frac{\rho_0}{N}\right)^{\frac{N-2}{2}}}
\nwc{\rnNf}    {\left(\frac{\rho_0}{N}\right)^{\frac{N-4}{2}}}
\nwc{\rNs}    {\left(\frac{\rho_0}{N}\right)^{\frac{N-6}{2}}}
\nwc{\kNt}    {\left(\frac{\kappa}{N}\right)^{\frac{N-2}{2}}}
\nwc{\kNf}    {\left(\frac{\kappa}{N}\right)^{\frac{N-4}{2}}}
\nwc{\kNs}    {\left(\frac{\kappa}{N}\right)^{\frac{N-6}{2}}}

\newcounter{app}
\def\app{\par
 \addtocounter{app}{1}
 \def\thesection{\Alph{app}}
 \def\ksection{\Alph{app}}}

\def\appendix#1{\app\sect{#1}}

\renewcommand{\theequation}{\thesection.\arabic{equation}}

\newcommand{\sect}[1]{ \section{#1} \setcounter{equation}{0} }

\newcommand{\lx}{\lambda}
\newcommand{\Lx}{\Lambda}
\newcommand{\beq}{\begin{eqalignno}}
\newcommand{\eeq}{\end{eqalignno}}
\newcommand{\rhz}{\rho_0}
\newcommand{\rht}{\tilde{\rho}}
\newcommand{\kx}{\kappa}
\newcommand{\dkl}{\delta \kappa_{\Lambda}}
\newcommand{\ex}{\epsilon}

\newcommand{\tcr}{T_{cr}}
\newcommand{\lr}{{\lambda}_R}

\begin{document}

\clearpage
\thispagestyle{empty}
\begin{titlepage}

  \title{Universal Critical Equation of State\\
  and the\\
  Chiral Phase Transition in QCD\\ \vspace*{0.5cm}}

  \author{\vspace*{0.5cm} {\sc J{\"u}rgen Berges\thanks{Email:
        J.Berges@thphys.uni-heidelberg.de}}\\
    {\em Institut f\"ur Theoretische Physik} \\ {\em Universit\"at
      Heidelberg} \\ {\em Philosophenweg 16} \\ {\em D-69120 Heidelberg,
      Germany}\\ }

\date{}
\maketitle

\begin{picture}(5,2.5)
\put(300,300){HD--THEP--97--41}
\end{picture}

\vspace*{1cm}

\thispagestyle{empty}

\begin{abstract}
We employ non--perturbative flow equations to compute the equation of 
state for two flavor QCD within an effective quark meson model. This 
yields the temperature and quark mass dependence of quantities like
the chiral condensate or the pion mass. Our treatment covers both 
the chiral perturbation
theory domain of validity and the domain of validity of universality
associated with critical phenomena: We explicitly connect the
physics at zero temperature and realistic quark mass with the 
universal behavior near the critical temperature $T_c$ and the chiral
limit. For realistic quark masses the
pion correlation length near $T_c$
turns out to be smaller than its zero temperature value. 
In the vicinity of $T_c$ and zero quark mass we obtain a
precision estimate of the universal critical equation of state.
It belongs to the universality class of the three dimensional 
$O(4)$ symmetric Heisenberg model. We also investigate scalar field
theories directly in three dimensions
near second order and first order phase transitions. In particular, 
we obtain the universal 
equation of state for weak first order phase transitions in
scalar matrix models. We also study the coarse grained free energy.
Its dependence on the coarse graining scale gives a quantitative
criterion for the validity of Langer's theory of
spontaneous bubble nucleation.
\end{abstract}

\end{titlepage}

\clearpage
\thispagestyle{empty}
\newpage
\mbox{ }
\newpage

\setcounter{page}{1}
\vspace*{-2.cm}
\tableofcontents

\clearpage
\thispagestyle{empty}
\newpage
\mbox{ }
\newpage

\include{mainin}

\include{field}

\include{second}

\include{first}

\include{qcd}

\include{outlook}

\include{append}

\include{bib}
\newpage
\thispagestyle{empty}

\end{document}

%% file: mainin.tex
\sect{Introduction and overview \label{mainin}}

Quantum chromodynamics (QCD) describes qualitatively different physics
at different length scales. The theory is asymptotically free \cite{GWP} 
and at short distances or high energies
the strong interaction dynamics of quarks and gluons can be 
determined from perturbation theory. 
On the other hand, at scales of a few hundred $\MeV$ confinement sets 
in and the spectrum of the theory consists only of color neutral 
states. A further important aspect of strong interaction dynamics 
is spontaneous chiral symmetry breaking. In the limit of
vanishing current quark masses the classical or ``short distance'' QCD 
action does not couple left-- and right--handed quarks. As a consequence, 
for $N_f$ massless quark flavors the classical QCD action 
exhibits a global chiral invariance 
under $U_L(N_f) \times U_R(N_f) = SU_L(N_f) \times SU_R(N_f) \times 
U_V(1) \times U_A(1)$, with different unitary transformations
acting on the left-- and right--handed quark fields. However, this
symmetry is not fully observed in the hadron spectrum. It is 
phenomenologically well established that in the quantum theory 
the symmetry $SU_L(N_f) \times SU_R(N_f)$  
is spontaneously broken to the diagonal $SU_V(N_f)$ vector--like
subgroup. For $N_f=2$ the isospin symmetry is realized in the 
hadron spectrum to a very good approximation. The comparably stronger
explicit symmetry breaking due to non--zero current quark masses 
makes the realization of the $SU_V(3)$ symmetry less accurate.
In addition, the axial abelian subgroup $U_A(1)$ is broken in the quantum 
theory by an anomaly of the axial--vector current \cite{Ho}. 
(The abelian $U_V(1)$ subgroup corresponds to baryon number 
conservation.)

Extrapolating QCD from short distance to long distance scales 
is clearly a non--pertur\-ba\-tive problem. The strong gauge coupling
$\alpha_s$ changes with scale, growing with increased distance.
However, not only effective couplings but also the relevant 
degrees of freedom can change with scale.
Indeed, at low energies an essential part of strong interaction 
dynamics can be encoded in the masses and interactions of 
mesons. Prominent examples of systematic effective 
descriptions are the successes of chiral 
perturbation theory \cite{GL82-1,CPT2} based on the 
non--linear sigma model  
or a description within the linear sigma model \cite{GML60-1}
\footnote{It has
been demonstrated recently~\cite{QuMa,JW97-1} that the results of
chiral perturbation theory can be reproduced within the linear meson
model once certain higher dimensional operators in its effective
action are taken into account.}. 
The success of any analytical approach to bridge the   
gap between short distance and long distance QCD may depend on 
the ability to introduce field variables for composite objects
such as mesons at a certain ``compositeness'' scale 
(cf.\ section \ref{2qcd}).

An even more challenging subject is finite temperature QCD \cite{Sm}.
In view of present and future heavy--ion collision experiments 
\cite{HIC}, but also for its cosmological relevance \cite{KoTu}, there is
an increased interest in the thermal properties of QCD. It was 
realized early \cite{CP75} that in a thermal equilibrium situation
at sufficiently high temperature or density QCD may differ in
important aspects from the corresponding zero temperature or
vacuum properties. If one considers the phase structure of QCD
as a function of temperature one expects at a critical temperature
around $T_c \simeq 100-200 \MeV$ two types of phase 
transitions or crossover phenomena\footnote{Transitions are expected
for both increasing density and increasing temperature \cite{CGS86,Bar92}.
We will restrict the discussion here to QCD at non--zero temperature.}. 
First, there is the expectation
that the confined quarks and gluons at zero temperature become
deconfined at sufficiently high temperature where the hot state is 
conventionally called the quark--gluon plasma. 
In the pure gauge theory, or equivalently in the limit of infinite 
quark masses, a high temperature deconfinement phase transition is 
well--established \cite{SY82,DEC}. Second, the spontaneously broken
chiral symmetry in the limit of vanishing quark masses and 
zero temperature is expected to be restored in a high temperature phase. 
It is not yet clear whether these phase transitions occur at the same 
temperature or whether they persist for realistic quark masses or 
rather turn into smooth crossover.

Such a transition must have occured during the early stages of the 
evolution of the universe. According to the standard hot big bang
cosmology the transition took place about a microsecond after
the big bang, where the temperature dropped to the order of 
$100 \MeV$. Three orders of magnitude higher the previous transition 
occured, related to the dynamics of the electroweak 
interactions\footnote{Existence and properties of the electroweak 
phase transition depend strongly on the mass of the Higgs scalar
$M_H$. For Higgs boson masses around $M_H \gta 80 \GeV$ there is no
phase transition but rather a smooth crossover 
\cite{KLRS,RW1,BeW95,BP95}.},
and for most of its evolution the early universe was to a good 
approximation in thermal equilibrium \cite{KoTu}.

A promising prospect is to reproduce the QCD transition in  
heavy--ion collisions in the laboratory at sufficiently high energy.
Present fixed target heavy--ion collision experiments at the
AGS accelerator (BNL) and at the CERN SPS accelerator have 
produced heavy--ion beams with energies that could yield initial
temperatures above $T_c$ \cite{InT}. 
Here the initial temperature
refers to a frequently used scenario in which the central rapidity
region reaches a stage of local thermal equilibrium, preceded
by an intermediate pre--equilibrium stage after the first instant
of nuclear contact 
\cite{Sh78,Bj83}.
In this scenario the thermalized\footnote{Thermalization enters as
an assumption that remains to be justified. There are results from 
the parton cascade model 
\cite{ParKa,Ge95} 
which give some support to the idea that there will be enough time
in a RHIC collision for local thermal equilibrium to be established.} 
central region is characterized by a large number ratio of mesons 
to baryons \cite{Bj83}, thus QCD at finite temperature and zero 
baryon number density would be of relevance for its 
description. 
Whereas SPS realizes for very heavy ion Pb beams a center--of--mass
energy of $160 \GeV$ per nucleon, future colliders, RHIC with
$200 \GeV$ per nucleon (gold on gold) and the LHC with $6300 \GeV$
per nucleon (lead on lead), will allow to produce temperatures
substantially larger that the typical time scales of $10^{-23}$
sec.\ or distances of $1$ fm associated with the QCD transition
\cite{MO96-1}.

Of particular interest is the order of a possible phase transition
which may have far--reaching phenomenological consequences. One 
reason is that a first order phase transition may create an out of
equilibrium situation. For a first order transition the (coarse
grained) free energy as a function of a suitable order parameter 
exhibits two distinct minima near the critical temperature $T_c$.
If in a heavy--ion collision of sufficiently high energy a  
region of the high temperature phase is created, it then cools through 
the phase transition. The initial high temperature phase becomes 
a local minimum of the free energy and the lowest minimum corresponds to
the low temperature phase. However, a barrier separating both minima  
prevents a smooth transition at $T_c$. At a first order
phase transition, typically the conversion proceeds in a short dramatic
period through the nucleation and growth of ``bubbles'' 
(droplets) of the
low temperature phase (cf.\ section \ref{secfirst}). 

In contrast to the discontinuous behavior at a first order transition,
at a second order phase transition the order parameter changes 
continuously (but non--analytically) from one phase to the other. 
In particular, there is no mass scale present at a second order phase 
transition which corresponds to infinite correlation lengths.
Large correlation lengths may yield distinctive,
qualitative signatures in a relativistic heavy--ion collision. 
It has been argued \cite{RaWi93-1,Raj95-1} that in the central
rapidity region a large correlation length may be responsible for
the creation of large domains, in which the pion field has a 
non--zero expectation value pointing in a fixed direction in
isospin space (``disoriented chiral condensate'' \cite{Ans88-1}).
The observational consequences would be strong fluctuations
in the number ratio of neutral to charged pions. Such a 
phenomenon was even related \cite{RaWi93-1,Raj95-1} to the
prominent Centauro events \cite{LFH80-1} in high energy
cosmic ray experiments. Apart from the phenomenological
implications, a large correlation length opens the spectacular 
possibility that the QCD phase transition is characterized by 
universal properties! The notion of universality
for critical phenomena is well--established in statistical physics
\cite{RG}. Universal properties are independent of the details,
like short distance couplings, of the model under investigation.
They only depend on the symmetries, the dimensionality of space
and the number of field degrees of freedom. 
As a consequence a whole class of models 
is described by the same universal form  of the equation of state
in the vicinity of the critical temperature. The range of 
applicability typically covers very different physical systems
in condensed matter physics and high temperature quantum field
theory (cf.\ section \ref{secsec}).

Finally there is the possibility of no true phase 
transition at all. For such a
crossover phenomenon quantities may change rapidly in the 
transition region, and the phenomenological consequences
may be almost indistinguishable from a situation with a
true phase transition. However, for a crossover there occurs 
no non--analytical behavior and all correlation lengths remain 
finite.

Large correlation lengths near a second order  
(or weak first order) phase transition are appealing both from  
the theoretical and the experimental point of view. For the 
experimentalist they may yield a unique, distinctive signature that a QCD 
phase transition has occured. The theoretical challenge 
is to find a method that describes both the known low--energy zero 
temperature properties of QCD and the possible universal behavior 
associated with a large correlation length near the critical temperature
of a second order phase transition. The link between the zero temperature
and the universal critical properties is crucial.
One may think of separating the problem into 
two parts: First, one considers a tractable, easier model than the
original one where both models belong to the same universality class.
The simplified description can develop enormous predictive power
because of the universal properties of the equation of state near
the critical temperature. For instance, a number of stimulating 
contributions \cite{PW84-1,RaWi93-1,Raj95-1} 
pointed out that for sufficiently small
up and down quark masses and sufficiently large strange 
quark mass the chiral phase transition is expected to be in the universality
class of the three dimensional $O(4)$ symmetric Heisenberg 
model\footnote{We note that for two flavors the relevant chiral symmetry 
$SU(2) \times SU(2)$ is locally isomorphic to $O(4)$.
This does not hold in the case of a speculative 
``effective restoration'' \cite{PW84-1,Shu94-1} of the axial $U_A(1)$ at 
temperatures around $T_c$.}. 
This model is known to exhibit a second 
order phase transition, where universal properties can be observed. 
Apart from the expected observational consequences, a number 
of interesting conclusions have been drawn from universality arguments
\cite{RaWi93-1,Raj95-1}, e.g.\ that the pion mass increases with 
temperature around 
$T_c$. In particular, near the critical temperature only the pions and 
the sigma resonance play a role for the behavior of the chiral condensate 
and long--distance correlation functions. 
Here the chiral condensate, i.e.\ the expectation value of the 
quark bilinear $\langle \overline{\psi}\psi\rangle$,
plays the role of an order parameter for the chiral transition in 
the limit of vanishing current quark masses.
However, the questions how 
small the up and down quark mass $m_u$ and $m_d$ would have to be in order 
to obtain a large correlation length near $T_c$ and if the drawn conclusions
are valid for realistic quark mass values remained
unanswered. Nothing can be infered from universality arguments about the 
absolute magnitude of correlation lengths. Likewise the question of the 
value of the critical temperature is non--universal in nature. The 
non--universal amplitudes crucially depend on the zero temperature 
properties of the theory under investigation. 
In a second step
one therefore finally has to link the universal behavior near $T_c$
and zero current quark mass on one hand with the known physical
properties at $T=0$ for realistic quark masses on the other hand.

One reason for the ``missing link'' is the breakdown of chiral 
perturbation theory for temperatures exceeding about one third of the 
zero temperature pion mass $m_{\pi}=135 \MeV$ (cf.\ section \ref{2qcd}).
Exploring the universal region in lattice QCD, on the other hand,
is limited by present computer resources, though the use of improved 
actions seems promising \cite{Im79}. 
Since decreasing quark 
masses increase the computational cost, simulations are typically 
done with unphysically heavy quarks where extrapolation to realistic 
values or to the chiral limit is needed.\\

{\bf Equation of state for two flavor QCD} 
\vspace*{0.1cm}

It is a major result of this work to provide the equation of state
for two flavor QCD which covers {\em both} the chiral perturbation
theory domain of validity and the domain of validity of universality
associated with critical phenomena \cite{BJW}. 
The equation of state connects in 
a temperature range below $\simeq 170 \MeV$ the derivative of the 
free energy or the effective potential with the average light current 
quark mass $\hat{m}=(m_u+m_d)/2$. It therefore allows to study the 
temperature and quark mass dependence 
of various quantities, like the pion mass $m_{\pi}$ or the chiral 
condensate $\langle \overline{\psi}\psi \rangle$ or the pion decay 
constant $f_{\pi}$ etc. 
In addition, we obtain a precision estimate of the universal critical 
equation of state of the three dimensional $O(4)$ symmetric 
Heisenberg model in the vicinity of the critical temperature and the 
chiral limit. From the scaling form of the equation of state we
extract universal critical exponents, amplitude ratios and couplings.
Recent lattice simulations for the universal critical equation of state
of the three dimensional $O(4)$ model show agreement with our 
results at the few per cent level (cf.\ section \ref{2qcd}). 
In addition, we find all non--universal 
information to be encoded in two independent amplitudes that are properly 
computed for the two flavor case.

Though these results will be presented in the text in more detail
we point out two important answers one obtains from this study:
First of all, for a thermal equilibrium situation the chiral  
transition gives no indication for strong fluctuations of
pions with long wavelength\footnote{We have restricted the
discussion to the thermal equilibrium properties of QCD.
The question of strong fluctuations of pions with long
wavelength has been investigated in the context of a chiral
QCD phase transition far from equilibrium 
\cite{RaWi93-1,DCCR}.}.
In the presence of light quark masses the second order phase transition
is turned into a smooth crossover: The expectation value of 
the quark bilinear $\langle \overline{\psi}\psi \rangle$ 
decreases smoothly as 
the temperature is increased. The longest correlation
lengths near the ``crossover temperature'' $T_{pc}$, which turns 
out to be around $130 \MeV$ for realistic quark masses, even becomes smaller 
than at $T=0$. (It should be emphasized, however, that a tricritical behavior 
with a massless excitation remains possible for three flavors.
This would not be characterized by the universal behavior of the
$O(4)$ model; see the discussion below.)
The second answer concerns the 
applicability of universal, i.e. almost model independent, arguments 
for a description of the chiral 
phase transition. Though the quark masses are 
fixed in nature, varying 
the average current quark mass to zero allows us to observe critical 
behavior independent from the realistic physical situation. In this way  
one can probe to what extent universality approximately is realized 
for realistic or even heavier quark masses. Universality 
has been assumed e.g.\
in two flavor lattice QCD simulations to guide extrapolation 
to realistic quark
mass values or even to the chiral limit \cite{MILC97-1}.
Despite the observed comparably short 
correlation lengths at non--zero temperature, we find the approximate
validity of the $O(4)$ scaling behavior over a large temperature interval
near and above $T_c=100.7 \MeV$ even for quark masses slightly
larger than the realistic ones (cf.\ section \ref{2qcd}).

Our approach is based on the use of an exact non--perturbative 
flow equation for a scale dependent effective action $\Gamma_k$
\cite{Wet91-1,Wet93-2}, which is the generating 
functional of the $1 PI$ Green
functions in the presence of an infrared cutoff $\sim k$. In a thermal
equilibrium context the so-called effective average action $\Gamma_k$
becomes temperature dependent and describes a coarse 
grained free energy with
a coarse graining length scale $\sim k^{-1}$. The coarse grained description
allows us to consider the relevant physics at a given momentum--like 
scale $k$. The approach is closely connected to the Wilsonian
renormalization group \cite{RG,WH73-1,NC77,Wei76-1,Pol84-1,Has86-1}, 
often also called exact renormalization 
group. An introduction to the method is given in section 
\ref{fieldtheory}. 

The flow equation for $\Gamma_k$ can be supplemented by an exact 
formalism for the introduction of composite field variables or,
more generally, a change of degrees of freedom at some given scale
$k$ \cite{EW94-1}. We employ for scales below a ``compositeness scale'' of 
$k_{\Phi} \simeq 600 \MeV$ a description in terms of quark and 
scalar mesonic degrees of freedom. This effective quark meson model 
can be obtained from QCD in principle by ``integrating out'' the gluon
degrees of freedom and by introducing fields for composite operators 
\cite{EW94-1,Wet95-2} (cf.\ the discussion in sections \ref{2qcd}
and \ref{outlook}).
In this picture the scale $k_{\Phi}$ is associated
to the scale at which the formation of mesonic bound states can be
observed \cite{EW94-1} in the flow of the momentum dependent four--quark 
interaction. Though considerable progress has been made, a reliable
quantitative derivation of the effective quark meson model from 
QCD is still missing. We emphasize, however, that the quantitative 
aspects of this derivation will be of minor relevance for our practical 
calculations in the mesonic sector: If the effective Yukawa coupling 
between the quarks and the mesons turns out to be strong at the 
compositeness scale we observe a fast 
approach of the scale dependent effective couplings to approximate
partial infrared fixed points \cite{Ju95-7,BJW}.
This behavior is discussed in section 
\ref{FlowEquationsAndInfraredStability}.
As a consequence, the detailed form of the meson potential at 
$k_{\Phi}$ becomes unimportant, except for the value of one relevant 
scalar mass term $\overline{m}_{k_{\Phi}}$. In this work we fix
$\overline{m}_{k_{\Phi}}$ from phenomenological input such that
$f_{\pi}=92.4 \MeV$ (for $m_{\pi}=135 \MeV$) which sets our unit of
mass for two flavor QCD. The only other input parameter we use is the 
constituent quark mass $M_q$ to determine the scale $k_{\phi}$. 
We consider a range $300 \MeV \lta M_q \lta 350 \MeV$ and find a
rather weak dependence of our results on the precise value of $M_q$.
We point out that though a strong Yukawa coupling at $k_{\Phi}$
is phenomenologically suggested by the comparably large value of 
the constituent quark mass $M_q$ it enters our description as a 
(consistent) assumption. On the present level of approximation
the quark meson model for two flavors can be reduced to the 
well known $O(4)$ symmetric Gell-Mann--Levy linear sigma 
model~\cite{GML60-1} for the three pions
and the ``sigma resonance'', however, it is
coupled to quarks now. The meson degrees of freedom can be 
understood as quark--antiquark bound states. Spontaneous chiral
symmetry breaking is described by a non--vanishing expectation
value of the meson field in absence of quark masses.
The non--perturbative flow equations for the quark--meson model
are obtained from a suitable truncation of the exact flow
equation for $\Gamma_k$. We solve them numerically. 
Further details about the model and the 
applied approximations can be found in section \ref{2qcd}.\\

{\bf Quark mass dependent phase structure}
\vspace*{0.1cm}

The details of a possible QCD phase transition, like the order
or the ``strength'' of the transition, crucially depend on the 
number of flavors and the current quark masses. Though the number  
of flavors and the quark masses are fixed in nature, on the 
theoretical side it is possible to adjust them in order to get more 
insight into the phase structure of QCD. 
A considerable amount of information about the phase structure
of non--zero temperature QCD has been provided by lattice Monte 
Carlo simulations whose results will be instructive in the 
following. The majority of lattice
simulations incorporating dynamical fermions have been done
with two or four flavors of equal mass quarks. There are a few
simulations carried out for two light and one heavier quark
which comes closest to the physical light quark mass ratios.
To save computational cost, however, 
simulations are typically done with unphysically heavy quarks.

Lattice QCD simulations in the limit of two light quark flavors 
show no signs of a first order deconfinement phase 
transition \cite{LatDec,MILC97-1}. The expectation value of the
Polyakov loop\footnote{We note that 
the Polyakov loop  
represents no order parameter for a possible deconfinement phase 
transition in the presence of light quarks. Only in the limit of 
infinite quark masses, or equivalently
in the pure gauge theory, it is an order parameter
for the spontaneous breaking of the center $Z(N)$ of the gauge group 
$SU(N)$ at high temperature \cite{SY82}.}  increases 
smoothly with increased temperature when dynamical fermions are 
present. There is no sharply defined 
transition, but rather a 
smooth crossover from one regime to the other. In accordance
with our results, the chiral transition for two small, non--zero quark 
masses is found to be described by a crossover phenomenon 
\cite{LatDec,MILC97-1}. The lattice results indicate that both 
crossover, associated to the deconfinement and to the chiral 
transition, occur at the same temperature approximately 
around\footnote{Here the zero temperature $\rho$ meson
mass is used to set the energy scale.} 
$140-150 \MeV$. 
This qualitative picture almost remains the same for the more
realistic case of two light and one heavier quark.
Simulations with ``2+1'' flavors carried out by the Columbia
group in the staggered fermion scheme find a first order signal
from the chiral condensate for three light flavors. However,
the first order transition disappears for an increased strange quark 
mass, even before the strange quark adopts its physical mass value 
\cite{Br90}.
In contrast, from a Wilson fermion approach the Tsukuba group finds 
results that indicate
a first order chiral phase transition for realistic 
quark masses\cite{Tsu}. However, simulations with Wilson 
fermions in this context are difficult
because they break chiral symmetry completely away from the 
continuum limit. Though no final
conclusion can be drawn from the lattice results by now, 
if there is a first order phase transition at all 
it is very likely to be weak. In addition, lattice data seem to 
indicate that QCD exhibits a (smooth) transition which is dominated 
by the approximate restoration of chiral symmetry.

Our results and the results from lattice simulations are 
compatible with the following picture proposed from effective sigma
model considerations \cite{PW84-1,Wil92,RaWi93-1,GGP2}: 
For three massless quark 
flavors the chiral phase transition is first order in nature. If the 
strange quark mass $m_s$ is raised the phase space shows a line of first 
order transitions which ends in a tricritical point with a massless 
excitation. For larger strange quark masses the transition remains of 
second order. If one takes into account that the up 
and down quark masses are non--zero this second order phase transition 
is smoothed into a crossover phenomenon. However, there remains a line 
of tricritical points separating the first order region for small
$m_s$ and the crossover region for large $m_s$.
It has been pointed out \cite{GGP2,mes} that the QCD 
chiral phase transition 
may be close to a tricritical point with a massless excitation.\\

{\bf Effective three dimensional behavior at high temperature}
\vspace*{0.1cm}

Our investigations have ruled out the appearance of a large correlation 
length at the equilibrium chiral transition in the two flavor case, 
corresponding 
to an infinite mass limit for the remaining quark flavors. However, 
we have seen that a chiral transition with a large 
correlation length remains possible through the influence of the
strange quark.
In particular, along the mentioned line of tricritical points the 
transition is expected to be in the universality class of the 
Ising model \cite{GGP2}. In this case, as has been discussed 
above, in a first step one may learn as much as possible about the 
phase transition only relying on universality arguments. 
This alone typically poses a complicated non--perturbative problem.
The problems are related to the effectively three dimensional behavior of 
high temperature quantum field theory in four dimensions. 
Quantum field theory at non--zero temperature $T$ can be formulated
in terms of an Euclidean functional integral where the ``time''
dimension is compactified on a torus with 
radius $T^{-1}$ \cite{Kap}.
Accordingly, non--zero temperature results in (anti--) periodic
boundary conditions for (fermionic) bosonic fields in the Euclidean
``time'' direction with periodicity $T^{-1}$. As a consequence, 
the zeroth component of the Euclidean four--momentum becomes 
discrete, i.e.\ $q_0=2 l \pi T$ for bosons and $q_0=(2l+1)\pi T$
for fermions with integers\footnote{These are commonly called
Matsubara frequencies.} $l$. It is suggestive that
if the relevant correlation length becomes much larger than the 
inverse temperature the compactified ``time'' dimension cannot be 
resolved anymore. This phenomenon is known as ``dimensional reduction'' 
\cite{DR}. The effective three dimensional behavior of the theory
can also be understood from the observation that for sufficiently
high temperature the theory is dominated by classical statistical
fluctuations. In particular, the critical exponents which describe 
the singular behavior of various quantities near a second order phase 
transition are those of the corresponding classical 
system\footnote{In the vicinity of the critical temperature of 
a second order phase transition the correlation length of the
relevant bosonic field is $\xi \sim T_c^{-1}(|T-T_c|/T_c)^{-\nu}$
with a (positive) exponent $\nu$, 
and one {\em always} faces a high temperature situation for
$T \to T_c$, i.e.\ $\xi \gg T^{-1}$.} \cite{TetWet}
(cf.\ section \ref{2qcd}). An important particularity of the 
dimensionally reduced theory is the absence of fermionic degrees
of freedom. This can be seen as a consequence of the fact that 
the ``lightest'' fermionic excitation is of the order $\pi T$,
whereas there exists a zero mode $(l=0)$ for a bosonic field. 
We have explicitly verified the 
phenomenon of dimensional reduction and the absence of fermions
within the effective quark meson model at  
high temperature (cf.\ section \ref{2qcd}). 

Near a second order (or a sufficiently weak first order) phase
transition we 
therefore have to deal with a purely bosonic three dimensional
theory. Standard high temperature perturbation theory can give a reliable 
description only if the relevant dimensionless couplings remain small
near the transition. In three dimensions the (effective) scalar quartic 
coupling $\bar{\lambda}$ has positive canonical mass dimension. If one 
denotes by $m_R$ the relevant mass at the critical temperature, 
standard perturbation theory corresponds to an expansion in the
dimensionless effective coupling $\bar{\lambda}/m_R$. For a
given fixed coupling $\bar{\lambda}$ a perturbative expansion will
only converge if $m_R$ is sufficiently large, 
$m_R \gg \bar{\lambda}$. In consequence, standard perturbation theory 
is not applicable to second order ($m_R = 0$) or weak first 
order ($m_R \ll \bar{\lambda}$) phase transitions due to the
apparent infrared divergences in a perturbative treatment. 
Obviously in a suitable non--perturbative treatment a ``running''
effective renormalized coupling $\lambda_R$ has to be introduced. 
The running coupling approaches zero at a second order phase transition, 
with $\lambda_R/m_R$ going to a constant as the correlation length 
$m_R^{-1}$ tends to infinity. Typically no small
effective coupling characterizes the interactions at the phase
transition (cf.\ section \ref{secsec}).

The three dimensional systems may be treated within the 
$\epsilon$--expansion \cite{WilFis72}
or by the use of perturbation series at fixed dimension \cite{Parisi}.
Most of the lattice results
come from the analysis of high temperature expansions
\cite{Gu89} or from Monte Carlo simulations \cite{MCS}. 
For the critical equation of state of the three dimensional $O(4)$
model a result from the $\epsilon$--expansion up to second order in 
$\epsilon=4-d$ ($d=3$) is available \cite{BWW73-1}. The 
reliability of this expansion is not guaranteed a priori and
results from a lattice Monte Carlo simulation \cite{Tou} that has 
recently been obtained show substantial deviations from the
$\epsilon$--expansion results
(cf.\ section \ref{2qcd}).
The lattice simulations are quite demanding in the vicinity of the
critical temperature of a second order (or weak first order)
phase transition and require careful correction for finite size 
effects. As has been pointed out above our results agree quite well
with the lattice study.

The use of non--perturbative flow equations within the framework of the 
effective average action is particularly suitable for dealing with
theories which are plagued by infrared problems in perturbation theory.
The employed formulation is both infrared and ultraviolet finite 
in arbitrary
dimensions. In particular, the method is not restricted to an 
expansion in small couplings which has been crucial for the treatment
of the effective quark meson model.\\

{\bf Equation of state for second order and (weak) first order
transitions}
\vspace*{0.1cm}

The first two parts of this work are devoted to the study of scalar
field theories near the critical temperature of second order and
first order phase transitions. These investigations are performed
directly in three dimensions. We obtain a detailed quantitative
picture of the critical equation of state in the vicinity of the 
second order phase transition of the $O(N)$ symmetric $N$-component
model \cite{BTW95,ABBTW95,JB}. 
Similar to the case $N=4$, for $N=1$ our results 
have been confirmed by lattice simulations \cite{Tsy2,Tsy1} while there
remains a considerable discrepancy to the $\epsilon$--expansion 
results\footnote{For $N=1$ the $\epsilon$--expansion 
for the scaling equation of state is available
up to order $\epsilon^3$ \cite{ZJ}.}.   
We point out that our approximations are based on a derivative 
expansion for the effective average action that takes into account
the {\em most general} non--derivative, i.e.\ potential, term 
consistent with the symmetries. It does not rely on any restriction
of the potential to a polynomial form.
It therefore allows to study the complete non--analytic behavior
of the effective potential, or equivalently the free energy, near the 
critical temperature of the second order phase transition.
For $N=4$ our universal results in three dimensions are found to 
correspond to the results from the quark meson model near the
critical temperature and the chiral limit. This 
correspondence is a manifestation of the phenomenon of dimensional 
reduction. 
The $O(N)$ model has a wide range of different applications in 
statistical physics and high temperature quantum field theory. 
For $N=4$ the model also describes the scalar
sector of the electroweak standard model in the limit of
vanishing gauge and Yukawa couplings. In condensed matter physics 
$N=3$ corresponds to the well known Heisenberg model used to describe 
the ferromagnetic phase transition. There are other applications like the
helium superfluid transition ($N=2$), liquid-vapor transition
($N=1$) or statistical properties of long polymer chains
($N=0$) \cite{ZJ}. The critical properties of these systems
are presented in section \ref{secsec}.

The second part of this work provides a detailed investigation 
of first order phase transitions. In particular weak first order
phase transitions are a much less thoroughly studied subject in the
literature than second order transitions. This becomes especially
desirable in the context of a possible weak first order chiral
phase transition in QCD. We consider \cite{BW97-1,BTW97-1}
models with $U(N) \times U(N)$
symmetry with a scalar field in the $(\bar{N},N)$ 
representation, described by an arbitrary complex $N \times N$
matrix. We compute the equation of state for $N=2$ which can be
used for an effective description of two flavor QCD in the
case of a speculative ``effective restoration'' 
\cite{PW84-1,Shu94-1} of the chiral 
anomaly at high temperature. 
The considered theory also has a relation to the electroweak
phase transition in models with two Higgs doublets. In certain limiting
cases the model describes a non--linear matrix model for unitary
matrices or one for singular $2\times 2$ matrices 
(cf.\ section~\ref{secfirst}).

For a large part of the parameter space the $U(2) \times U(2)$
symmetric matrix model exhibits a weak first
order phase transition. Though the corresponding classical or short 
distance action indicates a second order phase transition, the 
transition becomes first order once fluctuations are taken into 
account. The fluctuation induced first order phase transition is
known in four dimensions as the Coleman--Weinberg phenomenon
\cite{Col73}. A reliable method for a description of this 
phenomenon in three dimensions may be of relevance for the
QCD chiral phase transition. For the realistic case of three flavors
it has been argued that the chiral transition can be mainly
fluctuation induced \cite{GGP2,mes}.

We put special emphasis on the investigation of  
(approximate) universal properties near weak first order
phase transitions. 
Whereas for second order transitions the universal equation 
of state can be expressed as a function of only one scaling 
variable (cf.\ section \ref{secsec}), 
due to an additional mass scale at a first order 
transition there is a second dimensionless ratio. 
The equation of state or, equivalently,
the effective potential or free energy therefore
depends on two independent scaling variables in the
universal region. We have succeeded to describe this situation
\cite{BW97-1} for the present matrix models and present
the universal equation of state in section \ref{sce}.\\

{\bf Coarse graining and first order transitions}
\vspace*{0.1cm}

Another challenging problem is the discussion of 
the dynamics of a first order phase transition \cite{Langer}
which usually relies on the study
of a non-convex potential or free energy. 
The decay of metastable states 
is associated either with tunneling fluctuations through 
barriers in the potential \cite{ColCal}, or, at non-zero temperature, 
with thermal fluctuations above them
\cite{Lin}. However, the 
effective potential \cite{Sch51-1}, which seems 
at first sight a natural tool for 
such studies, 
is expected to be a convex quantity with no
barrier. The resolution of this paradox lies in the 
realization that the effective potential is convex because
the tunneling or thermal fluctuations are incorporated in it.
These fluctuations are associated with low frequency modes, while the
non-convex part of the potential is related to the 
classical potential and
the integration of high frequency modes. 
A natural approach to the study of first order phase transitions
separates the problem in two parts. First, the
high frequency modes are integrated out, with the possible
generation of new minima of the potential through radiative symmetry 
breaking \cite{Col73}. 
Subsequently, the decay of metastable states is discussed with
semiclassical techniques \cite{ColCal,Lin}, 
using the non-convex potential that has 
resulted from the first step. 
This leads one to the notion of the coarse grained free energy, which
is fundamental in statistical physics. Every physical system
has a characteristic length scale associated with it. The dynamics
of smaller length scales is integrated out, and
is incorporated in the parameters of the 
free energy one uses for the study of the behavior at larger length scales.
Here we note that by construction, i.e.\ the
inclusion of fluctuations with characteristic momenta larger
than a given infrared cutoff $\sim k$, the effective average
action $\Gamma_k$ is the appropriate quantity 
for the study of physics at
a scale $k$ (cf.\ section \ref{fieldtheory}). 
It therefore realizes the concept of a coarse
grained free energy in the sense of Langer \cite{Langer}.

The use of the effective average action will allow us to 
address the old question of the validity \cite{BW97-1,BTW97-1} 
of the standard theory of spontaneous bubble nucleation proposed
by Langer \cite{Langer}.
This requires first
a meaningful definition of a coarse grained free energy
with a coarse graining scale $k$ and second the 
validity of a saddle point approximation for the 
treatment of fluctuations around the ``critical bubble''
(cf.\ section \ref{coarse}).
Here only fluctuations with momenta smaller than $k$
must be included. Both issues turn out to be 
closely related. 
The validity of the saddle point 
approximation typically requires small dimensionless
couplings. On the other hand we observe
for large effective couplings that the form of the 
relevant coarse grained
effective potential $U_k$ depends strongly on the coarse graining
scale $k$. This means that
the lowest order in the saddle point approximation
(classical contribution) depends strongly on the details of
the coarse graining procedure. Since the final results
as nucleation rates etc.\ must be independent of the
coarse graining prescription this is only compatible
with a large contribution from the higher orders of the
saddle point expansion. We will consider 
this issue in section \ref{coarse} in a quantitative way.\\

This work is organized as follows. We provide an introduction into the
non--perturbative effective average action method in section 
\ref{fieldtheory}. The two following sections concern our studies
of second order and first order phase transitions in three
dimensional scalar field theories. These parts are based on refs.\ 
\cite{BTW95,ABBTW95,JB} and \cite{BW97-1,BTW97-1} respectively. 
Section \ref{2qcd} is devoted
to the chiral phase transition for two flavor QCD within the
effective quark meson model. This part is based on refs.\
\cite{BJW,BJW97-2}.
Each section is supposed to be
rather self--contained with a short introduction and conclusions
to allow for a selective approach to the presented material.
Section \ref{outlook} presents an outlook. 
The acknowledgments are found in section \ref{ack}.
The three appendices A--C contain a discussion and definitions
of threshold functions used in preceding sections.
In appendix D the connection of the average current quark 
mass with the effective source term of the quark meson model
is established. This is used in section~\ref{2qcd}.

%% file: field.tex
\sect{Non--perturbative flow equations \label{fieldtheory}}

\subsection{Effective average action}

We will introduce here the
effective average action~\cite{Wet91-1} $\Gamma_k$ with an infrared 
cutoff $\sim k$. The effective average action is based on
the quantum field theoretical concept of the effective
action~\cite{Sch51-1} $\Gamma$, i.e.\
the generating functional of the $1 PI$ Green functions. The field
equations derived from $\Gamma$ include all quantum
effects. For a field theoretical description of thermal equilibrium
this concept is easily generalized to a temperature dependent
effective action which includes in addition the thermal fluctuations. In
statistical physics $\Gamma$ describes the free energy as a functional
of some (space dependent) order parameter. In particular, for a
constant order parameter, $\Gamma$ yields the effective potential
which encodes the equation of
state. The effective average action $\Gamma_k$ is a simple
generalization of the effective action, with the distinction that only
fluctuations with momenta $q^2\gta k^2$ are included. In the
language of statistical physics, $\Gamma_k$ is a coarse grained free 
energy with a coarse graining scale $k$. Lowering $k$ results in 
a successive inclusion of fluctuations with momenta
$q^2\, \gtap\,\, k^2$ and therefore permits to explore the theory on 
larger and larger length scales.
We note that $\Gamma_k$
is closely related to an effective action for averages of
fields~\cite{Wet91-1}, where the average is taken over a volume
of size $\sim k^{-d}$ and which is similar
in spirit to the block--spin action 
\cite{Ka,RG} in lattice theories. In a theory with a physical 
ultraviolet cutoff $\Lambda$ we
can associate $\Gamma_\Lambda$ with the microscopic or 
classical action $S_{}$ 
since no fluctuations below $\Lambda$ are effectively included. 
For example, in 
statistical systems on a lattice the scale $\Lambda$ plays
the role of the inverse lattice spacing. In the context of dimensional
reduction in high temperature quantum field theory
$\La$ appears as the ultraviolet cutoff scale for the effective 
three dimensional models (cf.\ section 
\ref{FiniteTemperatureFormalism}).
By definition, the effective
average action equals the effective action for $k=0$, i.e.\
$\Gamma_{0}=\Gamma$, since the infrared cutoff is absent. Thus
$\Gamma_k$ interpolates between the classical action $S_{}$ and 
the effective action $\Gamma$ as $k$ is lowered from $\Lambda$ to zero.
The ability to follow the evolution to $k\ra0$ is equivalent to the
ability to solve the theory. The dependence of the
effective average action on the scale $k$ is described by an exact
flow equation which is presented in section \ref{ExactFlowEquation}. 

To be explicit 
we consider the path integral representation of the 
generating functional for the connected Green functions 
in $d$ Euclidean dimensions
with $N$ real scalar fields 
$\chi_a$ $(a=1 \ldots N)$, classical action $S_{}$ and sources
$J_{a}$. We introduce a $k$--dependent generating functional
\be
W_k[J]=\ln \int D \chi \exp\left(-S_{}[\chi]-\Dt_k S[\chi]+
\int d^dx J_a(x)\chi^a(x) \right)\; ,
\label{genfunc}
\ee 
with an additional 
infrared cutoff term which is quadratic in the fields and reads
in momentum space\footnote{In order to avoid a proliferation
of symbols we distinguish functions or operators in position space
and their Fourier transforms according to their arguments,
with the convention $\chi_a(q)=\int d^dx \exp(-iq_{\mu}x^{\mu})
\chi_a(x)$.}
\be
\Dt_k S[\chi]=\hal \int 
\frac{d^dq}{(2\pi)^d} R_k(q)\chi_a(-q)\chi^a(q).
\ee
Without this term $W_k$ equals the usual generating functional
for the connected Green functions. Here the 
infrared cutoff function $R_k$ is required to vanish
for $k \to 0$ and to diverge for $k \to \Lambda$
and fixed $q^2$.
For $\Lambda \to \infty$ this can be achieved, for example, by the choice
\be
 R_k(q)=\frac{Z_k q^2 e^{-q^2/k^2}}{1- e^{-q^2/k^2}}
 \label{Rk(q)}.
\ee
Here $Z_k$ denotes an appropriate wave function renormalization
constant which will be defined later. 
For fluctuations with small momenta
$q^2\ll k^2$ the cutoff $R_k\simeq Z_k k^2$ acts like an
additional mass term. 
For $q^2\gg k^2$ the infrared cutoff vanishes 
such that the functional integration of the high momentum modes
is not disturbed. 
The expectation value of $\chi$ in the presence of $\Dt_k S[\chi]$
and $J$ reads 
\be
\label{classicfield}
\phi^a(x) \equiv \langle\chi^a(x)\rangle = 
\frac{\dt W_k[J]}{\dt J_a(x)}.
\ee
In terms of $W_k$ the effective average action is
defined via a modified Legendre transform
\be
\label{GaDef}
\Gm_k[\phi]=-W_k[J]+\int d^dx J_a(x)\phi^a(x)-\Dt_k S[\phi]. 
\ee
The subtraction of the infrared cutoff piece in (\ref{GaDef}) guarantees
that the only difference between $\Gamma_k$ and $\Gamma$ is the
effective infrared cutoff in the fluctuations. For $k>0$ this allows the
interpretation of $\Gamma_k$ as a reasonable coarse grained
free energy (cf.\ sections \ref{ExactFlowEquation} and \ref{coarse}). 
Furthermore, one observes that $\Gamma_k$ does not 
need to be convex whereas 
the standard effective action $\Gamma=\lim_{k\to 0} \Gamma_k$
is a convex functional by its definition as a Legendre transform. 
To establish the property $\Gamma_{\Lambda}=S$ one may consider an
alternative integral equation for $\Gamma_k$.
Using
\be 
J_a(x)=\frac{\dt (\Gamma_k+\Dt_k S)[\phi]}{\dt \phi^a(x)} 
\label{sou}
\ee
and eqs.\ (\ref{genfunc}), (\ref{GaDef}) one 
finds 
\be
\label{IntegralEquation}
\exp(-\Gamma_k[\phi])=\ds{\int D \chi \exp\left(-S_{}[\chi]
-\Dt_kS[\chi-\phi]+\int d^dx \frac{\dt \Gamma_k[\phi]}{\dt \phi^a}
[\chi^a-\phi^a]\right)}\; .
\ee
For $k > 0$ the term $\exp(-\Dt_kS[\chi-\phi])$ constrains the 
functional integral and behaves $\sim \dt[\chi-\phi]$ for
$k\to \Lambda$ thus leading to the property $\Gamma_k \to S$
in this limit \cite{Wet93-2}.
One will often not consider the limit $\La \to \infty$,
but rather keep this scale fixed and associate it with a physical
ultraviolet cutoff for the validity of the investigated field
theory. 
In particular, near second order or weak first order phase
transitions the universal properties of the theory become independent 
of the precise form of $\Gm_{\La}$.

The formulation of the effective average action can be easily 
generalized to fermionic
degrees of freedom. In particular, it is possible 
to incorporate chiral fermions since a
chirally invariant cutoff $R_k$ can be formulated~\cite{Wet90-1}.
Possible local gauge symmetries can be treated along similar
lines \cite{RW93-1,RW1,Bec96-1,BAM94-1,EHW94-1,Wet95-2,EHW96-1,Ell97-1}. 
Here $\Delta_k S$ may
not be gauge invariant and the usual Ward identities
receive corrections \cite{EHW94-1} which vanish for $k\ra0$.

\subsection{Exact flow equation\label{ExactFlowEquation}}

The dependence of the effective average action $\Gamma_k$ on the 
coarse graining scale $k$ is
described by an exact flow equation which can be 
derived in a straightforward way~\cite{Wet93-2}. From eq.\
(\ref{GaDef}) and (\ref{classicfield}) one finds
\begin{eqnarray}
  \label{abl1}
  \ds{\prl_t \left.\Gamma_k\right|_\phi} 
  \equiv k {\ds \frac{\prl}{\prl k}\left.\Gamma_k\right|_\phi} &=& 
  \ds{
  -\prl_t \left.W_k\right|_J-
  \prl_t\Delta_kS[\phi]}\nnn
  &=& \ds{
  \frac{1}{2}\Tr\left\{
   W^{(2)}_k \prl_t R_k \right\} }\; .
\end{eqnarray}
The trace in momentum space reads 
$\Tr=\sum_a \int d^dq/(2\pi)^d$ and $W^{(2)}_k$ denotes
the $k$-dependent connected two-point function
\begin{eqnarray}
  \label{abl2}
  \ds{(W_{k}^{(2)})_{a b}(q,q^\prime)} &=& \ds{
  \frac{\delta^2 W_k[J]}
  {\delta J^a(-q)\delta J^b(q^\prime)} }\nnn
  &=& \ds{\langle \chi_a(q)\chi_b(-q^\prime)\rangle
- \langle \chi_a(q)\rangle\langle \chi_b(-q^\prime)\rangle} .
\end{eqnarray}
With the help of eq.\ (\ref{classicfield}) differentiated 
with respect to $\phi$ 
and (\ref{sou}) with respect to $J$ one finds that eq.\ (\ref{abl1}) 
can be rewritten as
\begin{equation}
  \prl_t\Gm_k[\phi] =  \hal\Tr\left\{\left[
  \Gm_k^{(2)}[\phi]+R_k\right]^{-1}\prl_t R_k\right\} \; . 
  \label{ERGE}
\end{equation}
The exact flow equation (\ref{ERGE}) describes
the scale dependence of $\Gamma_k$ in terms of the inverse
average propagator $\Gm_k^{(2)}$ as given by
the second functional derivative of $\Gm_k$ with respect 
to the field components
\begin{equation}
  \left(\Gamma_k^{(2)}\right)_{a b}(q,q^\prime)=
  \frac{\delta^2\Gamma_k}
  {\delta\phi^a(-q)\delta\phi^b(q^\prime)}\; . 
\end{equation} 
The trace in eq.\ (\ref{ERGE}) involves only one
momentum integration as well as the summation
over the internal indices. 
The additional cutoff function $R_k$ with a form like
the one given in eq.\ (\ref{Rk(q)}) renders the momentum integration both 
infrared and ultraviolet finite. In particular, the direct 
implementation of
the additional mass--like term $R_k \simeq Z_k k^2$ for $q^2 \ll k^2$
into the inverse average propagator makes the formulation suitable
for dealing with theories which are plagued by infrared problems
in perturbation theory. Up to exponentially small
corrections the integration of the high momentum modes with $q^2\gg
k^2$ is not affected by the infrared cutoff. We note that
the derivation of the exact flow equation does not 
depend on the particular choice of the cutoff function and 
it may sometimes be technically easier to use
a different infrared cutoff function, as e.g.\
$R_k \sim k^2$. Ultraviolet finiteness, however, is related
to a fast decay of $\prl_t R_k$ for $q^2\gg k^2$. 
For a rapidly decaying cutoff
function $R_k$ like the one given in eq.\ (\ref{Rk(q)}) only 
fluctuations with momenta in a small ``window'' 
around $q^2 \simeq k^2$ effectively contribute to the
trace on the r.h.s.\ of eq.\ (\ref{ERGE}) for given $k$. 
Accordingly, the large momentum
fluctuations with $q^2 \gta k^2$, which are not affected by the 
infrared cutoff, have already been integrated out.
Since the momentum integrations for $q^2 \lta k^2$ are effectively 
cut off, $\Gamma_k$ realizes the concept of a coarse
grained free energy in the sense of Langer \cite{Langer}
(cf.\ section \ref{coarse}).
Of course, the particular choice for the infrared cutoff function 
should have no effect on the physical results for $k \to 0$.
Different choices of $R_k$ correspond
to different trajectories in the space of effective actions along
which the unique infrared limit $\Gamma_0$ is reached.
Nevertheless, once approximations are applied not
only the trajectory but also its end point may depend on the precise
definition of the function $R_k$. This dependence may be used
to study the robustness of the approximation.

The flow
equation (\ref{ERGE}) closely resembles a one--loop equation:
Replacing $\Gamma_k^{(2)}$ by the second functional derivative of the
classical action, $S^{(2)}$, one obtains the corresponding one--loop
result. The ``renormalization
group improvement'' $S^{(2)}\ra\Gamma_k^{(2)}$ turns the one--loop
flow equation into an {\em exact} non-perturbative flow equation.
It also turns the equation into a functional 
differential equation. Possible methods for its solution include
standard perturbation theory in the case of a small coupling,
the $1/N$--expansion or the $\epsilon$-expansion.
Particularly suitable for our purposes is the 
derivative expansion. We 
will allow for an arbitrary field dependence of the non-derivative
term, i.e.\ the effective potential. 
This enables us to study the
non-analytic behavior of the equation of state near second order
phase transitions or the simultaneous description
of degenerate minima at first order transitions.
With the help of the derivative expansion the functional differential
equation (\ref{ERGE}) can be transformed into a coupled set
of non--linear partial differential equations. 
If one does not want to resort to further approximations
at this level, one needs appropriate tools for
solving this type of partial differential equations.
Analytical solutions can be found only in certain limiting
cases and for most purposes numerical solutions seem the
adequate tool. We have developed suitable numerical algorithms  
which will be addressed
to in section \ref{solv}.

We emphasize that the flow equation (\ref{ERGE}) is closely connected 
to the Wilsonian (exact) renormalization group 
equation\footnote{The exact flow equation may also be interpreted 
as a differential form of the Schwinger-Dyson 
equations \cite{SD}.} \cite{RG,WH73-1,NC77,Wei76-1,Pol84-1,Has86-1}. 
The latter describes how
the Wilsonian effective action $S_\Lambda^{\rm W}$ changes with the 
ultraviolet cutoff $\Lambda$. Polchinski's continuum
version of the Wilsonian flow
equation~\cite{Pol84-1} can be transformed into eq.\ (\ref{ERGE}) 
by means of a Legendre transform and a suitable variable
redefinition~\cite{BAM,BAM93-1}. In contrast to the effective
average action the Wilsonian effective action
does not generate the $1PI$ Green functions~\cite{KKS92-1}.
We also note that eq.\ (\ref{ERGE}) is compatible with 
all symmetries of the model. Extensions of the
flow equations to gauge fields 
\cite{RW93-1,RW1,Bec96-1,BAM94-1,EHW94-1,Wet95-2,EHW96-1,Ell97-1} and 
fermions~\cite{Wet90-1,CKM97-1} (cf.\ section \ref{2qcd}) are
available. The exact flow
equation can be supplemented by an exact formalism for the
introduction of composite field variables or, 
more generally, a change of variables~\cite{EW94-1}. 
We will exploit this crucial feature for a description of 
low--energy QCD in section \ref{2qcd}.

%% file: second.tex
\sect{Equation of state and second order phase transitions
\label{secsec}}

\subsection{Introduction}

The prototype for investigations concerning the
restoration of a spontaneously broken symmetry at 
high temperature is the $N$-component
scalar field theory with $O(N)$ symmetry.
For $N=4$ it describes the scalar
sector of the electroweak standard model in the limit of
vanishing gauge and Yukawa couplings. It is
also used as an effective model for the chiral
phase transition in QCD in the limit of two quark flavors 
\cite{PW84-1,Wil92,RaWi93-1,Raj95-1}.
In condensed matter physics $N=3$ corresponds to the well
known Heisenberg model used to describe the ferromagnetic
phase transition. There are other applications like the
helium superfluid transition ($N=2$), liquid-vapor transition
($N=1$) or statistical properties of long polymer chains
($N=0$) \cite{ZJ}.

The equation of state (EOS)
for a magnetic system is specified by the
free energy density (here denoted by $U$) as a function
of arbitrary magnetization $\phi$ and temperature $T$.
All thermodynamic quantities can be derived from the function
$U(\phi,T)$. For example, the response of the system to
a homogeneous magnetic field $H$ follows from 
$\partial U/ \partial \phi = H$. This permits the computation
of $\phi$ for arbitrary $H$ and $T$.
There is a close 
analogy to quantum field theory at non-vanishing temperature.
Here $U$ corresponds to the temperature dependent effective
potential as a function of a scalar field $\phi$.
For instance, in the $O(4)$ symmetric model for the chiral
phase transition in two flavor QCD the meson field $\phi$
has four components. In this picture, the average light 
quark mass $\hat{m}$ is associated
with the source $H \sim \hat{m}$ 
and one is interested in the behavior during the phase 
transition (or crossover) for $H \not= 0$.
The temperature and source
dependent meson masses and zero momentum interactions 
are given by derivatives of $U$ (cf.\ section \ref{2qcd}).

The applicability of the $O(N)$ symmetric
scalar model to a wide class of very different
physical systems in the vicinity of the critical temperature
$T_{c}$ at which the phase transition occurs
is a manifestation of universality of critical phenomena.
There exists a universal scaling form of the EOS
in the vicinity of the second order phase transition
which is not accessible to 
standard perturbation theory. 
The main reason is the effective three dimensional behavior 
of high temperature quantum field theory already discussed
in section~\ref{mainin}. 
In particular, the critical exponents which describe the
singular behavior of various quantities near the phase
transition are those of the corresponding classical statistical
system (cf.\ section \ref{2qcd}).  

The quantitative description of the scaling
form of the EOS will be the main topic of this section
\cite{BTW95,ABBTW95,JB}.
The calculation of the effective potential $U(\phi,T)$ in the
vicinity of the critical
temperature of a second order phase transition is an 
old problem. One can prove 
through a general
renormalization group analysis \cite{RG} the Widom
scaling form \cite{Widom} of the EOS\footnote{We frequently suppress
in our notation an appropriate power of a suitable ``microscopic'' 
length scale $\Lambda^{-1}$ which is used to render quantities 
dimensionless.}    
\be
H = \phi^{\delta} \tilde{f}
\left((T-T_{c})/\phi^{1/\beta}\right) \label{wid}.
\ee
Only the 
limiting cases $\phi \rightarrow 0$ and $\phi \rightarrow \infty$
are quantitatively well described by 
critical exponents and amplitudes.
The critical exponents $\beta$ and $\delta$ have been
computed with high accuracy \cite{ZJ,BC95-1,Rei95-1} 
but the scaling function
$\tilde{f}$ is more difficult to access.
A particular difficulty for a perturbative computation in three 
dimensions arises from the singularities induced by 
massless Goldstone modes in the phase with spontaneous symmetry 
breaking for models with continuous symmetry ($N \neq 1$). 
The scaling function $\tilde{f}$ has been computed within
the $\eps$--expansion up to second order in $\eps$ \cite{BWW73-1}
(third order for $N=1$ \cite{ZJ}). Recently, the scaling form of
the EOS has also been studied for $N=1$ and $N=4$ using 
lattice Monte-Carlo simulations \cite{Tsy1,Tsy2,Tou}. 
We will observe that the lattice results 
agree quite well with those
obtained from the effective average action method, while there
appears a considerable discrepancy to the $\eps$--expansion results.

In the following 
we compute the scaling EOS for $O(N)$ symmetric 
scalar $N$-component models directly in three dimensions 
\cite{BTW95,ABBTW95,JB}. 
Section \ref{onpot} presents a computation of the effective potential 
$U=\lim_{k\to 0} U_k$ from a derivative expansion
of the effective average action. The approximation 
takes into account 
the most general field dependence of the potential term. 
This will allow us to compute the non--analytic behavior
of $U$ in the vicinity of the second order phase transition.
From $U$ the universal scaling form of the equation of state
is extracted in section \ref{uceos}. We conclude in section
\ref{oncon}.

\subsection{Scale dependence of the effective average 
potential\label{onpot}}

In this section we compute the
effective average potential $U_k(\rho)$
for an $O(N)$ symmetric scalar field
theory directly in three dimensions \cite{BTW95,ABBTW95,JB}. Here 
$\rho = \frac{1}{2} \phi^a \phi_a$ and $\phi^a$ denotes the 
$N$-component real scalar field. For
$k \rightarrow 0$ one obtains the effective potential 
(Helmholtz free energy)
$U_0(\rho)\equiv U(\rho)$. 
In the phase with spontaneous
symmetry breaking the minimum of the potential occurs for $k=0$ at
$\rhz \not= 0$. In the symmetric phase
the minimum of $U_k(\rho)$ ends at $\rhz=0$ for $k=0$. 
The two phases are separated by a scaling solution for which
$U_k/k^3$ becomes independent of $k$ once expressed in terms
of a suitably rescaled field variable and the
corresponding phase transition is of second order.

Though the evolution equation (\ref{ERGE})
for the effective average
action is exact, it remains a complicated functional
differential equation. In practice, one has to find
a truncation for $\Gm_k$ in order to obtain approximate solutions.
An important feature of the exact flow equation is therefore
its simple and intuitive form which helps to find a 
non-perturbative approximation scheme. The r.h.s.\ of eq.\
(\ref{ERGE}) expresses the scale dependence of $\Gm_k$
in terms of the exact propagator. Known properties
of the propagator can be used as a guide to find an
appropriate truncation for the effective average action.
For a scalar theory the propagator is a matrix 
characterized by mass terms and kinetic terms $\sim Z q^2$.
The mass matrix is given by the second derivative of 
the potential $U_k$ with respect to the fields. In general
$Z$ can be a complicated function of the fields and momenta.
We may exploit the fact that the function $Z$ plays the role of
a field and momentum dependent wave function 
renormalization factor. For second order phase transitions
and approximately for weak first order phase transitions
the behavior of $Z$ is governed by the anomalous 
dimension $\eta$. Typically for three dimensional
scalar theories $\eta$ is small. One therefore expects a weak
dependence of $Z$ on the fields and momenta (cf.\ also
the discussion in section \ref{outlook}).

Our truncation is the lowest order in a systematic derivative
expansion of $\Gamma_k$ 
\cite{Wet91-1,TetWet,TW94-1,Mor,morris,BTW95,ABBTW95,MT97-1}
\be
\Gamma_k = \int d^dx \bigl\{
U_k(\rho) + \frac{1}{2} Z_k \partial^{\mu} \phi_a
\partial_{\mu} \phi^a \bigr\}.
\label{three} \ee
We keep for the
potential term the most general $O(N)$ symmetric
form $U_k(\rho)$, whereas the wave function renormalization
is approximated by one $k$-dependent parameter.
Next order in the derivative expansion would be the
generalization to a $\rho$-dependent wavefunction
renormalization $Z_k(\rho)$ plus a function
$Y_k(\rho)$ accounting for a possible different
index structure of the kinetic term for $N \geq 2$
\cite{Wet91-1,TW94-1,MT97-1} (cf.\ 
section \ref{outlook}).
Going further would require the consideration of
terms with four derivatives and so on.
The weak $\rho$-dependence of the wave function 
renormalization factor employed in the ansatz (\ref{three}) has
been established explicitly for the scaling solution \cite{Mor,MT97-1}
and also recently for the equation of state ($N=1$) \cite{BSW}.

If the ansatz (\ref{Ansatz}) is inserted into the 
evolution equation
for the effective average action (\ref{ERGE}) one can extract 
flow equations  
for the effective average potential $U_k(\rho)$ and for the
wave function renormalization constant $Z_k$ (or equivalently the 
anomalous dimension $\eta$)\footnote{
Details of the calculation may be found 
in section \ref{secfirst} for the discussion of scalar matrix models
which covers the considerations involved here.}.
For a study of the behavior in the vicinity of the phase transition
it is convenient to work
with dimensionless renormalized fields
\footnote{We keep the number of dimensions $d$ arbitrary and specialize
only later to $d=3$.}
\ben
\rht &=& Z_k k^{2-d} \rho \; , \nonumber \\
u_k(\rht) &=& k^{-d} U_k(\rho).
\label{four} \een
With the truncation of eq. (\ref{three}) the exact
evolution equation for $u'_k \equiv \partial u_k/\partial \rht$
\cite{TetWet,TW94-1,ABBTW95}
reduces then to the partial differential equation
\ben
\frac{\partial u'_k}{\partial t} =~&&(-2 + \eta) u'_k +(d-2+ \eta)\rht u_k''
\nonumber \\
&&- 2 v_d (N-1) u_k'' l^d_1(u_k';\eta)
- 2 v_d (3 u_k'' + 2 \rht u_k''') l^d_1(u_k'+2 \rht u_k'';\eta),
\label{five}
\een
where $t = \ln \left( k/\Lx \right)$, with $\Lx$
the ultraviolet cutoff of the theory.
The anomalous dimension $\eta$ is defined by
\be
\eta = - \frac{\partial}{\partial t} \ln Z_k
\label{six} \ee
and
\be
v_d^{-1} = 2^{d+1} \pi^{\frac{d}{2}} \Gamma \left( \frac{d}{2}
\right),
\label{seven} \ee
with $v_3 = 1/8 \pi^2$.
The ``threshold'' functions $l^d_n(w;\eta)$
result from the momentum integration on the
r.h.s. of eq. (\ref{ERGE}), and read for $n \geq 1$, with
$y=q^2/k^2$
\ben
l^d_n(w;\eta) =
{}~&&- n \int^{\infty}_0 dy y^{\frac{d}{2}+1}
\frac{\partial r(y)}{\partial y}
\left[ y(1+r(y)) + w \right]^{-(n+1)}
\nonumber \\
&&- \frac{n}{2} \eta \int^{\infty}_0 dy y^{\frac{d}{2}}
r(y)
\left[ y(1+r(y)) + w \right]^{-(n+1)}.
\label{eight} \een
Here $r(y)$ depends on the choice of the momentum dependence
of the infrared cutoff and we employ according to eq.\
(\ref{Rk(q)})
\be
r(y) = \frac{e^{-y}}{1 - e^{-y}}.
\label{nine} \ee
This choice has the property
$\lim_{q^2 \rightarrow 0} R_k/Z_k k^2 =
\lim_{y \rightarrow 0} y r(y) = 1$, whereas for $q^2 \gg k^2$
the effect of the infrared cutoff is exponentially suppressed.
The ``threshold'' functions $l^d_n$ decrease rapidly for
increased $w$ and account for the decoupling
of modes with mass term $w \gg 1$. We evaluate these
functions numerically. Finally, the anomalous dimension is given
in our truncation by\footnote{We neglect here for
simplicity the implicit, linear $\eta$-dependence of the
function $m^d_{2,2}$. We have numerically verified this approximation
to have only a minor effect on the value of $\eta$.} 
\cite{TetWet,TW94-1,ABBTW95}
\be
\eta(k) =  \frac{16 v_d}{d} \kx \lx^{2} m^d_{2,2}(2 \lx \kx),
\label{ten} \ee
with $\kx$ the location of the minimum of the potential
and $\lx$ the quartic coupling
\ben
u'_k(\kx)&=&0 \nonumber\; , \\
u''_k(\kx)&=&\lambda\; .
\label{eleven} \een
The function $m^d_{2,2}$ is given by 
\ben
m^d_{2,2}(w) &= & \int_0^{\infty} dy y^{\frac{d}{2}-2}
\frac{1 + r +y \frac{\partial r}{ \partial y}}{(1+r)^2
\left[(1+r)y +w \right]^2 }
\nonumber \\
&&\biggl\{
2 y \frac{\partial r}{ \partial y}
+ 2 \left(y \frac{\partial }{ \partial y} \right)^2 r
- 2 y^2 \left( 1 +r + y \frac{\partial r}{ \partial y} \right)
\frac{\partial r}{ \partial y}
\left[ \frac{1}{(1+r)y} + \frac{1}{(1+r)y + w} \right]
\biggr\}.
\nonumber \\
{}~&~
\label{twelve} \een
We point out that the argument
$2 \lx \kx$ turns out generically to
be of order one for the scaling solution. Therefore,
$\kx \sim \lx^{-1}$ and the mass effects are important,
in contrast to perturbation theory where they are
treated as small quantities $\sim \lx$.

At a second order phase transition 
there is no mass scale present in the theory.
In particular, one 
expects a scaling behavior of the rescaled effective 
average potential $u_k(\tilde{\rho})$. This can be studied
by following the trajectory describing the scale dependence
of $u_k(\tilde{\rho})$ as $k$ is lowered from $\Lambda$ to zero.  
Near the phase transition the   
trajectory spends most of the ``time'' $t$ in the vicinity of
the $k$-independent scaling solution of eq.\ (\ref{five}) given by
$\partial_t u'_* (\rht) = 0$.\footnote{The resulting ordinary 
differential
equation for $u_*(\rht)$ has already been solved numerically
for a somewhat different choice of the infrared cutoff
\cite{Mor,morris,MT97-1}.}
Only at the end of the running the ``near-critical''
trajectories deviate from the scaling solution. For
$k \rightarrow 0$ they either end up in the symmetric phase with
$\kx =0$ and positive constant mass term $m^2$ such that
$u'_k(0) \sim m^2/k^2$; or they lead to a non-vanishing
constant $\rhz$ indicating spontaneous symmetry breaking with
$\kx \rightarrow Z_0 k^{2-d} \rhz$. The equation of state
involves the potential $U_0(\rho)$ for
temperatures away from the critical temperature.
Its computation requires the
solution for the running 
away from the critical trajectory which involves the
full partial differential equation (\ref{five}).
We have developed two alternative algorithms for its numerical 
solution which will be discussed in section \ref{solv}.

In fig. 1 we present the results of the numerical integration
of eq. (\ref{five}) for $d=3$ and $N=1$.
\begin{figure}
\leavevmode
\centering  
\epsfxsize=4in
\epsffile{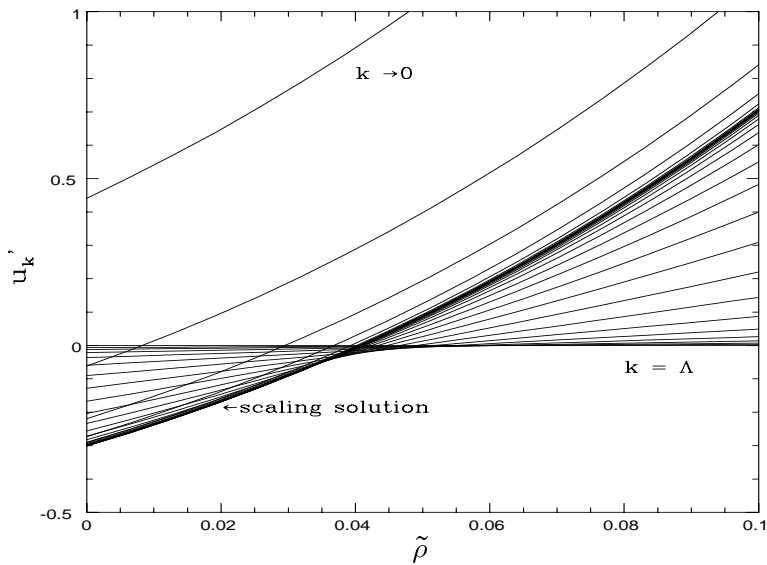}                                   
\caption{\footnotesize The evolution of $u'_k(\rht)$ 
as $k$ is lowered from $\Lx$ to zero for $N=1$.
The initial conditions (bare couplings)
have been chosen such that the scaling solution is approached
before the system evolves towards the symmetric phase
with $u'_k(0) > 0$.}
\end{figure}
The function $u'_k(\rht)$ is plotted
for various values of $t= \ln(k/\Lx)$. The evolution starts
at $k=\Lx$ ($t=0$)  where the average potential is equal to
the classical potential (no effective integration of modes
has been performed). We start with a quartic classical potential
parameterized as
\be
u'_{\Lx}(\rht) = \lx_\Lx (\rht - \kx_\Lx).
\label{eighteen} \ee
We arbitrarily choose $\lx_\Lx=0.1$ and fine tune $\kx_\Lx$
so that a scaling solution is approached at later stages
of the evolution. There is a critical value
$\kx_{cr} \simeq 6.396 \times 10^{-2}$
for which the evolution leads to the scaling solution.
For the results in fig. 1 a value
$\kx_\Lx$ slightly smaller than $\kx_{cr}$
is used. As $k$ is lowered
(and $t$ turns negative), $u'_k(\rht)$ deviates from its initial
linear shape. Subsequently it evolves towards a form which is
independent of $k$ and corresponds to the scaling
solution $\partial_t u'_* (\rht) = 0$. It spends a long ``time''
$t$ - which can be rendered arbitrarily long through appropriate
fine tuning of $\kx_\Lx$ -
in the vicinity of the scaling solution. During this
``time'', the minimum of the potential $u'_k(\rht)$
takes a fixed value $\kx_*$,
while the minimum of $U_k(\rho)$ evolves towards zero according to
\be
\rhz(k) = k \kx_* / Z_k.
\label{nineteen} \ee
The longer $u'_k(\rht)$ stays near the scaling solution, the
smaller the resulting value of $\rhz(k)$ when the
system deviates from it.
As this value determines the mass scale for the
renormalized theory at $k=0$, the
scaling solution governs the behavior of the system
very close to the phase transition, where the characteristic
mass scale goes to zero.
Another important property of the ``near-critical''
trajectories, which spend a long ``time'' $t$ near
the scaling solution, is that they become insensitive
to the details of the classical theory which determine the
initial conditions for the evolution. After $u'_k(\rht)$ has
evolved away from its scaling form $u'_*(\rht)$, its
shape is independent of the choice of $\lx_\Lx$ for
the classical theory.
This property gives rise to the universal behavior
near second order phase transitions.
For the solution depicted in
fig.\ 1, $u_k(\rht)$ evolves in such a way that its minimum
runs to zero with $u'_k(0)$ subsequently increasing.
Eventually the theory settles down in the
symmetric phase with a positive constant renormalized mass term
$m^2 = k^2 u'_k(0) $ as $k \rightarrow 0$.
Another possibility is that the system ends up in the
phase with spontaneous symmetry breaking. In this case
$\kx$ grows in such a way that
$\rhz(k)$ approaches a constant value
for $k \rightarrow 0$.

The approach to the scaling solution and the deviation
from it can also be seen in fig. 2. 
\begin{figure}
\unitlength1.0cm
\begin{center}
\begin{picture}(13.,8.)
\put(-4,-8){
\epsfysize=20.cm
\epsfxsize=18.cm
\epsffile{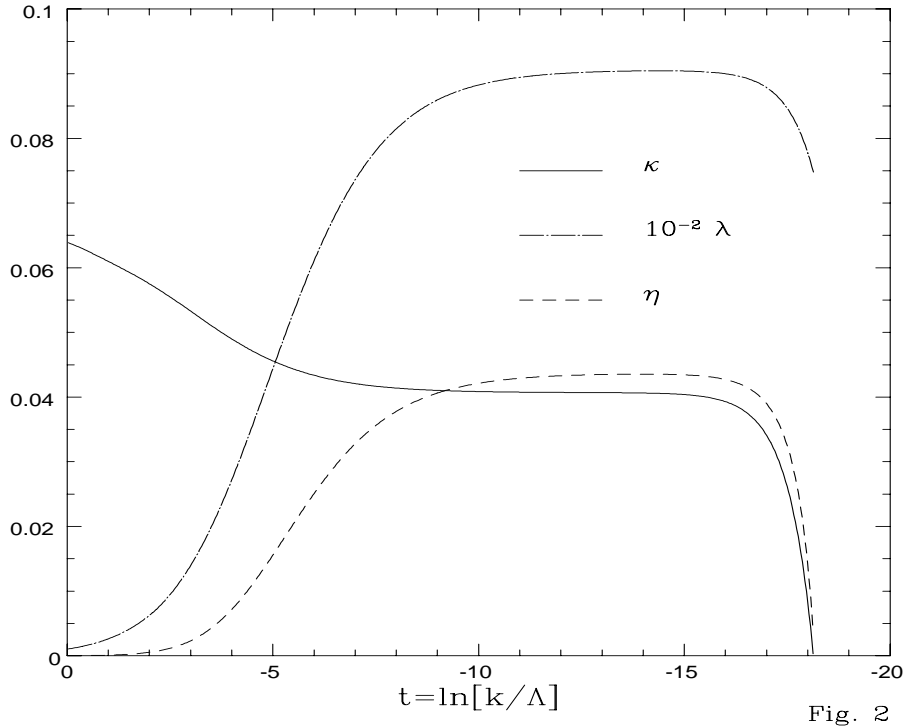}}
\end{picture}
\end{center}
\caption{\footnotesize The evolution of $\kx$, $\lx$ and $\eta$
for the solution of fig. 1.}
\end{figure}
The evolution of the
running parameters $\kx(t)$, $\lx(t)$ starts
with their initial classical values, leads to
fixed point values $\kx_*$, $\lx_*$ near the scaling solution,
and finally ends up in the symmetric phase
($\kx$ runs to zero).
Similarly the anomalous dimension $\eta(k)$, which is given
by eq. (\ref{ten}), takes a fixed point value $\eta_*$
when the scaling solution is approached.
During this part of the evolution the wave function
renormalization is given by
\be
Z_k \sim k^{-\eta_*}
\label{twenty} \ee
according to eq. (\ref{six}). When the parts of the evolution towards
and away from the fixed point become negligible
compared to the evolution near the fixed point
- that is, very close to the phase transition -
eq. (\ref{twenty}) becomes a very good approximation for
sufficiently low $k$.
This indicates that $\eta_*$ can be identified with the
critical exponent $\eta$.
For the solution of fig. 2 ($N=1$) we find
$\kx_*=4.07 \times 10^{-2}$, $\lx_*=9.04$ and
$\eta_*=4.4 \times 10^{-2}$.

As we have already mentioned the details of
the renormalized theory in the vicinity of the phase
transition are independent of the classical coupling
$\lx_\Lx$. Also the
initial form of the potential does not have
to be of the quartic form of eq.\ (\ref{eighteen})
as long as the symmetries are respected.
Moreover, the critical theory can
be parameterized in terms of critical exponents
\cite{WilFis72}, an example of
which is the anomalous dimension $\eta$.
These exponents are universal quantities which depend
only on the dimensionality of the system and its internal
symmetries. For our three-dimensional theory they depend only on the
value of $N$ and
can be easily extracted from our results. We concentrate here on
the exponent $\nu$, which parameterizes the behavior of the
renormalized mass in the critical region. Other exponents are
computed in sections \ref{uceos} and \ref{CriticalBehavior}
along with the critical equation of state. 
The other exponents
are not independent quantities, but
can be determined from $\eta$ and $\nu$ through universal
scaling laws \cite{WilFis72}. We define the exponent $\nu$ through
the renormalized mass term in the symmetric phase
\be
m^2 =  \frac{1}{Z_k} \frac{d U_k(0)}{d\rho}
= k^2 u'_k(0)~~~~~~~{\rm for}~~k \rightarrow 0.
\label{twentyone} \ee
The behavior of $m^2$ in the critical region
depends only on the distance from the phase transition, which
can be expressed in terms of the difference of
$\kx_\Lx$ from the critical value $\kx_{cr}$ for which
the renormalized theory has exactly $m^2 =0$.
The exponent $\nu$ is determined from the relation
\be
m^2 \sim |\dkl|^{2 \nu} = |\kx_\Lx - \kx_{cr}|^{2 \nu}.
\label{twentytwo} \ee
For a determination of $\nu$ from our results we calculate
$m^2$ for various values of $\kx_\Lx$ near $\kx_{cr}$.
We subsequently plot $\ln(m^2)$ as a function
of $\ln|\dkl|$. This curve becomes linear for
$\dkl \rightarrow 0$ and we
obtain $\nu$ from the constant slope.
In the past the critical exponents of the $O(N)$ symmetric
theory were calculated from truncated versions of
the partial differential equation (\ref{five}) \cite{TetWet,TW94-1}.
The strategy was to turn eq. (\ref{five}) into an infinite
system of ordinary differential equations for the
coefficients of a Taylor expansion 
around the ``running'' minimum of the potential.
This infinite system was approximately
solved by neglecting $\rht$-derivatives
of $u_k(\rht)$ higher
than a given order. The apparent convergence of the procedure
was checked by enlarging the level of truncation.
We now have an alternative way of estimating the accuracy of this
method. Our numerical solution of the partial differential
equation (\ref{five}) corresponds to an infinite level of
truncation where all the higher derivatives are taken into
account.
In table 1 we present results obtained through the procedure
of successive truncations and through our numerical solution
for $N=3$. 
\begin{table} [h]
\renewcommand{\arraystretch}{1.5}
\hspace*{\fill}
\begin{tabular}{|c|c|c|c|c|c|}  \hline

& $\kx_*$
& $\lx_*$
& $u^{(3)}_*$
& $\eta$
& $\nu$
\\ \hline \hline
a
& $6.57 \times 10^{-2}$
& 11.5
&
&
& 0.745
\\ \hline
b
& $8.01 \times 10^{-2}$
& 7.27
& 52.8
&
& 0.794
\\ \hline
c
& $7.86 \times 10^{-2}$
& 6.64
& 42.0
& $3.6 \times 10^{-2}$
& 0.760
\\ \hline
d
& $7.75 \times 10^{-2}$
& 6.94
& 43.5
& $3.8 \times 10^{-2}$
& 0.753
\\ \hline
e
& $7.71 \times 10^{-2}$
& 7.03
& 43.4
& $3.8 \times 10^{-2}$
& 0.752
\\ \hline
f
& $7.64 \times 10^{-2}$
& 7.07
& 44.2
& $3.8 \times 10^{-2}$
& 0.747
\\ \hline
\end{tabular}
\hspace*{\fill}
\renewcommand{\arraystretch}{1}
\caption[y]
{
The minimum $\kx$ of the
potential $u_k(\rht)$, the derivatives
$\lx=u''(\kx)$, $u_k^{(3)}(\kx)$
for the scaling solution, and
the critical exponents $\eta$ and $\nu$,
in various approximations: (a)-(e) from
refs. \cite{TetWet,TW94-1}
and (f) from the present
section \cite{ABBTW95,BTW95}. $N=3$. \\
a) Truncation where only the evolution of $\kx$ and $\lx$ is
considered and
higher derivatives of the potential and the anomalous
dimension are neglected. \\
b) $\kx$, $\lx$, $u^{(3)}_k(\kx)$ are included. \\
c) $\kx$, $\lx$, $u^{(3)}_k(\kx)$ are included and $\eta$ is approximated
by eq. (\ref{ten}). \\
d) with five parameters: $\kx$, $\lx$, $u^{(3)}_k(\kx)$, $u^{(4)}_k(\kx)$
and $\eta$. \\
e) as in d) and in addition
$u^{(5)}_k(\kx)$, $u^{(6)}_k(\kx)$ are estimated. \\
f) The partial differential equation (\ref{five})
for $u'_k(\rht)$ is solved numerically
and $\eta$ is approximated by eq. (\ref{ten}). \\
}
\end{table}
We give the values of $\kx$, $\lx$, $u^{(3)}_k(\kx)$
for the scaling solution and the critical exponents
$\eta$, $\nu$. We observe how the results stabilize as more
$\rht$-derivatives of $u_k(\rht)$ at $\rht=\kx$ and the
anomalous dimension are taken into account. The last line
gives the results of our numerical solution of eq. (\ref{five}).
By comparing with the previous line we conclude
that the inclusion of all the $\rht$-derivatives higher than
$u_k^{(6)}(\kx)$ and the term $\sim \eta$ in the
``threshold'' function of eq. (\ref{eight})
generates an improvement of less than
1 \% for the results. This is smaller than the error
induced by the omission of the higher derivative terms
in the average action, which typically generates an uncertainty
of the order of the anomalous dimension.
In table 2 we compare our values for the critical exponents
with results obtained from other methods\footnote{See also
  ref.~\cite{MT97-1} and references therein for a calculation
  of critical exponents using similar methods as in this work.}
(such as the $\epsilon$-expansion, perturbation series
at fixed dimension, lattice high temperature expansions 
and the $1/N$-expansion).
\begin{table} [h]
\renewcommand{\arraystretch}{1.}
\hspace*{\fill}
\begin{tabular}{|c||c|c||c|c|}
\hline
$N$
&\multicolumn{2}{c||}{$\nu$}
&\multicolumn{2}{c|}{$\eta$}
\\
\hline \hline

&
&$0.5880(15)^a$
&
&$0.027(4)^a$
\\
0
&$0.589$
&$0.5880(15)^{b}$
&$0.040$
&$0.0320(25)^{b}$
\\

&
&$0.5878(6)^c$
&
&
\\ \hline
&
&$0.6300(15)^a$
&
&$0.032(3)^{a}$
\\
1
&0.643
&$0.6310(15)^{b}$
&0.044
&$0.0375(25)^{b}$
\\

&
&$0.6315(8)^{c}$
&
&
\\ \hline

&
&$0.6695(20)^{a}$
&
&$0.033(4)^{a}$
\\
2
&0.697
&$0.671(5)^{b}$
&0.042
&$0.040(3)^{b}$
\\

&
&$0.675(2)^{c}$
&
&
\\ \hline

&
&$0.705(3)^{a}$
&
&$0.033(4)^{a}$
\\
3
&0.747
&$0.710(7)^{b}$
&0.038
&$0.040(3)^{b}$
\\

&
&$0.716(2)^{c}$
&
&
\\ \hline
4
&0.787
&$0.73(2)^{a*}$
&0.034
&$0.03(1)^{a*}$
\\

&
&$0.759(3)^{c}$
&
&\\  \hline
10
&0.904
&$0.894(4)^{c}$
&0.019
&
\\

&
&$0.877^{d}$
&
&$0.025^{d}$
\\ \hline
100
&0.990
&$0.989^{d}$
&0.002
&$0.003^{d}$
\\ \hline
\end{tabular}
\hspace*{\fill}
\renewcommand{\arraystretch}{1}
\caption[y]
{
Critical exponents $\nu$ and $\eta$
for various values of $N$.
For comparison we list results obtained with other methods as summarized in
\cite{ZJ}, \cite{journal} and \cite{BC95-1}: \\
a) From perturbation series at fixed dimension to six--loop 
( $^*$seven--loop \cite{BMN78-1}) order. \\
b) From the $\ex$-expansion at order $\ex^5$. \\
c) From lattice high temperature expansions. \\
d) From the $1/N$-expansion at order $1/N^2$. 
}
\end{table}
As expected $\eta$ is rather poorly determined since it is
the quantity most seriously affected by the omission of the
higher derivative terms in the average action. The exponent
$\nu$ is in agreement with the known results at the
1-5 \% level, with a discrepancy roughly equal to the
value of $\eta$ for various $N$.

In conclusion, the shape of the average potential is under
good quantitative control for every scale $k$. This
permits a quantitative understanding of the most
important properties of the system at every length
scale. We will exploit this in the following to extract the scaling
form of the equation of state which is the main
result here.

\subsection{Universal critical equation of state\label{uceos}}

In this section we extract the Widom scaling form
of the equation of state (EOS) 
from a solution \cite{BTW95,JB} of eqs.\ (\ref{five}),
(\ref{ten}) for the three dimensional $O(N)$ model. 
Its asymptotic behavior yields the universal critical
exponents and amplitude ratios.
We also present fits for the scaling function for $N=1$ and $N=3$.
(See in addition section \ref{CriticalBehavior} for $N=4$.)
An alternative parameterization of the equation
of state in terms of renormalized quantities is used in order
to compute universal couplings. 

Eq.\ (\ref{wid}) establishes the scaling properties of the
EOS. The external field $H$ is related to the derivative
of the effective potential $U'= \partial U/\partial \rho$ by
$H_a = U' \phi_a$.
The critical equation of state, relating
the temperature, the external field and the
order parameter, can then be written in the scaling
form ($\phi=\sqrt{2 \rho}$) 
\be
\frac{U'}{\phi^{\delta-1}} =  f(x),
\qquad x=\frac{-\dkl} {\phi^{1/\beta}},
\label{sixb}
\ee
with critical exponents $\delta$ and $\beta$.
A measure of the distance from the 
phase transition is the difference
$\dkl = \kx_{\Lambda} - \kx_{cr}$.
If $\kx_{\Lambda}$ is interpreted as a function of temperature, 
the deviation $\dkl$ 
is proportional to the deviation from the critical temperature, i.e.\
$\dkl = A(T) (\tcr- T)$ with $A(\tcr) > 0$.
For $\phi \to \infty$
our numerical solution for $U'$ obeys $U' \sim \phi^{\delta-1}$ with
high accuracy. The inferred value of $\delta$ is displayed in 
table \ref{tableeos}, and we have checked the scaling relation 
$\delta=(5-\eta)/(1+\eta)$.
The value of the critical exponent $\eta$ is obtained from eq. 
(\ref{five}) for the scaling solution.
We have also verified explicitly that 
$f$ depends only
on the scaling variable $x$ for the value of $\beta$ given in table
\ref{tableeos}. 
In figs.\ 1 and 2 we plot log$(f)$
and log$(df/dx)$ 
as a function of log$|x|$ for $N=1$ and $N=3$.
Fig.\ 1 corresponds to the symmetric phase $(x > 0)$, and 
fig.\ 2 to the phase with spontaneous symmetry breaking
$(x < 0)$. This is our main result of this section.
\begin{figure}
\leavevmode
\centering  
\epsfxsize=4in
\epsffile{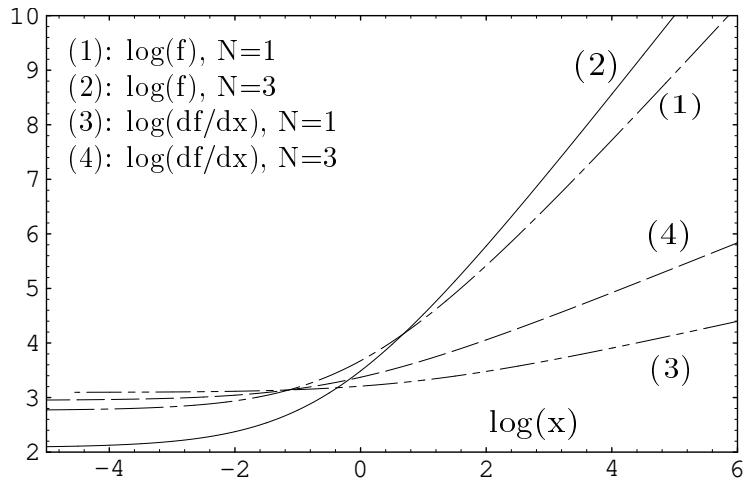}                                   
\caption{Logarithmic plot of $f$ and $df/dx$ for $x > 0$.}
\end{figure}
\begin{figure}
\leavevmode
\centering  
\epsfxsize=4in
\epsffile{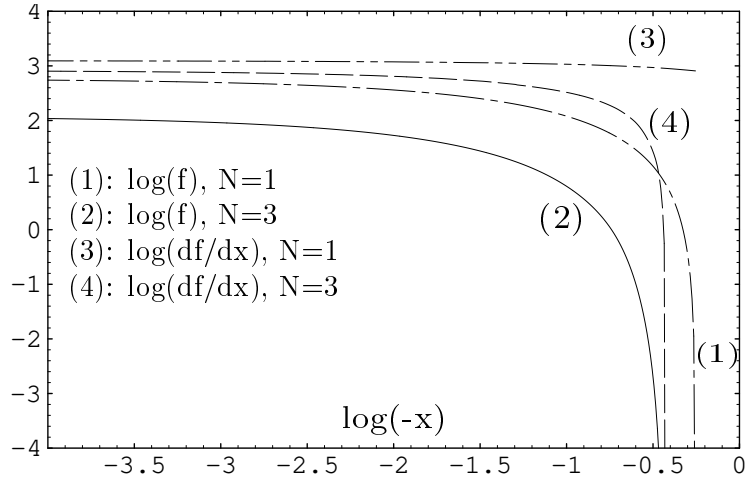}                                   
\caption{Logarithmic plot of $f$ and $df/dx$ for $x < 0$.}
\end{figure}
 
One can easily extract the asymptotic behavior from the logarithmic
plots and compare with known values of critical exponents and
amplitudes.
The curves become constant, both for 
$x \to 0^+$ and $x \to 0^-$ with the same value, consistently with the
regularity of $f(x)$ at $x=0$. For the universal function one
obtains
\be
\lim\limits_{x \to 0} f(x) = D,  \label{hd}
\ee
and $H = D \phi^{\delta}$ on the critical isotherm.
For $x \to \infty$ one 
observes that log$(f)$ becomes a linear function of log$(x)$  
with constant slope $\gamma$.
In this limit the universal function takes the form
\be
\lim\limits_{x \to \infty} f(x) = (C^+)^{-1} x^{\gamma} .
\ee     
The amplitude $C^+$ and the critical exponent $\gamma$
characterize the behavior of the 'unrenormalized' squared mass
or inverse susceptibility
\be 
\bar{m}^2=\chi^{-1}=\lim_{\phi \to 0}
\left(\frac{\prl^2 U}{\prl \phi^2}=
(C^+)^{-1}|\dt \kp_{\La}|^{\gamma}\phi^{\delta-1-\gamma/\beta}
\right).
\ee   
We have verified the scaling relation $\gamma/\beta=\delta-1$
that connects $\gamma$ with the exponents $\beta$ and $\delta$
appearing in the Widom scaling form (\ref{wid}). 
One observes that the zero-field magnetic susceptibility,
or equivalently the inverse unrenormalized squared mass
$\bar{m}^{-2}=\chi$, is 
non-analytic for $\dkl \to 0$ in the symmetric phase:
$\chi = C^+ |\dkl|^{-\gamma}$. In this phase 
we find that the correlation length
$\xi = (Z_0 \chi)^{1/2}$, which is
equal to the inverse of the renormalized mass $m_R$, 
behaves as $\xi=\xi^+|\dkl|^{-\nu}$
with $\nu=\gamma/(2-\eta)$.

The spontaneously broken phase is characterized by a nonzero
value $\phi_0$ of the minimum of the effective potential 
$U$ with
$H=(\prl U/\prl \phi) (\phi_0)=0$. The appearance
of spontaneous symmetry breaking below $T_{c}$ implies
that $f(x)$ has a zero $x=-B^{-1/\beta}$ and one observes
a singularity of the logarithmic plot in fig.\ 2. In
particular, according to eq.\ (\ref{wid}) the minimum
behaves as $\phi_0 = B (\dt \kp_{\La})^{\beta}$.
Below the critical temperature,
the longitudinal and transversal
susceptibilities $\chi_L$ and $\chi_T$ are different for $N > 1$
\be
\chi_L^{-1} = \frac{\partial^2 U}{\partial \phi^2}
= \phi^{\delta-1} \bigl(\delta f(x) - \frac{x}{\beta}f'(x)\bigr),
\qquad
\chi_T^{-1} = \frac{1}{\phi}\frac{\partial U}{\partial \phi}
= \phi^{\delta-1} f(x) \label{eightb}
\ee
(with $f'=df/dx$).
This is related to the existence of massless Goldstone modes in the 
$(N-1)$ transverse directions, which causes the transversal susceptibility
to diverge for vanishing external field.
Fluctuations of these modes 
induce the divergence of the zero-field longitudinal
susceptibility.
This can be concluded 
from the singularity 
of log$(f')$ for $N=3$
in fig.\ 2. The first $x$-derivative of the
universal function vanishes as $H \to 0$, i.e.\
$f'(x=-B^{-1/\beta})=0$ for $N > 1$.
For $N=1$ there is a non-vanishing constant value for
$f'(x=-B^{-1/\beta})$ with a finite zero-field susceptibility
$\chi = C^- (\dkl)^{-\gamma}$, where
$(C^-)^{-1}=B^{\delta-1-1/\beta}f'(-B^{-1/\beta})/\beta$.
For a non-vanishing physical infrared cutoff
$k$, the longitudinal susceptibility remains finite also
for $N > 1$: $\chi_L \sim (k\rho_0)^{-1/2}$.
For $N=1$ in the ordered phase, 
the correlation length behaves as
$\xi=\xi^-(\dkl)^{-\nu}$, and
the renormalized minimum $\rho_{0R}= Z_0 \rho_0$ of the potential $U$
scales as 
$\rho_{0R}=E (\dkl)^{\nu}$.

The amplitudes of singularities near the phase transition $D$, $C^{\pm}$,
$\xi^{\pm}$, $B$ and $E$ are given in table \ref{tableeos}. They are not
universal. There are two independent
scales in the vicinity of the transition point which can be related
to the deviation from the critical temperature and to the external
source. All models in the
same universality class can be related by a multiplicative
rescaling of $\phi$ and $\dt \kp_{\La}$ or $(T_{c}-T)$.
Accordingly there are only two independent amplitudes
and exponents respectively. 
Ratios of amplitudes which are invariant
under this rescaling are universal. We display the universal
combinations $C^+/ C^-$, $\xi^+/\xi^-$,
$R_{\chi}=C^+ D B^{\delta-1}$, 
$\tilde{R}_{\xi}=(\xi^+)^{\beta/\nu} D^{1/(\delta+1)} B$
and $\xi^+ E$ in table \ref{tableeos}.
\begin{table} [h]
\renewcommand{\arraystretch}{1.5}
\hspace*{\fill}
\begin{tabular}{|c|c|c|c|c|c||c|c|c|c|c|c|}     \hline

$N$
& $\beta$ 
& $\gamma$ 
& $\delta$
& $\nu$
& $\eta$
& $\lr / m_R$ 
& $\nu_R$ 
& $\hat{\lambda}_R / \rho_{0R}$
& $\hat{\nu}_R$
&
\\ \hline 
1
& 0.336
& 1.258
& 4.75
& 0.643
& 0.044
& 9.69 
& 108
& 61.6
& 107
&
\\ \hline
3
& 0.388
& 1.465
& 4.78
& 0.747
& 0.038
& 7.45
& 57.4
& 0
& $\simeq$ 250
&
\\ \hline \hline
$$
& $C^+$ 
& $D$ 
& $B$
& $\xi^+$
& $E$
& $C^+/C^-$
& $\xi^+/\xi^-$
& $R_{\chi}$
& $\tilde{R}_{\xi}$
& $\xi^+ E$
\\ \hline 
1
& 0.0742 
& 15.88
& 1.087
& 0.257
& 0.652
& 4.29
& 1.86
& 1.61
& 0.865
& 0.168
\\ \hline
3
& 0.0743
& 8.02
& 1.180
& 0.263
& 0.746
& -
& -
& 1.11
& 0.845
& 0.196
\\ \hline 
\end{tabular}
\hspace*{\fill}
\renewcommand{\arraystretch}{1}
\caption[]%
{Parameters for the equation of state.}
\label{tableeos}
\end{table}

The asymptotic behavior observed for the universal function can be
used in order to obtain a semi-analytical expression for $f(x)$.
We find that the following
fit reproduces the numerical values for both
$f$ and $df/dx$ with a 1\% accuracy
(apart from
the immediate vicinity of the zero of $f$
for $N=3$):
\be
f_{fit}(x)=D \bigl(1+B^{1/\beta} x \bigr)^a
\bigl(1+\Theta x \bigr)^{\Delta} 
\bigl(1+c x \bigr)^{\gamma-a-\Delta} , \label{nineb}
\ee
with $c=(C^+ D B^{a/\beta} \Theta^{\Delta} )^{-1/(\gamma-a-\Delta)}$
and $a=1$ $(a=2)$ for $N=1$ $(N>1)$. The fitting
parameters are chosen as $\Theta = 0.569 $ ($1.312$) and $\Delta = 0.180$
$(-0.595)$ for $N=1$ $(3)$. 

There is an alternative parameterization of the 
equation of state in terms
of renormalized quantities. In the symmetric phase ($\dkl < 0$)
we consider the dimensionless quantity
\be
F(s) = \frac{U_R'}{m_R^2} = C^+ x^{-\gamma} f(x), \qquad 
s=\frac{\rho_R}{m_R} = \frac{1}{2}(\xi^+)^3(C^+)^{-1} x^{-2\beta},
\label{tenb}
\ee
with
$\rho_R = Z_0 \rho$ and $U_R^{(n)} = Z_0^{-n} U^{(n)}$.
The derivatives of $F$ at $s=0$ yield the 
universal couplings
\be
\frac{d F}{d s}(0) = \frac{U_{R}''(0)}{m_R} \equiv \frac{\lambda_R}{m_R},
\qquad
\frac{d^2 F}{d s^2}(0) = U_{R}'''(0) \equiv \nu_R, \label{elev}
\ee 
and similarly for higher derivatives.
We confirm that the potential $U$ is well approximated by a 
$\phi^6$ (or $\rho^3$) polynomial \cite{TetWet,TW94-1,Tsy1}.
For $N=1$, 
our value for $\nu_R$ in the symmetric phase 
agrees with the results of a numerical
simulation \cite{Tsy1} within the expected accuracy (related to 
the size of the anomalous dimension $\eta$).
However, $\lx_R/m_R$ seems to be too small. This may be connected to
the more crude approximation for the anomalous dimension in the
symmetric phase, as compared to the ordered one. For a
recent detailed comparison of the dimensionless renormalized couplings
in the symmetric phase
with results obtained from other methods we refer the reader 
to ref.\ \cite{BC97-3}. Small field expansions of the
effective potential have also been used to reconstruct 
the equation of state $(N=1)$ within summed perturbation series
at fixed dimension or from high temperature expansions \cite{GZJ96-1}.
A comparison including the results obtained here
can be found in ref.\ \cite{GZJ96-1}. 

In the ordered 
phase $(\dkl > 0)$ we consider the ratio
\be
G(\tilde{s}) = \frac{U_{R}'}{\rho_{0R}^2} = \frac{1}{2} B^2 E^{-3}
(-x)^{-\gamma} f(x), 
\qquad
\tilde{s} = \frac{\rho_R}{\rho_{0R}} = B^{-2} (-x)^{-2\beta}.\label{fourt}
\ee
The values for the universal couplings
\be
\frac{d G}{d \tilde{s}}(1) = \frac{U_{R}''(\rho_{0R})}{\rho_{0R}}\equiv
\frac{\hat{\lambda}_R}{\rho_{0R}},
\qquad
\frac{d^2 G}{d \tilde{s}^2}(1)=U_{R}'''(\rho_{0R})\equiv
\hat{\nu}_R, \label{fivet}
\ee
as well as $\lambda_R/m_R$ and $\nu_R$ 
are given in table \ref{tableeos}. 
One observes that for $N>1$ the renormalized quartic coupling
$\hat{\lambda}_R$ vanishes in the ordered phase. 
The coupling  $\hat{\nu}_R$ takes a finite value and higher
derivatives $d^nG/d\tilde{s}^n(1)$ (for $n \le 3$) diverge.
This is due to  
the presence of massless fluctuations.

For comparison, 
\begin{figure}[h]
\unitlength1.0cm
\begin{center}
\begin{picture}(13.,8.)
\put(-3,-15){
\epsfysize=26cm
\epsfxsize=20.cm
\epsffile{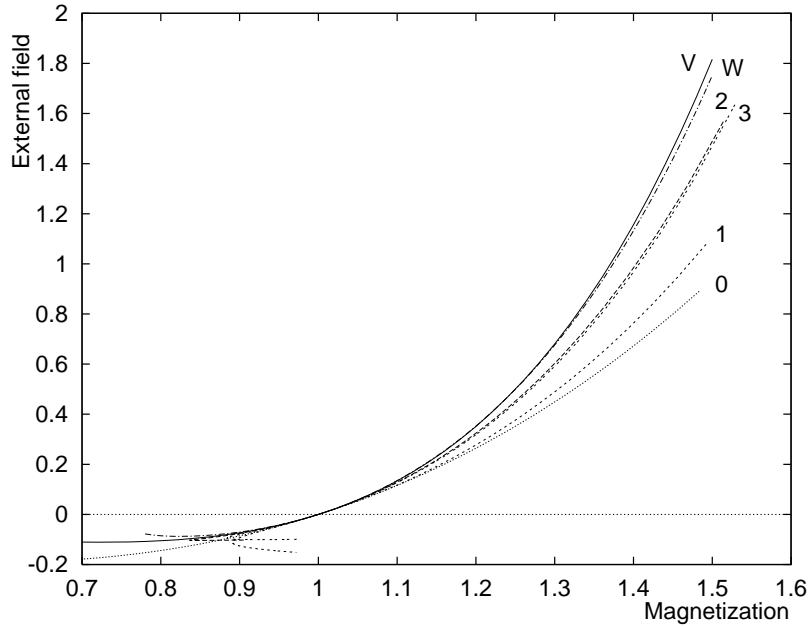}}
\end{picture}
\end{center}
\caption[]{\footnotesize The plot shows the ``Ising equation of
state'' ($N=1$) computed by different methods. The external source 
$\hat{\jmath}=\partial \hat{U}/\partial \hat{\phi}$ 
is plotted as a function of
the magnetization $\hat{\phi}$. In contrast to the parameters
used in the text here all curves are normalized by the conditions
$\partial \hat{U}/\partial \hat{\phi}(1)=0$ and 
$\partial^2 \hat{U}/\partial (\hat{\phi})^2(1)=1$. The curves 
$0-3$ are from the $\epsilon$-expansion in the parametric 
representation of the equation of state 
\cite{ZJ}, in orders
$\epsilon^0,\ldots,\epsilon^3$, respectively. The curve
$W$ represents our result and the curve $V$ denotes the
lattice result of ref.\ 
\cite{Tsy2} 
where this plot is taken from.}     
\label{figtsy}
\end{figure} 
figure 
\ref{figtsy} is taken from ref.\ \cite{Tsy2} which presents results
from a lattice Monte Carlo investigation
of the equation of state for $N=1$ (curve V), along with
our results (curve W) and those from the $\epsilon$--expansion
to order $\epsilon^0,\ldots ,\epsilon^3$ (curves $0-3$). 
The plot shows very good agreement of the lattice study 
with our results and the fit of eq. (\ref{nineb}) in the phase with
spontaneous symmetry breaking, whereas it deviates substantially 
from the results of the $\epsilon$-expansion. We also have computed
the equation of state for $N=4$. We present
these results in section \ref{CriticalBehavior} where we 
obtain the universal critical
equation of state within a study of an effective
quark meson model at non--zero temperature. 
Comparison of our $N=4$ results with a recent lattice
Monte-Carlo simulation \cite{Tou} shows agreement for the scaling 
function within a few percent in the symmetric and
the spontaneously broken phase (cf.\ section \ref{CriticalBehavior}).

\subsection{Conclusions\label{oncon}}

In summary, our numerical solution of eq.\ (\ref{four}) gives a very
detailed picture of the critical equation of state
of the three dimensional $O(N)$ model. The numerical 
uncertainties are estimated by 
comparison of results obtained through two independent integration
algorithms \cite{ABBTW95,BW97-1} (cf.\ section \ref{solv}).
They are small, typically less than $0.3 \%$ for critical exponents
and  $(1$ -- $3) \%$ for amplitudes. The scaling relations between the
critical exponents are fulfilled within a deviation of $2 \times 10^{-4}$.
The dominant quantitative error stems from the 
truncation of the exact flow equation and is related to the size of 
the anomalous dimension $\eta \simeq 4 \%$ (cf.\ also the discussion
in section \ref{outlook}). 

We have obtained a precise description of the universal equation of 
state in the vicinity of the critical
temperature of a second order phase transition. 
We will see in the next section that universal properties
can also be observed for weak first order
phase transitions. In particular, we will compute
the universal equation of state for weak
first order phase transitions in scalar matrix
models in section \ref{sce}.

%% file: first.tex
\sect{Equation of state and first order phase transitions \label{secfirst}}

\subsection{Introduction \label{intro}}

Matrix models are extensively discussed in statistical
physics. Beyond the $O(N)$ symmetric
Heisenberg models (``vector models''), which we have discussed
in the previous section, they correspond
to the simplest scalar field theories. 
There is a wide set of different applications as 
the metal insulator transition \cite{Weg}
or liquid crystals \cite{deG} or strings and
random surfaces \cite{FGZ}. 
The universal behavior of these models in the vicinity
of a second order or weak first order phase transition
is determined by the symmetries and the field content
of the corresponding field theories. We will consider
here \cite{BW97-1} models with $U(N) \times U(N)$ symmetry with
a scalar field in the $(\bar{N},N)$ representation,
described by an arbitrary complex $N \times N$ matrix
$\vp$. We do not impose nonlinear constraints for
$\vp$ a priori  but rather use a ``classical''
potential.  This enforces nonlinear constraints
in certain limiting cases. Among those, our model
describes a nonlinear matrix model for unitary matrices
or one for singular $2 \times 2$ matrices. The universal
critical behavior does not depend on the details of the 
classical potential and there is no difference
between the linear and nonlinear models in the 
vicinity of  the limiting cases. We concentrate here
on three dimensions, relevant for statistical
physics and critical phenomena in high temperature field 
theory.

The cases $N=2$, $3$ have a relation to high
temperature strong interaction physics. At vanishing
temperature the four dimensional models can be used
for a description of the pseudoscalar and scalar mesons
for $N$ quark flavors (cf.\ section \ref{2qcd}). 
For $N=3$ the antihermitean
part of $\vp$ describes here the pions, kaons, $\eta$
and $\eta^{\pri}$ whereas the hermitean part accounts for the 
nonet of scalar $0^{++}$ mesons.\footnote{See ref.\ \cite{JuWet96}
for a recent phenomenological analysis.} For nonzero
temperature $T$ the effects of fluctuations with momenta
$p^2 \, \ltap \, (\pi T)^2$ are described by the 
corresponding three dimensional models. These models
account for the long distance physics and obtain after
integrating out the short distance fluctuations by
virtue of dimensional reduction which will be treated
in detail in section \ref{FiniteTemperatureFormalism}.
In particular, the three dimensional models embody
the essential dynamics in the immediate vicinity
of a second order or weak first order chiral phase transition
\cite{PW84-1,Wil92,RaWi93-1,Raj95-1}. 
The four dimensional models at nonvanishing
temperature have also been used for investigations of
the temperature dependence of meson masses \cite{mes2,mes}.
The simple model investigated in this section is not yet 
realistic -- it neglects the effect of the axial anomaly
which reduces the chiral flavor symmetry to 
$SU(N) \times SU(N)$. For simplicity we will concentrate here on
$N=2$, but our methods can be generalized to $N=3$
and the inclusion of the axial anomaly.

The case $N=2$ also has a relation to the 
electroweak phase transition in models with two
Higgs doublets. Our model corresponds here
to the critical behavior in a special class
of left-right symmetric theories in the limit where 
the gauge couplings are neglected. Even though
vanishing gauge couplings are not a good 
approximation for typical realistic models
one would like to understand this limiting
case reliably.

For the present matrix models
one wants to know if the phase transition becomes second
order in certain regions of parameter space. In the context of 
flow equations this is equivalent 
to the question if the system of running couplings admits
a fixed point which is infrared stable (except one
relevant direction corresponding to $T-T_{c}$). 
We will employ here
the average action method introduced in section \ref{fieldtheory}.
We find that the phase transition for the investigated
matrix models with $N=2$ and symmetry breaking
pattern $U(2) \times U(2) \to U(2)$ is always 
(fluctuation induced) first order,
except for a boundary case with enhanced 
$O(8)$ symmetry. For a large part
of parameter space the transition is weak and one finds 
large renormalized dimensionless couplings near the critical
temperature. If the running of the couplings towards approximate
fixed points (there are no exact fixed points) is 
sufficiently fast the large distance physics looses 
memory of the details of the short distance or classical
action. In this case the physics near the phase transition
is described by an universal equation of state.

Besides the possible practical applications it is
a theoretical challenge to find the universal equation of 
state for weak first order phase transitions. 
Whereas for second order transitions the universal equation 
of state can be expressed as a function of only one scaling 
variable (cf.\ section \ref{uceos}), 
the additional mass scale at a first order 
transition induces a second dimensionless ratio. 
The equation of state or, equivalently,
the effective potential or free energy therefore
depends on two different scaling variables in the
universal region. We have succeeded to describe this situation
for the present matrix models and give
the universal equation of state in section \ref{sce}.

An important aspect in the study of first order phase transitions
is the possible creation of an out of equilibrium situation,
as has been discussed in section
\ref{mainin}. An old and challenging problem
in this context concerns the question of the validity 
of Langer's theory \cite{Langer} of spontaneous bubble nucleation.
In Langer's approach the decay of metastable
states is described in terms of a steady state solution of the 
associated Fokker--Planck equation. The nucleation rate is
obtained from the integrated probability current flowing
across the saddle point (``critical bubble'') of a suitably
defined coarse grained free energy. The treatment requires
first a meaningful definition of a coarse grained free energy
with a coarse graining scale $k$. Second, the 
validity of a saddle point approximation for the 
treatment of fluctuations around the critical bubble
has to be ensured.
Here only fluctuations with momenta smaller than $k$
must be included. We will see that the two issues
are closely related. 
The validity of the saddle point 
approximation typically requires small dimensionless
couplings. On the other hand we observe
for large couplings that the form of the coarse grained
effective potential $U_k$ depends strongly on $k$
even for scales where the location of the minima
of $U_k$
is essentially independent of $k$. This means that
the lowest order in the saddle point approximation
(classical contribution) depends strongly on the details of
the coarse graining procedure. Since the final results
as nucleation rates etc.\ must be independent of the
coarse graining prescription this is only compatible
with a large contribution from the higher orders of the
saddle point expansion. Section \ref{coarse} deals with
this issue in a quantitative way.

The text is organized as follows. In section \ref{model}
we define the $U(2) \times U(2)$ symmetric matrix model
and we establish the connection to a matrix model
for unitary matrices and to one for
singular complex $2 \times 2$ matrices. There we also give
an interpretation of the model as the coupled system
of two $SU(2)$-doublets for the weak interaction Higgs sector.
Section \ref{ps} is devoted to an overview over the phase 
structure and the coarse grained effective potential $U_k$
for the three dimensional theory. The evolution equation
for $U_k$ and its scaling form is computed in section 
\ref{scale}. A method for its numerical solution is discussed 
in section \ref{solv}
which also contains the flow equation for the wave 
function renormalization constant or the anomalous dimension
$\eta$. A detailed account on the renormalization group flow
is presented in section \ref{rg}. We compute the 
universal form of the equation of state 
for weak first order phase transitions 
in section \ref{sce} and we extract critical exponents
and the corresponding index relations.
The dependence of the coarse grained effective potential
on the coarse graining scale is studied in detail
in section \ref{coarse}. The quantitative analysis
is applied to Langer's theory of bubble formation.
Section \ref{con} contains the conclusions.

\subsection{Scalar matrix model
\label{model}}

We consider a $U(2) \times U(2)$ symmetric effective 
action for a scalar field
$\vp$ which transforms in the $(2,2)$ representation
with respect to the subgroup $SU(2) \times SU(2)$. Here $\vp$
is represented by a complex $2 \times 2$ matrix
and the transformations are
\bea
 \vp &\ra& U^{ }\vp V^\dagger\,\, , \nnn
 \vp^\dagger &\ra& V^{ }\vp^\dagger U^\dagger
 \label{Transformations}
\eea
where $U$ and $V$ are unitary $2 \times 2$ matrices 
corresponding to the two distinct $U(2)$ factors.

We classify the invariants for the construction of the effective 
average action by the number of derivatives. The lowest order
in a systematic derivative expansion \cite{TW94-1,Mor}
of $\Gm_k$ is given by
\be
 \Gm_k = \ds{\int d^d x\left\{U_k(\vp,\vp^\dagger )+
 Z_k \prl_\mu \vp^*_{ab} \prl^\mu \vp^{ab}
  \right\}}\qquad  (a,b=1,2).
 \label{Ansatz}
\ee
The term with no derivatives defines the scalar potential $U_k$
which is an arbitrary function of traces of powers of 
$\vp^\dagger \vp$. The most general $U(2) \times U(2)$ symmetric
scalar potential can be expressed as a function of only two 
independent invariants,
\bea
 \rho &=& \ds{\tr\left(\vp^\dagger \vp \right) }\non\\
 \tau &=& \ds{2 \tr\left(\vp^\dagger \vp - \frac{1}{2} \rho \right)^2}
 = \ds{2 \tr\left(\vp^\dagger \vp \right)^2 - \rho^2 }.
 \label{Invaria}
\eea
Here we have used for later convenience the traceless matrix 
$\vp^\dagger \vp - \frac{1}{2} \rho$ to construct the second
invariant.
Higher invariants, $\tr\left(\vp^\dagger \vp 
- \frac{1}{2} \rho \right)^n$ for
$n > 2$, can be expressed as functions of $\rho$ and $\tau$ 
\cite{Ju95-7}.

For the derivative part we consider a standard kinetic term with a 
scale dependent wave function renormalization constant $Z_k$. 
The first correction to the kinetic term would include field 
dependent wave function
renormalizations $Z_k(\rho,\tau)$ plus functions not
specified in eq.\ (\ref{Ansatz}) which account for a
different index structure of invariants with two
derivatives. These wave function renormalizations may be
defined at zero momentum. The next level involves 
invariants with four derivatives and so on.
We define $Z_k$ 
at the minimum $\rho_0$, $\tau_0$ of $U_k$ and
at vanishing momenta $q^2$,
\be
Z_k=Z_k(\rho=\rho_0,\tau=\tau_0;q^2=0). \label{zet}
\ee
The factor $Z_k$ appearing
in the definition of the infrared cutoff $R_k$ in eq.\ (\ref{Rk(q)})
is identified with (\ref{zet}).
The $k$-dependence of this function is given by the anomalous
dimension
\be
\eta(k)=-\frac{\mbox{d}}{\mbox{d}t}\mbox{ln} Z_k .
\label{Eta} 
\ee 

If the ansatz (\ref{Ansatz}) is inserted into the flow equation
for the effective average action (\ref{ERGE}) one obtains flow equations  
for the effective average potential $U_k(\rho,\tau)$ and for the
wave function renormalization constant $Z_k$ (or equivalently the 
anomalous dimension $\eta$).
This is done in sections
\ref{scale} and \ref{solv}. These flow equations 
have to be integrated starting from some short distance scale 
$\La$ and one has to specify $U_{\La}$ and $Z_{\La}$
as initial conditions. At the scale $\La$, where 
$\Gm_{\La}$ can be taken as the classical 
or short distance action, no integration 
of fluctuations 
has been performed.
The short distance potential is taken to be a quartic potential
which is parametrized by two quartic couplings $\bar{\la}_{1\La}$,
$\bar{\la}_{2\La}$ and a mass term. We start
in the spontaneously broken regime where the minimum
of the potential occurs at a nonvanishing field value 
and there is a negative mass term at
the origin of the potential $(\bar{\mu}_{\La}^2 > 0)$, 
\be
U_{\La}(\rho,\tau)=-\bar{\mu}_{\La}^2 \rho + \hal \bar{\la}_{1\La}
\rho^2 +\frac{1}{4} \bar{\la}_{2\La} \tau \quad
\label{uinitial}
\ee 
and $Z_{\La}=1$. 
The potential is bounded from
below provided 
$\bar{\la}_{1\La} > 0$ and 
$\bar{\la}_{2\La} > - 2 \bar{\la}_{1\La}$.
For $\bar{\la}_{2\La} > 0$ one observes 
the potential minimum for the configuration
$\vp_{ab}=\vp \delta_{ab}$ corresponding
to the spontaneous symmetry breaking 
down to the diagonal $U(2)$ subgroup of $U(2) \times U(2)$.
For negative $\bar{\la}_{2\La}$ the potential is minimized
by the configuration $\vp_{ab}=\vp \delta_{a1} \delta_{ab}$
which corresponds to the symmetry breaking pattern 
$U(2) \times U(2) \lra U(1) \times U(1) \times U(1)$. 
In the special case $\bar{\la}_{2\La}=0$ the theory
exhibits an enhanced $O(8)$ symmetry. This constitutes
the boundary between two phases with different symmetry
breaking patterns.

The limits of infinite couplings correspond to nonlinear
constraints in the matrix model. For 
$\bar{\la}_{1\La} \to \infty$ with fixed ratio 
$\bar{\mu}_{\La}^2/\bar{\la}_{1\La}$ one finds the constraint
$\tr(\vp^{\dagger}\vp)=\bar{\mu}_{\La}^2/\bar{\la}_{1\La}$.
By a convenient choice of $Z_{\La}$ (rescaling of $\vp$)
this can be brought to the form $\tr(\vp^{\dagger}\vp)=2$.
On the other hand, the limit $\bar{\la}_{2\La} \to +\infty$
enforces the constraint 
$\vp^{\dagger}\vp=\hal \tr(\vp^{\dagger}\vp)$.
Combining the limits $\bar{\la}_{1\La} \to \infty$,
$\bar{\la}_{2\La} \to \infty$ the constraint reads
$\vp^{\dagger}\vp=1$ and we deal with a matrix model for 
unitary matrices. (These considerations generalize to arbitrary
$N$.) Another interesting limit obtains for 
$\bar{\la}_{1\La}=-\hal \bar{\la}_{2\La} + \Dt_{\la}$,
$\Dt_{\la} > 0$ if $\bar{\la}_{2\La} \to - \infty$. In this 
case the nonlinear constraint reads
$(\tr\vp^{\dagger}\vp)^2=\tr(\vp^{\dagger}\vp)^2$ which
implies for $N=2$ that $\det \vp =0$. This is a matrix model
for singular complex $2 \times 2$ matrices.

One can also interpret our model as the coupled 
system of two
$SU(2)$-doublets for the weak interaction Higgs
sector. This is simply done by decomposing the 
matrix $\vp_{ab}$ into two two-component complex 
fundamental representations of one of the $SU(2)$
subgroups, $\vp_{ab} \to \vp_{1b},\vp_{2b}$. The
present model corresponds to a particular
left-right symmetric model with interactions specified
by
\bea
\rho&=&\vp_1^{\dagger}\vp_1+\vp_2^{\dagger}\vp_2\\
\tau&=&\left(\vp_1^{\dagger}\vp_1-\vp_2^{\dagger}\vp_2
\right)^2+4\left(\vp_1^{\dagger}\vp_2\right)
\left(\vp_2^{\dagger}\vp_1\right) \label{th}\, .
\eea
We observe that for a typical weak interaction symmetry
breaking pattern the expectation values of $\vp_1$ and
$\vp_2$ should be aligned in the same direction or one
of them should vanish. In the present model this 
corresponds to the choice $\bar{\la}_{2\La} < 0$.
The phase structure of a related model without the term
$\sim (\vp_1^{\dagger}\vp_2)(\vp_2^{\dagger}\vp_1)$
has been investigated previously \cite{TwoHig}
and shows second or first order 
transitions\footnote{First order phase transitions and 
coarse graining have also been discussed in a 
multi-scalar model with $Z_2$ symmetry \cite{AlMr}.
}. Combining
these results with the outcome of this work leads 
already to a detailed qualitative overview over the phase 
pattern in a more general setting with three independent
couplings for the quartic invariants 
$(\vp_1^{\dagger}\vp_1+\vp_2^{\dagger}\vp_2)^2$, 
$(\vp_1^{\dagger}\vp_1-\vp_2^{\dagger}\vp_2)^2$
and $(\vp_1^{\dagger}\vp_2)(\vp_2^{\dagger}\vp_1)$. 
We also note that the special case 
$\bar{\la}_{2\La}=2\bar{\la}_{1\La}$ corresponds to two
Heisenberg models interacting only by a term sensitive
to the alignment between $\vp_1$ and $\vp_2$, i.e.\
a quartic interaction of the form 
$(\vp_1^{\dagger}\vp_1)^2+(\vp_2^{\dagger}\vp_2)^2
+2(\vp_1^{\dagger}\vp_2)(\vp_2^{\dagger}\vp_1)$.

The model is now completely specified and it remains to extract
the flow equations for $U_k$ and $Z_k$. Before this is carried
out in sections \ref{scale} and \ref{solv}
we present an overview over the phase structure in the next
section. These results are obtained from a numerical solution
of the evolution equations.

\subsection{Phase structure \label{ps}}

In this section we consider the
$U(2) \times U(2)$ symmetric model in three space dimensions.
The aim is to give an overview of our results concerning
the phase structure and
the effective average potential.  
We concentrate here on the spontaneous 
symmetry breaking with a residual $U(2)$ symmetry
group. 
This symmetry breaking can be observed for a
configuration which is proportional to the 
identity and with (\ref{Invaria}) one finds $\tau = 0$. 
In this case we shall use
an expansion of $U_k(\rho,\tau)$ around $\tau=0$ keeping only the
linear term in $\tau$. This amounts to 
assuming
\be
\ds{\frac{\prl^n U_k}{\prl \tau^n}}
(\rho,\tau=0)=0 \quad \mbox{for}
\quad n \ge 2.
\label{Truncationdim}
\ee
We will motivate this truncation in section \ref{solv} where
we present a more detailed analysis. 
We make no expansion of $U_k(\rho,\tau)$ in terms of $\rho$.
This allows the description
of a first order phase transition where a second local minimum
of $U_k(\rho) \equiv U_k(\rho,\tau=0)$ appears. The 
$\rho$-dependence also gives information about the
equation of state of the system. 
 
For the considered symmetry breaking pattern 
the short distance potential $U_{\La}$
given in eq.\ (\ref{uinitial}) is parametrized by positive 
quartic couplings, 
\be
\bar{\la}_{1\La},\bar{\la}_{2\La}>0
\ee 
and the location of its minimum is given by 
\be
\rho_{0\La}=\bar{\mu}_{\La}^2/\bar{\la}_{1\La}. 
\ee
To study the phase structure of the model we integrate the
flow equation for the effective average potential $U_k$
(cf.\ sect.\ \ref{scale}, \ref{solv})
for a variety of initial 
conditions $\rho_{0\La},\bar{\la}_{1\La}$ and
$\bar{\la}_{2\La}$. In particular, for general 
$\bar{\la}_{1\La},\bar{\la}_{2\La}>0$ we are able to find a critical
value $\rho_{0\La}=\rho_{0 c}$ for which the system exhibits
a first order phase transition. In this case the evolution of
$U_k$ leads at some scale $k_2 < \La$ to the appearance 
of a second local
minimum at the origin of the effective average potential and both
minima become degenerate in the limit $k \to 0$. If $\rho_0(k)>0$
denotes the $k$-dependent outer minimum of the potential 
($U_k^{\pri}(\rho_0)=0$, where the prime on $U_k$ denotes the
derivative with respect to $\rho$ at fixed $k$) at
a first order phase transition one has
\be
\lim\limits_{k \to 0}( U_k(0)-U_k(\rho_0))=0.
\label{critcon}
\ee
A measure of the distance from the phase transition
is the difference $\dt\kp_{\La}=(\rho_{0\La}-\rho_{0 c})/\La$.
If $\bar{\mu}_{\La}^2$ and therefore
$\rho_{0\La}$ is interpreted as a function of 
temperature, the deviation $\dt\kp_{\La}$ is 
proportional to the deviation from the critical 
temperature $T_{c}$, i.e.\ $\dt\kp_{\La}=A(T)
(T_{c}-T)$ with $A(T_{c}) > 0$. 

\begin{figure}[h]
\unitlength1.0cm
\begin{center}
\begin{picture}(13.,9.)
\put(-0.6,4.4){$\ds{\frac{U_{k_f}}{\vp_{0R}^6}}$}
\put(6.4,-0.5){$\ds{\frac{\vp_R}{\vp_{0R}}}$}
\put(-0.5,0.){
\epsfysize=13.cm
\epsfxsize=9.cm
\rotate[r]{\epsffile{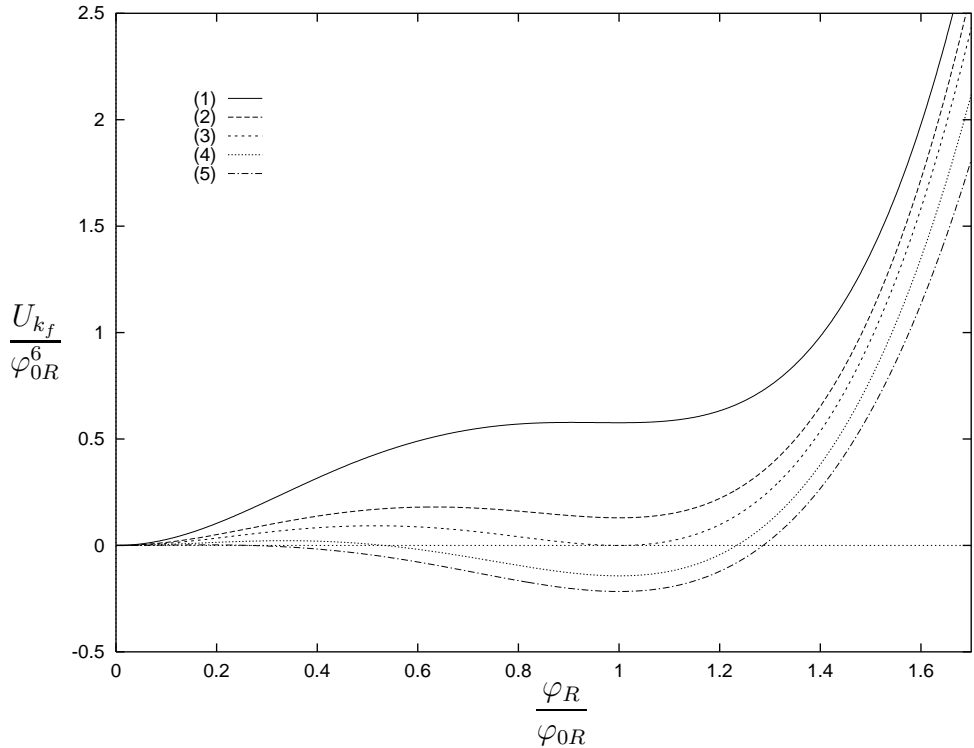
}}
}
\end{picture}
\end{center}
\vspace{0.3cm}
\caption{\footnotesize The effective average potential
$U_{k=k_f}$ as a function of the renormalized field 
$\vp_R$. The potential
is shown for various values of $\dt \kp_{\La} \sim T_c-T$.
The parameters for the short distance potential $U_{\La}$ are
(1) $\dt \kp_{\La}=-0.03$, (2) $\dt \kp_{\La}=-0.015$,
(3) $\dt \kp_{\La}=0$, (4) $\dt \kp_{\La}=0.04$,
(5) $\dt \kp_{\La}=0.1$ and
$\la_{1\La}=0.1$, $\la_{2\La}=2$.
\label{tempot}
}
\end{figure}

\begin{figure}[h]
\unitlength1.0cm
\begin{center}
\begin{picture}(13.,9.)
\put(4.5,3.5){\footnotesize $\ds{\frac{m_R}{\La}}$}
\put(4.5,7.){\footnotesize $\ds{\frac{\rho_{0R}}{\La}}$}
\put(6.4,-0.5){$-\dt\kp_{\La}$}
\put(-0.5,0.){
\epsfysize=13.cm
\epsfxsize=9.cm
\rotate[r]{\epsffile{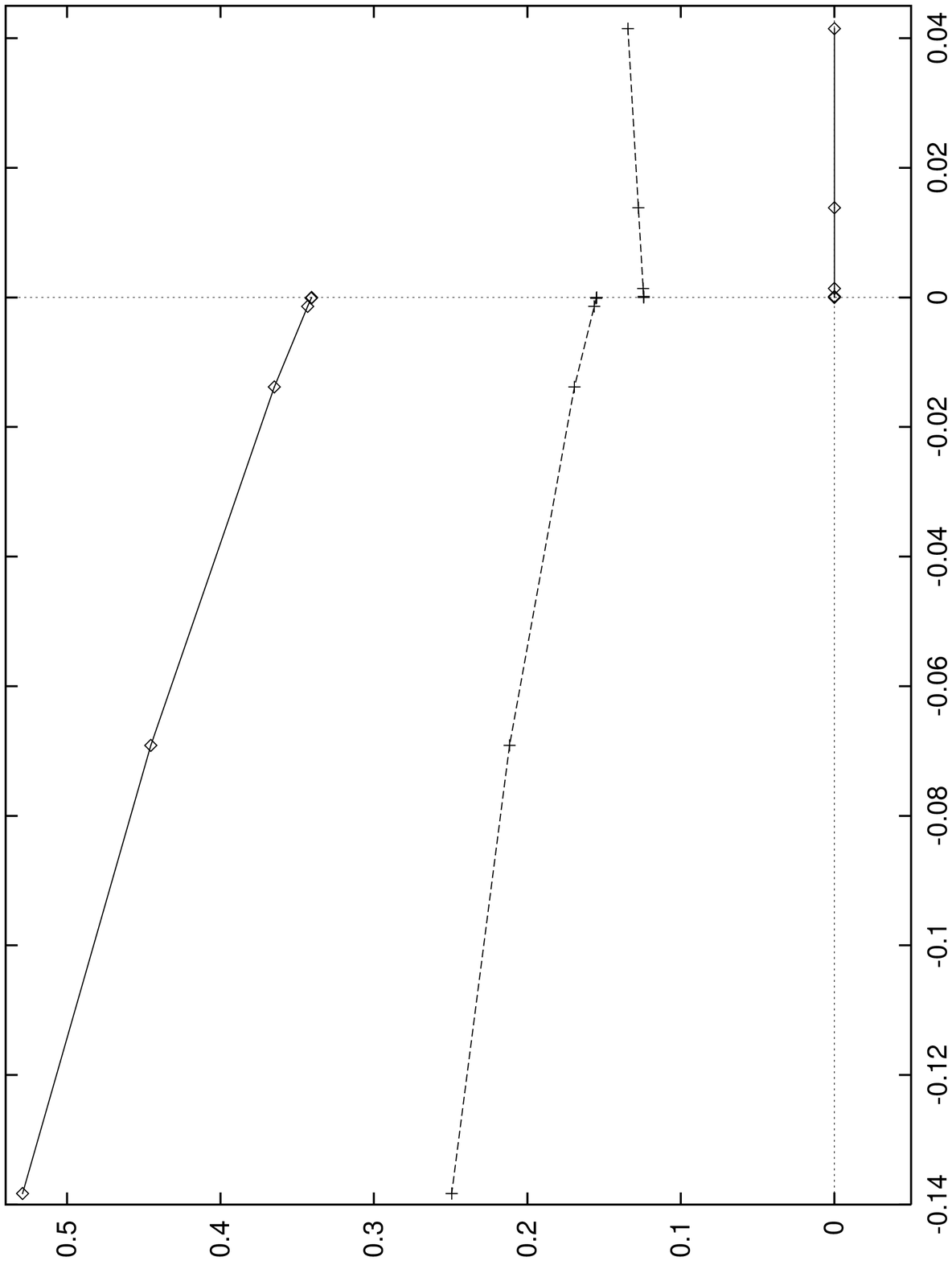
}}
}
\end{picture}
\end{center}
\caption[]{\label{temp} 
\footnotesize The
minimum $\rho_{0R}$ 
and the mass term 
$m_R$ in units of the momentum scale $\La$
as a function of $-\dt\kp_{\La}$ or temperature
($\la_{1\La}=0.1$, $\la_{2\La}=2$, $k=k_f$). For 
$\dt\kp_{\La}=0$ one observes the jump in the
renormalized order parameter $\Dt \rho_{0R}$
and mass $\Dt m_R$. 
}
\end{figure}

We consider in the
following the effective average potential $U_k$
for a nonzero scale $k$. This allows
to observe the nonconvex part of the potential (cf.\
sect.\ \ref{coarse}). 
As an example we show in 
fig.\ \ref{tempot} the effective average potential
$U_{k=k_f}$ for $\la_{1\La}= \bar{\la}_{1\La}/\La=0.1$
and $\la_{2\La}= \bar{\la}_{2\La}/\La=2$
as a function of the renormalized field 
$\vp_R=(\rho_R/2)^{1/2}$ with $\rho_{R}=Z_{k=k_f} \rho$. 
The scale $k_f$ is some 
characteristic scale 
below which the location of the minimum $\rho_0(k)$
becomes essentially independent of $k$.  
Its precise definition is given below. We have normalized 
$U_{k_f}$ and $\vp_R$
to powers of the renormalized minimum 
$\vp_{0R}(k_f)= (\rho_{0R}(k_f)/2)^{1/2}$ with
$\rho_{0R}(k_f)=Z_{k_f}\rho_0(k_f)$.
The potential
is shown for various values of deviations from the critical
temperature or $\dt\kp_{\La}$. 
For the given examples $\dt \kp_{\La}=-0.03$, $-0.015$
the minimum at the origin becomes
the absolute minimum and the system is in the
symmetric (disordered) phase. Here $\vp_{0R}$
denotes the minimum in the metastable ordered phase.
In contrast, for $\dt \kp_{\La}=0.04$, $0.1$ the 
absolute minimum is located at $\vp_R/\vp_{0R}=1$ which
characterizes the spontaneously broken phase.
For large enough $\dt \kp_{\La}$ the local minimum at the 
origin vanishes.
For $\dt\kp_{\La}=0$ the two distinct minima are degenerate in 
height\footnote{
We note that the critical 
temperature is determined by condition (\ref{critcon})
in the limit $k \to 0$. Nevertheless for the employed
nonvanishing scale $k=k_f$ the minima of $U_k$
become almost degenerate at the critical temperature.
}.
As a consequence the order parameter makes a discontinuous jump
at the phase transition which characterizes the transition to be 
first order.
It is instructive to consider some characteristic values
of the effective average potential.
In fig.\ \ref{temp} we consider for
$\la_{1\La}=0.1,\la_{2\La}=2$ the value of the renormalized
minimum $\rho_{0R}(k_f)$ and the radial mass term 
as a function of $-\dt\kp_{\La}$ or temperature.
In the spontaneously
broken phase the renormalized radial mass squared is given by
(cf.\ section \ref{scale})
\be
m^2_R(k_f)=2 Z_{k_f}^{-1} \rho_0 U^{\pri\pri}_{k_f}(\rho_0),
\label{radialmass}
\ee 
in the symmetric phase the renormalized mass term reads
\be
m^2_{0R}(k_f)=Z_{k_f}^{-1} U^{\pri}_{k_f}(0).
\label{originmass}
\ee
At the critical temperature ($\dt\kp_{\La}=0$) one observes
the discontinuity $\Dt \rho_{0R}=\rho_{0R}(k_f)$ and the jump
in the mass term $\Dt m_R=m_R(k_f)-m_{0R}(k_f)
=m_R^c-m_{0R}^c$. (Here the index ``c'' denotes
$\dt\kp_{\La}=0$).
The ratio $\Dt \rho_{0R}/\La$ is a rough measure for 
the ``strength'' of the first order transition.
For $\Dt \rho_{0R}/\La \ll 1$ the phase transition is
weak in the sense that typical masses are small compared to
$\La$. In consequence, the long-wavelength fluctuations
play a dominant role and the system exhibits universal 
behavior, i.e.\ it becomes largely independent of the
details at the short distance scale $\La^{-1}$.
We will discuss the universal behavior in more 
detail below.     

\begin{figure}[h]
\unitlength1.0cm
\begin{center}
\begin{picture}(13.,9.)
\put(-0.9,4.6){$\ds{\ln\left(\la_{2\La}\right)}$}
\put(6.4,-0.5){$\ds{\ln\left(\la_{1\La}\right)}$}
\put(9.5,6.5){\footnotesize $(1)$}
\put(9.5,5.4){\footnotesize $(2)$}
\put(9.5,3.){\footnotesize $(3)$}
\put(9.5,1.5){\footnotesize $(4)$}
\put(-0.5,0.){
\epsfysize=13.cm
\epsfxsize=9.cm
\rotate[r]{\epsffile{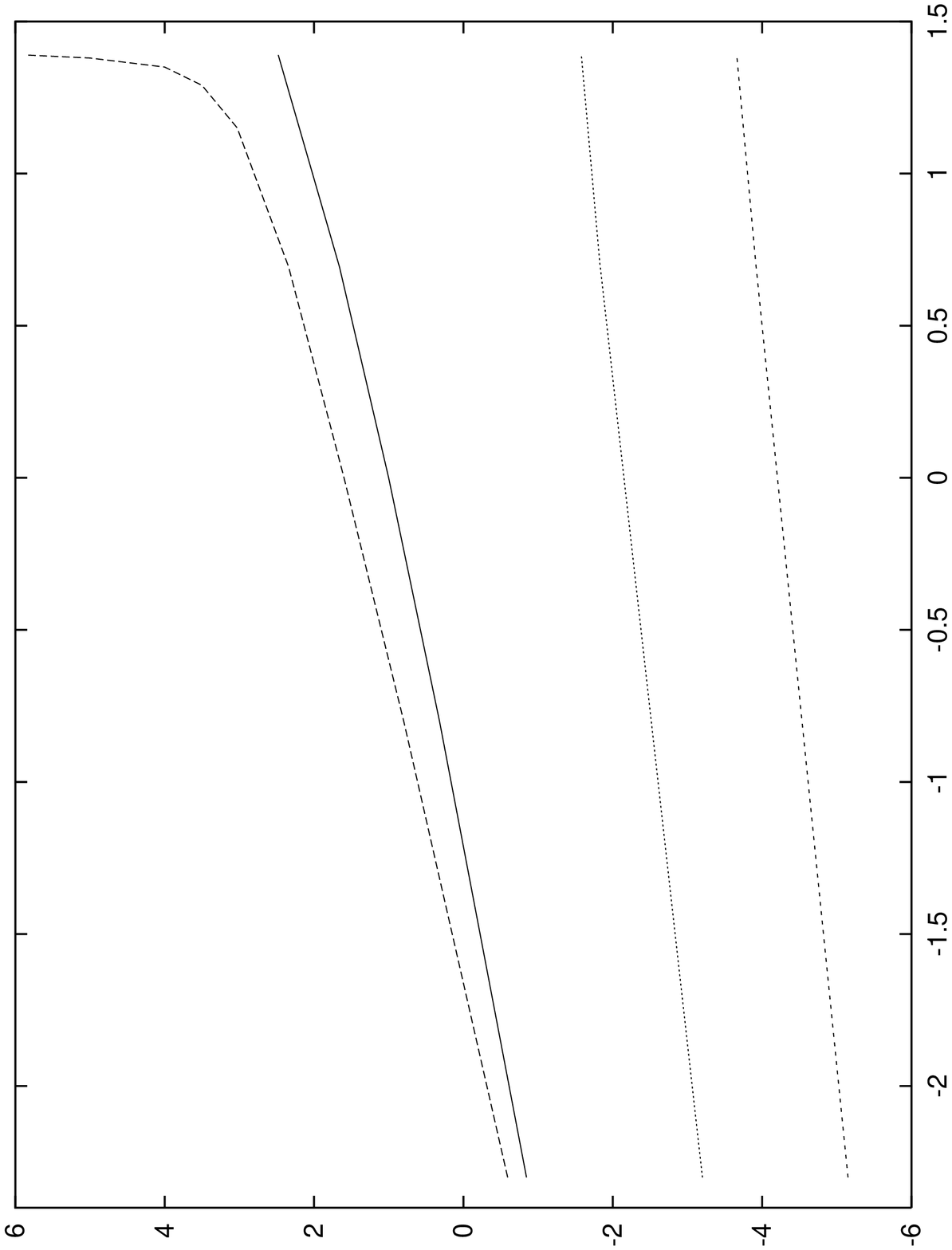
}}
}
\end{picture}
\end{center}
\caption[]{\label{la1la2} \footnotesize
Lines of constant jump of the 
renormalized order parameter
$\Dt \rho_{0R}/\La$ at the phase transition in the
$\ln(\la_{1\La}) , \ln(\la_{2\La})$ plane. The curves
correspond to (1) $\ln(\Dt \rho_{0R}/\La)=-4.0$,
(2) $\ln(\Dt \rho_{0R}/\La)=-4.4$,
(3) $\ln(\Dt \rho_{0R}/\La)=-10.2$,
(4) $\ln(\Dt \rho_{0R}/\La)=-14.3$.    
}
\end{figure}

In order to characterize the strength of 
the phase transition for 
arbitrary positive values of $\la_{1\La}$ and $\la_{2\La}$
we consider lines of constant $\Dt \rho_{0R}/\La$ in the
$\la_{1\La} , \la_{2\La}$ plane. In fig.\ \ref{la1la2} this is
done for the logarithms of these quantities. 
For fixed
$\la_{2\La}$ one observes that the discontinuity
at the phase transition weakens with increased $\la_{1\La}$.
On the other hand for given $\la_{1\La}$ one finds a larger jump in the
order parameter for increased $\la_{2\La}$. This is true up to a 
saturation point where $\Dt \rho_{0R}/\La$ becomes independent
of $\la_{2\La}$. In the plot this can be observed from the
vertical part of the line of constant ln$(\Dt \rho_{0R}/\La)$.
This phenomenon occurs for arbitrary nonvanishing   
$\Dt \rho_{0R}/\La$ in the strong $\la_{2\La}$ coupling limit
and is discussed in section \ref{rg}. 

\begin{figure}[h]
\unitlength1.0cm
\begin{center}
\begin{picture}(13.,9.)
\put(-1.2,4.6){$\ds{\ln\left(\frac{\Dt \rho_{0R}}{\La}\right)}$}
\put(6.2,-0.5){$\ds{\ln\left(\la_{2\La}\right)}$}
\put(8.9,8.5){\footnotesize $(1)$}
\put(10.5,7.7){\footnotesize $(2)$}
\put(10.5,6.7){\footnotesize $(3)$}
\put(-0.2,0.){
\epsfysize=13.cm
\epsfxsize=9.cm
\rotate[r]{\epsffile{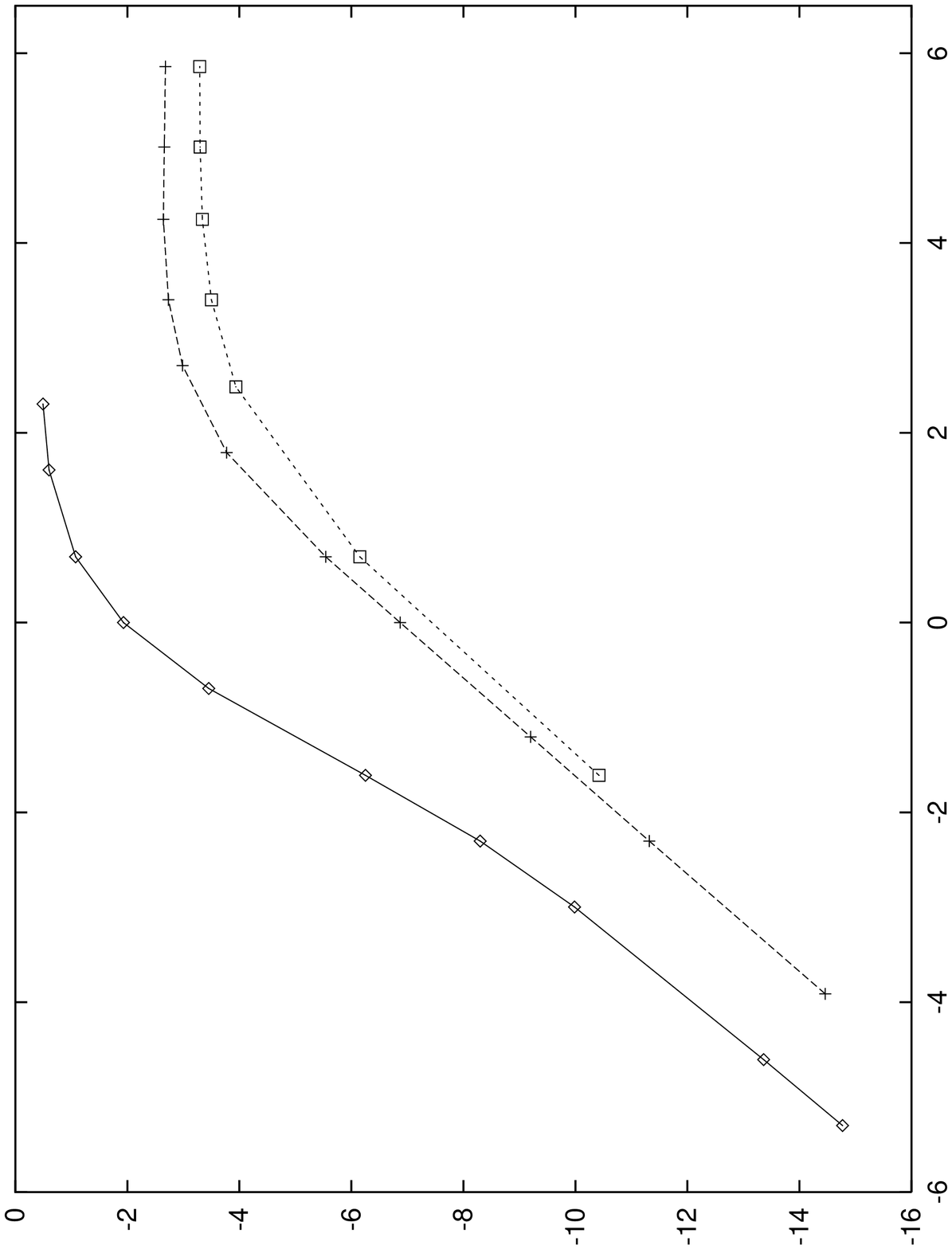
}}
}
\end{picture}
\end{center}
\caption[]{\footnotesize The logarithm of the discontinuity 
of the renormalized order parameter
$\Dt \rho_{0 R}/\La$ 
as a function of $\ln(\la_{2\La})$.
Data points
for fixed (1) $\la_{1\La}=0.1$,
(2) $\la_{1\La}=2$, (3) $\la_{1\La}=4$ are
connected by straight lines. 
\label{phase}
}
\end{figure}

In the following we give a detailed quantitative
description of the first order phase
transitions and a separation in weak and strong transitions.
We consider some characteristic quantities for the effective average 
potential in dependence on the short distance parameters 
$\la_{1\La}$ and
$\la_{2\La}$ for $\dt \kp_{\La}=0$.  
We consider the discontinuity
in the renormalized order parameter $\Dt \rho_{0R}$ 
and the inverse correlation
lengths (mass terms) 
$m_R^c$ and $m_{0R}^c$ in the ordered and the 
disordered phase respectively. 
Fig.\ \ref{phase}
shows the logarithm of  
$\Dt \rho_{0 R}$ in units of 
$\La$ as a function of the logarithm of the initial 
coupling $\la_{2\La}$. 
We have connected the calculated values
obtained for various fixed $\la_{1\La}=0.1$, 
$2$ and $\la_{1\La}=4$
by straight lines. The values are listed in table \ref{table1}. 
\begin{table} [h]
\renewcommand{\arraystretch}{1.5}
\hspace*{\fill}
\begin{tabular}{|c|c|c|c|c||c|c|c|c|c|}     \hline

$\la_{1 \La}$
&$\la_{2 \La}$
&$\ds{\frac{\Dt \rho_{0 R}}{\La}}$
&$\ds{\frac{m_R^c}{\Dt \rho_{0 R}}}$
&$\ds{\frac{m_{0R}^c}{\Dt \rho_{0 R}}}$
&$\la_{1 \La}$
&$\la_{2 \La}$
&$\ds{\frac{\Dt \rho_{0 R}}{\La}}$
&$\ds{\frac{m_R^c}{\Dt \rho_{0 R}}}$
&$\ds{\frac{m_{0R}^c}{\Dt \rho_{0 R}}}$
\\ \hline \hline
$0.1$
&0.005
&$0.386 \times 10^{-6}$
&$1.69$
&$1.26$
&$2$
&2
&$ 0.392\times 10^{-2}$
&1.66
&1.24
\\ \hline
$0.1$
&0.01
&$0.158 \times 10^{-5}$
& $1.68$
&1.26
&2
&6
&$0.230\times 10^{-1}$
&1.68
&1.25 
\\ \hline
$0.1$
&0.05
&$0.461 \times 10^{-4}$
& $1.66$
&1.23
&2
&15 
&$0.505 \times 10^{-1}$
&1.80
&1.35 
\\ \hline
$0.1$
&0.1
&$0.249 \times 10^{-3}$
& $1.58$
&1.17
&2
&30
&$0.649 \times 10^{-1}$
&1.91
&1.45
\\ \hline
$0.1$
&0.2
&$0.193 \times 10^{-2}$
& $1.34$
&0.992
&2
&70
&$0.712 \times 10^{-1}$
&2.01
&1.60
\\ \hline
$0.1$
&0.5
&$0.316 \times 10^{-1}$
& $0.772$
&0.571
&2
&150
&$0.699 \times 10^{-1}$
&2.02 
&1.65
\\ \hline
$0.1$
&1
&0.145
& $0.527$
&0.395
&2
&350
&$0.685 \times 10^{-1}$
&2.03
&1.68
\\ \hline
$0.1$
&2
&0.341
& $0.455$
&0.360
&4
&0.2
&$0.298 \times 10^{-4}$
&1.69 
&1.26
\\ \hline
$0.1$
&5
&0.547
& $0.450$
&0.414
&4
&2
&$0.213 \times 10^{-2}$
&1.70
&1.27 
\\ \hline
$0.1$
&10
&0.610
& $0.462$
&0.490
&4
&12
&$0.195 \times 10^{-1}$ 
&1.80
&1.35
\\ \hline
$2$
&0.02
&$ 0.523 \times 10^{-6}$
&$ 1.69 $
&1.26
&4
&30
&$0.302 \times 10^{-1}$
&1.89 
&1.43
\\ \hline
$2$
&0.1
&$0.121 \times 10^{-4}$
&$ 1.69 $
&1.26
&4
&70
&$0.355 \times 10^{-1}$
&1.96
&1.49
\\ \hline
$2$
&0.3
&$0.101 \times 10^{-3}$
&$ 1.69 $
&1.25
&4
&150
&$0.369 \times 10^{-1}$
&1.98
&1.55
\\ \hline
$2$
&1
&$ 0.104\times 10^{-2}$
&1.66
&1.25
&4
&350
&$0.372 \times 10^{-1}$
&1.97
&1.57
\\ \hline 
\end{tabular}
\hspace*{\fill}
\caption{\footnotesize
The discontinuity in the renormalized order parameter 
$\Dt \rho_{0R}$ and the critical inverse correlation
lengths $m_R^c$ and $m_{0R}^c$ in the ordered and the 
disordered phase respectively. 
For small $\la_{2\La}/\la_{1\La}$ the ratios 
$m_R^c/\Dt \rho_{0R}$ and $m_{0R}^c/\Dt \rho_{0R}$ become
universal. 
\label{table1}
}
\end{table}
For 
$\la_{2\La}/\la_{1\La}\,\, \ltap \,\,1$ the curves show constant 
positive slope. In this range $\Dt \rho_{0 R}$ follows a 
power law behavior
\be
\Dt \rho_{0 R}=R \, (\la_{2\La})^{\th}, \quad  \th=1.93 
\label{Powrho}.
\ee
The critical exponent $\th$ is obtained from the slope
of the curve in fig.\ \ref{phase} for 
$\la_{2\La}/\la_{1\La}\ll 1$. The exponent is
universal and, therefore, does not depend on the specific 
value for $\la_{1\La}$.
On the other hand, the amplitude $R$ grows with
decreasing $\la_{1\La}$. For vanishing $\la_{2\La}$
the order parameter changes continuously at the transition
point and one observes a second order phase transition
as expected for the $O(8)$ symmetric vector model. As 
$\la_{2\La}/\la_{1\La}$ becomes larger than one the
curves deviate substantially
from the linear behavior. The deviation
depends on the specific choice of the short distance
potential. For $\la_{2\La}/\la_{1\La}\gg 1$ the curves
flatten. In this range
$\Dt \rho_{0 R}$ becomes insensitive to a variation
of the quartic coupling $\la_{2\La}$.

\begin{figure}[h]
\unitlength1.0cm
\begin{center}
\begin{picture}(13.,9.)
\put(5.5,-0.8){$\ds{\ln\left(\frac{\la_{2 \La}}
{\la_{1 \La}}\right)}$}
\put(10.6,1.3){\footnotesize $(1)$}
\put(10.6,8.28){\footnotesize $(2)$}
\put(10.6,7.55){\footnotesize $(3)$}
\put(10.6,6.65){\footnotesize $(2)$}
\put(10.6,5.7){\footnotesize $(3)$}
\put(2.2,6.05){\footnotesize $\ds{\frac{m_R^c}
{\Dt \rho_{0R}}}$}
\put(2.2,4.1){\footnotesize $\ds{\frac{m_{0R}^c}
{\Dt \rho_{0R}}}$}
\put(-0.5,0.){
\epsfysize=13.cm
\epsfxsize=9.cm
\rotate[r]{\epsffile{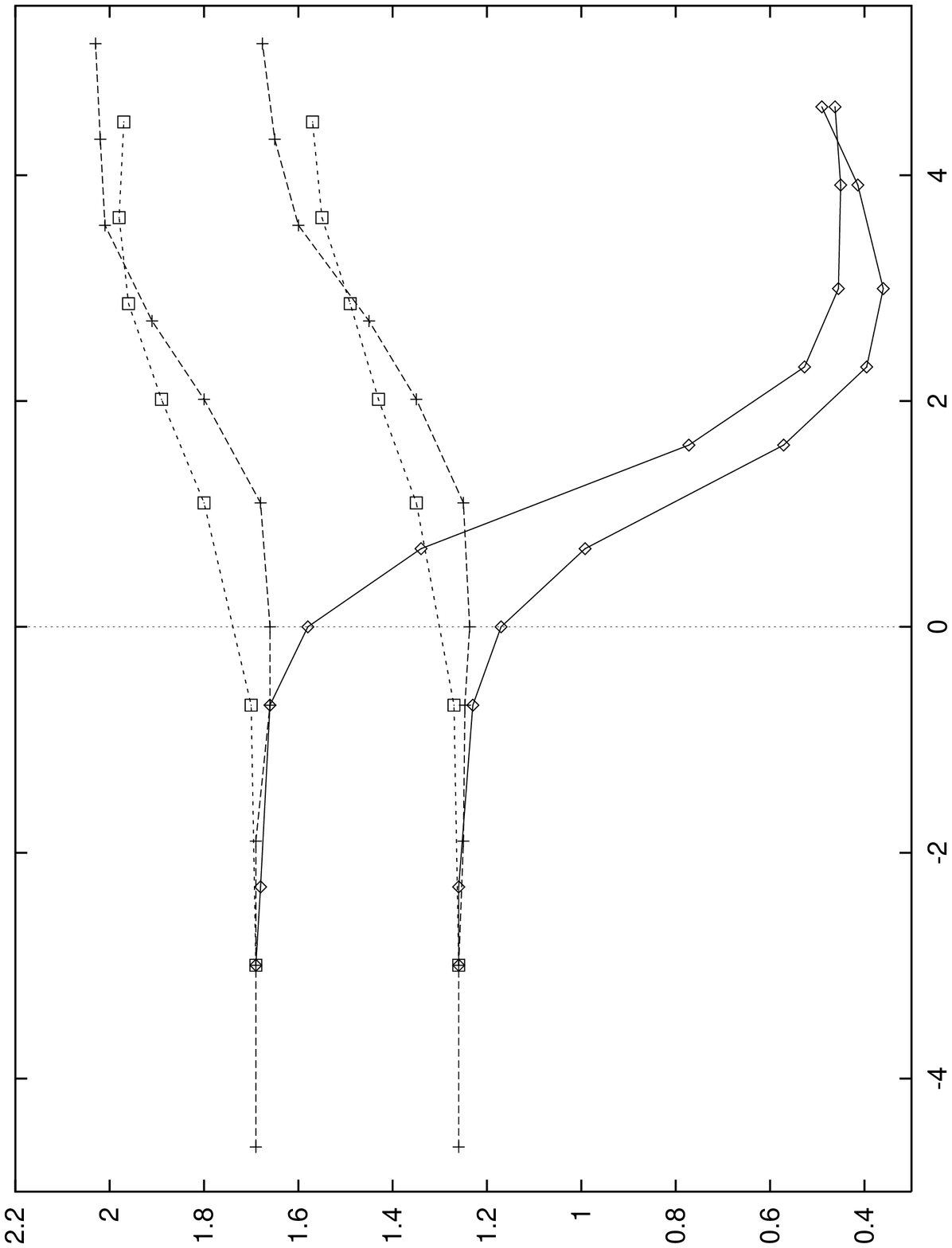
}}
}
\end{picture}
\end{center}
\vspace{0.5cm}
\caption[]{\label{ratio} \footnotesize
The inverse correlation
lengths $m_R^c$ and $m_{0R}^c$ in the ordered and the 
disordered phase respectively. They are normalized to
$\Dt \rho_{0R}$ and given as a function
of $\ln(\la_{2\La}/\la_{1\La})$. Data points
for fixed (1) $\la_{1\La}=0.1$,
(2) $\la_{1\La}=2$, (3) $\la_{1\La}=4$ are
connected by straight lines. 
}
\end{figure}

In addition to the jump in the order parameter 
we present the mass terms $m_R^c$ 
and $m_{0R}^c$ which we
normalize to $\Dt \rho_{0 R}$.
In fig.\ \ref{ratio} these ratios are
plotted versus the logarithm of the ratio of the initial
quartic couplings $\la_{2\La}/\la_{1\La}$. 
Again values
obtained for fixed $\la_{1\La}=0.1$, $2$ and $\la_{1\La}=4$
are connected by straight lines. The universal range is set
by the condition $m_R^c/ \Dt \rho_{0 R} \simeq \mbox{const}$
(equivalently for $m_{0R}^c/ \Dt \rho_{0 R}$).  
The universal ratios are 
$m_R^c/ \Dt \rho_{0 R}=1.69$
and $m_{0R}^c/ \Dt \rho_{0 R}=1.26$
as can be seen from table \ref{table1}. 
For the given curves universality holds 
approximately for
$\la_{2\La}/\la_{1\La} \, \ltap \, 1/2$ 
and becomes ``exact'' in the limit 
$\la_{2\La}/\la_{1\La} \to 0$. In this range we obtain
\be
m_R^c = S (\la_{2\La})^{\th},\qquad 
m_{0R}^c = \tilde{S} (\la_{2\La})^{\th} 
\label{Powm}.
\ee
The 
universal features of the system are not restricted to the weak coupling 
region of $\la_{2\La}$. This is demonstrated in fig.\ \ref{ratio}
for values up to  $\la_{2\La} \simeq 2$. 
The ratios $m_R^c/ \Dt \rho_{0 R}$ and $m_{0R}^c/ \Dt \rho_{0 R}$
deviate from the universal
values as $\la_{2\La}/\la_{1\La}$ is increased. For fixed
$\la_{2\La}$
a larger $\la_{1\La}$ results in a weaker transition
concerning $\Dt \rho_{0 R}/\La$.
The ratio $m_R^c/ \Dt \rho_{0 R}$ increases with $\la_{1\La}$
for small fixed $\la_{2\La}$ whereas in the asymptotic
region, $\la_{2\La}/\la_{1\La} \gg 1$, one observes from
fig.\ \ref{ratio} that this tendency is reversed 
and $m_R^c/ \Dt \rho_{0 R}$,
$m_{0R}^c/ \Dt \rho_{0 R}$ start to decrease at about 
$\la_{1\La} \simeq 2$.  

In summary,
the above results show that though the short distance 
potential $U_{\La}$ indicates a second order phase transition,
the transition becomes first order once fluctuations are taken into
account. This fluctuation induced first order phase transition
is known in four dimensions as the Coleman-Weinberg phenomenon
\cite{Col73}. Previous studies of the three dimensional 
$U(2) \times U(2)$ symmetric model using
the $\eps$-expansion \cite{Pisa84,PW84-1} already indicated that the
phase transition should be a fluctuation induced first order
transition. The validity of the $\eps$-expansion for weak 
first order transitions is, however, not clear a priori
since the expansion parameter is not small
-- there are known cases where it fails 
to predict correctly the order of the transition \cite{AbeHig}.
The question of the order of the phase transition has been
addressed also in lattice studies \cite{Dreher92Shen94} and in
high-temperature expansion \cite{Khl95}. All studies are 
consistent with the first order nature of the 
transition and with the absence of nonperturbative infrared
stable fixed points. It is, however, notoriously
difficult to distinguish by these methods between weak
first order and second order transitions. Our method
gives here a clear and unambiguous answer and allows
a detailed quantitative
description of the phase transition. The universal form
of the equation of state for weak first order phase
transitions is presented in section \ref{sce}.

In the following we specify the scale $k_f$ 
for which we have given
the effective average potential $U_{k}$.
We observe that $U_k$ depends
strongly on the infrared cutoff $k$ as long as $k$ is larger
than the scale $k_2$ where 
the second minimum of the potential
appears. Below $k_2$ the two minima start to become
almost degenerate for $T$ near $T_{c}$ and the running
of $\rho_0(k)$ stops rather soon. The
nonvanishing value of $k_2$ induces
a physical infrared cutoff and represents a characteristic
scale for the first order phase transition.  
We stop the integration of the flow equation
for the effective average potential at a scale
$k_f < k_2$ which is determined in terms of the 
curvature (mass term) at the top
of the potential barrier that separates the two local minima
of $U_k$ at the origin and at $\rho_{0}(k)$.
The top of the potential barrier at
$\rho_B(k)$ is determined
by
\be
U_k^{\pri}(\rho_B)=0 \label{bar}
\ee
for $0 < \rho_B(k) < \rho_0(k)$ and for the 
renormalized mass term at 
$\rho_B(k)$ one obtains 
\be
m_{B,R}^2(k)=2 Z_k^{-1} \rho_B U_k^{\pri\pri}(\rho_B) < 0. 
\label{mbr}
\ee  
We fix our final value for the running by 
\be
\ds{\frac{k_f^2-|m_{B,R}^2(k_f)|}{k_f^2}}=0.01
\label{fix}
\ee 
For this choice the coarse grained effective potential $U_{k_f}$
essentially includes all fluctuations with momenta larger than the
mass $|m_{B,R}|$ at the top of the potential barrier. It is a
nonconvex function which is the appropriate quantity for the 
study of physical
processes such as tunneling or inflation.
The nonconvex part of $U_k$ is
considered in detail in section \ref{coarse}. 
There we will also discuss that 
the appropriate choice for a coarse graining
scale $k$ is often far from obvious.\\

\subsection{Scale dependence of the effective average potential 
\label{scale}}

In this section we derive the flow equation for
the effective average potential $U_k$.
We point out a related
investigation of the four dimensional
$SU(N) \times SU(N)$ symmetric
linear sigma model coupled to fermions that has been
studied previously within the framework of the effective
average action \cite{Ju95-7}. 
There the flow equations for a polynomial approximation of
$U_k$ can be found for arbitrary dimension $d$. 
To study the equation of state of the three dimensional
theory we keep here the most general form for $U_k$.

We define $U_k$ by evaluating the average action 
for a constant field with $\Gm_k=\Omega U_k$ where
$\Omega$ denotes the  
volume. To evaluate the r.h.s.\ of (\ref{ERGE}) with the
ansatz (\ref{Ansatz}) we expand $\Gm_k$  
around a constant background configuration.
With the help of the $U(2)\times
U(2)$ transformations the matrix field $\vp$ can be turned  
into a standard diagonal form with real nonnegative eigenvalues.
Without loss of generality the evolution
equation for the effective
potential can therefore be obtained by calculating the trace 
in (\ref{ERGE}) for small field fluctuations 
$\chi_{ab}$ around a constant
background configuration which is real and
diagonal, 
\be
\vp_{ab}=\vp_{a} \delta_{ab} 
\,\, , \quad \vp^*_{a}=\vp_{a}\label{ConstConfig}.
\ee
We separate the fluctuation field into its real and
imaginary part, $\chi_{ab}=\frac{1}{\sqrt{2}}
(\chi_{Rab}+i\chi_{Iab})$ and perform the second functional
derivatives of $\Gm_k$ with respect to the
eight real components.
For the constant configuration
(\ref{ConstConfig}) it turns out that $\Gm_k^{(2)}$ has a 
block diagonal form because mixed derivatives with 
respect to real and imaginary parts of the field vanish.
The remaining submatrices 
$\dt^2\Gm_k/\dt\chi_R^{ab}\dt\chi_R^{cd}$ and 
$\dt^2\Gm_k/\dt\chi_I^{ab}\dt\chi_I^{cd}$ can be
diagonalized in order to find
the inverse of $\Gm_k^{(2)} + R_k$ under the trace
occuring in eq.\ (\ref{ERGE}).
Here the momentum independent part
of $\Gm_k^{(2)}$ defines the mass matrix
by the second functional derivatives
of $U_k$. 
The eight eigenvalues of the mass matrix are
\bea
(M_1^{\pm})^2 &=& U_k^{\pri}+2\left(\rho \pm
(\rho^2-\tau)^{1/2}\right)\prl_{\tau}U_k \,,\nnn
(M_2^{\pm})^2 &=& U_k^{\pri}\pm 2 \tau^{1/2} \prl_{\tau}U_k
\label{MassEigena} 
\eea
corresponding to second derivatives with respect to $\chi_I$
and
\bea
(M_3^{\pm})^2 &=& (M_1^{\pm})^2 \,,\nnn
(M_4^{\pm})^2 &=& U_k^{\pri}+\rho U_k^{\pri\pri}
+2 \rho \prl_{\tau} U_k
+4 \tau \prl_{\tau} U_k^{\pri} + 4 \rho \tau \prl_{\tau}^2 U_k 
\nnn &&
\pm \left\{ \tau \left( U_k^{\pri\pri} + 4 \prl_{\tau} U_k
+ 4 \rho \prl_{\tau} U_k^{\pri} + 4 \tau \prl_{\tau}^2 U_k \right)^2 
\right.\nnn
&& \left.
+ \left(\rho^2-\tau\right) \left( U_k^{\pri\pri} 
- 2 \prl_{\tau} U_k 
-4 \tau \prl_{\tau}^2 U_k \right)^2 \right\}^{1/2}
\label{MassEigen}
\eea
corresponding to second derivatives with respect to $\chi_R$.
Here the eigenvalues are expressed in terms of the 
invariants $\rho$ and $\tau$ using
\be
\vp_1^2=\hal(\rho+\tau^{1/2}),\quad
 \vp_2^2=\hal(\rho-\tau^{1/2})
\ee
and we adopt the convention that a prime on $U_k(\rho,\tau)$
denotes the derivative with respect to $\rho$ at fixed $\tau$ 
and $k$ and $\prl_{\tau}^n U_k \equiv \prl^nU_k/(\prl\tau)^n$.

The flow equation for the effective average potential is
simply expressed in terms of the mass eigenvalues
\bea
\lefteqn{ \ds{\prlt U_k(\rho,\tau)} = \ds{
 \hal\iddq\prlt R_k(q)}}\nnn
 &&\ds{ \left\{\frac{2}{P_k(q)+(M_1^+(\rho,\tau))^2}+
 \frac{2}{P_k(q)+(M_1^-(\rho,\tau))^2}
+\frac{1}{P_k(q)+(M_2^+(\rho,\tau))^2}\right.} \nnn
 &&+ \ds{ \left.
 \frac{1}{P_k(q)+(M_2^-(\rho,\tau))^2}+
 \frac{1}{P_k(q)+(M_4^+(\rho,\tau))^2}+
 \frac{1}{P_k(q)+(M_4^-(\rho,\tau))^2} \right\}  }.
 \label{UkEvol}
\eea
On the right hand side of the evolution equation appears the 
(massless) inverse average propagator 
\be
 P_k(q)=Z_k q^2+R_k(q)=\frac{Z_k q^2}{1-e^{-q^2/k^2}}
 \label{Propagator}
\ee
which incorporates the infrared cutoff function $R_k$ given
by eq.\ (\ref{Rk(q)}).
The only approximation so far is due to the derivative 
expansion (\ref{Ansatz}) of $\Gm_k$ which
enters into the flow equation 
(\ref{UkEvol}) through the form of $P_k$.
The mass eigenvalues (\ref{MassEigena}) and (\ref{MassEigen}) 
appearing in the above flow equation are exact since we have kept
for the potential the most general form $U_k(\rho,\tau)$.\\

{\bf Spontaneous symmetry breaking and mass spectra}

In the following we consider spontaneous symmetry 
breaking patterns
and the corresponding mass spectra for
a few special cases.
For the origin at $\vp_{ab}=0$ all eigenvalues equal 
$U_k^{\pri}(0,0)$. If the origin is the absolute minimum of the
potential we are in the symmetric regime where all excitations have
mass squared $U_k^{\pri}(0,0)$. 

Spontaneous symmetry breaking to the diagonal $U(2)$ subgroup
of $U(2) \times U(2)$ 
can be observed for a field configuration which is proportional
to the identity matrix, i.e. $\vp_{ab}=\vp \delta_{ab}$.
The invariants $(\ref{Invaria})$ take on values  
$\rho=2 \vp^2$ and $\tau=0$. The relevant
information for this symmetry breaking pattern is contained
in $U_k(\rho) \equiv U_k(\rho,\tau=0)$.
In case of spontaneous symmetry breaking there is a 
nonvanishing value for the minimum $\rho_0$ of the potential.
With $U_k^{\pri}(\rho_0)=0$ one finds the expected four 
massless Goldstone bosons with 
$(M_1^-)^2=(M_2^{\pm})^2=(M_3^-)^2=0$.
In addition there are three massive scalars in the 
adjoint representation of the unbroken diagonal
$SU(2)$ with mass squared 
$(M_1^+)^2=(M_3^+)^2=(M_4^-)^2=
4 \rho_0 \prl_{\tau} U_k$ and one singlet with mass squared
$(M_4^+)^2=2 \rho_0 U_k^{\pri\pri}$. The situation
corresponds to chiral symmetry breaking in two flavor
QCD in absence of quark masses and the chiral anomaly.
The Goldstone modes are the pseudoscalar pions and the
$\eta$ (or $\eta^{\prime}$), the scalar triplet has the
quantum numbers of $a_0$ and the singlet is the so-called
$\sigma$-field.

Another interesting case is the spontaneous symmetry breaking
down to a residual
$U(1) \times U(1) \times U(1)$ subgroup of $U(2) \times U(2)$
which can be observed for the configuration 
$\vp_{ab} =\vp \dt_{a1} \dt_{ab}$ ($\rho=\vp^2$, 
$\tau=\vp^4=\rho^2$). 
Corresponding to the number of broken generators 
one observes the five massless Goldstone bosons 
$(M_1^{\pm})^2=(M_2^+)^2=(M_3^{\pm})^2=0$ for the minimum 
of the potential at 
$U_k^{\pri}+2 \rho_0 \prl_{\tau} U_k = 0$. In addition there are two
scalars with mass squared 
$(M_2^-)^2=(M_4^-)^2=U_k^{\pri}-2 \rho_0 \prl_{\tau} U_k$
and one with 
$(M_4^+)^2=U_k^{\pri}+2 \rho_0 U_k^{\pri\pri} + 6 \rho_0
\prl_{\tau} U_k + 8 \rho_0^2 \prl_{\tau} U_k^{\pri} + 8 \rho_0^3
\prl_{\tau}^2 U_k$. 

We finally point out the special case where 
the potential is independent of the second invariant
$\tau$. In this case there is an enhanced 
$O(8)$ symmetry instead of 
$U(2) \times U(2)$. With $\prl^n_{\tau}U_k \equiv 0$
and $U_k^{\pri}(\rho_0)=0$ one observes the expected
seven massless Goldstone bosons and one massive mode with mass
squared $2 \rho_0 U_k^{\pri\pri}$.\\

{\bf Scaling form of the flow equation}

For the $O(8)$ symmetric  
model in the limit $\bar{\la}_{2\La}=0$
one expects a region of
the parameter space which is characterized by renormalized
masses much smaller than the ultraviolet cutoff
or inverse microscopic length scale of the theory. 
In particular, in the absence of a mass scale one
expects a scaling behavior of the effective average
potential $U_k$. The behavior of $U_k$ at or near a second
order phase transition is most conveniently studied
using the scaling form of the evolution equation.
This form is also appropriate for an investigation that has 
to deal with weak first order phase transitions as 
encountered in the present model for 
$\bar{\la}_{2\La} > 0$. The remaining
part of this section is devoted to the derivation of
the scaling form (\ref{DlessEvol}) of 
the flow equation (\ref{UkEvol}).

In the present form of eq.\ (\ref{UkEvol}) the r.h.s.\
shows an explicit dependence on the scale $k$ once the
momentum integration is performed.
By a proper choice of 
variables we cast the evolution equation into a form
where the scale no longer appears explicitly. We
introduce a dimensionless potential $u_k=k^{-d}  U_k$
and express it in terms of dimensionless 
renormalized fields 
\bea
\trho &=& Z_k k^{2-d} \rho\,,\nnn
\ttau &=& Z_k^2 k^{4-2 d} \tau\,.  
\label{dlessfield}
\eea
The derivatives of $u_k$ are given by
\be
\prl_{\ttau}^n u_k^{(m)}(\trho,\ttau)=Z_k^{-2 n-m} 
k^{(2 n+m-1) d-4 n-2 m} \prl_{\tau}^n U_k^{(m)}
\left(\rho,\tau\right)\,.
\label{dlessder}
\ee  
(Note that $u_k^{(m)}$ denotes $m$ derivatives
with respect to $\trho$ at fixed $\ttau$ and $k$, while 
$U_k^{(m)}$ denotes $m$ derivatives
with respect to $\rho$ at fixed $\tau$ and $k$). With
\bea
\ds{\prlt u_k(\trho,\ttau)_{|\trho,\ttau}} &=& 
\ds{-d u_k(\trho,\ttau) +
(d-2+\eta) \trho u_k^{\pri}(\trho,\ttau)
 + (2 d-4+2 \eta) \ttau
\prl_{\ttau} u_k(\trho,\ttau)}\nnn 
&&+ \ds{k^{-d} \prlt 
U_k\left(\rho(\trho),\tau(\ttau)\right)_{|\rho,\tau}}
\label{dlesstrans}    
\eea
one obtains from (\ref{UkEvol})
the evolution equation for the 
dimensionless potential. Here the anomalous dimension 
$\eta$ arises from the $t$-derivative acting on $Z_k$ and 
is given by eq.\ (\ref{Eta}). It is convenient to introduce
dimensionless integrals by 
\be
\ds{\hal\iddq\frac{\prl R_k}{\prl t}}\frac{1}{P_k+Z_k k^2 \om}
=2 v_d k^d \left[ l^d_0(\om) - \eta \hat{l}^d_0(\om) \right]
\ee
where 
\be
v_d^{-1} = 2^{d+1} \pi^\frac{d}{2} \Gm\left(\frac{d}{2}\right). 
\ee
The explicit form of $l^d_0$ and $\hat{l}^d_0$
reads
\bea
\ds{l^d_0(\om)} &=&
\ds{-  \int^{\infty}_0 dy y^{\frac{d}{2}+1} 
\frac{\partial r(y)}{\partial y}
\left[ y(1+r(y)) + \om \right]^{-1}}, 
\nnn 
\ds{\hat{l}^d_0(\om)}
&= &\ds{ \frac{1}{2} \int^{\infty}_0 dy y^{\frac{d}{2}} }
r(y)
\left[ y(1+r(y)) + \om \right]^{-1}
\label{threshld0} 
\eea
with the dimensionless infrared cutoff function
\be
r(y) = \frac{e^{-y}}{1 - e^{-y}}.
\label{ry} 
\ee
The behavior of these functions is discussed in section 
\ref{solv}
where we also consider
their derivatives with respect to $\om$. Using the notation
\be
l^d_0(\om;\eta )=l^d_0(\om) - \eta \hat{l}^d_0(\om)
\ee  
one obtains the flow equation for the dimensionless potential
\bea
\lefteqn{\ds{\prlt u_k(\trho,\ttau)} = 
\ds{-d u_k(\trho,\ttau) +
(d-2+\eta) \trho u_k^{\pri}(\trho,\ttau)
 + (2 d-4+2 \eta) \ttau
\prl_{\ttau} u_k(\trho,\ttau)}}\nnn
&&+ \ds{4 v_d l^d_0\left((m_1^+(\trho,\ttau))^2;\eta\right)
+ 4 v_d l^d_0\left((m_1^-(\trho,\ttau))^2;\eta\right)
+ 2 v_d l^d_0\left((m_2^+(\trho,\ttau))^2;\eta\right)}\nnn
&&+ \ds{2 v_d l^d_0\left((m_2^-(\trho,\ttau))^2;\eta\right)
+2 v_d l^d_0\left((m_4^+(\trho,\ttau))^2;\eta\right)
+2 v_d l^d_0\left((m_4^-(\trho,\ttau))^2;\eta\right)
}
\label{DlessEvol}    
\eea
where the dimensionless mass terms are related to 
(\ref{MassEigen}) according to 
\be
(m_i^{\pm}(\trho,\ttau))^2=\frac
{\left(M_i^{\pm}(\rho(\trho),\tau(\ttau))\right)^2}{Z_k k^2}.
\label{massrel}
\ee
Eq.\ (\ref{DlessEvol}) is the scaling form of the
flow equation we are looking for. For a 
$\ttau$-independent potential it reduces to the 
evolution equation for the $O(8)$ symmetric model
\cite{TW94-1,Mor}. The potential $u_k$ at
a second order phase transition is given by
a $k$-independent (scaling) solution 
$\prl u_k/\prl t=0$ \cite{TW94-1,Mor}. For this solution 
all the $k$-dependent functions
on the r.h.s.\ of eq.\
(\ref{DlessEvol}) become independent
of $k$. For a weak
first order phase transition these functions
will show a weak $k$-dependence for $k$
larger than the inherent mass scale of the
system (cf.\ section \ref{rg}). There is no
particular advantage of the scaling form of
the flow equation for strong first order phase transitions.\\

\subsection{Solving the flow equation \label{solv}}

Eq.\ (\ref{DlessEvol}) describes the scale dependence
of the effective average potential $u_k$
by a nonlinear partial differential equation for the
three variables $t$, $\trho$ and $\ttau$. 
A particular difficulty for its analytical
study is that the integral $l^d_0(\om;\eta)$
(cf.\ eq.\ (\ref{threshld0}))
can be done analytically only for certain limits
of the arguments. The complicated form of the equation
therefore suggests a numerical solution. 
We will use a method that relies on a simultaneous
expansion of the potential around a number 
of field values $\trho_i,\ttau_j$, $i=1,\ldots,l$,
$j=1,\ldots,l^{\pri}$ for given numbers $l,l^{\pri}$. 
The expansions around
different points are matched to obtain the general field
dependence of the potential. As a consequence we
cast the partial differential equation (\ref{DlessEvol})
into a system of ordinary differential equations.
The method is developed in \cite{ABBTW95} for a computation of 
the critical equation of state for $O(N)$ symmetric
Heisenberg models \cite{BTW95} (cf.\ section \ref{secsec}). 
The generalization to the present model
is described below. This approach has some
favorable aspects. The main advantage is that it 
allows a very efficient integration of the 
differential equations using
a standard Runge-Kutta algorithm without the occurance
of numerical instabilities\footnote{These instabilities
arise through the inevitable small round-off errors
at every integration step. The standard discretised
version of eq.\ (\ref{DlessEvol}),
using finite difference quotients for the derivatives,
becomes strongly oscillating after a few integration steps.
Already for the most simple linear partial differential 
equations one finds that the resulting error may grow 
exponentially with the number of steps (see e.g.\ ref.\
\cite{John} for an analytically solved example). 
A possible finite difference scheme uses 
``backward'' difference quotients as we have exemplified 
in ref.\ \cite{ABBTW95}.}. The coupled set of
ordinary differential equations describes the
flow of couplings defined as derivatives of the potential
at given points, e.g.\ at the minimum of the potential.
These couplings often allow direct physical interpretation 
and some of their properties can be read off from the
analytic structure of their flow equations.
We will exploit this fact to explain the 
results we obtain from the numerical solution
in section \ref{rg}.

We concentrate in the following on spontaneous 
symmetry breaking with a residual $U(2)$ symmetry
group. As we have already pointed out in section \ref{model}
this symmetry breaking can be observed for a
configuration which is proportional to the 
identity and we have $\ttau=0$. In this case
the eigenvalues (\ref{MassEigena}) and
(\ref{MassEigen}) of the mass matrix 
with (\ref{massrel}) are given by
\bea
(m_1^-)^2 &=& (m_2^{\pm})^2 = (m_3^-)^2 = u_k^{\pri}\,,\nnn 
(m_1^+)^2 &=& (m_3^+)^2 = (m_4^-)^2 = 
u_k^{\pri}+4 \trho \prl_{\ttau}u_k\,,\nnn
(m_4^+)^2 &=& u_k^{\pri} + 2 \trho u_k^{\pri\pri}\,
\eea
and on the r.h.s.\ of the partial differential equation 
(\ref{DlessEvol}) for $u_k(\trho) \equiv u_k(\trho,\ttau=0)$
only the functions $u_k^{\pri}(\trho)$, $u_k^{\pri\pri}(\trho)$
and $\prl_{\ttau}u_k(\trho)$ appear. At fixed 
$\trho=\trho_i$ the $k$-dependence of $u_k$ is
then determined by the couplings 
$u_k^{\pri}(\trho_i)$, $u_k^{\pri\pri}(\trho_i)$
and $\prl_{\ttau}u_k(\trho_i)$. We determine these
couplings through the use of flow equations which
are obtained by taking the derivative in eq.\ 
(\ref{DlessEvol}) with respect to $\trho$ and
$\ttau$ evaluated at $\trho=\trho_i$, $\ttau=0$.
These flow equations for $u_k^{\pri}$, $u_k^{\pri\pri}$
and $\prl_{\ttau}u_k$ involve also
higher derivatives of the
potential as $u_k^{\pri\pri\pri}$,
$u_k^{(4)}$, $\prl_{\ttau}u_k^{\pri}$
and $\prl_{\ttau}u_k^{\pri\pri}$. The procedure
will be to evaluate the flow
equations for $u_k^{\pri}$ and $u_k^{\pri\pri}$
at different points $\trho_i,\ttau=0$ for $i= 1,\ldots ,l$
and to estimate the higher $\trho$-derivatives 
appearing on the right hand side of the flow equations
by imposing matching conditions. The same procedure
can be applied to $\prl_{\ttau}u_k$ and in order to obtain
an equivalent matching we
also consider the flow equation for $\prl_{\ttau}u_k^{\pri}$.
One could proceed in a similar way for $\prl_{\ttau}^2u_k$ which
appears on the right hand side of the evolution equation
of $\prl_{\ttau}u_k$ and so on.
However, since we are 
interested in the $\trho$-dependence of the potential at 
$\ttau=0$ we shall use a truncated expansion\footnote{
In principle one could also consider points with $\ttau \not = 0$ 
in the neighborhood
of $\ttau=0$ and use the additional information to estimate
the higher $\ttau$-derivatives as it is done for the
higher $\trho$-derivatives.} in $\ttau$ with 
\be
\prl_{\ttau}^n u_k(\trho,\ttau=0)=0 \quad \mbox{for}
\quad n \ge 2.
\label{Truncation}
\ee
In three space dimensions the neglected  
($\trho$-dependent) operators have 
negative canonical mass dimension.
We make no expansion in terms of $\trho$
since the general $\trho$-dependence allows a description
of a first order phase transition where a second local minimum
of $u_k(\trho)$ appears. 
The approximation (\ref{Truncation}) only affects  
the flow equations for $\prl_{\ttau}u_k$ and 
$\prl_{\ttau}u_k^{\pri}$ (cf.\ eqs.\ (\ref{La2}) and 
(\ref{La2Pri})). The form of the flow equations for
$u_k^{\pri}$ and $u_k^{\pri\pri}$ is not affected by the 
truncation (cf.\ eqs.\ (\ref{Epsi}) and 
(\ref{La1})). 
From $u_k^{\pri}$ we obtain the effective average
potential $u_k$ by simple integration.
We have tested the sensitivity of our results
for $u_k^{\pri}$ to a change in $\prl_{\ttau}u_k$ by neglecting
the $\trho$-dependence of the $\ttau$-derivative. We observed
no qualitative change of the results. We expect that the main
truncation error is due to the derivative expansion (\ref{Ansatz}) for
the effective average action.  

To simplify notation we introduce 
\bea
\eps(\trho)&=&u_k^{\pri}(\trho,\ttau=0),\nnn
\la_1(\trho)&=&u_k^{\pri\pri}(\trho,\ttau=0),\nnn
\la_2(\trho)&=&4 \prl_{\ttau} u_k(\trho,\ttau=0).
\label{notation}
\eea
Higher derivatives are denoted by primes on the 
$\trho$-dependent quartic
``couplings'',
i.e.\ $\la_1^{\pri}=u_k^{\pri\pri\pri}$,
$\la_2^{\pri}=\prl_{\ttau}u_k^{\pri}$ etc. It is
convenient to introduce functions $l^d_n(\om;\eta)$
that can be related
to $l^d_0(\om;\eta)$
(\ref{threshld0}) by differentiation with respect to the
mass argument: 
\bea
\ds{l^d_1(\om;\eta)} 
&=& l^d_1(\om) - \eta \hat{l}^d_1(\om)\nnn
&=& \ds{- \frac{\prl}{\prl \om} l^d_0(\om;\eta),} \nnn
l^d_{n+1}(\om;\eta) &=& \ds{-\frac{1}{n} 
\frac{\prl}{\prl \om} l^d_n(\om;\eta)}\quad
\mbox{for}\quad n \ge 1.
\eea
The explicit form of $l^d_n$ and $\hat{l}^d_n$ is given in
eq.\ (\ref{threshldn}) in the appendix. 
For $\om \ge -1$ they are positive,
monotonically decreasing functions of $\om$.
In leading order $l^d_n$ and $\hat{l}^d_n$ vanish
$\sim \om^{-(n+1)}$ for arguments $\om \gg 1$. They 
introduce a ``threshold'' behavior that accounts for
the decoupling of modes with mass squared larger
than the infrared cutoff $Z_k k^2$. For 
vanishing argument they are of order unity. As $\om \to -1$
the functions $l^d_n$, $\hat{l}^d_n$ exhibit
a pole for $d < 2(n+1)$. The pole structure is discussed
in the appendix. We also use two-parameter functions
$l^d_{n_1,n_2}(\om_1,\om_2;\eta)$ \cite{Ju95-7}. 
For $n_1=n_2=1$ their relation to the functions
$l^d_n(\om;\eta)$ can be expressed as
\bea
l^d_{1,1}(\om_1,\om_2;\eta) &=&
\ds{\frac{1}{\om_2 - \om_1} \left[ l^d_1(\om_1;\eta)
- l^d_1(\om_2;\eta) \right]} \quad 
\mbox{for}\quad \om_1 \ne \om_2, \nnn
l^d_{1,1}(\om ,\om ;\eta)&=& l^d_2(\om ;\eta)
\eea
and
\be
l^d_{n_1 + 1, n_2}(\om_1,\om_2;\eta) =\ds{
-\frac{1}{n_1}  \frac{\prl}{\prl \om_1}
l^d_{n_1,n_2}(\om_1,\om_2;\eta)},\quad
l^d_{n_1,n_2}(\om_1,\om_2;\eta) =  
l^d_{n_2,n_1}(\om_2,\om_1;\eta).                         
\ee

With the help of these functions the scale dependence of 
$\eps$ is described by
\bea
\ds{\frac{\prl \eps}{\prl t}} &=& 
\ds{(-2+\eta) \eps +
(d-2+\eta) \trho \la_1 - 6 v_d (\la_1+\la_2+\trho \la_2^{\pri}) 
l^d_1(\eps+\trho \la_2;\eta)}\nnn
&&- \ds{2 v_d (3 \la_1+2 \trho \la_1^{\pri}) 
l^d_1(\eps+2\trho \la_1;\eta)   
-8 v_d \la_1 l^d_1(\eps ;\eta)}
\label{Epsi}
\eea
and for $\la_1$ one finds
\bea
\ds{\frac{\prl \la_1}{\prl t}} &=& 
\ds{(d-4+2\eta) \la_1 +
(d-2+\eta) \trho \la_1^{\pri} }\nnn
&&+\ds{6 v_d \left[ (\la_1+\la_2+\trho \la_2^{\pri})^2
l^d_2(\eps+\trho \la_2;\eta)
-(\la_1^{\pri}+2\la_2^{\pri}+\trho \la_2^{\pri\pri})
l^d_1(\eps+\trho \la_2;\eta)\right]}\nnn
&&+\ds{2 v_d \left[(3\la_1+2\trho \la_1^{\pri})^2
l^d_2(\eps+2\trho \la_1;\eta)
-(5\la_1^{\pri}+2\trho \la_1^{\pri\pri})
l^d_1(\eps+2\trho \la_1;\eta)\right] }\nnn
&&+\ds{8 v_d \left[(\la_1)^2 l^d_2(\eps;\eta)
-\la_1^{\pri}l^d_1(\eps;\eta)\right]}.
\label{La1}                      
\eea
Similarly the scale dependence of $\la_2$ 
is given by
\bea
\ds{\frac{\prl \la_2}{\prl t}} &=& 
\ds{(d-4+2\eta) \la_2 +
(d-2+\eta) \trho \la_2^{\pri}       
- 4 v_d (\la_2)^2 l^d_{1,1}(\eps+\trho \la_2,\eps;\eta) }\nnn
&&+\ds{2 v_d \left[ 3 (\la_2)^2+12\la_1\la_2+8 \trho
\la_2^{\pri}(\la_1+\la_2)+4\trho^2 (\la_2^{\pri})^2\right]
l^d_{1,1}(\eps+\trho \la_2,\eps+2\trho \la_1;\eta) }\nnn
&&- \ds{14 v_d \la_2^{\pri} l^d_1(\eps+\trho \la_2;\eta)
-2 v_d (5 \la_2^{\pri}+2\trho \la_2^{\pri\pri})
l^d_1(\eps+2\trho \la_1;\eta) }\nnn
&&+\ds{2 v_d \left[ (\la_2)^2 l^d_2(\eps;\eta)-4 \la_2^{\pri}
l^d_1(\eps;\eta)\right] }
\label{La2} 
\eea
and for $\la_2^{\pri}$ it reads
\bea
\ds{\frac{\prl \la_2^{\pri}}{\prl t}} &=&
(2 d-6+3\eta) \la_2^{\pri} + 
(d-2+\eta) \trho \la_2^{\pri \pri} \nnn
&&
+4 v_d (\la_2)^2 \Big[(\la_1+\la_2+\trho \la_2^{\pri})
l^d_{2,1}(\eps+\trho \la_2,\eps;\eta)+
\la_1 l^d_{1,2}(\eps+\trho \la_2,\eps;\eta)\Big]\nnn
&&
-2 v_d \Big[3 \la_2 (4 \la_1+\la_2) + 4\trho \la_2^{\pri}
(2\la_1+2\la_2+\trho \la_2^{\pri})\Big]
\Big[(\la_1+\la_2+\trho \la_2^{\pri})\nnn
&&
l^d_{2,1}(\eps+\trho \la_2,\eps+2\trho \la_1;\eta)+
(3 \la_1+ 2 \trho \la_1^{\pri})
l^d_{1,2}(\eps+\trho \la_2,\eps+2\trho \la_1;\eta)\Big]\nnn
&&
+4 v_d \Big[\la_2^{\pri}(7 \la_2  +10 \la_1) + 
6 \la_2 \la_1^{\pri} + 4 \trho \Big(\la_2^{\pri \pri}
(\la_1+\la_2+\trho \la_2^{\pri})
+\la_2^{\pri} (2\la_2^{\pri}+\la_1^{\pri})\Big)\Big]\nnn
&&
l^d_{1,1}(\eps+\trho \la_2,\eps+2\trho \la_1;\eta)
-8 v_d \la_2 \la_2^{\pri} l^d_{1,1}(\eps+\trho \la_2,\eps;\eta)
\nnn &&
+2 v_d (3 \la_1 + 2\trho \la_1^{\pri})(5 \la_2^{\pri}+
2\trho\la_2^{\pri \pri})
l^d_{2}(\eps+2\trho \la_1;\eta)
+14 v_d \la_2^{\pri} (\la_1+\la_2+\trho \la_2^{\pri})\nnn
&&
l^d_{2}(\eps+\trho \la_2;\eta)-2 v_d
(7 \la_2^{\pri \pri} + 2 \trho \la_2^{\pri \pri \pri})
l^d_{1}(\eps+2\trho \la_1;\eta)-14 v_d
\la_2^{\pri \pri} l^d_{1}(\eps+\trho \la_2;\eta)\nnn
&&
-4 v_d (\la_2)^2 \la_1 l^d_{3}(\eps;\eta)
+4 v_d \la_2^{\pri} (2 \la_1+\la_2) l^d_{2}(\eps;\eta)
-8 v_d \la_2^{\pri \pri} l^d_{1}(\eps;\eta)
\label{La2Pri} .
\eea
We evaluate the above flow equations
at different points $\trho_i$ for $i= 1,\ldots ,l$ 
and use a set of 
matching conditions that has been proposed in ref.\ \cite{ABBTW95}.
The generalization
of these conditions to the present model is obtained by
considering fourth order polynomial expansions of 
$\eps (\trho)$ and $\la_2 (\trho)$ around some arbitrary
point $\trho_i$,
\bea
(\eps)_i (\trho) &=& \ds{ \eps_i+\la_{1,i}
(\trho-\trho_i)+\hal \la_{1,i}^{\pri} (\trho-\trho_i)^2
+\frac{1}{6} \la_{1,i}^{\pri \pri} (\trho-\trho_i)^3}
,\nnn
(\la_2)_i (\trho) &=&\ds{ \la_{2,i} +\la_{2,i}^{\pri}
(\trho-\trho_i) + \hal \la_{2,i}^{\pri \pri} (\trho-\trho_i)^2
+\frac{1}{6} \la_{2,i}^{\pri \pri \pri} (\trho-\trho_i)^3}
\eea
with $\eps_i=\eps (\trho_i)$, $\la_{2,i}=\la_2 (\trho_i)$ etc.
Using similar expressions for
\be
(\la_1)_i (\trho) = \frac{\prl}{\prl \trho}
(\eps)_i (\trho)\quad,
\qquad  (\la_2^{\pri})_i (\trho) = \frac{\prl}{\prl \trho} 
(\la_2)_i (\trho)
\ee
the matching is done by imposing continuity 
at half distance between neighboring expansion points,
\bea 
\ds{ (\eps)_i \left( \frac{\trho_i+\trho_{i+1}}{2} \right)} =   
\ds{ (\eps)_{i+1} \left( \frac{\trho_i+\trho_{i+1}}{2} \right) 
}&,&
\ds{ (\la_1)_i \left( \frac{\trho_i+\trho_{i+1}}{2} \right)} =
\ds{ (\la_1)_{i+1} \left( \frac{\trho_i+\trho_{i+1}}{2} \right), 
}\nnn
\ds{ (\la_2)_i \left( \frac{\trho_i+\trho_{i+1}}{2} \right)} =   
\ds{ (\la_2)_{i+1} \left( \frac{\trho_i+\trho_{i+1}}{2} \right) 
}&,& 
\ds{ (\la_2^{\pri})_i \left( \frac{\trho_i+\trho_{i+1}}{2} \right)} =
\ds{ (\la_2^{\pri})_{i+1} \left( \frac{\trho_i+\trho_{i+1}}{2} \right) 
} \label{match1}
\eea 
for $i=1,\ldots ,l-1$ and 
\bea
\ds{ (\la_1^{\pri})_j \left( \frac{\trho_j+\trho_{j+1}}{2} \right)} =  
\ds{ (\la_1^{\pri})_{j+1} \left( \frac{\trho_j+\trho_{j+1}}{2} \right) 
}&,&
\ds{ (\la_2^{\pri \pri})_j \left( \frac{\trho_j+\trho_{j+1}}{2}
\right)} =
\ds{ (\la_2^{\pri \pri})_{j+1} \left( \frac{\trho_j+\trho_{j+1}}{2} 
\right) 
}\label{match2}
\eea
for the initial and end points, $j=1$ and $j=l-1$. Together these
$4 (l-1)$ conditions (\ref{match1})
for all $l-1 \ge 2$ intermediate points and the four conditions
(\ref{match2})
make up two independent algebraic systems of each
$2 l$ equations. From the first set of equations
one obtains a unique solution 
for $\la_{1,i}^{\pri}$ and $\la_{1,i}^{\pri \pri}$.
The second set is identical in structure and 
$\la_{2,i}^{\pri \pri}$, $\la_{2,i}^{\pri \pri \pri}$
can be obtained
from the solutions for $\la_{1,i}^{\pri}$ and 
$\la_{1,i}^{\pri \pri}$
with the substitutions $\eps_j \to \la_{2,j}$, $\la_{1,j} \to   
\la_{2,j}^{\pri}$, $\la_{1,j}^{\pri} \to \la_{2,j}^{\pri \pri}$,
$\la_{1,j}^{\pri \pri} \to \la_{2,j}^{\pri \pri \pri}$ for 
$j=1,\ldots,l$. With the help of these algebraic solutions
we eliminate $\la_1^{\pri}(\trho)$, $\la_1^{\pri\pri}(\trho)$,
$\la_2^{\pri\pri}(\trho)$ and $\la_2^{\pri\pri\pri}(\trho)$
in the flow equations (\ref{Epsi}) - (\ref{La2Pri}) for all
$l$ points $\trho_i$.\footnote{
The algebraic solutions $\la_{1,i}^{\pri}$ and $\la_{1,i}^{\pri \pri}$
(equivalently $\la_{2,i}^{\pri \pri}$ and 
$\la_{2,i}^{\pri \pri \pri}$) do incorporate information from
the whole range of points $\trho_j$ with $j=1,\ldots,l$. It is
a feature of the matching conditions (\ref{match1}),
(\ref{match2}) that the contributions from points $\trho_{j \not = i}$
to $\la_{1,i}^{\pri}$, $\la_{1,i}^{\pri \pri}$ rapidly
decrease with increasing $|i-j|$. For equal spacings between
neighboring expansion points contributions from points
$\trho_j$ with $j > i+1$ $(j < i-1)$ are typically suppressed
by a factor $\ltap 10^{-|i-j|+1}$ as compared to the contribution
from the nearest neighbor point $\trho_{i+1}$ ($\trho_{i-1}$).
As a consequence solutions $\la_{1,i}^{\pri}$,
$\la_{1,i}^{\pri \pri}$ for inner points with $1 \ll i \ll l$
become independent from boundary points. We observe approximate
translational invariance for inner point solutions, i.e.
$\la_{1,i \pm n}^{\pri}$ and $\la_{1,i \pm n}^{\pri \pri}$
are approximately obtained from the solutions 
$\la_{1,i}^{\pri}$ and $\la_{1,i}^{\pri \pri}$ with the 
substitutions $\eps_j \to \eps_{j \pm n}$, 
$\la_{1,j} \to \la_{1,j \pm n}$ for $1 \le j \le l$
if $i$ and $i\pm n$ are sufficiently far away from
the boundaries.
The decoupling from distant points and the translational 
invariance for inner points can be used to obtain
approximate expressions which become useful if a large
number of expansion points is considered. 
We use the exact algebraic solution 
for $l=10$ points. For $l > 10$
we apply the approximate translational 
invariance to 
generate from $\la_{1,5}^{\pri}$, $\la_{1,6}^{\pri}$
additional solutions $\la_{1,5 + 2i}^{\pri}$, 
$\la_{1,6 + 2i}^{\pri}$ for $i=1,\ldots,(l-10)/2$ 
with $l$ even and equivalently  
for $\la_{1,5}^{\pri\pri}$, $\la_{1,6}^{\pri\pri}$.
With $\la_{1,l-3}^{\pri},\ldots,\la_{1,l}^{\pri}$
and $\la_{1,l-3}^{\pri \pri},\ldots,\la_{1,l}^{\pri \pri}$
from the calculation with 10 points one obtains the 
desired generalization.
We have used runs with different choices of $l$
in order to check the stability of the numerical
results.}
Therefore, equations (\ref{Epsi}) - (\ref{La2Pri})
are turned 
into a closed system of $4l$ ordinary differential equations
for the unknowns $\eps(\trho_i)$, $\la_1(\trho_i)$,
$\la_2(\trho_i)$ and 
$\la_2^{\pri}(\trho_i)$. 

If there is a 
minimum of the potential at nonvanishing
$\kp \equiv \trho_0$
we use expansion points that are proportional to
the minimum, i.e.\ $\trho_i = \frac{i-1}{n} \kappa$ with 
$i=1,\ldots,l$ and fixed integer $n$. 
The condition $\eps(\kp)=0$ can be used to 
obtain the  scale dependence of $\kp(k)$:
\bea
\ds{\frac{\mbox{d} \kp}{\mbox{d} t}}
&=&\ds{-[\la_1(\kp)]^{-1}\frac{\prl \eps}{\prl t}}
|_{\trho=\kp}\nnn
&=&\ds{ -(d-2+\eta) \kp 
+ 6 v_d \left(1+\frac{\la_2(\kp) + \kp \la_2^{\pri}(\kp)}
{\la_1(\kp)} \right) l^d_1\left(\kp \la_2(\kp);\eta \right) }\nnn
&+& \ds{2 v_d \left( 3+\frac{2\kp \la_1^{\pri}(\kp)}{\la_1(\kp)}
\right) l^d_1\left(2\kp \la_1(\kp);\eta \right) 
+8 v_d l^d_1\left(0;\eta\right) }.
\label{kappa}
\eea
To make contact with $\beta$-functions for the couplings 
at the potential minimum $\kp$ we point out the relation
\be
\ds{\frac{\mbox{d} \la_{1,2}^{(m)}(\kp)}{\mbox{d} t}}=
\ds{\frac{\prl \la_{1,2}^{(m)} }{\prl t}|_{\trho=\kp}
+ \la_{1,2}^{(m+1)}(\kp) \frac{\mbox{d} \kp}{\mbox{d} t} }.
\label{lakappa}
\ee
Similar relations hold for $\eps(\trho_i)$,
$\la_{1}(\trho_i)$ etc., e.g.\
\bea
\ds{\frac{\mbox{d} \eps(\trho_i)}{\mbox{d} t}}
&=& \ds{\frac{\prl \eps}{\prl t}|_{\trho=\trho_i}
+\frac{i-1}{n} \la_{1}(\trho_i) \frac{\mbox{d} \kp}
{\mbox{d} t}}, \nnn
\ds{\frac{\mbox{d} \la_1(\trho_i)}{\mbox{d} t}}
&=& \ds{\frac{\prl \la_1}{\prl t}|_{\trho=\trho_i}
+\frac{i-1}{n} \la_{1}^{\pri}(\trho_i) \frac{\mbox{d} \kp}
{\mbox{d} t}}. \label{prop}
\eea
We integrate the $4 l-1$ differential equations
(\ref{Epsi}) - (\ref{La2Pri})
for the couplings $\eps(\trho_i)$, $\la_1(\trho_i)$,
$\la_2(\trho_i)$ and 
$\la_2^{\pri}(\trho_i)$ (with $\prl/\prl t$ replaced by
$\mbox{d} / \mbox{d} t$ according to (\ref{prop}))  
and the one for $\kp$ (\ref{kappa})
with a fifth-order Runge-Kutta algorithm using the
embedded fourth-order method for precision control.
The general $\trho$-dependence is recovered
by patching the simultaneous expansions around
different points at half distance between neighboring 
expansion points. The polynomial patching improves with 
decreasing distance which is used to check the
stability of numerical results. 
In order to check for systematic errors the algorithm has been
verified by comparison with a conventional ``backward'' finite 
difference method \cite{ABBTW95}.   

It remains
to compute the anomalous dimension $\eta$ defined in (\ref{Eta})
which describes the scale
dependence of the wave function renormalization $Z_k$. 
We consider a space dependent distortion of the constant 
background field configuration (\ref{ConstConfig})
of the form
\be
 \vp_{ab}(x) = \vp_a\dt_{ab} +
 \left[\dt\vp e^{-iQx} + \dt\vp^* e^{iQx}\right] 
\Si_{ab}.
 \label{ConfAnDi}
\ee
Insertion of the above configuration into the parametrization 
(\ref{Ansatz}) of $\Gm_k$ yields
\be
 \ds{
 Z_k }=
\ds{
  Z_k(\rho,\tau,Q^2=0) }
 = \ds{
 \hal\frac{1}{\Si_{ab}^*\Si_{ab}}
 \lim_{Q^2 \ra 0}\frac{\prl}{\prl Q^2}
 \frac{\dt \Gm_k}{\dt (\dt\vp \dt\vp^*)}|_{\dt\vp=0}} .
 \label{Z}
\ee 
\\
To obtain 
the flow equation of the wave function renormalization
one expands the effective average action around a configuration
of the form (\ref{ConfAnDi}) and evaluates the r.h.s.\ of eq.\
(\ref{ERGE}). This computation has been done in ref.\ 
\cite{Ju95-7} for a ``Goldstone'' configuration with   
\be
 \Si_{ab}=\dt_{a1}\dt_{b2}-\dt_{a2}\dt_{b1}
\ee
and $\vp_a \dt_{ab}=\vp\dt_{ab}$ corresponding to a 
symmetry breaking pattern
with residual $U(2)$ symmetry. The result of ref.\ \cite{Ju95-7}
can be easily generalized to arbitrary fixed field values 
of $\trho$
and we find
\bea
 \ds{ \eta(k)}
&=&\ds{
 4 \frac{v_d}{d}\trho \left[
 4(\la_1)^2 
m_{2,2}^d(\eps,\eps+
2\trho \la_1;\eta) 
  +(\la_2)^2 
m_{2,2}^d(\eps,\eps+
\trho \la_2;\eta) 
 \right] }.
 \label{EtaSkale}
\eea  
The definition of the threshold function
\be 
m_{2,2}^d(\om_1,\om_2;\eta)=
m_{2,2}^d(\om_1,\om_2)-\eta\hat{m}_{2,2}^d(\om_1,\om_2)
\ee
can be found in appendix \ref{ThresholdFunctions}. 
For vanishing arguments the 
functions $m_{2,2}^d$ and $\hat{m}_{2,2}^d$ are of order
unity. They are symmetric with respect to their arguments
and in leading order $m_{2,2}^d(0,\om) \sim 
\hat{m}_{2,2}^d(0,\om) \sim \om^{-2}$ for $\om \gg 1$. 
According to eq.\ (\ref{zet})
we use $\trho=\kp$ to define the uniform
wave function renormalization
\be
Z_{k} \equiv Z_{k}(\kp).
\label{wfr}
\ee 
We point out that 
according to our truncation of the
effective average action with eq.\ (\ref{EtaSkale}) 
the anomalous
dimension $\eta$ is exactly zero at $\trho = 0$. 
This is an artefact of the truncation and
we expect the symmetric phase
to be more affected by truncation errors than the
spontaneously broken phase. We typically observe
small values for 
$\eta(k)=-\mbox{d} (\ln Z_k)/ \mbox{d} t$ (of the order of 
a few per cent). The smallness of $\eta$ is crucial
for our approximation of a uniform wave function
renormalization to give quantitatively reliable
results for the equation of state. For the universal
equation of state given in sect.\ \ref{sce}
one has $\eta=0.022$ as given by the corresponding
index of the $O(8)$ symmetric ``vector'' model.

\subsection{Renormalization group flow \label{rg}}

To understand the detailed picture of the phase structure
presented in section \ref{ps} we will consider 
the flow of some characteristic quantities for the effective
average potential as the infrared
cutoff $k$ is lowered. We will always consider in this section
the trajectories for the critical ``temperature'', i.e.\
$\dt \kp_{\La}=0$,
and we follow the flow for different values of the
short distance parameters $\la_{1\La}$
and $\la_{2\La}$. The discussion for sufficiently
small $\dt \kp_{\La}$ is analogous.
In particular, we compare the
renormalization group flow of these quantities for a weak and a
strong first order phase transition. In some limiting cases
their behavior can be studied analytically.
For the discussion we will frequently consider the flow 
equations for the quartic ``couplings'' $\la_1(\trho)$,
$\la_2(\trho)$ eqs.\ (\ref{La1}), (\ref{La2}) and for the
minimum $\kp$ eq.\ (\ref{kappa}).    

\begin{figure}[h]
\unitlength1.0cm
\begin{center}
\begin{picture}(17.,10.)
\put(8.5,2.8){\footnotesize $\ds{\frac{U_B}{\La^3}}\, 
2\times 10^{15}$}
\put(2.2,2.3){\footnotesize $\ds{\frac{U_0}{\La^3}}\, 
2\times 10^{15}$}
\put(0.7,6.9){\footnotesize $\ds{\frac{
\rho_{0R}}{\La}} 10^{5}$}
\put(7.2,9.1){\footnotesize $\ds{\frac{
m_{R}^2}{\La^2}} 10^{9}$}
\put(12.2,8.2){\footnotesize $\tilde{m}_2^2 $}
\put(10.1,5.8){\footnotesize $\tilde{m}^2 $}
\put(10.2,4.4){\footnotesize $\kp$}
\put(6.4,-0.5){$t=\ln(k/\La)$}
\put(-0.7,0.){
\epsfysize=11.7cm
\epsfxsize=10.cm
\rotate[r]{\epsffile{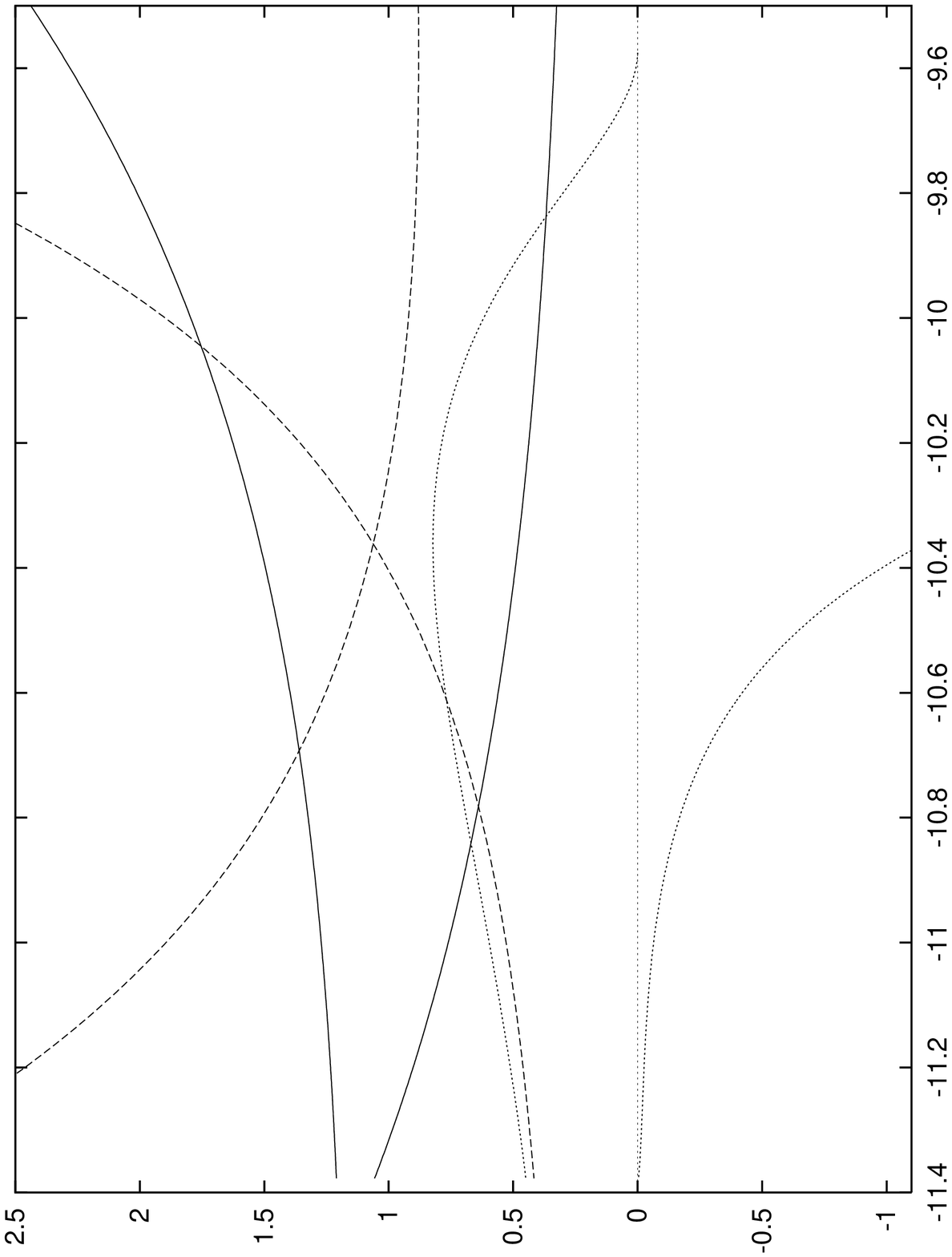
}}
}
\put(10.4,0.){
\epsfysize=5.5cm
\epsfxsize=10.cm
\rotate[r]{\epsffile{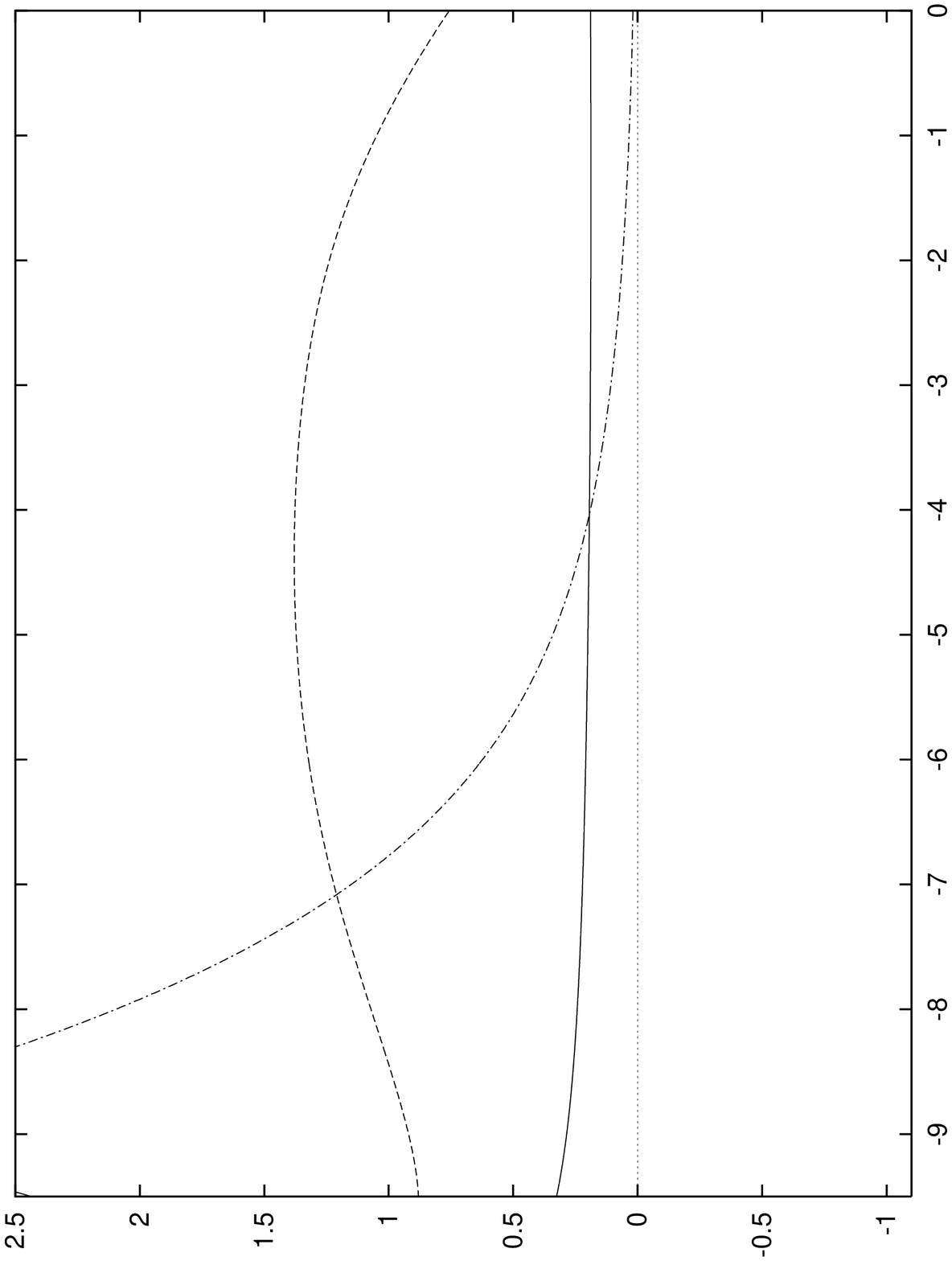
}}
}
\end{picture}
\end{center}
\caption[]{\footnotesize
Scale dependence of 
the dimensionless renormalized 
masses $\tilde{m}^2$, $\tilde{m}_2^2$, minimum $\kp$
and dimensionful counterparts 
$m_R^2=k^2 \tilde{m}^2$, $\rho_{0 R}=k \kp$ 
in units of $\La$. We also 
show $U_B(k)$ and
$U_0(k)$, the value of the potential at the top
of the potential barrier and at the minimum
$\rho_{0R}$, respectively. The short distance parameters are  
$\la_{1\La}=2$, $\la_{2\La}=0.1$ and $\dt \kp_{\La}=0$. 
\label{scaledep1}
}
\end{figure}

\begin{figure}[h]
\unitlength1.0cm
\begin{center}
\begin{picture}(15.,10.)
\put(8.,2.15){\footnotesize $\ds{\frac{U_B}{\La^3}}\, 
2\times 10^{3}$}
\put(3.6,1.5){\footnotesize $\ds{\frac{U_0}{\La^3}}\, 
2\times 10^{3}$}
\put(1.1,3.73){\footnotesize $\ds{\frac{
\rho_{0R}}{\La}}$}
\put(6.4,8.6){\footnotesize $\ds{\frac{
m_{R}^2}{\La^2}} 10^{2}$}
\put(11.5,7.8){\footnotesize $\tilde{m}_2^2 $}
\put(13.,3.15){\footnotesize $\tilde{m}^2 $}
\put(13.,4.4){\footnotesize $\kp$}
\put(6.4,-0.5){$t=\ln(k/\La)$}
\put(-0.7,0.){
\epsfysize=15.cm
\epsfxsize=10.cm
\rotate[r]{\epsffile{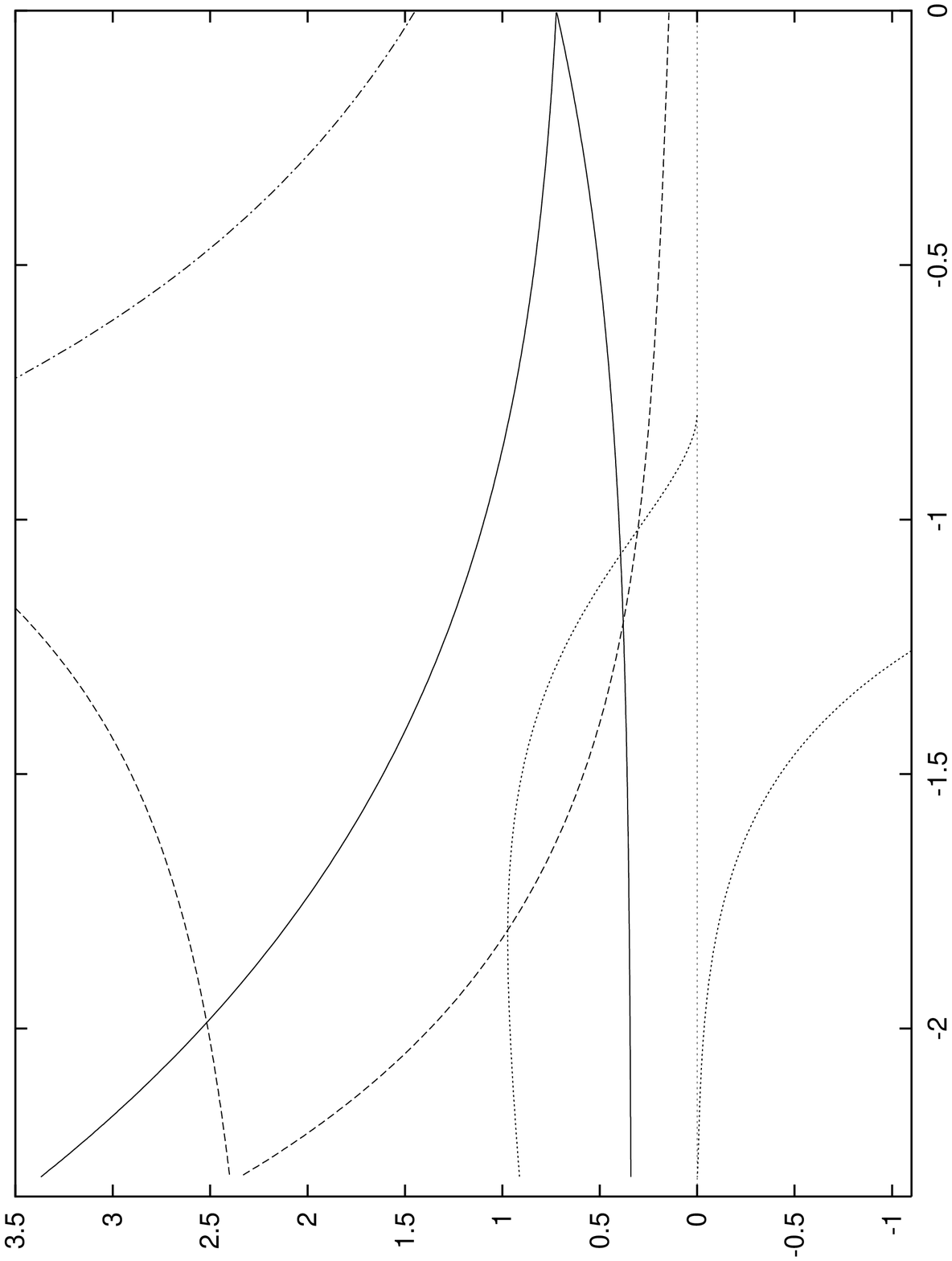
}}
}
\end{picture}
\end{center}
\caption[]{\label{scaledep2} \footnotesize
Same as fig.\ \ref{scaledep1}
for $\la_{1\La}=0.1$ and $\la_{2\La}=2$.
}
\end{figure}

In fig.\ \ref{scaledep1}, \ref{scaledep2} we follow the flow
of the dimensionless renormalized minimum $\kp$ and
the radial mass term $\tilde{m}^2=2 \kp \la_{1}(\kp)$ in
comparison to their dimensionful
counterparts $\rho_{0 R}=k \kp$
and $m_R^2=k^2 \tilde{m}^2$ in units of the momentum scale $\La$.
We also consider the dimensionless renormalized mass term
$\tilde{m}_2^2=\kp \la_{2} (\kp)$ corresponding to the curvature of the
potential in the direction of the second
invariant $\ttau$. The height of the potential 
barrier $U_B(k)=k^3 u_k(\trho_B)$ 
with $u_k^{\pri}(\trho_B)=0$, $0 < \trho_B < \kp$,
and the height
of the outer minimum $U_0(k)=k^3 u_k(\kp)$ 
is also displayed and will be discussed
in section \ref{coarse}. Fig.\ \ref{scaledep1} shows these quantities
as a function of $t=\mbox{ln}(k/\La)$ for $\la_{1 \La}=2$, 
$\la_{2 \La}=0.1$.
One observes that
the flow can be separated into two parts.
The first part ranging from $t=0$ to $t \simeq -6$
is characterized by $\kp \simeq \mbox{const}$ and small
$\tilde{m}_2^2$. 
It is instructive to consider what happens in the case
$\tilde{m}_2^2 \equiv 0$. In this case $\la_{2}\equiv 0$
and the flow is governed by the Wilson-Fisher fixed
point of the O(8) symmetric theory.
At the corresponding second order phase transition
the evolution of $u_k$ leads to the scaling
solution of (\ref{DlessEvol}) which obtains for
$\prl u_k/ \prl t=0$. As a consequence $u_k$ becomes a
$k$-independent function that takes on constant (fixed
point) values \cite{TW94-1,Mor}.
In particular, the minimum $\kp$ of the potential  
takes on its fixed point value
$\kp(k)= \kp_{\star}$. 
The fixed point is 
not attractive in the $U(2) \times U(2)$ symmetric theory and
$\la_{2\La}$ is an additional relevant parameter for the system.
For small $\la_2$ the evolution is governed by an anomalous
dimension $d\la_2/ d t=A \la_2$ with $A < 0$, leading to the
increasing $\tilde{m}_2^2$ as $k$ is lowered.

The system exhibits 
scaling behavior only for sufficiently small $\la_{2}$.
As $\tilde{m}_2^2$ increases the quartic coupling $\la_1$ and
therefore the radial mass term $\tilde{m}^2$ is driven to 
smaller values as can be observed from fig.\ \ref{scaledep1}.
For nonvanishing $\la_{2}$ the corresponding 
qualitative change in the flow equation 
(\ref{La1}) for $\la_1$ is the occurance of a term 
$\sim \la_2^2$.
It allows to drive $\la_1$ to negative values in a certain range
of $\trho<\kp$ and, therefore,
to create a potential barrier inducing a first order phase
transition. We observe from the plot that at $t \ltap -9.5$
a second minimum arises $(U_B \not = 0)$. The corresponding value
of $k=\La e^t=k_2$ sets a characteristic scale for the first order
phase transition. Below this scale the dimensionless,
renormalized  quantities approximately scale 
according to their canonical dimension.
The dimensionful quantities like $\rho_{0R}$ or
$m_R^2$ show only a weak scale dependence in this range. 
In contrast to the
above example of a weak first order phase transition 
with characteristic renormalized masses much smaller
than $\La$ fig.\
\ref{scaledep2} shows the flow of the corresponding quantities for
a strong first order phase transition. 
The short distance 
parameters employed are $\la_{1\La}=0.1$,
$\la_{2\La}=2$.
Here the range with 
$\kp \simeq \mbox{const}$ is almost absent and one observes
no approximate scaling behavior.   

With the above examples it becomes easy to understand
the phase structure presented in section \ref{ps}.
For the given curves of figs.\ \ref{phase} and \ref{ratio} we 
distinguished between the range $\la_{2\La}/\la_{1\La} \ll 1$
and $\la_{2\La}/\la_{1\La} \gg 1$ to denote the weak
and the strong first order region.
For $\la_{2\La}/\la_{1\La} \ll 1$ the initial   
renormalization group flow is dominated by the 
Wilson-Fisher fixed point of the $O(8)$ symmetric theory.
In this range 
the irrelevant couplings are driven close to 
the fixed point for some ``time'' $|t|=-\mbox{ln}(k/\La)$, loosing
their memory on the initial conditions given by the short
distance potential $u_{\La}$. As a consequence we are able
to observe universal behavior as is 
demonstrated in fig.\ \ref{ratio}.
 
To discuss the case $\la_{2\La}/\la_{1\La} \gg 1$ we consider the
flow equations for the couplings
at the minimum $\kp \not= 0$ of the potential
given by (\ref{kappa}) and (\ref{lakappa})
with (\ref{La1}), (\ref{La2}). 
In the limit of an infinite mass term 
$\tilde{m}_2^2=\kp \la_{2}(\kp) \to \infty$ the $\bt$-functions for
$\la_{1}(\kp)$ and $\kp$ become independent from $\la_{2}(\kp)$ due to
the threshold functions, with 
$l^3_n(\kp \la_{2}) \sim (\kp \la_{2})^{-(n+1)}$ for large
$\kp \la_{2}(\kp)$. 
As a consequence
$\bt_{\la_1}$ and $\bt_{\kp}$ equal the $\beta$-functions for an 
$O(5)$ symmetric model. 
We argue in the following that in this large
coupling limit fluctuations of massless Goldstone bosons
lead to an attractive fixed point for $\la_{2}(\kp)$.
We take the flow equation (\ref{lakappa}),
(\ref{La2}) for $\la_{2} (\kp)$ 
keeping only terms with positive canonical mass dimension
for a qualitative discussion. (This amounts to the 
approximation 
$\la_1^{(n)}(\kp)=\la_2^{(n)}(\kp)=0$ for $n \ge 1$.) 
To be explicit, one may consider the
case for given $\la_{1\La}=2$. The critical cutoff value
for the potential minimum is $\kp_{\La}\simeq 0.2$ for
$\la_{2\La}\gg 1$. For $\kp \la_2(\kp) \gg 1$ and taking 
$\eta \simeq 0$ the $\bt$-function for $\la_2 (\kp)$
is to a good approximation given by $(d=3)$
\be
\frac{\mbox{d} \la_{2} (\kp)}{\mbox{d} t} =
-\la_{2} (\kp)+2 v_3 (\la_{2} (\kp))^2 l^3_2(0).
\label{fpla2}
\ee
The second term on the r.h.s.\ of eq.\ (\ref{fpla2})
is due to massless Goldstone modes which give the
dominant contribution in the considered range. 
The solution of (\ref{fpla2})
implies an attractive fixed point for $\la_{2} (\kp)$ with a value
\be
\la_{2\star} (\kp)=\ds{\frac{1}{2 v_3 l^3_2(0)} } \simeq 4 \pi^2.
\ee 
Starting from $\la_{2\La}$ one finds for the ``time''
$|t|$ necessary
to reach a given $\la_{2} (\kp) > \la_{2\star} (\kp)$ 
\be
|t|= - \mbox{ln}
\ds{\frac{\la_{2} (\kp)-\la_{2\star} (\kp)}
{\la_{2} (\kp) \left(1-\frac{\ds{\la_{2\star} (\kp)}}{
\ds{\la_{2 \La}}} 
\right)} }\quad .
\ee
This converges to a 
finite value for $\la_{2 \La} \to \infty$.  
The further evolution therefore 
becomes insensitive
to the initial value for $\la_{2\La}$ in the large coupling
limit. The flow of $\la_{1}(\kp)$ and $\kp$ is not
affected by the initial running of $\la_{2}(\kp)$
and quantities like $\Dt \rho_{0R}/\La$ or 
$m_R/\Dt \rho_{0R}$ become independent of $\la_{2\La}$
if the coupling is sufficiently large. 
This qualitative
discussion is confirmed by the numerical solution of the full
set of equations presented in figs.\ \ref{phase} and 
\ref{ratio}.
For the fixed point value we obtain $\la_{2\star} (\kp)=38.02$.
We point out that an analogous discussion for the large
coupling region of $\la_{1 \La}$ cannot be made. This can be seen by
considering the mass term at the origin of the short
distance potential 
(\ref{uinitial}) given by 
$u^{\pri}_{\La}(0,0)=-\kp_{\La} \la_{1\La}$. Due to the
pole of $l^3_n(\om,\eta)$ at $\om = -1$ for $n > 1/2$
(cf.\ appendix \ref{poles}) one obtains the constraint
\be
\kp_{\La} \la_{1\La} < 1 \label{constraint}\quad.
\ee  
In the limit $\la_{1\La} \to \infty$
the mass term $2 \kp_{\La} \la_{1\La}$
at the minimum $\kp$ of the potential
at the critical temperature therefore remains finite.

\subsection{Universal equation of state for 
weak first order phase
transitions \label{sce}}

We presented in section \ref{ps} some characteristic
quantities for the effective average potential which
become universal at the phase transition 
for a sufficiently small 
quartic coupling $\la_{2\La}=\bar{\la}_{2\La}/\La$
of the short distance potential $U_{\La}$ (\ref{uinitial}). 
The aim of this section is to generalize this observation
and to find a universal scaling form of 
the equation of state for weak first order phase transitions.
The equation of state relates the derivative of the free energy
$U=\lim_{k\to 0}U_k$ to an external source, 
$\prl U/\prl \vp = j$.
Here the derivative has to be evaluated in the outer
convex region of the potential. For instance, for the meson model
of strong interactions the source $j$ is proportional
to the average quark mass \cite{PW84-1,QuMa} and the 
equation of state
permits to study the quark mass dependence of properties of the
chiral phase transition. We will compute the equation
of state for a nonzero coarse graining scale $k$. It 
therefore contains information for quantities
like the ``classical'' bubble surface tension in the
context of Langer's theory of bubble formation which will be
discussed in section \ref{coarse}.
  
In three dimensions the $U(2) \times U(2)$ symmetric 
model exhibits a second order phase transition in
the limit of a vanishing quartic coupling ${\la}_{2\La}$ due to
an enhanced $O(8)$ symmetry. In this case  
there is no scale present in the theory at the critical
temperature. 
In the vicinity of the critical temperature
(small $|\dt \kp_{\La}| \sim |T_c-T|$) and for
small enough $\la_{2\La}$
one therefore expects a scaling behavior of
the effective average potential $U_k$ and
accordingly a universal scaling form of the equation of state. 
At the second order phase transition in the 
$O(8)$ symmetric model there are only two independent
scales that can be related to the deviation from
the critical temperature and to the external source
or $\vp$. 
As a consequence the properly rescaled potential $U/ \rho_R^{3}$
or $U/ \rho^{(\dt +1)/2}$ (with the usual critical exponent
$\dt$) can only depend on one dimensionless ratio.
A possible set of variables
to represent the two independent scales are the renormalized
minimum of the potential $\vp_{0R}=(\rho_{0R}/2)^{1/2}$ 
(or the renormalized mass for the symmetric phase) and
the renormalized field $\vp_R=(\rho_{R}/2)^{1/2}$. The 
rescaled potential will then only depend on the 
scaling variable $z=\vp_R/\vp_{0R}$ \cite{BTW95}. Another
possible choice is the Widom scaling variable
$x=-\dt \kp_{\La}/\vp^{1/\bt}$ \cite{Widom}.  
In the $U(2) \times U(2)$ symmetric 
theory $\la_{2\La}$ is an additional relevant
parameter which renders the phase transition first
order and introduces a new scale, e.g.\ the nonvanishing
value for the jump in the renormalized order parameter
$\Dt \vp_{0R}=(\Dt \rho_{0R}/2)^{1/2}$ at the critical 
temperature or $\dt \kp_{\La}=0$.
In the universal range we therefore observe three
independent scales and the scaling form of the equation
of state will depend on two dimensionless ratios. 
The rescaled potential
$U/\vp_{0R}^6$ can then be written as a universal function $G$
\be
\frac{U}{\vp_{0R}^6}=G(z,v)
\label{scalingeos}
\ee
which depends on the two scaling variables
\be
z=\frac{\vp_R}{\vp_{0R}},\quad v= \frac{\Dt \vp_{0R}}{\vp_{0R}}\,\,.
\ee
The relation (\ref{scalingeos}) is the scaling 
form of the equation of
state we are looking for. At a second order phase
transition the variable $v$ vanishes identically and 
$G(z,0)$ describes the scaling equation of state
for the model with $O(8)$ symmetry \cite{BTW95}.
The variable $v$ accounts for the additional scale 
present at the first order phase transition.
We note that $z=1$ corresponds to a vanishing source
and $G(1,v)$ describes the temperature dependence
of the free energy for $j=0$. In this case $v=1$ denotes the 
critical temperature $T_c$ whereas for $T < T_c$
one has $v < 1$. Accordingly $v > 1$ obtains for
$T > T_c$ and $\vp_{0R}$ describes here the local 
minimum corresponding to the metastable ordered phase. 
The function $G(z,1)$ accounts for the
dependence of the free energy on $j$ for $T=T_c$.

\begin{figure}[h]
\unitlength1.0cm
\begin{center}
\begin{picture}(13.,9.)
\put(-0.6,4.4){$G_{k_f}$}
\put(6.4,-0.5){$z$}
\put(-0.5,0.){
\epsfysize=13.cm
\epsfxsize=9.cm
\rotate[r]{\epsffile{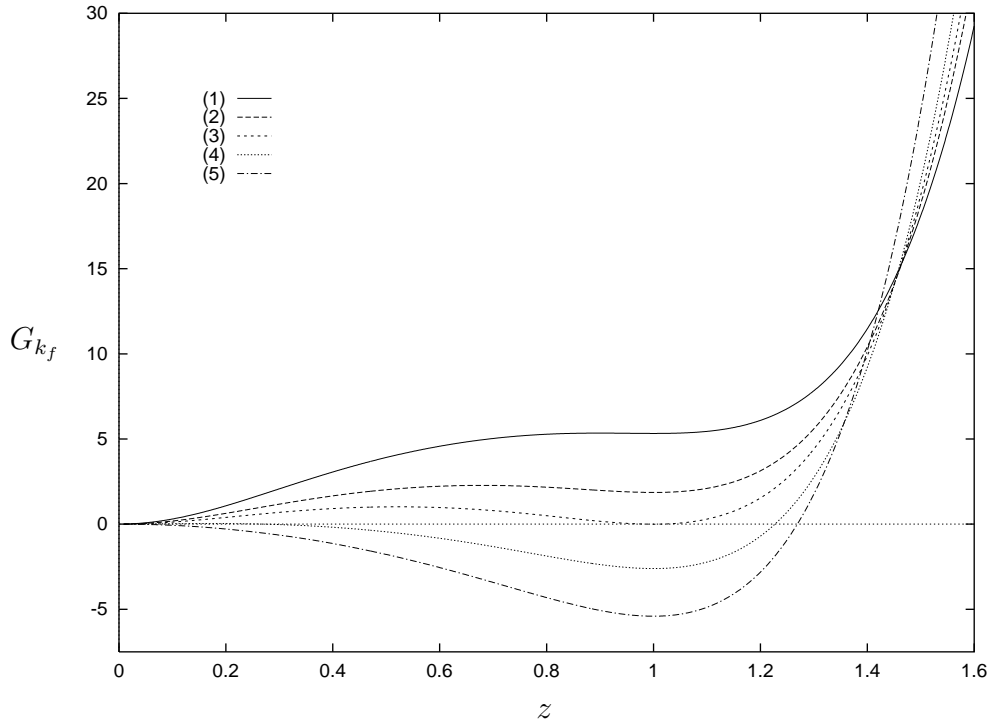
}}
}
\end{picture}
\end{center}
\caption[]{\label{scalfu} \footnotesize
Universal shape of the coarse grained potential $(k=k_f)$
as a function of the scaling variable $z=\vp_R/\vp_{0R}
=(\rho_R/\rho_{0R})^{1/2}$ for different values of
$v=\Dt \vp_{0R}/\vp_{0R}=(\Dt \rho_{0R}/\rho_{0R})^{1/2}$.
The employed values for $v$ are (1) $v=1.18$,
(2) $v=1.07$, (3) $v=1$, (4) $v=0.90$, (5) $v=0.74$. 
For vanishing sources one has $z=1$. In this case $v=1$
denotes the critical temperature $T_c$. Similarly
$v > 1$ corresponds to $T > T_c$ with $\vp_{0R}$
denoting the minimum in the metastable ordered phase.
}
\end{figure}

We consider the scaling form (\ref{scalingeos}) of the
equation of state for a nonzero coarse graining scale $k$.
The renormalized field is given by $\vp_R=Z_k^{1/2} \vp$.
We pointed out in section \ref{ps} that there is a
characteristic scale $k_2$ for the first order phase
transition where the second local minimum of the effective
average potential appears. For weak first order phase
transitions one finds $\rho_{0R} \sim k_2$. To observe
the scaling form of the equation of state the infrared cutoff
$k$ has to run below $k_2$ with $k \ll k_2$. For
the scale $k_f$ defined in eq.\ (\ref{fix})
we observe universal behavior to high
accuracy (cf.\ fig.\ \ref{ratio} and the corresponding universal 
ratios in table \ref{table1} for small $\la_{2\La}/\la_{1\La}$).
The result for the universal function 
$U_{k_f}/\vp_{0R}^6=G_{k_f}(z,v)$ is presented in fig.\
\ref{scalfu}. 
For $v=1$ one has $\vp_{0R}(k_f)=\Dt \vp_{0R}(k_f)$ 
which denotes the critical temperature.
Accordingly $v > 1$ denotes temperatures above 
and $v < 1$ temperatures below the critical temperature.
One observes that $G_{k_f}(z,1)$ 
shows two almost degenerate minima. (They become exactly
degenerate in the limit $k \to 0$).
For the given examples $v=1.18$, $1.07$
the minimum at the origin becomes
the absolute minimum and the system is in the
symmetric phase. In contrast, for $v = 0.90$, $0.74$ the 
absolute minimum is located at $z=1$ which
characterizes the spontaneously broken phase.
For small enough $v$ the local minimum at the 
origin vanishes.

We explicitly verified that the universal
function $G_{k_f}$ depends only 
on the scaling variables $z$ and $v$ by choosing various
values for $\dt \kp_{\La}$ and for the quartic couplings of
the short distance potential, $\la_{1\La}$ and $\la_{2\La}$.
In section \ref{ps} we observed that the model shows
universal behavior for a certain range of the parameter
space. For given $\la_{1\La}$ and small enough 
$\la_{2\La}$ one always observes universal behavior.
For $\la_{1\La}=0.1$, $2$ and $4$ it is 
demonstrated that (approximate) universality holds
for $\la_{2\La}/ \la_{1\La} \, \ltap \, 1/2$ 
(cf.\ fig.\ \ref{ratio}). 
For $\la_{1\La}$ around $2$ one observes 
from figs.\ \ref{phase}, \ref{ratio} and table \ref{table1} that
the system is to a good accuracy described by
its universal properties for even larger values 
of $\la_{2\La}$. The corresponding phase transitions
cannot be considered as particularly weak first order. 
The universal function $G_{k_f}$
therefore accounts 
for a quite large range of the parameter space.

We emphasize that the universal form of the 
effective potential given in fig.\ \ref{scalfu} depends on
the scale $k_f$ where the integration of the flow 
equations is stopped (cf.\ eq.\ (\ref{fix})).  
A different prescription
for $k_f$ will, in general, 
lead to a different
form of the effective potential. We may interpret
this as a scheme dependence describing the effect of 
different coarse graining procedures. This is fundamentally
different from nonuniversal corrections since $G_{k_f}$
is completely independent of details of the short distance
or classical action and in this sense universal.
A more quantitative discussion of this scheme
dependence will be presented in the next section.
We note that fluctuations on scales 
$k < k_f$ do not influence substantially the location of
the minima of the coarse grained potential and the
form of $U_k(\vp_R)$ for $\vp_R > \vp_{0R}$ remains almost
$k$-independent. Here
$\prl U_{k_f}/\prl \vp=j(k_f)$ with
$j(k_f) \simeq \lim_{k \to 0} j(k) = j$.\footnote{
The role of massless
Goldstone boson fluctuations for the 
universal form of the effective average potential
in the limit $k \to 0$ has been discussed previously
for the $O(8)$ symmetric model \cite{BTW95}.}

We have established the scaling equation of state using the
renormalized variables $\vp_R$, $\vp_{0R}$ and
$\Dt \vp_{0R}$. Alternatively the scaling equation
of state can be presented in terms of the short
distance parameters $\dt \kp_{\La}$, $\la_{2\La}$
and the unrenormalized field $\vp$. For the 
$O(8)$ symmetric model $(\la_{2\La}=0)$ this is
known as the Widom scaling form of the equation of state
\cite{Widom} and in this case the relation between
$\vp_R$, $\vp_{0R}$ and $\vp$, $\dt \kp_{\La}$
is solely determined by critical exponents 
and amplitudes \cite{BTW95}. For the 
$U(2) \times U(2)$ symmetric model the dependence
of $\vp_R$, $\vp_{0R}$ and $\Dt \vp_{0R}$
on the parameters $\dt \kp_{\La}$, $\la_{2\La}$
and the unrenormalized field $\vp$ can be expressed
in terms of scaling functions. 
In general, only for limiting
cases as $\dt \kp_{\La}=0$ or $\la_{2\La}=0$ the
relation is determined by critical exponents 
and amplitudes. We will consider here these limits. 

We consider the
renormalized minimum $\vp_{0R}$ in two limits which are
denoted by  $\Dt \vp_{0R} = \vp_{0R}(\dt \kp_{\La}=0)$
and $\vp^0_{0R} = \vp_{0R}(\la_{2\La}=0)$.  
The behavior of $\Dt \vp_{0R}$ is described in
terms of the exponent $\theta$ according to
eq.\ (\ref{Powrho}),
\be
\Dt \vp_{0R} \sim (\la_{2\La})^{\theta/2}, \qquad \theta=1.93.
\label{scath}
\ee
The dependence of the minimum $\vp^0_{0R}$ of the
$O(8)$ symmetric potential on the temperature is 
characterized by the critical exponent $\nu$,
\be
\vp^0_{0R} \sim (\dt \kp_{\La})^{\nu/2}, \qquad \nu=0.882.
\label{scanu}
\ee
The exponent $\nu$ for the $O(8)$ symmetric model
is determined analogously to $\theta$ as described in section
\ref{phase} \footnote{For the $O(8)$ symmetric model
($\la_{2\La}=0)$ we consider the 
minimum $\vp^0_{0R}$ at $k=0$.}. 
We can also introduce a critical exponent $\zeta$ for the
jump of the unrenormalized order parameter
\be
\Dt \vp_{0} \sim (\la_{2\La})^{\zeta}, 
\qquad \zeta=0.988 \, .
\ee
With
\be
\vp^0_{0} \sim (\dt \kp_{\La})^{\beta}, 
\qquad \beta=0.451
\ee
it is related to $\th$ and $\nu$ by the additional index
relation
\be
\frac{\theta}{\zeta}=\frac{\nu}{\beta}=1.95 \, . 
\ee   
We have verified this numerically. 
For the case $\dt \kp_{\La}=\la_{2\La}=0$ one obtains
\be
j \sim \vp^{\dt}.
\ee
The exponent $\dt$ is related to the 
anomalous dimension $\eta$ via the usual
index relation $\dt=(5-\eta)/(1+\eta)$. From the
scaling solution of eq.\ (\ref{EtaSkale}) we
obtain $\eta=0.0224$.

With the help of the above relations one immediately
verifies that for $\la_{2\La}=0$ 
\be
z \sim (-x)^{-\bt}\, , \qquad v=0
\ee
and for $\dt \kp_{\La}=0$
\be
z \sim y^{-\zeta} \, , \qquad v=1 \, .
\ee
Here we have used the Widom scaling variable $x$ and the 
new scaling variable $y$ given by
\be
x=\frac{-\dt \kp_{\La}}{\vp^{1/\bt}}\, , 
\qquad y=\frac{\la_{2\La}}{\vp^{1/\zeta}}\, .
\ee

\subsection{Coarse graining and Langer's theory of bubble nucleation
\label{coarse}}

The coarse grained effective potential $U_k$ results from
the integration of fluctuations with momenta larger
than $\sim k$. It is a nonconvex function whereas the 
standard effective potential $U=\lim_{k\to 0} U_k$
has to be convex by its definition as a Legendre transform
(cf.\ section \ref{ExactFlowEquation}).
The difference between $U$ and $U_k$
is due to fluctuations with
characteristic length scales larger than the inverse
coarse graining scale $\sim k^{-1}$. The role of these
fluctuations for the approach to convexity has been
established explicitly \cite{RTW,BW97-1,BTW97-1}.
The study of
first order phase transitions usually relies on the
nonconvex ``part'' of the potential. As an 
illustration we consider the change of the system
from the high temperature to the low temperature
phase by bubble nucleation as described by Langer's theory
\cite{Langer}. On the one hand, the approach relies on
the definition of a suitable
coarse grained free energy $\Gm_k$ with a coarse graining
scale $k$ and, on the other hand, a saddle point
approximation for the treatment of fluctuations around
the ``critical bubble'' is employed.
The problem is therefore separated into two parts:
One part concerns the treatment of fluctuations with 
momenta $q^2 \, \gtap \, k^2$ which are included in the
coarse grained free energy. The second part deals with an 
estimate of fluctuations around the bubble
for which only fluctuations with momenta smaller than
$k$ must be considered. These issues will be discussed 
in this section in a quantitative way. In particular, we will
give a criterion for the validity of Langer's 
formula. 

One may consider a system that starts at some high temperature
$T > T_c$ and investigate what happens as $T$ is lowered
as a function of time as for example during the evolution
of the early universe \cite{KoTu}. For large enough temperature
the origin of the potential $(\vp=0)$ is the only 
minimum and the system is therefore originally in the
symmetric phase. As $T$ approaches $T_c$ a second local
minimum develops at $\vp_0 > 0$. This becomes the
absolute minimum below $T_c$. Nevertheless, the potential
barrier prevents the system to change smoothly to the
ordered phase. For a short while where $T$ is in the vicinity
of $T_c$ but below $T_c$ the system remains therefore 
in a state with higher energy density as
compared to the state corresponding to the absolute minimum
away from the origin. This is the so-called ``false vacuum'' in
high energy physics or the metastable state in statistical 
physics. Such a state is unstable with respect to
fluctuations which  penetrate or cross the barrier.
The picture is familiar from the condensation of vapor.
The false vacuum corresponds to the supercooled
vapor phase and the true vacuum to the fluid phase.
Bubbles of the true vacuum (droplets) occur through thermal or
quantum fluctuations\footnote{
In the real world the condensation of vapor is triggered
by impurities but this is not the issue here.}. 
If a bubble is large enough 
so that the decrease in volume energy exceeds the 
surface energy it will grow. The phase transition is 
completed once the whole 
space is filled with the true vacuum. 
On the other hand, small bubbles shrink due to the 
surface tension. The
critical bubble is just large enough that it
does not shrink.
To be explicit we consider a spherical bubble
where the bubble wall with ``thickness'' $\Dt$
is thin as compared to the bubble radius $R$, i.e.
$\Dt \ll R$.
In leading order the coarse grained free energy $\Gm_k$ 
for such a bubble configuration
can be decomposed
in a volume and a surface term \cite{ColCal,Lin}
\be
\Gm_k^{(0)} = -\frac{4 \pi}{3}  R^3 \eps + 4 \pi R^2 \sigma_k.
\label{dec}
\ee
In the thin wall approximation one obtains for the 
surface tension $\sigma_k$ in our conventions
\be
\sigma_k=2\int\limits^{\vp_0}_0 d\vp
\sqrt{2 Z_k U_k(\vp)}.
\label{sur}
\ee
For the difference in the free energy density $\eps$ one 
has
\be
\eps=U(0)-U(\vp_0)=-\lim_{k \to 0} U_0(k).
\label{vo}
\ee
We include in $\eps$ fluctuations with arbitrarily
small momenta. In contrast, the long wavelength
contributions to the true surface tension\footnote{
The true surface tension is a ``measurable quantity''.
It is independent of $k$ and all fluctuations 
must be included. It therefore differs, in general,
from $\si_k$ which includes only part of the fluctuations.} 
are effectively cut off by the 
characteristic length scale of the bubble surface.
We include for the computation of the ``classical''
surface tension $\si_k$ only fluctuations with 
$q^2 \, \gtap \, k^2$. The modes with $q^2 \, \ltap \, k^2$ 
contribute to the ``fluctuation determinant''
$A_k$ (cf.\ eq.\ (\ref{nc})). 
The determination of a suitable
coarse graining scale $k$ for the computation of 
$\si_k$ is discussed below.

The critical bubble maximizes $\Gm_k^{(0)}$ with respect to the
radius. It minimizes the coarse grained free energy
with respect to other deformations because the spherical 
form is energetically favorable. The critical bubble
therefore represents a saddle point in the space of
possible ``bubble configurations''. 
In Langer's theory of bubble formation
one considers a saddle point expansion 
around the critical bubble. There is exactly one negative 
mode that corresponds to the shrinking or growth
of the bubble and there are infinitely many
positive modes (there are also translational
zero modes) 
\footnote{
Langer's theory is not restricted to the thin wall approximation
which is considered here for simplicity. In particular,
the property of the critical bubble to 
represent a saddle point with exactly one negative mode is 
independent from the thin wall approximation.
}. The bubble nucleation
rate $\bar{\Gm}$, which describes the probability
per unit volume per unit time for the transition
to the new vacuum, 
can be written in the form \cite{ColCal,Lin,Langer}
\be
\bar{\Gm} = A_k \, \exp (-\Gm^{(0)}_k [\vp^{(0)}_b]) 
=A_k \, \exp \left(-\frac{16 \pi}{3}\frac{\sigma_k^3}
{\eps^2}\right)
\label{nc}
\ee 
where $\Gm_k^{(0)}$ is evaluated for $\vp^{(0)}_b$ corresponding to
the critical bubble and approximated by (\ref{dec}).
The exponential term with the coarse grained free energy
$\Gm^{(0)}_k$ denotes the lowest order or classical
contribution.
The prefactor $A_k$ contains several
factors that depend on the details of the system
under investigation. 
In particular, $A_k$ accounts for the contribution
to the free energy from the fluctuations
with momenta smaller than $k$.
It depends on $k$ through the effective ultraviolet
cutoff for these fluctuations which is present
since fluctuations with momenta larger than $k$
are already included in 
$\Gm^{(0)}_k[\vp^{(0)}_b]$~\footnote{
The effective average action \cite{Wet91-1} also provides the
formal tool how the ultraviolet cutoff  $\sim k$
is implemented in the remaining functional
integral for large length scale fluctuations.}. 
Langer's formula for
bubble nucleation amounts essentially to a perturbative
one loop estimate of $A_k$.

For a determination of a useful choice of $k$
it is convenient to place the discussion in a more
general context which does not rely on the thin wall 
approximation or a saddle point approximation.
What one is finally interested in is the free energy
$\Gm[\vp_b]$ for bubble configurations of a given shape.
The ``true critical bubble'' $\vp_b^c$
corresponds to a saddle point
in the space of ``bubble configurations'' which are
characterized by boundary conditions connecting the
false and the true vacuum. The nucleation rate
is then proportional $\exp-\Gm[\vp_b^c]$. The coarse 
graining can be seen as a convenient strategy to
evaluate $\Gm[\vp_b]$ by separating contributions from 
different momentum scales. We propose to choose
the coarse graining scale $k$ somewhat above
but in the vicinity of the inverse thickness $\Dt^{-1}$
of the bubble wall. We will argue below that in
this case the corrections to the effective surface tension
from the prefactor $A_k$ should be best accessible.

In fact, we can write $\bar{\Gm} = B \, \exp-\Gm[\vp_b^c]$
where $B$ contains dynamical factors and $\Gm[\vp_b^c]$
does not include contributions from fluctuations of the
negative mode and the translational modes present for the
critical bubble. The prefactor in eq.\ (\ref{nc}) can then be
written as
\be
A_k=B \exp-(\dt_k+\eta_k)
\ee
where
\be
\dt_k= \Gm[\vp_b^c]-\Gm_k[\vp_b^c]\, ,\qquad 
\eta_k= \Gm_k[\vp_b^c]-\frac{16 \pi}{3}\frac{\sigma_k^3}
{\eps^2}\, .  
\ee
The term $\eta_k$ includes the difference between
the true critical bubble and the configuration used
to estimate $\sigma_k$ as well as a correction term 
to $\eps$ to be discussed below. We first concentrate 
on $\dt_k$ which describes the difference between the
free energy and the coarse grained free energy for
the critical bubble. As mentioned above this is due 
to fluctuations with momenta $q^2\, \ltap \, k^2$
and incorporates the dominant $k$-dependence of $A_k$.
Since the bubble provides for inherent effective
infrared cutoff scales  
$\sim R^{-1}$  
or $\Dt^{-1}$ the
contribution $\dt_k$ is both infrared and 
ultraviolet finite. The larger $k$, the more 
fluctuations are included in $\dt_k$ and from this
point of view one wants to take $k$ as low as possible.
On the other hand, $k$ should not be taken smaller than
$\Dt^{-1}$ if the approximation used for a computation 
of $\Gm_k$ relies on almost constant field configurations
rather than real bubbles, as is usually the case. Only
for $k$ sufficiently large compared to $\Dt^{-1}$ the
difference between an evaluation of the potential
and kinetic terms in $\Gm_k$ for almost constant field 
configurations (e.g.\ by a derivative expansion) 
rather than for bubbles remains small. In this way the 
technique of course graining combines a relatively simple 
treatment of the modes with $q^2 \, \gtap \, k^2$
for which the detailed properties of the bubble are 
irrelevant with an estimate of fluctuations around the
bubble for which the short distance physics
($q^2 \, \gtap \, k^2$) needs not to be considered anymore.
It is clear that $k$ is only a technical construct
and for physical quantities the $k$-dependence of
$\dt_k$ and $\Gm_k^{(0)}$ must cancel. More 
precisely, this concerns the sum 
$\dt_k+\eta_k+16 \pi \sigma_k^3/3 \eps^2$ .
For thin wall bubbles the most important contribution
to $\eta_k$ is easily identified: By our definition
of $\eps$ we have included contributions from fluctuations
with length scales $\gtap \, R$. They should not be
present in the effective action for a bubble with finite 
radius. Therefore $\eta_k$ contains a correction term
$(16 \pi/3)\sigma_k^3(\eps^{-2}(R)-\eps^{-2})$
which replaces effectively $\eps$ by $\eps(R)$ in eq.\ 
(\ref{nc}). We can evaluate $\eps(R)$ in terms of the
coarse grained free energy at a scale $k_R$
\be
\eps(R) \simeq U_{k_R}(0)-U_{k_R}(\vp_0)=-U_0(k_R)\, ,
\qquad
k_R = \ds{\frac{1}{R}} \,\, .
\ee
For $\Dt \ll R$ one should not confound $k_R$
with the coarse graining scale $k$ since one has the
inequality
\be
k_R \ll \frac{1}{\Dt} \,\, \ltap \,\, k\,\, .
\ee
Only for $\Dt \simeq R$ the clear separation between
$k_R$ and $k$ disappears. We note that at the critical
temperature one has $R \to \infty$ and therefore 
$\eps(R)=\eps$\, .
  
Since $\sigma_k$ enters the 
nucleation rate
(\ref{nc}) exponentially even small changes with $k$
will have large effects. If one finds a strong
dependence of $\si_k$ on the
coarse graining scale $k$ this is only compatible with
a large contribution from the higher orders
of the saddle point expansion. The $k$-dependence of
$\si_k$ therefore
gives direct information about the validity of 
Langer's formula. We find a strong scale
dependence of $\si_k$ if the phase transition is characterized
by large effective
dimensionless couplings.
A weak scale dependence is observed for small effective
couplings. This gives a very consistent
picture: The validity of the saddle point
approximation typically requires small dimensionless couplings.
In this case also the details of the
coarse graining are not of crucial importance
within an appropriate range of $k$. The
remaining part of this section provides a detailed
quantitative discussion.

We consider the dependence of
the effective average potential $U_k$ 
and the surface tension $\sigma_k$ on the coarse
graining scale $k$ near the critical temperature $T_{c}$
for three examples in detail.
They are 
distinguished by different choices for the quartic couplings
$\bar{\la}_{1\La}$ and $\bar{\la}_{2\La}$ of the short
distance potential $U_{\La}$ given by eq.\ (\ref{uinitial}).
The choice $\bar{\la}_{1\La}/\La=0.1$ and 
$\bar{\la}_{2\La}/\La=2$ corresponds to a strong first order
phase transition with renormalized masses
not much smaller than the cutoff 
scale $\La$.
The renormalized couplings will turn out
small enough such that the 
notion of a coarse grained potential $U_k$ 
and a surface tension $\si_k$ can be used
without detailed information on the coarse graining
scale within a certain range of $k$. In contrast
we give two examples where the dependence of $U_k$
and $\si_k$ on the coarse graining scale becomes of crucial
importance.
The choice $\bar{\la}_{1\La}/\La=2$ and 
$\bar{\la}_{2\La}/\La=0.1$ leads to a weak first
order phase transition with small renormalized masses
and the system shows universal behavior
(cf.\ sect. \ref{ps}). For 
$\bar{\la}_{1\La}/\La=4$ and 
$\bar{\la}_{2\La}/\La=70$ one observes a 
relatively strong
first order phase transition. Nevertheless for 
both examples the coarse grained potential 
and the surface tension show 
a similarly high sensitivity on the scale $k$.
 
\begin{figure}[h]
\unitlength1.0cm
\begin{center}
\begin{picture}(13.,9.)
\put(-0.8,4.6){$\ds{\frac{U_k}{\La^3}}\, 10^3$}
\put(6.3,-0.5){$\ds{\frac{\vp_R}{\La^{1/2}}}$}
\put(-0.4,0.){
\epsfysize=13.cm
\epsfxsize=9.cm
\rotate[r]{\epsffile{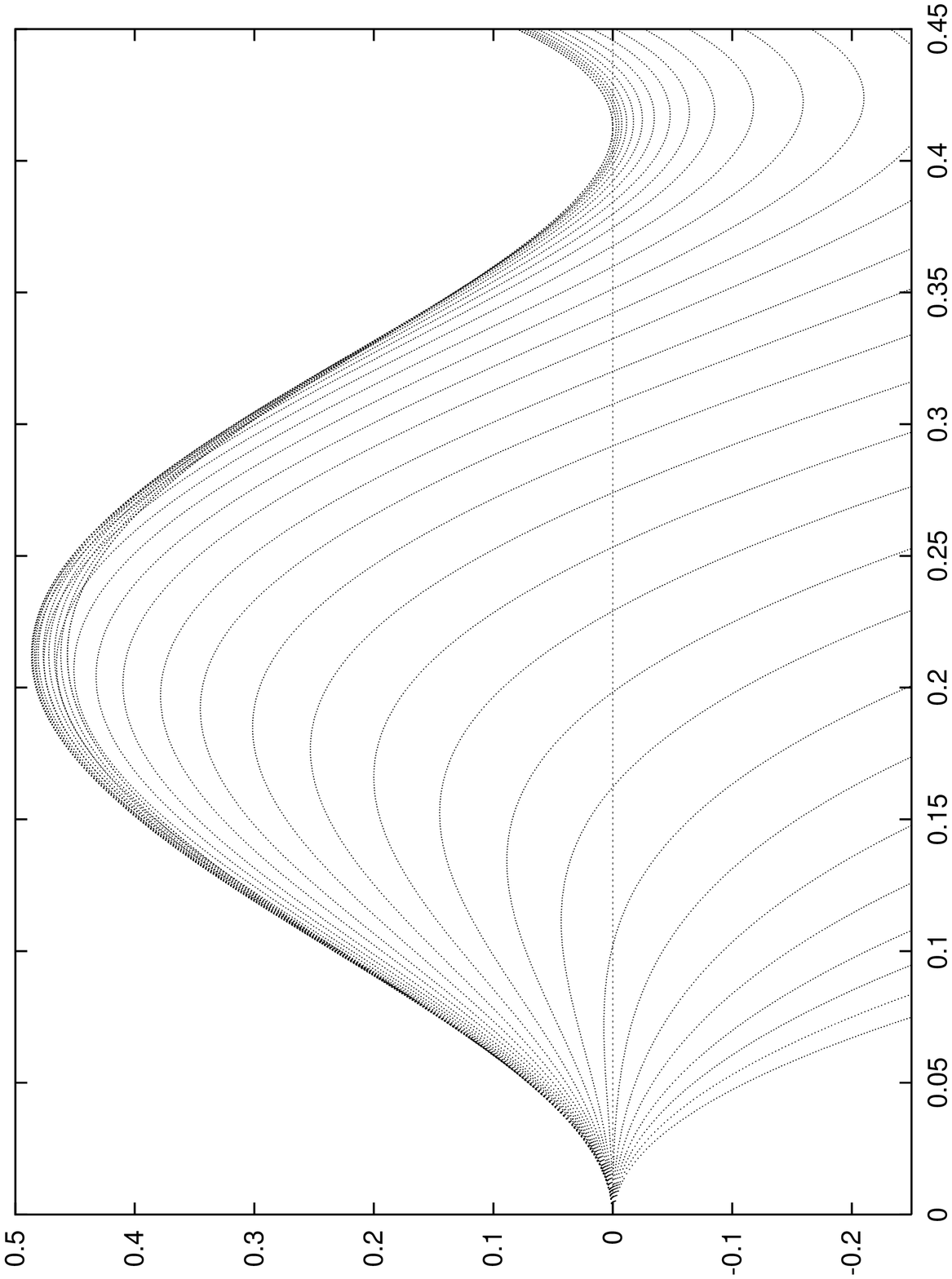
}}
}
\end{picture}
\end{center}
\caption[]{\footnotesize The coarse grained   
potential in dependence on the coarse graining scale $k$
for fixed (almost critical) temperature.
The example corresponds to
$\bar{\la}_{1\La}/\La=0.1,\bar{\la}_{2\La}/\La=2$.
\label{innercoarse}
}
\end{figure}

\begin{figure}[h]
\unitlength1.0cm
\begin{center}
\begin{picture}(13.,9.)
\put(-0.8,4.6){$\ds{\frac{U_k}{\La^3}}\, 10^{15}$}
\put(6.3,-0.5){$\ds{\frac{\vp_R}{\La^{1/2}}}$}
\put(-0.4,0.){
\epsfysize=13.cm
\epsfxsize=9.cm
\rotate[r]{\epsffile{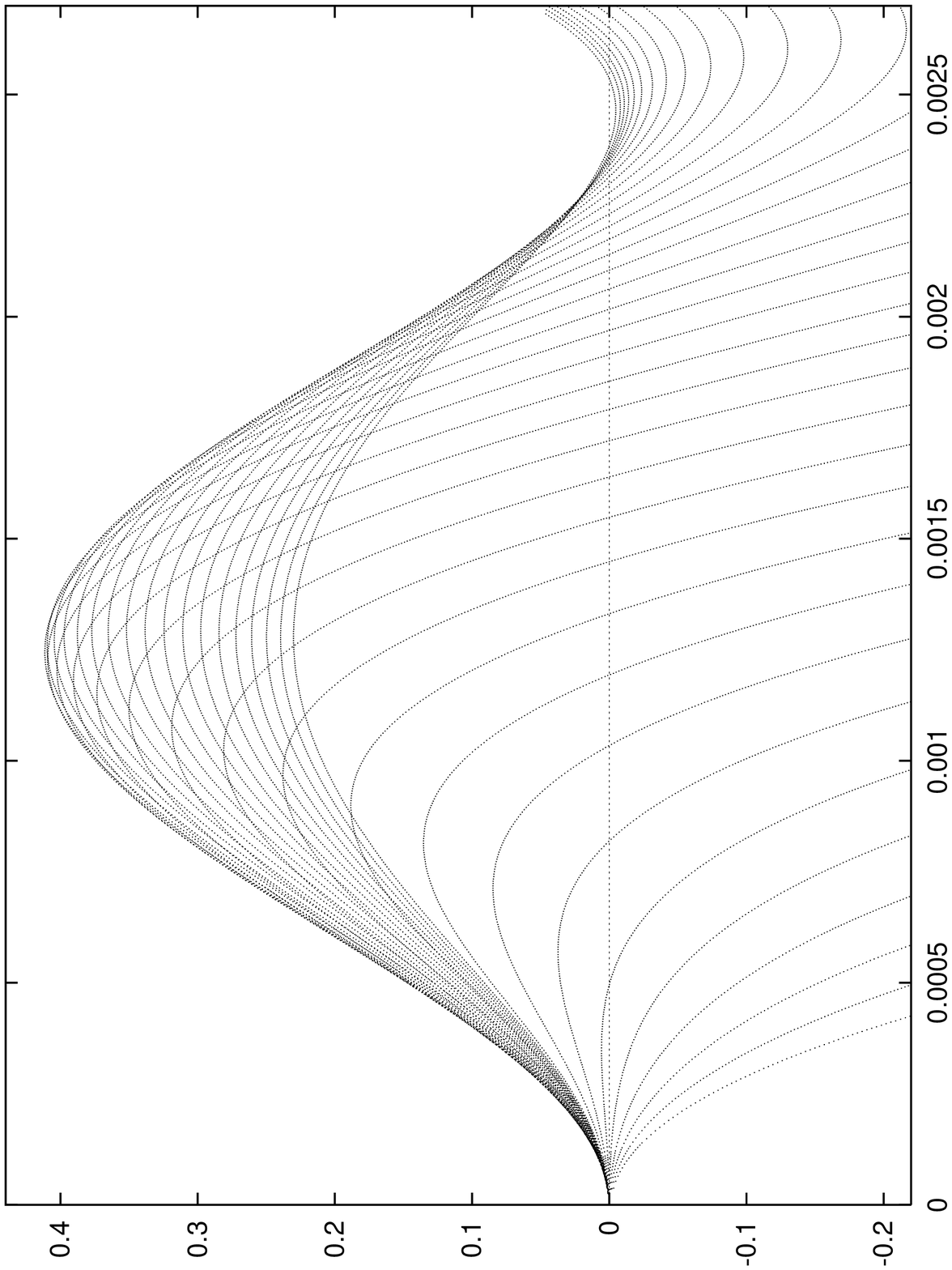
}}
}
\end{picture}
\end{center}
\caption[]{\footnotesize The coarse grained potential for
$\bar{\la}_{1\La}/\La=2,\bar{\la}_{2\La}/\La=0.1$.
\label{innercoarse2}
}
\end{figure}

In figs.\  \ref{innercoarse}, \ref{innercoarse2}
the scale dependence of $U_k$
is shown for a fixed temperature in the 
vicinity of $T=T_{c}$ or $\dt \kp_{\La} =0$.
We plot $U_k$ in units of $\La^3$ as a function of 
$\vp_R/\La^{1/2} = (\rho_{R}/2\La)^{1/2}
= (Z_k \rho/2\La)^{1/2}$ 
for different values of $k$. Each curve differs
in $k=\La e^{t}$ 
by $\Dt t=1/18$ and
the first curve to
the left with a negative curvature at the origin
corresponds to $t \simeq -0.40$ 
for fig.\ \ref{innercoarse} 
($t \simeq -9.3$ for fig.\ \ref{innercoarse2}).
For $0 \ge t \, \gtap \, -0.40$ ($0 \ge t \, \gtap \, -9.3$)
there is only one minimum of the potential
away from the origin which lies below the plotted
range in fig.\ \ref{innercoarse}
(\ref{innercoarse2}). Lowering $k$ 
results in a successive
inclusion of fluctuations on larger length scales
and one observes the appearance
of a second local minimum at the origin of the
potential. The barrier between both minima increases
and reaches its maximum value. 
As $k$ is lowered further the potential 
barrier starts to
decrease whereas the location of
the minima becomes almost $k$-independent
and degenerate in height. 
In this region also the outer convex part of the
potential shows an almost stable profile due to the 
decoupling of the massive modes. 
We stop the integration at the scale
$k=k_f$ given by eq.\ (\ref{fix}).
The coarse grained effective potential $U_k$
with $k=k_f$ essentially includes all fluctuations
with squared momenta larger than the scale $|m_{B,R}^2|$
given by the curvature at the top of the potential barrier
(cf.\ eq.\ (\ref{mbr})). Successive inclusion of fluctuations
with momenta smaller than $k_f$ would result in a further
decrease of the potential 
barrier. The flattening of the barrier
is induced by the pole structure of the
threshold functions $l^3_n(\om)$ appearing
in the flow equations for the couplings (e.g.\ $l^3_1(\om)$ 
appearing in eq.\ (\ref{La1}) exhibits a pole at
$\om=-1$ according to $l^3_1(\om) \sim (\om+1)^{-1/2}$
for $\om$ near $-1$ (cf.\ appendix \ref{poles})).
The argument $\om$ corresponds
to dimensionless mass terms as $U_k^{\pri}(\rho_R)/k^2$ or
$(U_k^{\pri}(\rho_R)+2\rho_R U_k^{\pri\pri}(\rho_R))/k^2$. 
In the nonconvex region of the potential the curvature 
is negative. Since the pole
cannot be crossed, negative $U_k^{\pri}$ or 
$U_k^{\pri\pri}$ must go to zero with $k^2$ and
as a consequence the potential barrier flattens.
For the scalar ``vector'' model the approach to convexity
in the limit $k\to 0$ has been studied analytically
previously \cite{RTW}. Here we are interested in the
potential for a nonzero coarse graining scale $k$.

\begin{figure}[h]
\unitlength1.0cm
\begin{center}
\begin{picture}(13.,9.)
\put(-0.7,4.6){$\ds{\frac{\si_k}{\si_{\max}}}$}
\put(6.3,-0.5){$\ds{\ln\left({k}/{k_f}\right)}$}
\put(6.9,2.){(1)}
\put(9.2,2.){(2)}
\put(11.3,2.){(3)}
\put(-0.4,0.){
\epsfysize=13.cm
\epsfxsize=9.cm
\rotate[r]{\epsffile{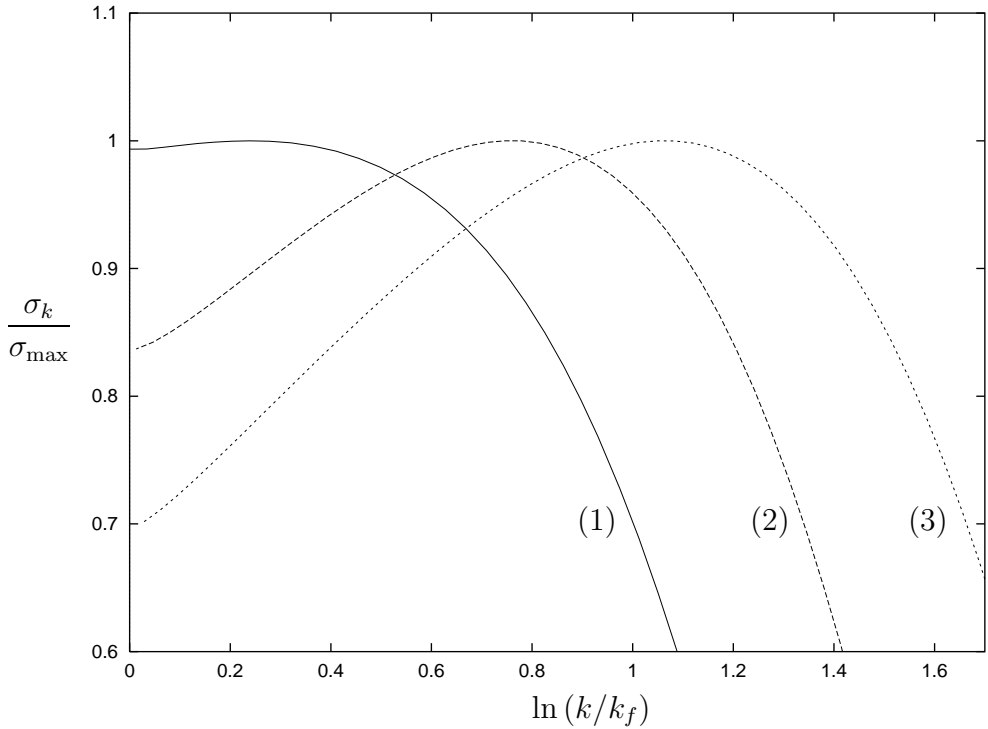
}}
}
\end{picture}
\end{center}
\caption[]{\footnotesize The normalized surface tension
$\si_k/\si_{\max}$ as a function of $\ln (k/k_f)$.
The short distance parameters are (1)
$\bar{\la}_{1\La}/\La=0.1$, $\bar{\la}_{2\La}/\La=2$,
(2) $\bar{\la}_{1\La}/\La=2$, $\bar{\la}_{2\La}/\La=0.1$,
(3) $\bar{\la}_{1\La}/\La=4$, $\bar{\la}_{2\La}/\La=70$. 
\label{surface}
}
\end{figure}

The most significant
difference between fig.\ \ref{innercoarse} and 
fig.\ \ref{innercoarse2}
is the $k$-dependence of the potential barrier
in a region where the minima become degenerate
and almost independent of $k$. Fig.\ \ref{innercoarse}
shows a barrier with a weak scale dependence
in the range of $k$ where the location of the minima stabilizes.
In contrast, in fig.\ \ref{innercoarse2} one observes a barrier 
with a strong scale dependence in this region. 
Figs.\ \ref{scaledep1} and 
\ref{scaledep2} (cf.\ sect.\ \ref{rg}) show
the relative height $U_0(k)=U_k(\vp_0)-U_k(0)$ between the 
two local minima and the height of the
potential barrier $U_B(k)=U_k(\vp_B)-U_k(0)$ 
($(\prl U_k/\prl \vp)(\vp_B)=0$, $0 < \vp_B < \vp_0$) as a function
of $t=\ln(k/\La)$. Accordingly one observes that for
$\bar{\la}_{1\La}/\La=0.1$, $\bar{\la}_{2\La}/\La=2$ 
the top of the potential barrier shows a weak
scale dependence in a region of $k$ where $U_0(k)$ is small
whereas for
$\bar{\la}_{1\La}/\La=2$, $\bar{\la}_{2\La}/\La=0.1$
the top of the potential barrier
depends strongly on the coarse graining scale
in this region.
The surface tension $\si_k$ is displayed in
fig.\ \ref{surface} and shows a corresponding behavior.
Here we consider $\si_k$ also for the short distance 
parameters $\bar{\la}_{1\La}/\La=4$,
$\bar{\la}_{2\La}/\La=70$. In fig.\ \ref{surface}
the surface tension $\si_k$ is normalized to $\si_{\max}$
($\si_{\max}/\La^2=1.67 \times 10^{-2} 
(8.41 \times 10^{-11})
(1.01 \times 10^{-3})$ for 
$\bar{\la}_{1\La}/\La=0.1(2)(4)$, 
$\bar{\la}_{2\La}/\La=2(0.1)(70)$)
and given as a function of $\ln (k/k_f)$.\footnote{
The integration according to eq. (\ref{sur}) is performed  
between the two zeros $\vp=0$ and $\vp=\vp_0^{\pri} \, \ltap \vp_0$ 
of $U_k$.}
For $\bar{\la}_{1\La}/\La=0.1$,
$\bar{\la}_{2\La}/\La=2$ the curve exhibits a small curvature
around its maximum and $\si_{\max} \simeq \si_{k=k_f}$.
One observes for the second and the third example 
a comparably large curvature around $\si_{\max}$ and 
$\si_{k=k_f}$ becomes considerably smaller than the maximum.
One may consider what happens if this scale
dependence is not taken into account correctly.
If one takes the difference in $\sigma_k$
from its maximum value to $\sigma_{k=k_f}$ as a rough
measure for its uncertainty this is about 30\%
for the last example. Entering exponentially in eq.\
(\ref{nc}) this would lead to tremendous errors.

We have pointed out above that the coarse graining scale $k$ 
should not be taken smaller
than the inverse bubble wall thickness
$\Dt^{-1}$. This ensures that the
detailed properties of the bubble are irrelevant
for the computation of the ``classical'' 
surface tension $\si_k$. 
We estimate the bubble wall thickness $\Dt$ by
\be
\Dt(k)= 2 Z_k \frac{\rho_{0}(k)}{\si_k}
\ee
where we have 
taken the gradient energy as half the total 
surface energy and 
approximated the mean field gradient
at the bubble wall by $\vp_{0}/\Dt$.
For the given examples we observe
$\Dt^{-1}(k_f) \simeq k_f/4$. We choose
$k \simeq k_f$ as the coarse graining scale. 

In order to quantify the differences between the three 
examples we have displayed some characteristic 
quantities in table \ref{table2}. 
\begin{table} [h]
\renewcommand{\arraystretch}{2.0}
\hspace*{\fill}
\begin{tabular}{|c|c|c|c|c|c|c|c|}     \hline

$\ds{\frac{\bar{\la}_{1\La}}{\La}}$
&$\ds{\frac{\bar{\la}_{2\La}}{\La}}$
&$\ds{\frac{\la_{1R}}{m_R^c}}$
&$\ds{\frac{\la_{2R}}{m_R^c}}$
&$\ds{\frac{m_R^c}{m_{2R}^c}}$
&$\ds{\frac{m_R^c}{\La}}$
&$\ds{\frac{m_{2R}^c}{\La}}$
&$\ds{\frac{k_f}{\La}}$
\\ \hline \hline
0.1
& 2
&0.228
&8.26
&0.235
&$1.55 \times 10^{-1}$
&$6.62 \times 10^{-1}$
&$1.011 \times 10^{-1}$
\\ \hline 
$2$
&0.1
&0.845
&15.0
&0.335
&$2.04 \times 10^{-5}$
&$6.10 \times 10^{-5}$
&$1.145 \times 10^{-5}$
\\ \hline 
4
&70
&0.980
&16.8
&0.341
&$6.96 \times 10^{-2}$
&$2.04 \times 10^{-1}$
&$3.781 \times 10^{-2}$
\\ \hline 
\end{tabular}
\hspace*{\fill}
\renewcommand{\arraystretch}{1}
\caption{\footnotesize The effective dimensionless couplings
$\la_{1R}/m_R^c$ and $\la_{2R}/m_R^c$. The couplings and the mass
terms $m_R^c$, $m_{2R}^c$ are evaluated at the scale $k_f$
and $\dt\kp_{\La}=0$. 
\label{table2}
}
\end{table}
The renormalized couplings
\be
\la_{1R}=U_{k_f}^{\pri\pri}(\rho_{0R}),\qquad
\la_{2R}=4\prl_{\tau}U_{k_f}(\rho_{0R})
\ee
are normalized to the mass term
\be
m_{R}^c=(2\rho_{0R} \la_{1R})^{1/2} \, .
\ee
In addition we give
the mass term
\be
m_{2R}^c=(\rho_{0R} \la_{2R})^{1/2}
\ee
corresponding to the curvature of the potential in the direction
of the second invariant $\tau$. In comparison with 
figs.\ \ref{innercoarse}--\ref{surface} one observes
that the smaller the effective couplings the weaker the scale 
dependence of $U_k$ and $\sigma_k$. In particular, a
reasonably weak scale dependence of $U_k$ and $\si_k$ requires
\be
\frac{\la_{1R}}{m_R^c}=\hal \frac{m_R^c}{\Dt \rho_{0R}}
\ll 1 \label{weakk}\, \, . 
\ee
This establishes a quantitative criterion for the range
where Langer's theory can be used without paying too much
attention to the precise definition of the coarse graining.
Comparison with fig.\ \ref{ratio} (cf.\ sect.\ \ref{phase}) 
shows that this condition is 
not realized for the range of couplings leading to
universal behavior and for large $\bar{\la}_{1\La}/\La$.
The only region where the saddle point expansion 
is expected to converge 
reasonably well is for small $\bar{\la}_{1\La}/\La$
and large $\bar{\la}_{2\La}/\La$.
The second and the third example 
given in fig.\ \ref{surface}, which exhibit a strong
$k$-dependence, show similar values for the effective 
couplings. More precisely, 
for the relatively strong phase transition
with slightly larger effective couplings 
one observes an increased scale 
dependence as compared to the weak phase transition.
Therefore, the strength of the transition is not the
primary criterion for the convergence of the saddle
point expansion.
In table \ref{table2} also the renormalized 
masses in units of $\La$ 
which indicate the strength of the phase transition and $k_f/\La$
are presented.

In addition to the dependence on $k$, the coarse grained 
free energy depends also on the precise shape
of the infrared cutoff function $R_k(q)$ (cf.\ eq.\ \ref{Propagator}). 
Analytical studies in the Abelian Higgs model 
indicate \cite{litim} that this scheme dependence
is rather weak  
for the effective potential and the
surface tension. 

In summary, we have shown that 
the coarse grained free energy cannot be defined
without detailed information on the coarse graining scale
$k$ unless the effective dimensionless couplings are small. 
Only for small
couplings we observe a weak $k$-dependence 
of the surface tension in a range
where the location of the
minima of the potential remains almost fixed. There is
a close relation between the dependence of the coarse grained
free energy on the coarse graining scale and the reliability
of the saddle point approximation in Langer's theory
of bubble nucleation. For a strong $k$-dependence of $U_k$
a small variation in the coarse graining scale can induce 
large changes in the predicted nucleation rate
in lowest order in a saddle point approximation.
In this case the $k$-dependence of the pre\-factor
$A_k$ has also to be computed. For strong dimensionless
couplings a realistic estimate of the nucleation rate
therefore needs the capability to compute
$\ln A_k$ with the same accuracy as $16 \pi \sigma_k^3/3 \eps^2$
and a check of the cancelation of the $k$-dependence
in the combined expression (\ref{nc}). 
Our observation that the details of the coarse graining 
prescription become less important in the case of small
dimensionless couplings is consistent with the
fact that typically small couplings are needed for a
reliable saddle point approximation for $A_k$.
For the electroweak high temperature phase transition
a small $k$-dependence of $\sigma_k$ is found for a
small mass $M_H$ of the Higgs scalar whereas for 
$M_H$ near the W-boson mass the picture resembles our 
fig.\ \ref{innercoarse2} \cite{Tet}. This corresponds to
the observation \cite{La,Bu,Ba} that the saddle point approximation
around the critical bubble converges well only for a
small enough mass of the Higgs scalar.

\subsection{Conclusions \label{con}}

We have presented in this chapter a detailed investigation
of the phase transition in three dimensional models for 
complex $2 \times 2 $ matrices. They are 
characterized by two quartic couplings
$\bar{\la}_{1\La}$ and $\bar{\la}_{2\La}$. 
In the limit
$\bar{\la}_{1\La}\to \infty$, $\bar{\la}_{2\La}\to \infty$
this also covers the model of unitary matrices.
The picture arising from this study is unambiguous:

(1) One 
observes two symmetry breaking patterns for
$\bar{\la}_{2\La}>0$ and $\bar{\la}_{2\La}<0$
respectively. The case $\bar{\la}_{2\La}=0$
denotes the boundary between the two phases
with different symmetry breaking patterns. 
In this special case the theory
exhibits an enhanced $O(8)$ symmetry.  
The phase transition is always first order
for the investigated symmetry breaking
$U(2) \times U(2) \to U(2)$ ($\bar{\la}_{2 \La}>0$). 
For $\bar{\la}_{2 \La}=0$ 
the $O(8)$ symmetric Heisenberg model is recovered
and one finds a second order phase transition.

(2) The strength of the phase transition depends on the size
of the classical quartic couplings $\bar{\la}_{1\La}/\La$ and  
$\bar{\la}_{2\La}/\La$. They describe the short distance or
classical action at a momentum scale $\La$. The strength of the
transition can be parametrized by $m_R^c/\La$ with $m_R^c$ 
a characteristic inverse correlation length at the 
critical temperature.
For fixed $\bar{\la}_{2\La}$ the strength of the
transition decreases with increasing $\bar{\la}_{1\La}$.
This is analogous to the Coleman-Weinberg effect in 
four dimensions.

(3) For a wide range of classical couplings the critical
behavior near the phase transition is universal.
This means
that it becomes largely independent of the details of the 
classical action once everything is expressed in terms of the
relevant renormalized parameters. 
In particular, characteristic ratios like
$m_R^c/\Dt \rho_{0R}$ (critical inverse correlation length
in the ordered phase over discontinuity in the order parameter)
or $m_{0R}^c/\Dt \rho_{0R}$ (same for the disordered phase)
are not influenced by the addition of
new terms in the classical action
as far as the symmetries are respected.

(4) The range of short distance parameters
$\bar{\la}_{1\La}$, $\bar{\la}_{2\La}$ for 
which the phase transition exhibits universal behavior
is not only determined by the strength of the phase
transition as measured by $m_R^c/\La$.
For a given $\bar{\la}_{1\La}/\La$ and 
small enough $\bar{\la}_{2\La}/\La$ one always observes
universal behavior. In the range of small 
$\bar{\la}_{1\La}/\La$ the essential criterion for universal 
behavior is given by the size of 
$\bar{\la}_{2\La}/\bar{\la}_{1\La}$, with approximate 
universality for $\bar{\la}_{2\La} < \bar{\la}_{1\La}$.
For strong couplings universality extends to larger
$\bar{\la}_{2\La}/\bar{\la}_{1\La}$ and occurs for much larger
$m_R^c/\La$ (cf.\ table \ref{table1}).

(5) We have investigated how various characteristic 
quantities like  
the discontinuity in the order parameter
$\Dt \rho_{0}$ or the corresponding renormalized quantity
$\Dt \rho_{0R}$ or critical correlation lengths
depend on the classical parameters. 
In particular, at the critical temperature
one finds universal 
critical exponents for not too large $\bar{\la}_{2\La}$,
\bea
\Dt \rho_{0R} &\sim& (\bar{\la}_{2\La})^{\theta},
\qquad \theta=1.93\, ,\nnn
\Dt \rho_{0} &\sim& (\bar{\la}_{2\La})^{2\zeta},
\qquad \zeta=0.988 \, .
\eea 
These exponents are related by a scaling relation
to the critical 
correlation length and order parameter exponents
$\nu$ and $\beta$ of the $O(8)$ symmetric 
Heisenberg model according to 
$\theta/\zeta=\nu/\beta=1.95$ ($\nu=0.882$, 
$\beta=0.451$ in our calculation for 
$\bar{\la}_{2\La}=0$). Small values of $\bar{\la}_{2\La}$
can be associated with a perturbation of the 
$O(8)$ symmetric model and $\theta , \zeta$ are
related to the corresponding crossover exponents.
On the other hand, $\Dt \rho_{0R}$ ($\Dt \rho_{0}$)
becomes independent of $\bar{\la}_{2\La}$ in the 
infinite coupling limit. 

(6) We have computed the universal equation of state.
The equation of state relates the 
derivative of the free energy
$U$ to an external source, $\prl U/\prl \vp = j$.
From there one can extract universal ratios
e.g.\ for the jump in the order parameter
$(\Dt \rho_{0R}/m_R^c=0.592)$ or for the ratios of critical 
correlation lengths in the disordered (symmetric) and
ordered (spontaneously broken) phase
$(m_{0R}^c/m_{R}^c=0.746)$.
It specifies critical couplings
$(\la_{1R}/m_R^c=0.845,\la_{2R}/m_R^c=15.0)$. 
The universal behavior
of the potential for large field arguments 
$\rho \gg \rho_{0}$ is
$U \sim \rho^{3/(1+\eta)}$
provided $\rho_R$ is sufficiently 
small as compared to $\La$. Here the critical exponent
$\eta$ which characterizes the dependence of the potential
on the unrenormalized field $\rho$
is found to be $\eta=0.022$. For large $\rho$ the universal
equation of state equals the one for the $O(8)$ symmetric
Heisenberg model and $\eta$ specifies the anomalous
dimension or the critical exponent $\dt=(5-\eta)/(1+\eta)$.
The equation of state is computed for a nonzero coarse
graining scale $k$. It 
therefore contains information for quantities
like the ``classical'' bubble surface tension in the
context of Langer's theory of bubble formation.
  
(7) We have investigated the dependence of the coarse
grained
effective potential $U_k(\rho)$ and the ``classical''
surface tension $\sigma_k$
on the coarse graining 
scale $k$ with special emphasis on the question
of the validity of Langer's 
theory of bubble formation. 
We find a strong scale
dependence of 
$U_k$ and $\sigma_k$
if the phase transition is characterized
by large dimensionless couplings.
A weak scale dependence is observed for small effective
couplings. 
There is
a close relation between the dependence of the coarse grained
free energy on the coarse graining scale and the reliability
of the saddle point approximation in Langer's theory
of bubble nucleation. A strong $k$-dependence of $\sigma_k$
is only compatible
with a large contribution from the higher orders of the
expansion.
We obtain a very consistent
picture: The validity of the saddle point
approximation typically requires small dimensionless couplings.
In this case also the details of the
coarse graining are not of crucial importance
within an appropriate range of $k$. The 
quantitative criterion for the validity of Langer's
formula is in our case $\la_{1R}/m_R^c \ll 1$.

(8) Our method is not restricted to the study of the universal
behavior. We can compute the effective potential for 
arbitrary values of the initial parameters and have done this
for particular examples.

The uncertainties of our results 
induced by the numerical integration
of the flow equations are well under control
and small. They
are negligible compared to
the expected error induced by our truncation.
For a significant improvement of our treatment one
would have to include higher order terms in the 
derivative expansion employed for the effective average
action. For weak first order or second order
phase transitions  we expect the error
to be related to the anomalous dimension
$\eta=0.022$. For the special case of the enhanced 
$O(8)$ symmetry one can compare e.g.\ with known values
for critical exponents obtained by other methods
\cite{ZJ,BC95-1,Rei95-1}. A comparison of our results 
for the critical exponents $\beta$ and $\nu$ with the
results of the most sophisticated calculations show
agreement within a few per cent. The anomalous dimension
is also well determined, even though it is most affected 
by our truncation.  

Finally, we should mention that our approach can be extended
to systems with
reduced $SU(N) \times SU(N)$ symmetry. They obtain by adding to
the classical potential a term involving the invariant
$\xi=\det \vp + \det \vp^{\dagger}$. (Note that $\xi$
is not invariant with respect to $U(N) \times U(N)$).
This will give an even richer pattern of phase transitions
and permits a close contact to realistic meson models in
QCD where the axial anomaly is incorporated. Finally
one can extend the three dimensional treatment to a
four dimensional study of field theories at nonvanishing
temperature. How this can be used to approach
the chiral phase transition in QCD is presented in the next section.

%% file: qcd.tex
\sect{The equation of state for two flavor QCD \label{2qcd}}

\subsection{Introduction}
\label{Introduction}

We have pointed out in section \ref{mainin} that
strong interactions in thermal equilibrium at high temperature $T$ 
differ in important aspects from the well tested vacuum or zero
temperature properties. A phase transition at some critical
temperature $T_c$ or a relatively sharp crossover may separate the
high and low temperature physics~\cite{MO96-1}. Many experimental
activities at heavy ion colliders~\cite{QM96} search for signs of such
a transition.  It was realized early that the transition should be
closely related to a qualitative change in the chiral condensate
according to the general observation that spontaneous symmetry
breaking tends to be absent in a high temperature situation. A series
of stimulating contributions~\cite{PW84-1,RaWi93-1,Raj95-1} pointed
out that for sufficiently small up and down quark masses, $m_u$ and
$m_d$, and for a sufficiently large mass of the strange quark, $m_s$,
the chiral transition is expected to belong to the universality class
of the $O(4)$ Heisenberg model. It
was suggested~\cite{RaWi93-1,Raj95-1} that a large correlation length
may be responsible for important fluctuations or lead to a disoriented
chiral condensate~\cite{Ans88-1}. The question how
small $m_u$ and $m_d$ would have to be in order to see a large
correlation length near $T_c$ and if this scenario could be realized
for realistic values of the current quark masses remained, however,
unanswered. The reason was the missing link between the universal
behavior near $T_c$ and zero current quark mass on one hand and the
known physical properties at $T=0$ for realistic quark masses on the
other hand.

It is the purpose of the present section to provide this link
\cite{BJW}. We
present here the equation of state for two flavor QCD within an
effective quark meson model. The equation of state expresses the
chiral condensate $\VEV{\ol{\psi}\psi}$ as a function of temperature
and the average current quark mass $\hat{m}=(m_u+m_d)/2$.  This
connects explicitly the universal critical behavior for $T\ra T_c$ and
$\hat{m}\ra0$ with the temperature dependence for a realistic value
$\hat{m}_{\rm phys}$. Since our discussion covers the whole
temperature range $0\le T \,\ltap\, 1.7\, T_c$ we can fix
$\hat{m}_{\rm phys}$ such that the (zero temperature) pion mass is
$m_\pi=135\MeV$. The condensate $\VEV{\ol{\psi}\psi}$ plays here the
role of an order parameter. Its precise definition will be given in
section \ref{TheQuarkMesonModelAtT=0}.  Figure \ref{ccc_T} shows our
results for $\VEV{\ol{\psi}\psi}(T,\hat{m})$:
\begin{figure}
\unitlength1.0cm

\begin{center}
\begin{picture}(13.,7.0)

\put(0.0,0.0){
\epsfysize=11.cm
\rotate[r]{\epsffile{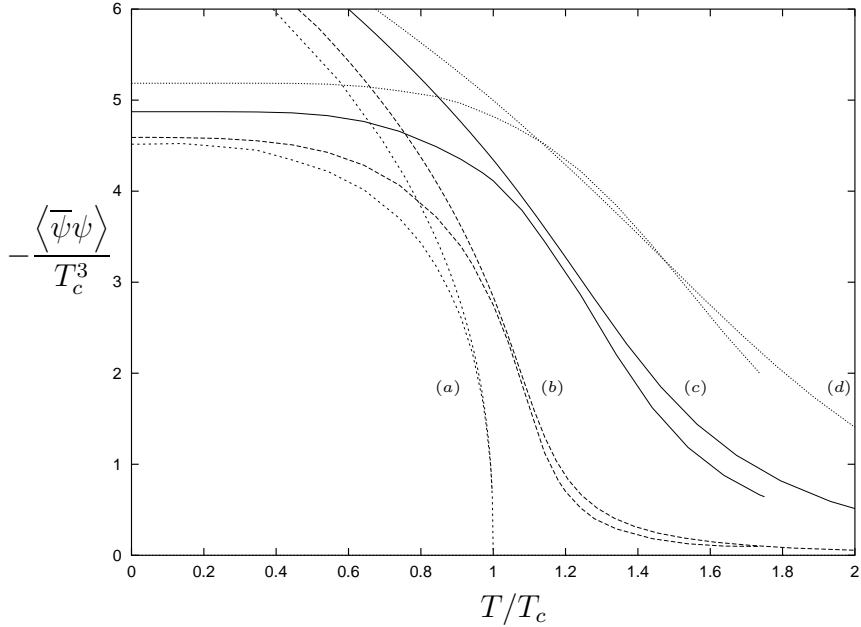}}
}
\put(-0.5,4.2){\bf $\ds{-\frac{\VEV{\ol{\psi}\psi}}{T_{c}^3}}$}
\put(5.8,-0.5){\bf $\ds{T/T_{c}}$}
\put(5.2,2.5){\tiny $(a)$}
\put(6.6,2.5){\tiny $(b)$}
\put(8.5,2.5){\tiny $(c)$}
\put(10.4,2.5){\tiny $(d)$}

\end{picture}
\end{center}
\caption{\footnotesize The plot shows the chiral condensate
  $\VEV{\ol{\psi}\psi}$ as a function of temperature $T$.  Lines
  $(a)$, $(b)$, $(c)$, $(d)$ correspond at zero temperature to
  $m_\pi=0,45\MeV,135\MeV,230\MeV$, respectively. For each pair of
  curves the lower one represents the full $T$--dependence of
  $\VEV{\ol{\psi}\psi}$ whereas the upper one shows for comparison the
  universal scaling form of the equation of state for the $O(4)$
  Heisenberg model. The critical temperature for zero quark mass is
  $T_c=100.7\MeV$. The chiral condensate is normalized at a scale
  $k_{\Phi}\simeq 620\MeV$.}
\label{ccc_T}
\end{figure}
Curve $(a)$ gives the temperature dependence of $\VEV{\ol{\psi}\psi}$
in the chiral limit $\hat{m}=0$. Here the lower curve is the full
result for arbitrary $T$ whereas the upper curve corresponds to the
universal scaling form of the equation of state for the $O(4)$
Heisenberg model.  We see perfect agreement of both curves for $T$
sufficiently close to $T_c=100.7 \MeV$. This demonstrates the
capability of our method to cover the critical behavior and, in
particular, to reproduce the critical exponents of the $O(4)$--model.
We have determined (cf.~section~\ref{CriticalBehavior}) the universal
critical equation of state as well as the non--universal amplitudes.
This provides the full functional dependence of $\VEV{\ol{\psi}\psi}
(T,\hat{m})$ for small $T-T_c$ and $\hat{m}$.  The curves $(b)$, $(c)$
and $(d)$ are for non--vanishing values of the average current quark
mass $\hat{m}$.  Curve $(c)$ corresponds to $\hat{m}_{\rm phys}$ or,
equivalently, $m_\pi(T=0)=135\MeV$. One observes a crossover in the
range $T=(1.2-1.5)T_c$. The $O(4)$ universal equation of state (upper
curve) gives a reasonable approximation in this temperature range. The
transition turns out to be much less dramatic than for $\hat{m}=0$. We
have also plotted in curve $(b)$ the results for comparably small
quark masses $\simeq1\MeV$, i.e.~$\hat{m}=\hat{m}_{\rm phys}/10$, for
which the $T=0$ value of $m_\pi$ equals $45\MeV$. The crossover is
considerably sharper but a substantial deviation from the chiral limit
remains even for such small values of $\hat{m}$. In order to
facilitate comparison with lattice simulations which are typically
performed for larger values of $m_\pi$ we also present results for
$m_\pi(T=0)=230\MeV$ in curve $(d)$. One may define a ``pseudocritical
temperature'' $T_{pc}$ associated to the smooth crossover as the
inflection point of $\VEV{\ol{\psi}\psi}(T)$ as usually done in
lattice simulations. Our results for this definition of $T_{pc}$ are
denoted by $T_{pc}^{(1)}$ and are presented in table \ref{tab11} for
the four different values of $\hat{m}$ or, equivalently, $m_\pi(T=0)$.
\begin{table}
\begin{center}
\begin{tabular}{|c||c|c|c|c|} \hline
  $\stackrel{ }{\frac{m_\pi}{\MeV}}$ &
  $0$ &
  $45$ &
  $135$ &
  $230$
  \\[1.0mm] \hline
  $\stackrel{ }{\frac{T_{pc}^{(1)}}{\MeV}}$ &
  $100.7$ &
  $\simeq110$ &
  $\simeq130$ &
  $\simeq150$
  \\[1mm] \hline
  $\stackrel{ }{\frac{T_{pc}^{(2)}}{\MeV}}$ &
  $100.7$ &
  $$113 &
  $$128 &
  $$---
  \\[1mm] \hline
\end{tabular}
\caption{\footnotesize The table shows the critical and
  ``pseudocritical'' temperatures for various values of the zero
  temperature pion mass. Here $T_{pc}^{(1)}$ is defined as the
  inflection point of $\VEV{\ol{\psi}\psi}(T)$ whereas $T_{pc}^{(2)}$
  is the location of the maximum of the sigma correlation length (see
  section \ref{TheQuarkMesonModelAtTNeq0}).}
\label{tab11}
\end{center}
\end{table}
The value for the pseudocritical temperature for $m_{\pi}=230 \MeV$
compares well with the lattice results for two flavor QCD
(cf.~section~\ref{CriticalBehavior}). One should mention, though, that
a determination of $T_{pc}$ according to this definition is subject to
sizeable numerical uncertainties for large pion masses as the curve in
figure \ref{ccc_T} is almost linear around the inflection point for
quite a large temperature range.  A problematic point in lattice
simulations is the extrapolation to realistic values of $m_\pi$ or
even to the chiral limit. Our results may serve here as an analytic
guide. The overall picture shows the approximate validity of the
$O(4)$ scaling behavior over a large temperature interval in the
vicinity of and above $T_c$ once the (non--universal) amplitudes are
properly computed.

\begin{figure}
\unitlength1.0cm
\begin{center}
\begin{picture}(13.,7.0)

\put(0.0,0.0){
\epsfysize=11.cm
\rotate[r]{\epsffile{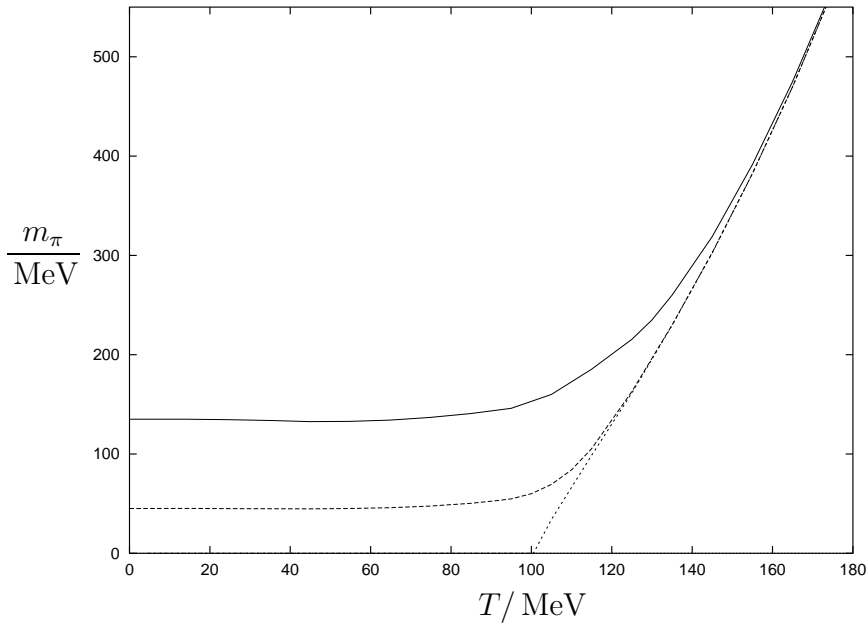}}
}
\put(-0.5,4.2){\bf $\ds{\frac{m_\pi}{\MeV}}$}
\put(5.8,-0.5){\bf $\ds{T/\MeV}$}

\end{picture}
\end{center}
\caption{\footnotesize The plot shows $m_\pi$ as a function of
  temperature $T$ for three different values of the average light
  current quark mass $\hat{m}$. The solid line corresponds to the
  realistic value $\hat{m}=\hat{m}_{\rm phys}$ whereas the dotted line
  represents the situation without explicit chiral symmetry breaking,
  i.e., $\hat{m}=0$. The intermediate, dashed line assumes 
  $\hat{m}=\hat{m}_{\rm phys}/10$.}
\label{mpi_T}
\end{figure}

A second important result of our investigations is the temperature
dependence of the space--like pion correlation length
$m_\pi^{-1}(T)$. (We will often call $m_\pi(T)$ the temperature
dependent pion mass since it coincides with the physical pion mass for
$T=0$.) The plot for $m_\pi(T)$ in figure \ref{mpi_T} again shows the
second order phase transition in the chiral limit $\hat{m}=0$. 
For $T<T_c$ the pions are massless Goldstone bosons whereas for
$T>T_c$ they form with the sigma particle a degenerate vector of
$O(4)$ with mass increasing as a function of temperature.  For
$\hat{m}=0$ the behavior for small positive $T-T_c$ is characterized
by the critical exponent $\nu$, i.e.
$m_\pi(T)=\left(\xi^+\right)^{-1}T_c \left( (T-T_c)/T_c\right)^\nu$
and we obtain $\nu=0.787$, $\xi^+=0.270$. For $\hat{m}>0$ we find that
$m_\pi(T)$ remains almost constant for $T\lta T_c$ with only a very
slight dip for $T$ near $T_c/2$. For $T>T_c$ the correlation length
decreases rapidly and for $T\gg T_c$ the precise value of $\hat{m}$
becomes irrelevant. We see that the universal critical behavior near
$T_c$ is quite smoothly connected to $T=0$.  The full functional
dependence of $m_\pi(T,\hat{m})$ allows us to compute the overall size
of the pion correlation length near the critical temperature and we
find $ m_\pi(T_{pc})\simeq 1.7 m_\pi(0)$ for the realistic value
$\hat{m}_{\rm phys}$. This correlation length is even smaller than the
vacuum ($T=0$) one and gives no indication for strong fluctuations of
pions with long wavelength. It would be interesting to see if a
decrease of the pion correlation length at and above $T_c$ is
experimentally observable.  It should be emphasized, however, that a
tricritical behavior with a massless excitation remains possible for
three flavors. This would not be characterized by the universal
behavior of the $O(4)$--model. We also point out that the present
investigation for the two flavor case does not take into account a
speculative ``effective restoration'' of the axial $U_A(1)$ symmetry
at high temperature \cite{PW84-1,Shu94-1}. We will comment on these
issues in section~\ref{AdditionalDegreesOfFreedom}.

Our method is based on the effective average action
$\Gamma_k$~\cite{Wet91-1} which has been introduced
in section \ref{fieldtheory}. 
Since in most cases the flow equation (\ref{ERGE}) for 
$\Gamma_k$ can not be solved exactly the
capacity to devise useful truncations in a non-perturbative context
becomes crucial. This requires first of all an identification of the
degrees of freedom which are most relevant for a given problem. In the
present work we concentrate on the chiral aspects of QCD\footnote{For
  a study of chiral symmetry breaking in QED using related exact
  renormalization group techniques see ref.~\cite{AMST97-1}.}. 
Spontaneous
chiral symmetry breaking occurs through the expectation value of a
(complex) scalar field $\Phi_{ab}$ which transforms as $(\ol{\bf
  N},\bf{N})$ under the chiral flavor group $SU_L(N)\times SU_R(N)$
with $N$ the number of light quark flavors. More precisely, the
expectation value
\begin{equation}
  \label{AAA50}
  \VEV{\Phi^{ab}}=\ol{\sigma}_0\dt^{ab} 
\end{equation} 
induces for $\ol{\sigma}_0\neq0$ a spontaneous breaking of the chiral
group to a vector--like subgroup, $SU_L(N)\times
SU_R(N)\longrightarrow SU_{L+R}(N)\equiv SU_V(N)$. In addition,
non--vanishing current quark masses $m_u,m_d,m_s$ break the chiral
group explicitly and also lift the $SU_V(N)$ degeneracy of the
spectrum if they are unequal. The physical degrees of freedom
contained in the field $\Phi_{ab}$ are pseudoscalar and scalar mesons
which can be understood as quark--antiquark bound states. It is
obvious that any analytical description of the chiral transition has
to include at least part of these (pseudo--)scalar fields as the most
relevant degrees of freedom.

In the present work we use for $k$ smaller than a ``compositeness
scale'' $k_\Phi\simeq600\MeV$ a description in terms of $\Phi_{ab}$
and quark degrees of freedom. The quarks acquire a constituent quark mass
$M_q$ through the chiral condensate $\ol{\sigma}_0$ which forms in our
picture for $k_{\chi SB}\simeq400\MeV$.  This effective quark meson
model can be obtained from QCD by ``integrating out'' the gluon
degrees of freedom and introducing fields for composite
operators~\cite{EW94-1,Wet95-2}. This will be explained in more detail
in the first part of section \ref{TheQuarkMesonModelAtT=0}. In this
picture the scale $k_\Phi$ is associated to the scale at which the
formation of mesonic bound states can be observed in the flow of the
effective (momentum dependent) four--quark interaction. We will
restrict our discussion here to two flavor QCD with equal
quark masses $m_u=m_d\equiv\hat{m}$.  Since in this case the scalar
triplet $a_0$ and the pseudoscalar singlet (associated with the
$\eta^\prime$) have typical masses around\footnote{More precisely,
  because of the anomalous $U_A(1)$ breaking in QCD these mesons are
  significantly heavier than the remaining degrees of freedom in the
  range of scales $k$ where the dynamics of the model is strongly
  influenced by mesonic fluctuations. The situation becomes more
  involved if the model is considered at high temperature which is
  discussed in section \ref{AdditionalDegreesOfFreedom}.} $1\GeV$ we
will neglect them for $k<k_\Phi$.  This reduces the scalar degrees of
freedom of our effective model to a four component vector of $O(4)$,
consisting of the three pions and the ``sigma resonance''.

We imagine that all other degrees of freedom besides the quarks $\psi$
and the scalars $\Phi$ are integrated out. This is reflected in the
precise form of the effective average action
$\Gamma_{k_\Phi}[\psi,\Phi]$ at the scale $k_\Phi$ which serves as an
initial value for the solution of the flow equation. The flow of
$\Gamma_{k}[\psi,\Phi]$ for $k<k_\Phi$ is then entirely due to the
quark and meson fluctuations which are not yet included in
$\Gamma_{k_{\Phi}}[\psi,\Phi]$. Obviously, the initial value
$\Gamma_{k_\Phi}$ may be a quite complicated functional of $\psi$ and
$\Phi$ containing, in particular, important non--local behavior. We
will nevertheless use a rather simple truncation in terms of standard
kinetic terms and a most general form of the scalar potential $U_k$,
i.e.\footnote{Our Euclidean conventions ($\ol{h}_k$ is real) are
  specified in refs.~\cite{Ju95-7,Wet90-1}.}
\begin{equation}
  \label{AAA60}
  \begin{array}{rcl}
  \ds{\hat{\Gamma}_k} &=& \ds{
    \Gamma_k-\frac{1}{2}\int d^4 x\tr
    \left(\Phi^\dagger\jmath+\jmath^\dagger\Phi\right)}\nnn
    \ds{\Gamma_{k}} &=& \ds{
      \int d^4x\Bigg\{
      Z_{\psi,k}\ol{\psi}_a i\slash{\prl}\psi^a+
      Z_{\Phi,k}\tr\left[\prl_\mu\Phi^\dagger\prl^\mu\Phi\right]+
      U_k(\Phi,\Phi^\dagger)
      }\nnn
    &+& \ds{
      \ol{h}_{k}\ol{\psi}^a\left(\frac{1+\gm_5}{2}\Phi_{ab}-
      \frac{1-\gm_5}{2}(\Phi^\dagger)_{ab}\right)\psi^b
      \Bigg\} }\; .
  \end{array}
\end{equation}
Here $\Gamma_k$ is invariant under the chiral flavor symmetry
$SU_L(2)\times SU_R(2)$ and the only explicit symmetry breaking arises
through the source term $\jmath\sim\hat{m}$. We will consider the flow
of the most general form of $U_k$ consistent with the symmetries
(without any restriction to a polynomial form as typically used in a
perturbative context). On the other hand, our approximations for the
kinetic terms are rather crude and parameterized by only two running
wave function renormalization constants, $Z_{\Phi,k}$ and
$Z_{\psi,k}$. The same holds for the effective Yukawa coupling
$\ol{h}_k$. The main approximations in this work concern
\begin{description}
\item [(i)] the simple form of the derivative terms and the Yukawa
  coupling, in particular, the neglect of higher derivative terms (and
  terms with two derivatives and higher powers of $\Phi$). This is
  partly motivated by the observation that at the scale $k_\Phi$ and
  for small temperatures the possible strong non--localities related
  to confinement affect most likely only the quarks in a momentum
  range $q^2\lta(300\MeV)^2$.  Details of the quark propagator and
  interactions in this momentum range are not very important in our
  context (see section~\ref{TheQuarkMesonModelAtT=0}).
\item[(ii)] the neglect of interactions involving more than two quark
  fields. This is motivated by the fact that the dominant multi--quark
  interactions are already incorporated in the mesonic description.
  Six--quark interactions beyond those contained effectively in $U_k$
  could be related to baryons and play probably only a minor role for
  the meson physics considered here.
\end{description}

We will choose a normalization of $\psi,\Phi$ such that
$Z_{\psi,k_\Phi}=\ol{h}_{k_\Phi}=1$. We therefore need as initial
values at the scale $k_\Phi$ the scalar wave function renormalization
$Z_{\Phi,k_\Phi}$ and the shape of the potential $U_{k_\Phi}$. We will
make here the important assumption that $Z_{\Phi,k}$ is small at the
compositeness scale $k_\Phi$ (similarly to what is usually assumed in
Nambu--Jona-Lasinio--like models). This results in a large value of
the renormalized Yukawa coupling
$h_k=Z_{\Phi,k}^{-1/2}Z_{\psi,k}^{-1}\ol{h}_k$. A large value of
$h_{k_\Phi}$ is phenomenologically suggested by the comparably large
value of the constituent quark mass $M_q$. The latter is related to
the value of the Yukawa coupling for $k\ra0$ and the pion decay
constant $f_\pi=92.4\MeV$ by $M_q=h f_\pi/2$ (with $h=h_{k=0}$), and
$M_q\simeq300\MeV$ implies $h^2/4\pi\simeq3.4$. For increasing $k$ the
value of the Yukawa coupling grows rapidly for $k\gta M_q$.  Our
assumption of a large initial value for $h_{k_\Phi}$ is therefore
equivalent to the assumption that the truncation (\ref{AAA60}) can be
used up to the vicinity of the Landau pole of $h_k$. The existence of
a strong Yukawa coupling enhances the predictive power of our approach
considerably.  It implies a fast approach of the running couplings to
partial infrared fixed points~\cite{Ju95-7}. In consequence, the
detailed form of $U_{k_\Phi}$ becomes unimportant, except for the
value of one relevant parameter corresponding to the scalar mass term
$\ol{m}^2_{k_\Phi}$. In this work we fix $\ol{m}^2_{k_\Phi}$ such
that $f_\pi=92.4\MeV$ for $m_\pi=135\MeV$. The possibility of such a
choice is highly non--trivial since $f_\pi$ can actually be
predicted~\cite{Ju95-7} in our setting within a relatively narrow
range. The value$f_\pi=92.4\MeV$ (for $m_\pi=135\MeV$) sets our unit
of mass for two flavor QCD which is, of course, not directly
accessible by observation. Besides $\ol{m}^2_{k_\Phi}$ (or $f_\pi$)
the other input parameter used in this work is the constituent quark
mass $M_q$ which determines the scale $k_\Phi$ at which $h_{k_\Phi}$
becomes very large. We consider a range $300\MeV\lta M_q\lta350\MeV$
and find a rather weak dependence of our results on the precise value
of $M_q$. We also observe that the limit $h_{k_\Phi}\ra\infty$ can be
considered as the lowest order of a systematic expansion in
$h_{k_\Phi}^{-1}$ which is obviously highly non-perturbative.

A generalization of our method to the realistic case of three light
flavors is possible and work in this direction is in progress. For the
time being we expect that many features found for $N=2$ will carry
over to the realistic case, especially the critical behavior for $T\ra
T_c$ and $\hat{m}\ra0$ (for fixed $m_s\neq0$).  Nevertheless, some
quantities like $\VEV{\ol{\psi}\psi}(T=0)$, the difference between
$f_\pi$ for realistic quark masses and $\hat{m}=0$ or the mass of the
sigma resonance at $T=0$ may be modified. This will also affect the
non--universal amplitudes in the critical equation of state and, in
particular, the value of $T_c$.  In the picture of the two flavor
quark meson model these changes occur through an effective temperature
dependence of the initial values of couplings at the scale $k_\Phi$.
This effect, which is due to the temperature dependence of effects
from fluctuations not considered in the present work is discussed
briefly in section \ref{AdditionalDegreesOfFreedom}.  It remains
perfectly conceivable that this additional temperature dependence may
result in a first order phase transition or a tricritical behavior for
realistic values of $\hat{m}$ for the three flavor case. Details will
depend on the strange quark mass.  We observe, however, that the
temperature dependence in the limit $\hat{m}\ra0$ involves for $T\le
T_c$ only information from the running of couplings in the range
$k\lta300\MeV$.  (The running for $k\gta3T$ effectively drops out in
the comparison between the thermal equilibrium results and those for
$T=0$.)  In this range of temperatures our model should be quite
reliable.

Finally, we mention that we have concentrated here only on the
$\Phi$--dependent part of the effective action which is related to
chiral symmetry breaking. The $\Phi$--independent part of the free
energy also depends on $T$ and only part of this temperature
dependence is induced by the scalar and quark fluctuations considered
in the present work. Most likely, the gluon degrees of freedom cannot
be neglected for this purpose. This is the reason why we do not give
results for ``overall quantities'' like energy density or pressure as
a function of $T$.

The text is organized as follows: In
section~\ref{TheQuarkMesonModelAtT=0} we review the linear quark meson
model at vanishing temperature. We begin with an overview of the
different scales appearing in strong interaction physics.
Subsequently, the flow equations for the linear quark meson model are
introduced and their approximate partial fixed point behavior is
discussed in detail leading to a ``prediction'' of the chiral
condensate $\VEV{\ol{\psi}\psi}$. In
section~\ref{FiniteTemperatureFormalism} the exact renormalization
group formulation of field theories in thermal equilibrium is given.
It is demonstrated how mass threshold functions in the flow equations
smoothly decouple all massive Matsubara modes as the temperature
increases, therefore leading to a ``dimensional reduction'' of the
model.  Section~\ref{TheQuarkMesonModelAtTNeq0} contains our results
for the linear quark meson model at non--vanishing temperature. Here
we discuss the $T$--dependences of the parameters and physical
observables of the linear quark meson model in detail for a
temperature range $0\le T\lta170\MeV$ including the (pseudo)critical
temperature $T_c$ of the chiral transition. The critical behavior of
the model near $T_c$ and $\hat{m}=0$, where $\hat{m}$ denotes the
light average current quark mass, is carefully analyzed in
section~\ref{CriticalBehavior}.  There we present the universal
scaling form of the equation of state including a fit
for the corresponding scaling function. Also the universal critical
exponents and amplitude ratios are given there.  The effects of
additional degrees of freedom of strong interaction physics not
included in the linear $O(4)$ symmetric quark meson model are
addressed in section~\ref{AdditionalDegreesOfFreedom}. Here we also
comment on differences between the linear quark meson model and chiral
perturbation theory. Some technical details concerning the quark mass
term and the definition of threshold functions at vanishing and
non--vanishing temperature are presented in three appendices.

\subsection{The quark meson model}
\label{TheQuarkMesonModelAtT=0}

Before discussing the finite temperature behavior of strong
interaction physics we will review some of its zero temperature
features. This will be done within the framework of a linear quark
meson model as an effective description for QCD for scales below the
mesonic compositeness scale of approximately $k_\Phi\simeq600\MeV$.
Relating this model to QCD in a semi--quantitative way in subsection
\ref{ASemiQuantitativePicture} will allow us to gain some information
on the initial value for the effective average action at the
compositeness scale $k_\Phi$.  We emphasize, however, that the
quantitative aspects of the derivation of the effective quark meson
model from QCD will not be relevant for our practical calculations in
the mesonic sector. This is related to the ``infrared stability'' for
large Yukawa coupling $h_{k_\Phi}$ as discussed in the introduction
and which will be made quantitative in subsection
\ref{FlowEquationsAndInfraredStability}.

\subsection{A short (scale) history of QCD}
\label{ASemiQuantitativePicture}

For an evaluation of the trace on the right hand side 
of the flow equation
(\ref{ERGE}) only a small momentum range $q^2\simeq k^2$ contributes
substantially. One therefore only needs to take into account
those fluctuations which are
important in this momentum interval. Here we are interested
in the description of chiral symmetry breaking.
The relevant fluctuations in relation to this phenomenon may
change with the scale $k$ and we
begin by summarizing the qualitatively different scale intervals
which appear for meson physics in QCD. 
Some of this will be explained
in more detail in the remainder of this section whereas other aspects
are well known. Details of this discussion may also be
found in refs.~\cite{EW94-1,Ju95-7,JW96-4}.
We will distinguish five qualitatively different
ranges of scales:
\begin{enumerate}
\item At sufficiently high momentum scales, say,
  \begin{displaymath}
    k\gta k_p\simeq1.5\GeV
  \end{displaymath}
  the relevant degrees of freedom of strong interactions are quarks
  and gluons and their dynamics is well described by perturbative QCD.
\item For decreasing momentum scales in the range
  \begin{displaymath}
    k_\Phi\simeq600\MeV\lta k\lta k_p\simeq1.5\GeV
  \end{displaymath}
  the dynamical degrees of freedom are still quarks and gluons. Yet,
  as $k$ is lowered part of their dynamics becomes dominated by
  effective non--local four quark interactions which cannot be fully
  accessed perturbatively.
\item At still lower scales this situation changes dramatically.
  Quarks and gluons are supplemented by mesonic bound states as
  additional degrees of freedom which are formed at a scale
  $k_\Phi\simeq600\MeV$. We emphasize that $k_\Phi$ is well separated
  from $\Lambda_{\rm QCD}\simeq200\MeV$ where confinement sets in and
  from the constituent masses of the quarks $M_q\simeq(300-350)\MeV$.
  This implies that below the compositeness scale $k_\Phi$ there
  exists a hybrid description in term of quarks {\em and} mesons! It
  is important to note that for scales not too much smaller than
  $k_\Phi$ chiral symmetry remains unbroken. This situation holds down
  to a scale $k_{\chi SB}\simeq400\MeV$ at which the scalar meson
  potential develops a non--trivial minimum thus breaking chiral
  symmetry spontaneously. The meson dynamics within the range
  \begin{displaymath}
    k_{\chi SB}\simeq400\MeV\lta k\lta k_\Phi\simeq600\MeV
  \end{displaymath}
  is dominated by light current quarks with a strong Yukawa coupling
  $h^2_k/(4\pi)\gg\alpha_s(k)$ to mesons. We will thus assume that
  the leading gluon effects are included below $k_\Phi$ already in the
  formation of mesons.  Near $k_{\chi SB}$ also fluctuations of the
  light scalar mesons become important as their initially large
  renormalized mass approaches zero. Other hadronic bound states like
  vector mesons or baryons should have masses larger than those of the
  lightest scalar mesons, in particular near $k_{\chi SB}$, and give
  therefore only subleading contributions to the dynamics. This leads
  us to a simple linear model of quarks and scalar mesons as an
  effective description of QCD for scales below $k_\Phi$.
\item As one evolves to scales below $k_{\chi SB}$ the Yukawa coupling
  decreases whereas $\alpha_s$ increases. Of course, getting closer to
  $\Lambda_{\rm QCD}$ it is no longer justified to neglect in the
  quark sector the QCD effects which go beyond the dynamics of the
  effective quark meson model in our truncation (\ref{AAA60}).  On the
  other hand, the final IR value of the Yukawa coupling $h$ is fixed
  by the typical values of constituent quark masses $M_q\simeq300\MeV$
  to be $h^2/(4\pi)\simeq3.4$. One may therefore speculate that the
  domination of the Yukawa interaction persists even for the interval
  \begin{displaymath}
    M_q\simeq300\MeV\lta k\lta k_{\rm \chi SB}\simeq400\MeV
  \end{displaymath}
  below which the quarks decouple from the evolution of the mesonic
  degrees of freedom altogether. Of course, details of the gluonic
  interactions are expected to be crucial for an understanding of
  quark and gluon confinement. Strong interaction effects may
  dramatically change the momentum dependence of the quark propagator
  for $k$ and $q^2$ around $\Lambda_{\rm QCD}$.  Yet, there is no
  coupling of the gluons to the color neutral mesons. As long as one
  is only interested in the dynamics of the mesons one is led to
  expect that confinement effects are quantitatively not too
  important.
\item Because of the effective decoupling of the quarks and therefore
  of the whole colored sector the details of confinement have only
  little influence on the mesonic dynamics for scales
  \begin{displaymath}
    k\lta M_q\simeq300\MeV\; .
  \end{displaymath}
  Here quarks and gluons disappear effectively from the spectrum and
  one is left with the pions. They are the only particles whose
  propagation is not suppressed by a large mass. For scales below the
  pion mass the flow of the couplings stops.
\end{enumerate}

Let us now try to understand these different ranges of scales in more
detail. We may start at $k_p=1.5\GeV$ where we assume that all gluonic
degrees of freedom have been integrated out while we have kept an
effective infrared cutoff $\sim k_p$ in the quark propagators.
Details of this procedure were outlined in ref.~\cite{Wet95-2}. This
results in a non--trivial momentum dependence of the quark propagator
and effective non--local four and higher quark interactions. Because
of the infrared cutoff the resulting effective action for the quarks
resembles closely the one for heavy quarks (at least for Euclidean
momenta). The dominant effect is the appearance of an effective quark
potential similar to the one for the charm quark which describes the
effective four quark interactions.

\begin{figure}
\unitlength1.0cm
\begin{picture}(13.,8.)
\put(1.7,0.3){$k_p=1.5\GeV$}
\put(9.6,0.3){$k_\Phi\simeq600\MeV$}
\put(10.5,3.4){$\Phi^{ba}$}

\put(0.7,6.2){$\ol{\psi}_a^i(p_1)$}
\put(3.9,6.2){$\psi_a^j(p_3)$}
\put(0.7,1.6){$\psi_b^i(p_2)$}
\put(3.9,1.6){$\ol{\psi}_b^j(p_4)$}

\put(7.2,5.8){$\ol{\psi}_a^i(p_1)$}
\put(13.2,5.8){$\psi_a^j(p_3)$}
\put(7.2,2.0){$\psi_b^i(p_2)$}
\put(13.2,2.0){$\ol{\psi}_b^j(p_4)$}

\put(1.5,1.3){
\epsffile{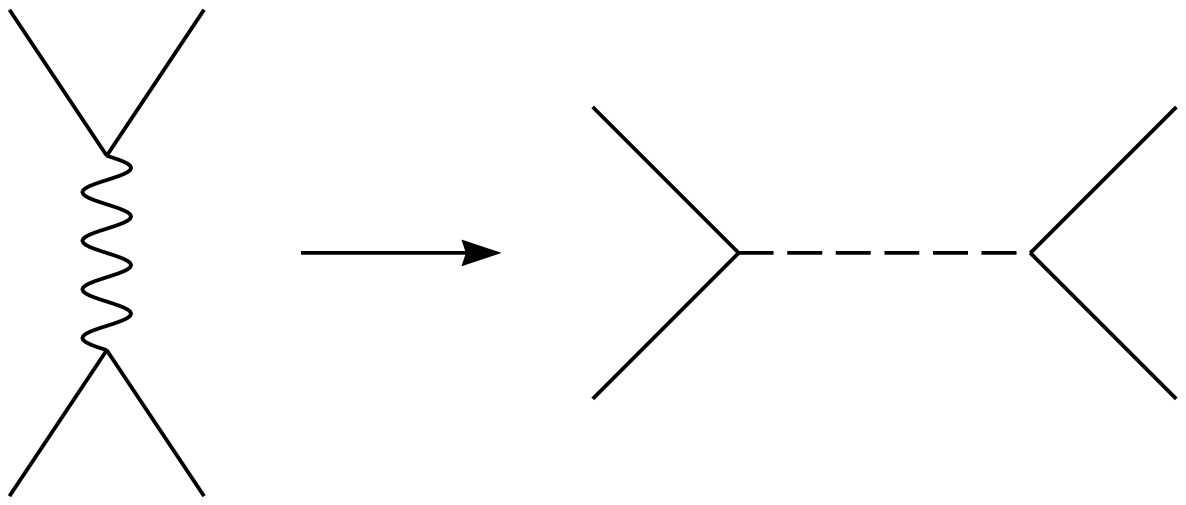}
}
\end{picture}
\caption{\footnotesize The left diagram represents the one gluon
  exchange $t$--channel contribution to the four quark vertex at the
  scale $k_p\simeq1.5\GeV$. It is assumed here that the gluon
  propagator is modified such that it accounts for the linearly rising
  term in the heavy quark potential. The right diagram displays the
  scalar meson $s$--channel exchange found at the compositeness scale
  $k_\Phi\simeq600\MeV$.}
\label{Feyn} 
\end{figure}

We next have to remove the infrared cutoff for the quarks, $k\ra0$.
This task can be carried out by means of the exact flow equation for
quarks only, starting at $k_p$ with an initial value
$\Gamma_{k_p}[\psi]$ as obtained after integrating out the gluons. A
first investigation in this direction~\cite{EW94-1} used a truncation
with a chirally invariant four quark interaction whose most general
momentum dependence was retained. A crucial point is, of course, the
initial value for this momentum dependence at $k_p$. The ansatz used
in ref.~\cite{EW94-1} is obtained by Fierz transforming the heavy
quark potential and keeping, for simplicity, only the scalar meson
channel while neglecting the $\rho$--meson and pomeron channels which
are also present.  The effective heavy quark potential was
approximated there by a one gluon exchange term $\sim\alpha_s(k_p)$
supplemented by a linearly rising string tension term.  This ansatz
corresponds to the four quark interaction generated by the flavor
neutral $t$--channel one gluon exchange depicted in figure \ref{Feyn}
with an appropriately modified gluon propagator and quark gluon vertex
in order to account for the linearly rising part of the potential.

The evolution equation for the four quark interaction can be derived
from the fermionic version of eq.~(\ref{ERGE}). It is by far not clear
that the evolution of the effective four quark vertex will lead at
lower scales to a momentum dependence representing the ($s$--channel)
exchange of colorless mesonic bound states. Yet, at the compositeness
scale
\begin{equation}
  \label{kphi}
  k_\Phi\simeq600\MeV
\end{equation}
one finds \cite{EW94-1} an approximate Bethe--Salpeter factorization
of the four quark amplitude with precisely this property.  This
situation is described by the right Feynman diagram in figure
\ref{Feyn}.  In particular, it was possible to extract the amputated
Bethe--Salpeter wave function as well as the mesonic bound state
propagator displaying a pole--like structure in the $s$--channel if it
is continued to negative $s=(p_1+p_2)^2$. In the limit where the
momentum dependence of the Bethe--Salpeter wave function and the bound
state propagator is neglected the effective action $\Gamma_{k_{\Phi}}$
resembles\footnote{Our solution of the flow equation for $\Gamma_k$
  with $Z_{\Phi,k_\Phi}=0$ (see below) may be considered as a solution
  of the NJL model with a particular form of the ultraviolet cutoff
  dictated by the shape of $R_k(q^2)$ as given in eq.~(\ref{Rk(q)}).}
the Nambu--Jona-Lasinio model~\cite{NJL61-1,Bij95-1}.  It is therefore
not surprising that our description of the dynamics for $k<k_\Phi$
will parallel certain aspects of the investigations of this model,
even though we are not bound to the approximations used typically in
such studies (large--$N_c$ expansion, perturbative renormalization
group, etc.).

It is clear that for scales $k\lta k_\Phi$ a description of strong
interaction physics in terms of quark fields alone would be rather
inefficient. Finding physically reasonable truncations of the
effective average action should be much easier once composite fields
for the mesons are introduced.  The exact renormalization group
equation can indeed be supplemented by an exact formalism for the
introduction of composite field variables or, more generally, a change
of variables~\cite{EW94-1}. In the context of QCD this amounts to the
replacement of the dominant part of the four quark interactions by
scalar meson fields with Yukawa couplings to the quarks.  In turn,
this substitutes the effective quark action at the scale $k_\Phi$ by
the effective quark meson action given in eq.\ (\ref{AAA60}) in the
introduction\footnote{We note that no double counting problem arises
  in this procedure.}.  The term in the effective potential
$U_{k_\Phi}$ which is quadratic in $\Phi$,
$U_{k_\Phi}=\ol{m}^2_{k_\Phi}\tr\Phi^\dagger\Phi+\ldots$, turns out to
be positive as a consequence of the attractiveness of the four quark
interaction inducing it. Its value was found for the simple
truncations used in ref.~\cite{EW94-1} to be
$\ol{m}_{k_\Phi}\simeq120\MeV$.  The higher order terms in
$U_{k_\Phi}$ cannot be determined in the four quark approximation
since they correspond to terms involving six or more quark fields.
(Their values will not be needed for our quantitative investigations
as is discussed in subsection
\ref{FlowEquationsAndInfraredStability}). The initial value of the
(bare) Yukawa coupling corresponds to the amputated Bethe--Salpeter
wave function. Neglecting its momentum dependence it can be normalized
to $\ol{h}_{k_\Phi}=1$.  Moreover, the quark wave function
renormalization $Z_{\psi,k}$ is normalized to one at the scale
$k_\Phi$ for convenience. One may add that we have refrained here for
simplicity from considering four quark operators with vector and
pseudo--vector spin structure.  Their inclusion is straightforward and
would lead to vector and pseudo--vector mesons in the effective
action.

In view of the possible large truncation errors made in
ref.~\cite{EW94-1}
we will take (\ref{kphi}) and the above value of 
$\ol{m}_{k_\Phi}$ only as order of magnitude
estimates. Furthermore, we will assume, as 
motivated in the introduction and usually done in
large--$N_c$ computations within the NJL model, that
\begin{equation}
  \label{AAA32}
  Z_{\Phi,k_\Phi}\ll1\; .
\end{equation}
As a consequence, the initial value of the renormalized Yukawa
coupling $h_{k_\Phi}$ $=$ $Z_{\Phi,k_\Phi}^{-1/2}$
$Z_{\psi,k_\Phi}^{-1}$ $\ol{h}_{k_\Phi}$ is much larger than one and
we will be able to exploit the infrared stable features of the flow
equations. As a typical coupling we take $h_{k_\Phi}=100$ in order to
simulate the limit $h_{k_\Phi}\to \infty$. The effective potential
$U_k(\Phi)$ must be invariant under the chiral $SU_L(N) \times
SU_R(N)$ flavor symmetry. In fact, the axial anomaly of QCD breaks the
Abelian $U_A(1)$ symmetry. The resulting $U_A(1)$ violating
multi--quark interactions\footnote{A first attempt for the computation
  of the anomaly term in the fermionic effective average action can be
  found in ref.~\cite{Paw96-1}.} lead to corresponding $U_A(1)$
violating terms in $U_k(\Phi)$.  Accordingly, the most general
effective potential $U_k$ is a function of the $N+1$ independent $C$
and $P$ conserving $SU_L(N)\times SU_R(N)$ invariants
\begin{equation}
  \begin{array}{rcl}
    \label{Invariants}
    \ds{\rho} &=&
    \ds{\tr\Phi^\dagger\Phi}\; ,\nnn
    \ds{\tau_i} &\sim& \ds{
      \tr\left(\Phi^\dagger\Phi- \frac{1}{N}\rho\right)^i\; ,\;\;\;
      i=2,\ldots,N}\; ,\nnn
    \ds{\xi} &=&
    \ds{\det\Phi+\det\Phi^\dagger}\; .
  \end{array}
\end{equation}
For a given initial form of $U_k$ all quantities in our truncation of
$\Gamma_k$ (\ref{AAA60}) are now fixed and we may follow the flow of
$\Gamma_k$ to $k \to 0$.  In this context it is important that the
formalism for composite fields~\cite{EW94-1} also induces an infrared
cutoff in the meson propagator. The flow equations are therefore
exactly of the form (\ref{ERGE}), with quarks and mesons treated on an
equal footing. At the compositeness scale the quadratic term of
$U_{k_\Phi}=\ol{m}^2_{k_\Phi}\Tr\Phi^\dagger\Phi+\ldots$ is positive
and the minimum of $U_{k_{\Phi}}$ therefore occurs for $\Phi=0$.
Spontaneous chiral symmetry breaking is described by a non--vanishing
expectation value $\VEV{\Phi}$ in absence of quark masses. This
follows from the change of the shape of the effective potential $U_k$
as $k$ flows from $k_\Phi$ to zero.  The large renormalized Yukawa
coupling rapidly drives the scalar mass term to negative values and
leads to a potential minimum away from the origin at some scale
$k_{\rm \chi SB}<k_\Phi$ such that finally
$\VEV{\Phi}=\ol{\sigma}_0\neq0$ for $k \to 0$ \cite{EW94-1,Ju95-7}.
This concludes our overview of the general features of chiral symmetry
breaking in the context of flow equations for QCD.

We will concentrate in this work on the two flavor case ($N=2$) and
comment on the effects of including the strange quark in
section~\ref{AdditionalDegreesOfFreedom}. Furthermore we will neglect
isospin violation and therefore consider a singlet source term
$\jmath$ proportional to the average light current quark mass
$\hat{m}\equiv\frac{1}{2}(m_u+m_d)$.  Due to the $U_A(1)$--anomaly
there is a mass split for the mesons described by $\Phi$.  The scalar
triplet $(a_0)$ and the pseudoscalar singlet $(\eta^\prime)$ receive a
large mass whereas the pseudoscalar triplet $(\pi)$ and the scalar
singlet $(\sigma)$ remains light. From the measured values
$m_{\eta^\prime},m_{a_0}\simeq1\GeV$ it is evident that a decoupling
of these mesons is presumably a very realistic limit\footnote{In
  thermal equilibrium at high temperature this decoupling is not
  obvious. We will comment on this point in section
  \ref{AdditionalDegreesOfFreedom}.}.  It can be achieved in a
chirally invariant way and leads to the well known $O(4)$ symmetric
Gell-Mann--Levy linear sigma model~\cite{GML60-1} which is, however,
coupled to quarks now. This is the two flavor linear quark meson model
which we will study in the remainder of this work.  For this model the
effective potential $U_k$ is a function of $\rho$ only.

It remains to determine the source $\jmath$ as a function of the
average current quark mass $\hat{m}$. This is carried out in appendix
\ref{Source} and we obtain in our normalization with
$Z_{\psi,k_{\Phi}}=1$, $\ol{h}_{k_{\Phi}}=1$,
\begin{equation}
  \label{AAA101a}
  \jmath=2 \ol{m}^2_{k_{\Phi}}\hat{m}\; .
\end{equation}
It is remarkable that higher order terms do not influence the relation
between $\jmath$ and $\hat{m}$.  Only the quadratic term
$\ol{m}^2_{k_\Phi}$ enters which is in our scenario the only relevant
coupling. This feature is an important ingredient for the predictive
power of the model as far as the absolute size of the current quark
mass is concerned.

The quantities which are directly connected to chiral symmetry
breaking depend on the $k$--dependent expectation value
$\VEV{\Phi}_k=\ol{\sigma}_{0,k}$ as given by
\begin{equation}
  \label{AAA101}
  \frac{\prl U_k}{\prl\rho}(\rho=2\ol{\sigma}_{0,k}^2)=
  \frac{\jmath}{2\ol{\sigma}_{0,k}}.
\end{equation}
In terms of the renormalized expectation value
\begin{equation}
  \label{AAA100}
  \sigma_{0,k}=Z_{\Phi,k}^{1/2}\ol{\sigma}_{0,k}\; 
\end{equation}
we obtain the following expressions for phenomenological observables
from (\ref{AAA60}) for\footnote{We note that the expressions
  (\ref{AAA65}) obey the well known Gell-Mann--Oakes--Renner relation
  $m_\pi^2
  f_\pi^2=-2\hat{m}\VEV{\ol{\psi}\psi}+\Oc(\hat{m}^2)$~\cite{GMOR68-1}.}
$d=4$
\begin{equation}
  \label{AAA65}
  \begin{array}{rcl}
    \ds{f_{\pi,k}} &=& \ds{2\sigma_{0,k}}\; ,\nnn
    \ds{\VEV{\ol{\psi}\psi}_k} &=& \ds{
      -2\ol{m}^2_{k_\Phi}\left[Z_{\Phi,k}^{-1/2}
      \sigma_{0,k}-\hat{m}\right]}\; ,\nnn
      \ds{M_{q,k}} &=& \ds{
        h_k\sigma_{0,k}}\; ,\nnn
      \ds{m^2_{\pi,k}} &=& \ds{
        Z_{\Phi,k}^{-1/2}
        \frac{\ol{m}^2_{k_\Phi}\hat{m}}{\sigma_{0,k}}=
        Z_{\Phi,k}^{-1/2}\frac{\jmath}{2\sigma_{0,k}}}\; ,\nnn
      \ds{m_{\sigma,k}^2} &=& \ds{
        Z_{\Phi,k}^{-1/2}
        \frac{\ol{m}^2_{k_\Phi}\hat{m}}{\sigma_{0,k}}+
        4\lambda_k\sigma_{0,k}^2}\; .
  \end{array}
\end{equation}
Here we have defined the dimensionless, renormalized couplings
\begin{equation}
  \label{AAA102}
  \begin{array}{rcl}
    \ds{\lambda}_k &=& \ds{Z_{\Phi,k}^{-2}
      \frac{\prl^2U_k}{\prl\rho^2}(\rho=2\ol{\sigma}_{0,k}^2)}\; ,\nnn
    \ds{h_k} &=& \ds{
      Z_{\Phi,k}^{-1/2}Z_{\psi,k}^{-1}\ol{h}_k}\; .
  \end{array}
\end{equation}
We will mainly be interested in the ``physical values'' of the
quantities (\ref{AAA65}) in the limit $k\ra0$ where the infrared
cutoff is removed, i.e.\ $f_{\pi}=f_{\pi,k=0}$,
$m_{\pi}^2=m_{\pi,k=0}^2$, etc.  We point out that the formalism of
composite fields provides the link \cite{EW94-1} to the chiral
condensate $\VEV{\ol{\psi}\psi}$ since the expectation value
$\ol{\sigma}_0$ is related to the expectation value of a composite
quark--antiquark operator.

\subsection{Flow equations and infrared stability}
\label{FlowEquationsAndInfraredStability}

At first sight, a reliable computation of $\Gamma_{k\ra0}$ seems a
very difficult task. Without a truncation $\Gamma_k$ is described by
an infinite number of parameters (couplings, wave function
renormalizations, etc.) as can be seen if $\Gamma_k$ is expanded in
powers of fields and derivatives. For instance, the sigma mass is
obtained as a zero of the exact inverse propagator,
$\lim_{k\ra0}\Gamma_k^{(2)}(q)|_{\Phi=\VEV{\Phi}}$, which formally
receives contributions from terms in $\Gamma_k$ with arbitrarily high
powers of derivatives and the expectation value $\sigma_0$.  Realistic
non-perturbative truncations of $\Gamma_k$ which reduce the problem to
a manageable size are crucial.  We will follow here a twofold
strategy:
\begin{itemize}
\item Physical observables like meson masses, decay constants, etc.,
  can be expanded in powers of (current) quark masses in a similar way
  as in chiral perturbation theory~\cite{GL82-1}. To a given finite
  order of this expansion only a finite number of terms of a
  simultaneous expansion of $\Gamma_k$ in powers of derivatives and
  $\Phi$ are required if the expansion point is chosen properly.
  Details of this procedure and some results can be found
  in~\cite{JuWet96,QuMa,JW97-1}.
\item Because of an approximate partial IR fixed point behavior of the
  flow equations in the symmetric regime, i.e. for $k_{\chi
    SB}<k<k_\Phi$, the values of many parameters of $\Gamma_k$ for
  $k\ra0$ will be almost independent of their initial values at the
  compositeness scale $k_{\Phi}$. For large enough $h_{k_\Phi}$ only a
  few relevant parameters need to be computed accurately from QCD.
  They can alternatively be determined from phenomenology. Because of
  the present lack of an explicit QCD computation we will pursue the
  latter approach.
\end{itemize}
In combination, these two points open the possibility for a perhaps
unexpected degree of predictive power within the linear quark meson
model.  We wish to stress, however, that a perturbative treatment of
the model at hand, e.g., using perturbative RG techniques, cannot be
expected to yield reliable results. The renormalized Yukawa coupling
is very large at the scale $k_\Phi$. Even the IR value of $h_k$ is
still relatively big
\begin{equation} 
  \label{IRh}
  h_{k=0}=\frac{2M_q}{f_\pi}\simeq6.5
\end{equation} 
and $h_k$ increases with $k$. The dynamics of the linear quark meson
model is therefore clearly non-perturbative for all scales $k\leq
k_\Phi$.

We will now turn to the flow equations for the linear quark meson
model.  We first note that the flow equations for $\Gamma_k$ and
$\Gamma_k-\frac{1}{2}\int d^4
x\tr\left(\jmath^\dagger\Phi+\Phi^\dagger\jmath\right)$ are identical.
The source term therefore does not need to be considered explicitly
and only appears in the condition (\ref{AAA101}) for $\VEV{\Phi}$.  It
is convenient to work with dimensionless and renormalized variables
therefore eliminating all explicit $k$--dependence. With
\begin{equation}
  \label{AAA190}
  u(t,\tilde{\rho})\equiv k^{-d}U_k(\rho)\; ,\;\;\;
  \tilde{\rho}\equiv Z_{\Phi,k} k^{2-d}\rho
\end{equation}
and using (\ref{AAA60}) as a first truncation of the effective average
action $\Gamma_k$ one obtains the flow equation ($t=\ln(k/k_\Phi)$)
\begin{equation}
  \label{AAA68}
  \begin{array}{rcl}
    \ds{\frac{\prl}{\prl t}u} &=& \ds{
      -d u+\left(d-2+\eta_\Phi\right)
      \tilde{\rho}u^\prime}\\[2mm]
    &+& \ds{
      2v_d\left\{
      3l_0^d(u^\prime;\eta_\Phi)+
      l_0^d(u^\prime+2\tilde{\rho}u^{\prime\prime};\eta_\Phi)-
      2^{\frac{d}{2}+1}N_c
      l_0^{(F)d}(\frac{1}{2}\tilde{\rho}h^2;\eta_\psi)
      \right\} }\; .
  \end{array}
\end{equation}
Here $v_d^{-1}\equiv2^{d+1}\pi^{d/2}\Gamma(d/2)$ and primes denote
derivatives with respect to $\tilde{\rho}$. The number of quark colors
is denoted as $N_c$. We will always use in the following $N_c=3$.
Eq.~(\ref{AAA68}) is a partial differential equation for the effective
potential $u(t,\tilde{\rho})$ which has to be supplemented by the flow
equation for the Yukawa coupling and expressions for the anomalous
dimensions $\eta_\Phi$, $\eta_\psi$. The symbols $l_n^d$, $l_n^{(F)d}$
denote bosonic and fermionic mass threshold functions, respectively,
which are defined in appendix \ref{ThresholdFunctions}.  They describe
the decoupling of massive modes and provide an important
non-perturbative ingredient. For instance, the bosonic threshold
functions
\begin{equation}
 \label{AAA85}
 l_n^d(w;\eta_\Phi)=\frac{n+\delta_{n,0}}{4}v_d^{-1}
 k^{2n-d}\int\frac{d^d q}{(2\pi)^d}
 \frac{1}{Z_{\Phi,k}}\frac{\prl R_k}{\prl t}
 \frac{1}{\left[ P(q^2)+k^2w\right]^{n+1}}
\end{equation}
involve the inverse average propagator
$P(q^2)=q^2+Z_{\Phi,k}^{-1}R_k(q^2)$ where the infrared cutoff is
manifest. These functions decrease $\sim w^{-(n+1)}$ for $w\gg1$.
Since typically $w=M^2/k^2$ with $M$ a mass of the model, the main
effect of the threshold functions is to cut off fluctuations of
particles with masses $M^2\gg k^2$. Once the scale $k$ is changed
below a certain mass threshold, the corresponding particle no longer
contributes to the evolution and decouples smoothly.

The dimensionless renormalized expectation value
$\kappa\equiv2k^{2-d}Z_{\Phi,k}\ol{\sigma}_{0,k}^2$, with
$\ol{\sigma}_{0,k}$ the $k$--dependent VEV of $\Phi$, may be computed
for each $k$ directly from the condition (\ref{AAA101})
\begin{equation}
  \label{AAA90}
  u^\prime(t,\kappa)=\frac{\jmath}{\sqrt{2\kappa}}
  k^{-\frac{d+2}{2}}Z_{\Phi,k}^{-1/2}\equiv
  \epsilon_g \; .
\end{equation}
Note that $\kappa\equiv0$ in the symmetric regime for vanishing source
term. Equation (\ref{AAA90}) allows us to follow the flow of $\kappa$
according to
\begin{eqnarray}
  \label{AAA91}
  \ds{\frac{d}{d t}\kappa} &=& \ds{
    \frac{\kappa}{\epsilon_g+2\kappa\lambda}
    \Bigg\{\left[\eta_\Phi-d-2\right]\epsilon_g-
    2\frac{\prl}{\prl t}u^\prime(t,\kappa)\Bigg\} }
\end{eqnarray}
with $\lambda\equiv u^{\prime\prime}(t,\kappa)$.  We define the Yukawa
coupling for $\tilde{\rho}=\kappa$ and its flow equation
reads~\cite{Ju95-7}
\begin{equation}
  \label{AAA70}
  \begin{array}{rcl}
  \ds{\frac{d}{d t}h^2} &=& \ds{
  \left(d-4+2\eta_\psi+\eta_\Phi\right)h^2-
    2v_d h^4\Bigg\{
    3l_{1,1}^{(F B)d}(\frac{1}{2}h^2\kappa,
    \epsilon_g;\eta_\psi,\eta_\Phi) }\\[2mm]
  &-& \ds{
    l_{1,1}^{(F B)d}(\frac{1}{2}h^2\kappa,
    \epsilon_g+2\lambda\kappa;
    \eta_\psi,\eta_\Phi)
  \Bigg\} }\; .
  \end{array}
\end{equation}
Similarly, the scalar and quark anomalous
dimensions are infered from
\begin{equation}
  \label{AAA69}
  \begin{array}{rcl}
    \ds{\eta_\Phi} &\equiv& \ds{
      -\frac{d}{d t}\ln Z_{\Phi,k}=
      4\frac{v_d}{d}\Bigg\{
      4\kappa\lambda^2
      m_{2,2}^d(\epsilon_g,\epsilon_g+2\lambda\kappa;
      \eta_\Phi) }\nnn
    &+& \ds{
      2^{\frac{d}{2}}N_c h^2
      m_4^{(F)d}(\frac{1}{2}h^2\kappa;
      \eta_\psi)
      \Bigg\}\; , }\nnn
    \ds{\eta_\psi} &\equiv& \ds{
      -\frac{d}{d t}\ln Z_{\psi,k}=
      2\frac{v_d}{d}h^2\Bigg\{
      3m_{1,2}^{(F B)d}(\frac{1}{2}h^2\kappa,
      \epsilon_g;\eta_\psi,\eta_\Phi) }\\[2mm]
    &+& \ds{
      m_{1,2}^{(F B)d}(\frac{1}{2}h^2\kappa,
      \epsilon_g+2\lambda\kappa;
      \eta_\psi,\eta_\Phi)
      \Bigg\}\; , }
  \end{array}
\end{equation}
which is a linear set of equations for the anomalous dimensions.  The
threshold functions $l_{n_1,n_2}^{(FB)d}$, $m_{n_1,n_2}^d$,
$m_4^{(F)d}$ and $m_{n_1,n_2}^{(FB)d}$ are also specified in appendix
\ref{ThresholdFunctions}.

The flow equations (\ref{AAA68}), (\ref{AAA91})---(\ref{AAA69}),
constitute a coupled system of ordinary and partial differential
equations which can be integrated numerically.  Here we take the
effective current quark mass dependence of $h_k$, $Z_{\Phi,k}$ and
$Z_{\psi,k}$ into account by stopping the evolution according to
eqs.~(\ref{AAA70}), (\ref{AAA69}), evaluated for the chiral limit,
below the pion mass $m_\pi$.  (For details of the algorithm used here
see refs.~\cite{ABBTW95,BW97-1}.)  One finds for $d=4$ that chiral
symmetry breaking indeed occurs for a wide range of initial values of
the parameters including the presumably realistic case of large
renormalized Yukawa coupling and a bare mass $\ol{m}_{k_\Phi}$ of
order $100\MeV$. Driven by the strong Yukawa coupling, the
renormalized mass term $u^\prime(t,\tilde{\rho}=0)$ decreases rapidly
and goes through zero at a scale $k_{\chi{\rm SB}}$ not far below
$k_\Phi$.  Here the system enters the spontaneously broken regime and
the effective average potential develops an absolute minimum away from
the origin.  The evolution of of the potential minimum
$\sigma_{0,k}^2=k^2\kappa/2$ turns out to be reasonably stable already
before $k\simeq m_\pi$ where it stops. We take this result as an
indication that our truncation of the effective action $\Gm_k$ leads
at least qualitatively to a satisfactory description of chiral
symmetry breaking. The reason for the relative stability of the IR
behavior of the VEV (and all other couplings) is that the quarks
acquire a constituent mass $M_q=h\sigma_0\simeq300\MeV$ in the
spontaneously broken regime. As a consequence they decouple once $k$
becomes smaller than $M_q$ and the evolution is then dominantly driven
by the light Goldstone bosons.  This is also important for our
approximation of neglecting the residual gluonic interactions in the
quark sector of the model as outlined in
subsection~\ref{ASemiQuantitativePicture}.

Most importantly, one finds that the system of flow equations exhibits
an approximate IR fixed point behavior in the symmetric
regime~\cite{Ju95-7}. To see this explicitly we study the flow
equations (\ref{AAA68}), (\ref{AAA91})---(\ref{AAA69}) subject to the
condition (\ref{AAA32}). For the relevant range of $\tilde{\rho}$ both
$u^\prime(t,\tilde{\rho})$ and $u^\prime(t,\tilde{\rho})+
2\tilde{\rho}u^{\prime\prime}(t,\tilde{\rho})$ are then much larger
than $\tilde{\rho}h^2(t)$ and we may therefore neglect in the flow
equations all scalar contributions with threshold functions involving
these large masses.  This yields the simplified equations
($d=4,v_4^{-1}=32\pi^2$)
\begin{equation}
  \label{AAA110}
  \begin{array}{rcl}
    \ds{\frac{\prl}{\prl t}u} &=& \ds{
      -4u+\left(2+\eta_\Phi\right)
      \tilde{\rho}u^\prime
      -\frac{N_c}{2\pi^2}
      l_0^{(F)4}(\frac{1}{2}\tilde{\rho}h^2)\; ,
      }\nnn
    \ds{\frac{d}{d t}h^2} &=& \ds{
      \eta_\Phi h^2 m\; ,
      }\nnn
      \ds{\eta_\Phi} &=& \ds{
        \frac{N_c}{8\pi^2}
        m_4^{(F)4}(0)h^2\; ,
        }\nnn
      \ds{\eta_\psi} &=& \ds{0}\; .
    \end{array}
\end{equation}
Of course, it should be clear that this approximation is only valid
for the initial range of running below $k_\Phi$ before the
(dimensionless) renormalized scalar mass squared
$u^\prime(t,\tilde{\rho}=0)$ approaches zero near the chiral symmetry
breaking scale.  The system (\ref{AAA110}) is exactly soluble. Using
$m_4^{(F)4}(0)=1$ which holds independently of the choice of the IR
cutoff we find
\begin{equation}
  \label{AAA113}
  \begin{array}{rcl}
    \ds{h^2(t)} &=& \ds{
      Z_\Phi^{-1}(t)=
      \frac{h_I^2}{1-\frac{N_c}{8\pi^2}h_I^2 t}\; ,
      }\nnn
    \ds{u(t,\tilde{\rho})} &=& \ds{
      e^{-4t}u_I(e^{2t}\tilde{\rho}\frac{h^2(t)}{h_I^2})-
      \frac{N_c}{2\pi^2}\int_0^t d r e^{-4r}
      l_0^{(F)4}(\frac{1}{2}h^2(t)\tilde{\rho}e^{2r}) }\; .
  \end{array}
\end{equation}
(The integration over $r$ on the right hand side of the solution for
$u$ can be carried out by first exchanging it with the one over
momentum implicit in the definition of the threshold function
$l_0^{(F)4}$ (see appendix \ref{ThresholdFunctions}).) Here
$u_I(\tilde{\rho})\equiv u(0,\tilde{\rho})$ denotes the effective
average potential at the compositeness scale and $h_I^2$ is the
initial value of $h^2$ at $k_\Phi$, i.e. for $t=0$.  For simplicity we
will use an expansion of the initial value effective potential
$u_I(\tilde{\rho})$ in powers of $\tilde{\rho}$ around
$\tilde{\rho}=0$
\begin{equation}
  \label{AAA140}
  u_I(\tilde{\rho})=
  \sum_{n=0}^\infty
  \frac{u_I^{(n)}(0)}{n!}\tilde{\rho}^n
\end{equation}
even though this is not essential for the forthcoming reasoning.
Expanding also $l_0^{(F)4}$ in eq.~(\ref{AAA113}) in powers of its
argument one finds for $n>2$
\begin{equation}
  \label{LLL00}
  \ds{\frac{u^{(n)}(t,0)}{h^{2n}(t)}} = \ds{
    e^{2(n-2)t}\frac{u_I^{(n)}(0)}{h_I^{2n}}+
    \frac{N_c}{\pi^2}
    \frac{(-1)^n (n-1)!}{2^{n+2}(n-2)}
    l_n^{(F)4}(0)
    \left[1-e^{2(n-2)t}\right]}\; .
\end{equation}
For decreasing $t\ra-\infty$ the initial values $u_I^{(n)}$ become
rapidly unimportant and $u^{(n)}/h^{2n}$ approaches a fixed point.
For $n=2$, i.e., for the quartic coupling, one finds
\begin{equation}
  \label{LLL01}
  \frac{u^{(2)}(t,0)}{h^2(t)}=
  1-\frac{1-\frac{u_I^{(2)}(0)}{h_I^2}}
  {1-\frac{N_c}{8\pi^2}h_I^2 t}
\end{equation}
leading to a fixed point value $(u^{(2)}/h^2)_*=1$. As a consequence
of this fixed point behavior the system looses all its ``memory'' on
the initial values $u_I^{(n\ge2)}$ at the compositeness scale
$k_\Phi$! This typically happens before the approximation
$u^\prime(t,\tilde{\rho}),u^\prime(t,\tilde{\rho})+
2\tilde{\rho}u^{\prime\prime}(t,\tilde{\rho})\gg\tilde{\rho}h^2(t)$
breaks down and the solution (\ref{AAA113}) becomes invalid.
Furthermore, the attraction to partial infrared fixed points continues
also for the range of $k$ where the scalar fluctuations cannot be
neglected anymore.  The initial value for the bare dimensionless mass
parameter
\begin{equation}
  \label{AAA142}
  \frac{u_I^\prime(0)}{h_I^2}=
  \frac{\ol{m}^2_{k_\Phi}}{k_\Phi^2}
\end{equation}
is never negligible. (In fact, using the values for
$\ol{m}^2_{k_\Phi}$ and $k_\Phi$ computed in ref.~\cite{EW94-1} one
obtains $\ol{m}^2_{k_\Phi}/k_\Phi^2\simeq0.036$.)  For large $h_I$
(and dropping the constant piece $u_I(0)$) the solution (\ref{AAA113})
therefore behaves with growing $\abs{t}$ as
\begin{equation}
  \label{AAA150}
  \begin{array}{rcl}
    \ds{Z_\Phi(t)} &\simeq& \ds{
      -\frac{N_c}{8\pi^2}t\; ,
      }\nnn
    \ds{h^2(t)} &\simeq& \ds{
      -\frac{8\pi^2}{N_c t}\; ,
      }\nnn
    \ds{u(t,\tilde{\rho})} &\simeq& \ds{
      \frac{u_I^\prime(0)}{h_I^2} e^{-2t} h^2(t)\tilde{\rho}
      -\frac{N_c}{2\pi^2}\int_0^t d r e^{-4r}
      l_0^{(F)4}(\frac{1}{2}h^2(t)\tilde{\rho}e^{2r}) }\; .  
  \end{array}
\end{equation}
In other words, for $h_I\ra\infty$ the IR behavior of the linear quark
meson model will depend (in addition to the value of the compositeness
scale $k_\Phi$ and the quark mass $\hat{m}$) only on one parameter,
$\ol{m}^2_{k_\Phi}$.  We have numerically verified this feature by
starting with different values for $u_I^{(2)}(0)$.  Indeed, the
differences in the physical observables were found to be small.  This
IR stability of the flow equations leads to a perhaps surprising
degree of predictive power!  For definiteness we will perform our
numerical analysis of the full system of flow equations (\ref{AAA68}),
(\ref{AAA91})---(\ref{AAA69}) with the idealized initial value
$u_I(\tilde{\rho})=u_I^\prime(0)\tilde{\rho}$ in the limit
$h_I^2\ra\infty$. It should be stressed, though, that deviations from
this idealization will lead only to small numerical deviations in the
IR behavior of the linear quark meson model as long as the condition
(\ref{AAA32}) holds, say for $h_I\gta15$~\cite{Ju95-7}.

With this knowledge at hand we may now fix the remaining three
parameters of our model, $k_\Phi$, $\ol{m}^2_{k_\Phi}$ and $\hat{m}$ by
using $f_\pi=92.4\MeV$, the pion mass $M_\pi=135\MeV$ and the
constituent quark mass $M_q$ as phenomenological input.  Because of
the uncertainty regarding the precise value of $M_q$ we give in table
\ref{tab1} the results for several values of $M_q$.
\begin{table}
\begin{center}
\begin{tabular}{|c|c||c|c|c||c|c|c|c|} \hline
  $\frac{M_q}{\MeV}$ &
  $\frac{\lambda_I}{h_I^2}$ &
  $\frac{k_\Phi}{\MeV}$ &
  $\frac{\ol{m}^2_{k_\Phi}}{k_\Phi^2}$ &
  $\frac{\jmath^{1/3}}{\MeV}$ &
  $\frac{\hat{m}(k_\Phi)}{\MeV}$ &
  $\frac{\hat{m}(1\GeV)}{\MeV}$ &
  $\frac{\VEV{\ol{\psi}\psi}(1\GeV)}{\MeV^3}$ &
  $\frac{f_\pi^{(0)}}{\MeV}$
  \\[0.5mm] \hline\hline
  $303$ &
  $1$ &
  $618$ &
  $0.0265$ &
  $66.8$ &
  $14.7$ &
  $11.4$ &
  $-(186)^3$ &
  $80.8$
  \\ \hline
  $300$ &
  $0$ &
  $602$ &
  $0.026$ &
  $66.8$ &
  $15.8$ &
  $12.0$ &
  $-(183)^3$ &
  $80.2$
  \\ \hline
  $310$ &
  $0$ &
  $585$ &
  $0.025$ &
  $66.1$ &
  $16.9$ &
  $12.5$ &
  $-(180)^3$ &
  $80.5$
  \\ \hline
  $339$ &
  $0$ &
  $552$ &
  $0.0225$ &
  $64.4$ &
  $19.5$ &
  $13.7$ &
  $-(174)^3$ &
  $81.4$
  \\ \hline
\end{tabular}
\caption{\footnotesize The table shows the dependence on the
  constituent quark mass $M_q$ of the input parameters $k_\Phi$,
  $\ol{m}^2_{k_\Phi}/k_\Phi^2$ and $\jmath$ as well as some of our
  ``predictions''. The phenomenological input used here besides $M_q$
  is $f_\pi=92.4\MeV$, $m_\pi=135\MeV$.The first line corresponds to
  the values for $M_q$ and $\lambda_I$ used in the remainder of this
  work. The other three lines demonstrate the insensitivity of our
  results with respect to the precise values of these
  parameters.}
\label{tab1}
\end{center}
\end{table}
The first line of table~\ref{tab1} corresponds to the choice of $M_q$
and $\lambda_I\equiv u_I^{\prime\prime}(\kappa)$ which we will use for
the forthcoming analysis of the model at finite temperature.  As
argued analytically above the dependence on the value of $\lambda_I$
is weak for large enough $h_I$ as demonstrated numerically by the
second line. Moreover, we notice that our results, and in particular
the value of $\jmath$, are rather insensitive with respect to the
precise value of $M_q$. It is remarkable that the values for $k_\Phi$
and $\ol{m}_{k_\Phi}$ are not very different from those computed in
ref.~\cite{EW94-1}.  As compared to the analysis of ref.~\cite{Ju95-7}
the present truncation of $\Gamma_k$ is of a higher level of accuracy:
We now consider an arbitrary form of the effective average potential
instead of a polynomial approximation and we have included the pieces
in the threshold functions which are proportional to the anomalous
dimensions. It is encouraging that the results are rather robust with
respect to these improvements of the truncation.

Once the parameters $k_\Phi$, $\ol{m}^2_{k_\Phi}$ and $\hat{m}$ are
fixed there are a number of ``predictions'' of the linear meson model
which can be compared with the results obtained by other methods or
direct experimental observation. First of all one may compute the
value of $\hat{m}$ at a scale of $1\GeV$ which is suitable for
comparison with results obtained from chiral perturbation
theory~\cite{Leu96-1} and sum rules~\cite{JM95-1}.  For this purpose
one has to account for the running of this quantity with the
normalization scale from $k_\Phi$ as given in table \ref{tab1} to the
commonly used value of $1\GeV$:
$\hat{m}(1\GeV)=A^{-1}\hat{m}(k_\Phi)$. A reasonable estimate of the
factor $A$ is obtained from the three loop running of $\hat{m}$ in the
$\ol{MS}$ scheme~\cite{JM95-1}. For $M_q\simeq300\MeV$ corresponding
to the first two lines in table~\ref{tab1} its value is $A\simeq1.3$.
The results for $\hat{m}(1\GeV)$ are in acceptable agreement with
recent results from other methods~\cite{Leu96-1,JM95-1} even though
they tend to be somewhat larger. Closely related to this is the value
of the chiral condensate
\begin{equation}
  \label{LLL04}
  \VEV{\ol{\psi}\psi}(1\GeV)\equiv
  -A\ol{m}^2_{k_\Phi}
  \left[f_\pi Z_{\Phi,k=0}^{-1/2}-
    2\hat{m}\right]
    \; .
\end{equation}
These results are quite non--trivial since not only $f_\pi$ and
$\ol{m}^2_{k_\Phi}$ enter but also the computed IR value
$Z_{\Phi,k=0}$.  We emphasize in this context that there may be
substantial corrections both in the extrapolation from $k_\Phi$ to
$1\GeV$ and in the factor $a_q$ (see~(\ref{PPF00})). The latter is due
to the neglected influence of the strange quark which may be important
near $k_\Phi$. These uncertainties have only little effect on the
physics at lower scales as relevant for our analysis of the
temperature effects. Only the value of $\jmath$ which is fixed by
$m_\pi$ enters here.

A further more qualitative test concerns the mass of the sigma
resonance or radial mode whose renormalized mass squared may be
computed according to (\ref{AAA65}) in the limit $k\ra0$. From our
numerical analysis we obtain $\lambda_{k=0}\simeq20$ which translates
into $m_\sigma\simeq 430\MeV$.  One should note, though, that this
result is presumably not very accurate as we have employed in this
work the approximation of using the Goldstone boson wave function
renormalization constant also for the radial mode. Furthermore, the
explicit chiral symmetry breaking contribution to $m_\sigma^2$ is
certainly underestimated as long as the strange quark is neglected.
In any case, we observe that the sigma meson is significantly heavier
than the pions. This is a crucial consistency check for the linear
quark meson model. A low sigma mass would be in conflict with the
numerous successes of chiral perturbation theory~\cite{GL82-1} which
requires the decoupling of all modes other than the Goldstone bosons
in the IR--limit of QCD. The decoupling of the sigma meson is, of
course, equivalent to the limit $\lambda\ra\infty$ which formally
describes the transition from the linear to the non--linear sigma
model and which appears to be reasonably well realized by the large
IR--values of $\lambda$ obtained in our analysis. We also note that
the issue of the sigma mass is closely connected to the value of
$f_\pi^{(0)}$, the value of $f_\pi$ in the chiral limit $\hat{m}=0$
also given in table~\ref{tab1}.  To lowest order in
$(f_\pi-f_\pi^{(0)})/f_\pi$ or, equivalently, in $\hat{m}$ one
has
\begin{equation}
  \label{ABC00}
  f_\pi-f_\pi^{(0)}=\frac{\jmath}{Z_\Phi^{1/2}m_\sigma^2}=
  \frac{f_\pi m_\pi^2}{m_\sigma^2}\; .
\end{equation}
A larger value of $m_\sigma$ would therefore reduce the difference
between $f_\pi^{(0)}$ and $f_\pi$.

\begin{figure}
\unitlength1.0cm
\begin{center}
\begin{picture}(13.,7.0)
\put(0.0,0.0){
\epsfysize=11.cm
\rotate[r]{\epsffile{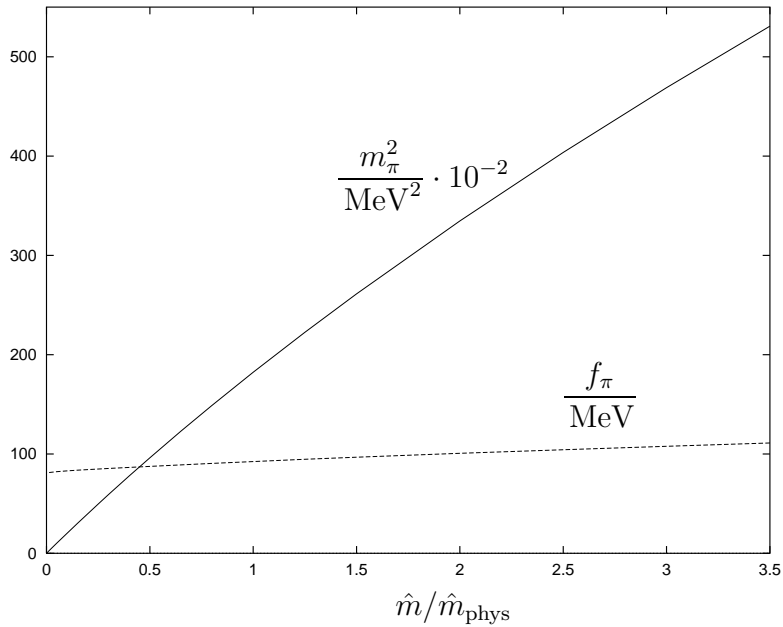}}
}
\put(8.0,2.3){\bf $\ds{\frac{f_\pi}{\MeV}}$}
\put(5.0,5.2){\bf $\ds{\frac{m^2_\pi}{\MeV^2}\cdot10^{-2}}$}
\put(5.8,-0.5){\bf $\ds{\hat{m}/\hat{m}_{\rm phys}}$}
\end{picture}
\end{center}
\caption{\footnotesize The plot shows $m_\pi^2$ (solid line) and
  $f_\pi$ (dashed line) as functions of the current quark mass
  $\hat{m}$ in units of the physical value $\hat{m}_{\rm phys}$.}
\label{mm}
\end{figure}

In figure~\ref{mm} we show the dependence of the pion mass and decay
constant on the average current quark mass $\hat{m}$.
These curves depend very little on the values of the initial
parameters as demonstrated in table~\ref{tab1} by $f_\pi^{(0)}$. We
observe a relatively large difference of $12\MeV$ between the pion
decay constants at $\hat{m}=\hat{m}_{\rm phys}$ and $\hat{m}=0$.
According to (\ref{ABC00}) this difference is related to the mass of
the sigma particle and will be modified in the three flavor case. We
will later find that the critical temperature $T_c$ for the second
order phase transition in the chiral limit is almost independent of
the initial conditions. The values of $f_\pi^{(0)}$ and $T_c$
essentially determine the non--universal amplitudes in the critical
scaling region (cf.~section~\ref{CriticalBehavior}). In summary, we
find that the behavior of our model for small $k$ is quite robust as
far as uncertainties in the initial conditions at the scale $k_\Phi$
are concerned. We will see that the difference of observables between
non--vanishing and vanishing temperature is entirely determined by the
flow of couplings in the range $0<k\lta3T$.

\subsection{Thermal equilibrium and dimensional reduction}
\label{FiniteTemperatureFormalism}

The extension of flow equations to thermal equilibrium situations at
non--vanishing temperature $T$ is straightforward~\cite{TetWet}. In
the Euclidean formalism non--zero temperature results in
(anti--)periodic boundary conditions for (fermionic) bosonic fields in
the Euclidean time direction with periodicity $1/T$~\cite{Kap}.
This leads to the replacement
\begin{equation}
  \label{AAA120}
  \int\frac{d^d q}{(2\pi)^d}f(q^2)\ra
  T\sum_{l\in\ZZZ}\int\frac{d^{d-1}\vec{q}}{(2\pi)^{d-1}}
  f(q_0^2(l)+\vec{q}^{\,2})
\end{equation}
in the trace in (\ref{ERGE}) when represented as a momentum
integration, with a discrete spectrum for the zero component
\begin{equation}
  \label{AAA121}
  q_0(l)=\left\{
  \begin{array}{lll}
    2l\pi T &{\rm for}& {\rm bosons}\\
    (2l+1)\pi T &{\rm for}& {\rm fermions}\; .
  \end{array}\right.
\end{equation}
Hence, for $T>0$ a four--dimensional QFT can be interpreted as a
three--dimensional model with each bosonic or fermionic degree of
freedom now coming in an infinite number of copies labeled by
$l\in\ZZ$ (Matsubara modes). Each mode acquires an additional
temperature dependent effective mass term $q_0^2(l)$. In a high
temperature situation where all massive Matsubara modes decouple from
the dynamics of the system one therefore expects to observe an
effective three--dimensional theory with the bosonic zero modes as the
only relevant degrees of freedom. In other words, if the
characteristic length scale associated with the physical system is
much larger than the inverse temperature the compactified Euclidean
``time'' dimension cannot be resolved anymore. This phenomenon is
known as ``dimensional reduction''~\cite{DR}.

\begin{figure}
\unitlength1.0cm
\begin{center}
\begin{picture}(13.,18.0)

\put(0.0,9.5){
\epsfysize=11.cm
\rotate[r]{\epsffile{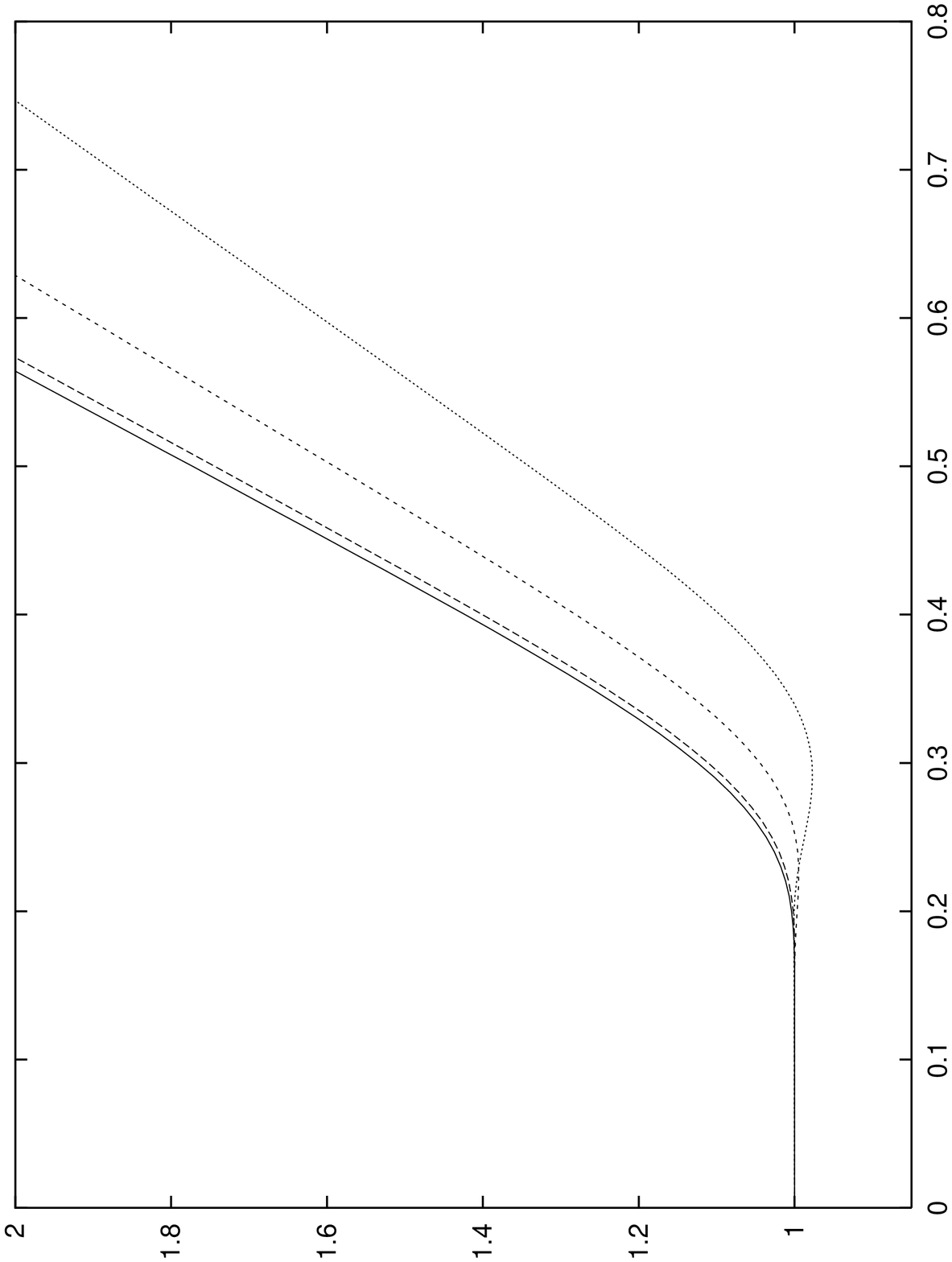}}
}
\put(-1.0,16.5){\bf $\ds{\frac{l_1^4\left(w,\ds{\frac{T}{k}}\right)}
    {l_1^4(w)}}$}
\put(1.5,16.5){\bf $\ds{(a)}$}

\put(0.0,0.5){
\epsfysize=11.cm
\rotate[r]{\epsffile{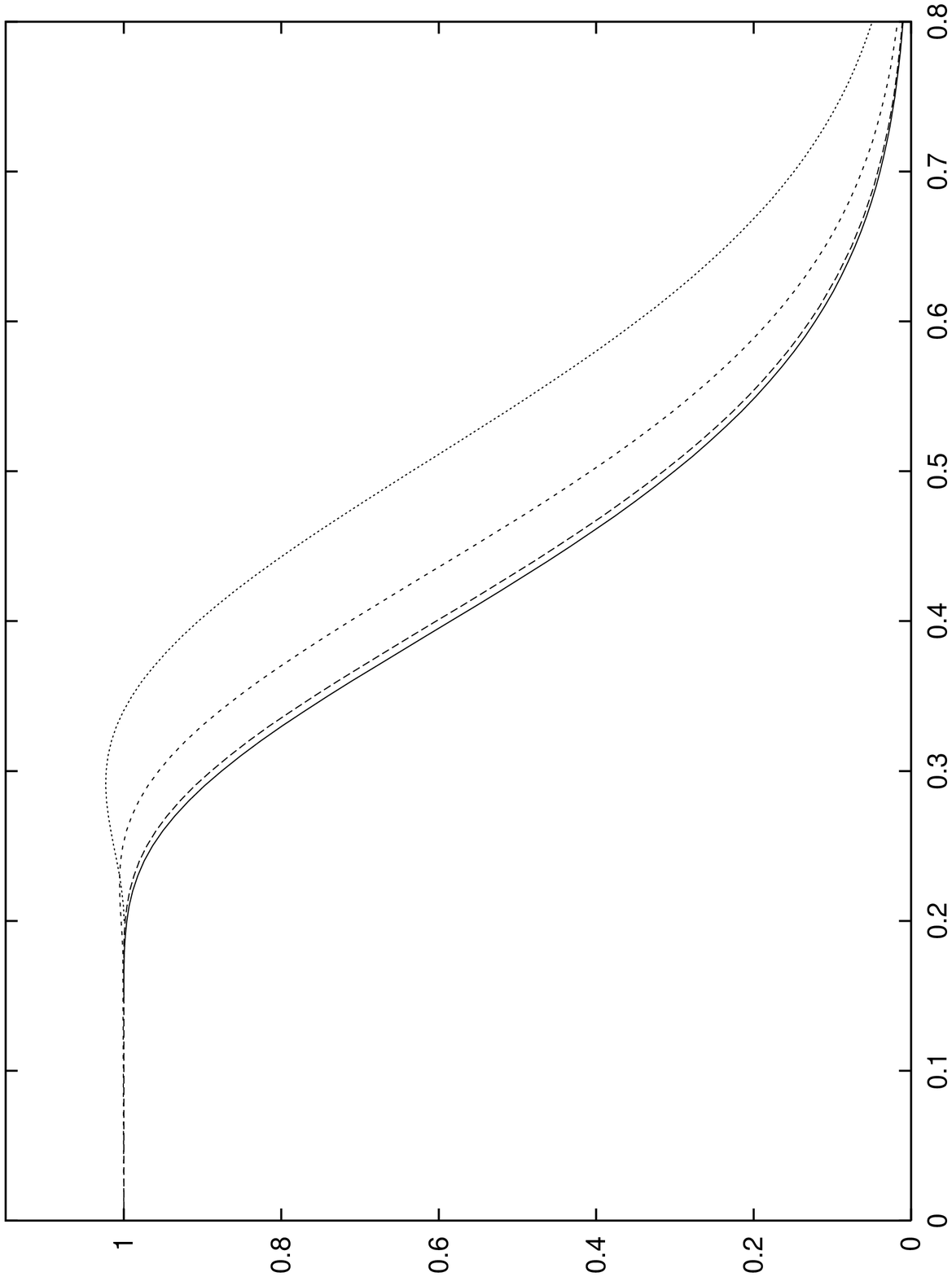}}
}
\put(-1.2,7.5){\bf
  $\ds{\frac{l_1^{(F)4}\left(w,\ds{\frac{T}{k}}\right)}
    {l_1^{(F)4}(w)}}$}
\put(6.1,-0.2){\bf $\ds{T/k}$}
\put(1.5,7.5){\bf $\ds{(b)}$}
\end{picture}
\end{center}
\caption{\footnotesize The plot shows the temperature 
  dependence of the bosonic (a) and the fermionic (b) threshold
  functions $l_1^4(w,T/k)$ and $l_1^{(F)4}(w,T/k)$, respectively, for
  different values of the dimensionless mass term $w$.  The solid line
  corresponds to $w=0$ whereas the dotted ones correspond to $w=0.1$,
  $w=1$ and $w=10$ with decreasing size of the dots.  For $T \gg k$
  the bosonic threshold function becomes proportional to $T/k$ whereas
  the fermionic one tends to zero.  In this range the theory with
  properly rescaled variables behaves as a classical
  three--dimensional theory.}
\label{Thresh}
\end{figure}

The formalism of the effective average action automatically provides
the tools for a smooth decoupling of the massive Matsubara modes as
the scale $k$ is lowered from $k\gg T$ to $k\ll T$.  It therefore
allows us to directly link the low--$T$, four--dimensional QFT to the
effective three--dimensional high--$T$ theory. The replacement
(\ref{AAA120}) in (\ref{ERGE}) manifests itself in the flow equations
(\ref{AAA68}), (\ref{AAA91})---(\ref{AAA69}) only through a change to
$T$--dependent threshold functions.  For instance, the dimensionless
functions $l_n^d(w;\eta_\Phi)$ defined in eq.~(\ref{AAA85}) are
replaced by
\begin{equation}
  \label{AAA200}
  l_n^d(w,\frac{T}{k};\eta_\Phi)\equiv
  \frac{n+\delta_{n,0}}{4}v_d^{-1}k^{2n-d}
  T\sum_{l\in\ZZZ}\int
  \frac{d^{d-1}\vec{q}}{(2\pi)^{d-1}}
  \left(\frac{1}{Z_{\Phi,k}}\frac{\prl R_k(q^2)}{\prl t}\right)
  \frac{1}{\left[P(q^2)+k^2 w\right]^{n+1}}
\end{equation}
where $q^2=q_0^2+\vec{q}^{\,2}$ and $q_0=2\pi l T$. A list of the
various temperature dependent threshold functions appearing in the
flow equations can be found in appendix~\ref{AnomalousDimensions}.
There we also discuss some subtleties regarding the definition of the
Yukawa coupling and the anomalous dimensions for $T\neq0$. In the
limit $k\gg T$ the sum over Matsubara modes approaches the integration
over a continuous range of $q_0$ and we recover the zero temperature
threshold function $l_n^d(w;\eta_\Phi)$.  In the opposite limit $k\ll
T$ the massive Matsubara modes ($l\neq0$) are suppressed and we expect
to find a $d-1$ dimensional behavior of $l_n^d$. In fact, one obtains
from~(\ref{AAA200})
\begin{equation}
  \label{AAA201}
  \begin{array}{rclcrcl}
    \ds{l_n^d(w,T/k;\eta_\Phi)} &\simeq& \ds{
      l_n^{d}(w;\eta_\Phi)}
    &{\rm for}& \ds{T\ll k}\; ,\nnn
    \ds{l_n^d(w,T/k;\eta_\Phi)} &\simeq& \ds{
      \frac{T}{k}\frac{v_{d-1}}{v_d}
      l_n^{d-1}(w;\eta_\Phi)}
    &{\rm for}& \ds{T\gg k}\; .
  \end{array}
\end{equation}
For our choice of the infrared cutoff function $R_k$,
eq.~(\ref{Rk(q)}), the temperature dependent Matsubara modes in
$l_n^d(w,T/k;\eta_\Phi)$ are exponentially suppressed for $T\ll k$
whereas the behavior is more complicated for other threshold functions
appearing in the flow equations (\ref{AAA68}),
(\ref{AAA91})---(\ref{AAA69}).  Nevertheless, all bosonic threshold
functions are proportional to $T/k$ for $T\gg k$ whereas those with
fermionic contributions vanish in this limit\footnote{For the present
  choice of $R_k$ the temperature dependence of the threshold
  functions is considerably smoother than in
  ref.~\cite{TetWet}.}. This behavior is demonstrated in figure
\ref{Thresh} where we have plotted the quotients
$l_1^4(w,T/k)/l_1^4(w)$ and $l_1^{(F)4}(w,T/k)/l_1^{(F)4}(w)$ of
bosonic and fermionic threshold functions, respectively.
One observes that for $k\gg T$ both threshold functions essentially
behave as for zero temperature. For growing $T$ or decreasing $k$ this
changes as more and more Matsubara modes decouple until finally all
massive modes are suppressed. The bosonic threshold function $l^4_1$
shows for $k \ll T$ the linear dependence on $T/k$ derived in
eq.~(\ref{AAA201}).  In particular, for the bosonic excitations the
threshold function for $w\ll1$ can be approximated with reasonable
accuracy by $l_n^4(w;\eta_\Phi)$ for $T/k<0.25$ and by
$(4T/k)l_n^3(w;\eta_\Phi)$ for $T/k>0.25$. The fermionic threshold
function $l_1^{(F)4}$ tends to zero for $k\ll T$ since there is no
massless fermionic zero mode, i.e.~in this limit all fermionic
contributions to the flow equations are suppressed.  On the other
hand, the fermions remain quantitatively relevant up to $T/k\simeq0.6$
because of the relatively long tail in figure~\ref{Thresh}b.  The
transition from four to three--dimensional threshold functions leads
to a {\em smooth dimensional reduction} as $k$ is lowered from $k\gg
T$ to $k\ll T$!  Whereas for $k\gg T$ the model is most efficiently
described in terms of standard four--dimensional fields $\Phi$ a
choice of rescaled three--dimensional variables
$\Phi_{3}=\Phi/\sqrt{T}$ becomes better adapted for $k\ll T$.
Accordingly, for high temperatures one will use a potential
\begin{equation}
  \label{CCC01}
  u_{3}(t,\tilde{\rho}_{3})=\frac{k}{T}
  u(t,\tilde{\rho})\; ;\;\;\;
  \tilde{\rho}_{3}=\frac{k}{T}\tilde{\rho}\; .
\end{equation}
In this regime $\Gamma_{k\ra0}$ corresponds to the free energy of
classical statistics and $\Gamma_{k>0}$ is a classical coarse grained
free energy.

For our numerical calculations at non--vanishing temperature we
exploit the discussed behavior of the threshold functions by using the
zero temperature flow equations in the range $k\ge10T$. For smaller
values of $k$ we approximate the infinite Matsubara sums
(cf.~eq.~(\ref{AAA200})) by a finite series such that the numerical
uncertainty at $k=10T$ is better than $10^{-4}$. This approximation
becomes exact in the limit $k\ll10T$.

\subsection{Equation of state and the chiral phase transition}
\label{TheQuarkMesonModelAtTNeq0}

In section \ref{ASemiQuantitativePicture} we have considered the
relevant fluctuations that contribute to the flow of $\Gamma_k$ in
dependence on the scale $k$. In a thermal equilibrium situation
$\Gamma_k$ also depends on the temperature $T$ and one may ask for the
relevance of thermal fluctuations at a given scale $k$.  In
particular, for not too high values of $T$
(cf.~sect.~\ref{AdditionalDegreesOfFreedom}) the ``initial condition''
$\Gamma_{k_\Phi}$ for the solution of the flow equations should
essentially be independent of temperature.  This will allow us to fix
$\Gamma_{k_\Phi}$ from phenomenological input at $T=0$ and to compute
the temperature dependent quantities in the infrared ($k \to 0$).  We
note that the thermal fluctuations which contribute to the r.h.s.\ of
the flow equation for the meson potential (\ref{AAA68}) are
effectively suppressed for $T \lta k/4$ (cf.~section
\ref{FiniteTemperatureFormalism}).  Clearly for $T \gta k_{\Phi}/3$
temperature effects become important at the compositeness scale. We
expect the linear quark meson model with a compositeness scale
$k_{\Phi} \simeq 600 \MeV$ to be a valid description for two flavor
QCD below a temperature of about\footnote{There will be an effective
  temperature dependence of $\Gamma_{k_{\Phi}}$ induced by the
  fluctuations of other degrees of freedom besides the quarks, the
  pions and the sigma which are taken into account here.  We will
  comment on this issue in section \ref{AdditionalDegreesOfFreedom}.
  For realistic three flavor QCD the thermal kaon fluctuations will
  become important for $T\gta170\MeV$.} $170 \MeV$.

We compute the quantities of interest for temperatures $T\lta170\MeV$
by solving numerically the $T$--dependent version of the flow
equations (\ref{AAA68}), (\ref{AAA91})---(\ref{AAA69})
(cf.~section~\ref{FiniteTemperatureFormalism} and
appendix~\ref{AnomalousDimensions}) by lowering $k$ from $k_\Phi$ to
zero. For this range of temperatures we use the initial values as
given in the first line of table \ref{tab1}.  This corresponds to
choosing the zero temperature pion mass and the pion decay constant
($f_{\pi}=92.4 \MeV$ for $m_{\pi}=135 \MeV$) as phenomenological
input. The only further input is the constituent quark mass $M_q$
which we vary in the range $M_q \simeq 300 - 350 \MeV$. We observe
only a minor dependence of our results on $M_q$ for the considered
range of values. In particular, the value for the critical temperature
$T_c$ of the model remains almost unaffected by this variation.

\begin{figure}
\unitlength1.0cm

\begin{center}
\begin{picture}(13.,7.0)

\put(0.0,0.0){
\epsfysize=11.cm
\rotate[r]{\epsffile{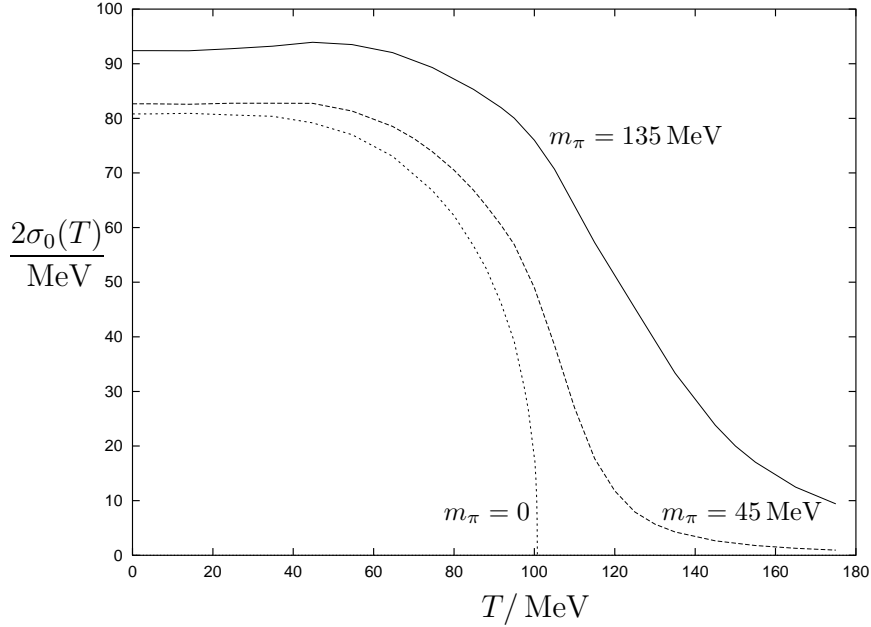}}
}
\put(-0.5,4.2){\bf $\ds{\frac{2\sigma_0(T)}{\MeV}}$}
\put(5.8,-0.5){\bf $\ds{T/\MeV}$}
\put(5.3,0.8){\footnotesize\bf $m_\pi=0$}
\put(8.2,0.8){\footnotesize\bf $m_\pi=45\MeV$}
\put(6.7,5.8){\footnotesize\bf $m_\pi=135\MeV$}

\end{picture}
\end{center}
\caption{\footnotesize The expectation value $2\sigma_0$ is shown as a
  function of temperature $T$ for three different values of the
  zero temperature pion mass.}
\label{fpi_T}
\end{figure}

We have plotted in figure \ref{fpi_T} the renormalized expectation
value $2\sigma_0$ of the scalar field as a function of temperature for
three different values of the average light current quark mass
$\hat{m}$. (We remind that $2\sigma_0(T=0)=f_{\pi}$.)
For $\hat{m}=0$ the order parameter $\sigma_0$ of chiral symmetry
breaking continuously goes to zero for $T\ra T_c = 100.7\MeV$
characterizing the phase transition to be of second order.  The
universal behavior of the model for small $T-T_c$ and small $\hat{m}$
is discussed in more detail in section \ref{CriticalBehavior}.  We
point out that the value of $T_c$ corresponds to
$f_\pi^{(0)}=80.8\MeV$, i.e.~the value of the pion decay constant for
$\hat{m}=0$, which is significantly lower than $f_\pi=92.4\MeV$
obtained for the realistic value $\hat{m}_{\rm phys}$.  If we would
fix the value of the pion decay constant to be $92.4\MeV$ also in the
chiral limit ($\hat{m}=0$), the value for the critical temperature
would raise to $115\MeV$.  The nature of the transition changes
qualitatively for $\hat{m}\neq0$ where the second order transition is
replaced by a smooth crossover.  The crossover for a realistic
$\hat{m}_{\rm phys}$ or $m_{\pi}(T=0)=135 \MeV$ takes place in a
temperature range $T \simeq(120-150)\MeV$.  The middle curve in figure
\ref{fpi_T} corresponds to a value of $\hat{m}$ which is only a tenth
of the physical value, leading to a zero temperature pion mass
$m_\pi=45\MeV$. Here the crossover becomes considerably sharper but
there remain substantial deviations from the chiral limit even for
such small quark masses $\hat{m}\simeq 1 \MeV$. The temperature
dependence of $m_\pi$ has already been mentioned in the introduction
(see fig.~\ref{mpi_T}) for the same three values of $\hat{m}$.  As
expected, the pions behave like true Goldstone bosons for $\hat{m}=0$,
i.e.~their mass vanishes for $T\le T_c$. Interestingly, $m_\pi$
remains almost constant as a function of $T$ for $T<T_c$ before it
starts to increase monotonically. We therefore find for two flavors no
indication for a substantial decrease of $m_\pi$ around the critical
temperature.

\begin{figure}
\unitlength1.0cm
\begin{center}
\begin{picture}(13.,7.0)

\put(0.0,0.0){
\epsfysize=11.cm
\rotate[r]{\epsffile{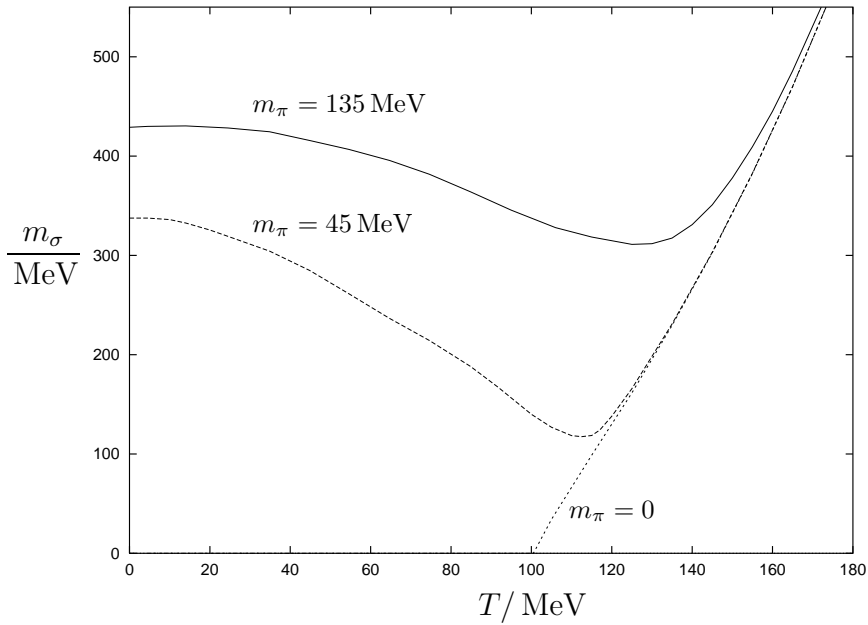}}
}
\put(-0.5,4.2){\bf $\ds{\frac{m_\sigma}{\MeV}}$}
\put(5.8,-0.5){\bf $\ds{T/\MeV}$}
\put(7.0,0.8){\footnotesize\bf $m_\pi=0$}
\put(2.8,4.6){\footnotesize\bf $m_\pi=45\MeV$}
\put(2.8,6.2){\footnotesize\bf $m_\pi=135\MeV$}

\end{picture}
\end{center}
\caption{\footnotesize The plot shows the $m_\sigma$ as a function of
  temperature $T$ for three different values of the 
  zero temperature pion mass.}
\label{ms_T}
\end{figure}

\begin{figure}
\unitlength1.0cm
\begin{center}
\begin{picture}(13.,7.0)
\put(0.0,0.0){
\epsfysize=11.cm
\rotate[r]{\epsffile{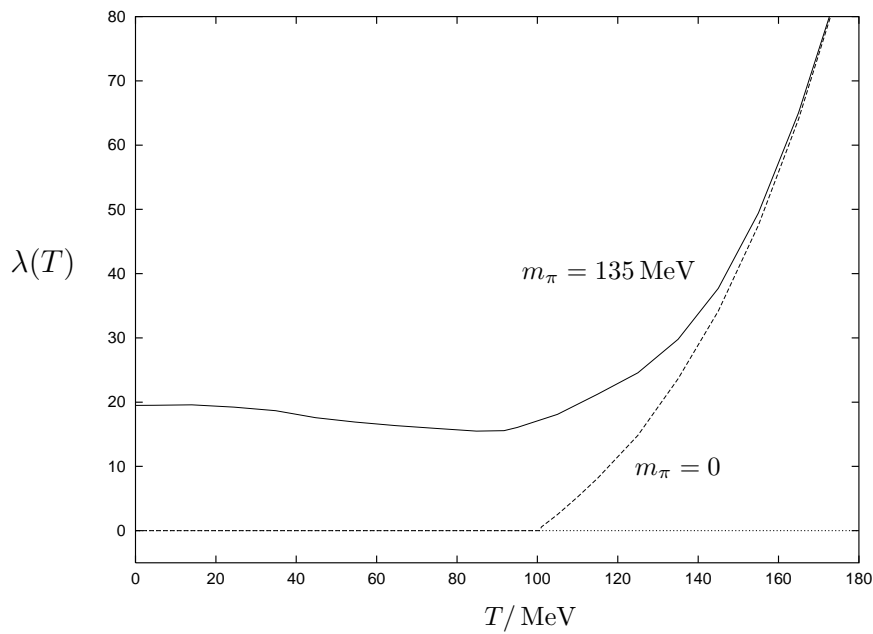}}
}
\put(-0.5,4.2){\bf $\ds{\lambda(T)}$}
\put(5.8,-0.5){\footnotesize\bf $\ds{T/\MeV}$}
\put(7.8,1.5){\footnotesize\bf $\ds{m_\pi=0}$}
\put(6.3,4.1){\footnotesize\bf $\ds{m_\pi=135\MeV}$}
\end{picture}
\end{center}
\caption{\footnotesize The plot shows the renormalized quartic scalar
  self coupling $\lambda$ as a function of temperature $T$ for the
  physical value of $\hat{m}$ (solid line) as well as for $\hat{m}=0$
  (dashed line).}
\label{lambda_T}
\end{figure}

\begin{figure}
\unitlength1.0cm
\begin{center}
\begin{picture}(13.,7.0)
\put(0.0,0.0){
\epsfysize=11.cm
\rotate[r]{\epsffile{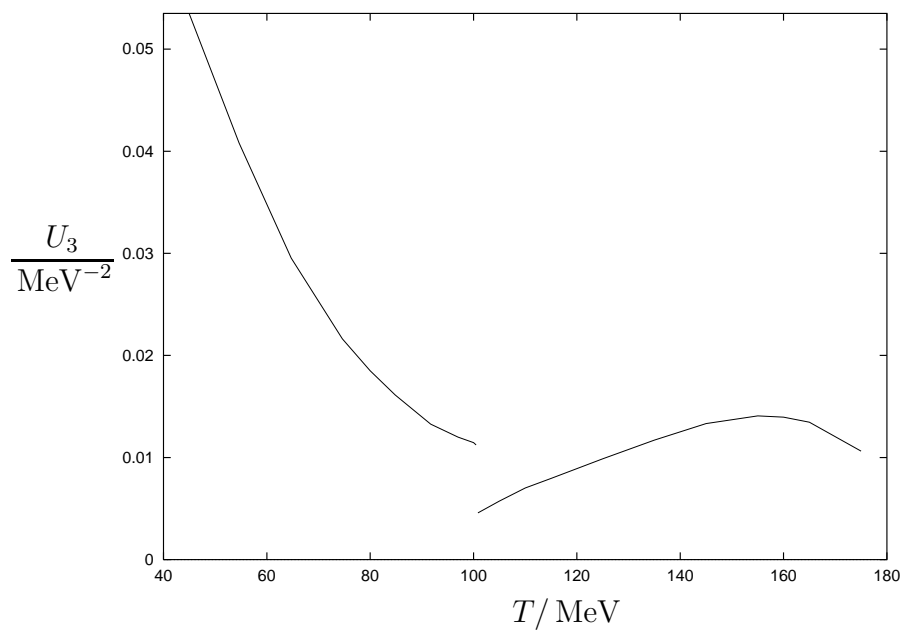}}
}
\put(-0.9,4.2){\bf $\ds{\frac{U_3}{\MeV^{-2}}}$}
\put(5.8,-0.5){\bf $\ds{T/\MeV}$}
\end{picture}
\end{center}
\caption{\footnotesize The plot shows the renormalized $\Phi^6$ scalar
  self coupling $U_3$ as a function of temperature $T$ in the chiral
  limit.}
\label{U3_T}
\end{figure}

The dependence of the mass of the sigma resonance $m_\sigma$ on the
temperature is displayed in figure~\ref{ms_T} for the above three
values of $\hat{m}$.
In the absence of explicit chiral symmetry breaking, $\hat{m}=0$, the
sigma mass vanishes for $T\le T_c$. For $T<T_c$ this is a consequence
of the presence of massless Goldstone bosons in the Higgs phase which
drive the renormalized quartic coupling $\lambda$ to zero.  In fact,
$\lambda$ runs linearly with $k$ for $T \gta k/4$ and only evolves
logarithmically for $T \lta k/4$.  Once $\hat{m}\neq0$ the pions
acquire a mass even in the spontaneously broken phase and the
evolution of $\lambda$ with $k$ is effectively stopped at $k\sim
m_\pi$. Because of the temperature dependence of $\sigma_{0,k=0}$
(cf.~figure~\ref{fpi_T}) this leads to a monotonically decreasing
behavior of $m_\sigma$ with $T$ for $T\lta T_c$. This changes into the
expected monotonic growth once the system reaches the symmetric phase
for\footnote{See section~\ref{FlowEquationsAndInfraredStability} for a
  discussion of the zero temperature sigma mass.} $T>T_c$. For low
enough $\hat{m}$ one may use the minimum of $m_{\sigma}(T)$ for an
alternative definition of the (pseudo-)critical temperature denoted as
$T_{pc}^{(2)}$. Table \ref{tab11} in the introduction shows our
results for the pseudocritical temperature for different values of
$\hat{m}$ or, equivalently, $m_{\pi}(T=0)$. For a zero temperature
pion mass $m_{\pi}=135 \MeV$ we find $T_{pc}^{(2)}=128 \MeV$. At
larger pion masses of about $230 \MeV$ we observe no longer a
characteristic minimum for $m_{\sigma}$ apart from a very broad,
slight dip at $T \simeq 90 \MeV$.  A comparison of our results with
lattice data is given in section \ref{CriticalBehavior}.  In
fig.~\ref{lambda_T} we show the renormalized quartic coupling
$\lambda$ as a function of temperature for two fixed values of the
average current quark mass.
The upper curve corresponds to
the physical value of $\hat{m}$ or, equivalently,  
$m_\pi(T=0)=135 \MeV$ whereas the lower
curve shows $\lambda$ for $\hat{m}=0$. One observes the 
vanishing of the renormalized quartic coupling in the chiral limit 
for $T \leq T_c$ as discussed above. The renormalized scalar $\Phi^6$
self interaction
\begin{equation}
  \label{ABC20}
  U_3(T)=Z_\Phi^{-3}
  \frac{\prl^3 U(\rho,T)}{\prl\rho^3}
  (\rho=2\ol{\sigma}_0^2(T))
\end{equation}
assumes a small negative value for realistic quark masses in the
temperature range $T\lta35\MeV$ with
$2U_3(T)\sigma_0^2(T)\simeq-0.5\ll\lambda(T)$ and
$2U_3(T)\sigma_0^2(T)\simeq8.0,8.5,1.5$ for $T=80,120,160\MeV$. We
display $U_3(T)$ in figure~\ref{U3_T} for the chiral limit where one
observes a discontinuity of $U_3(T)$ at the critical temperature
$T_c$.

\begin{figure}
\unitlength1.0cm
\begin{center}
\begin{picture}(13.,7.0)

\put(0.0,0.0){
\epsfysize=11.cm
\rotate[r]{\epsffile{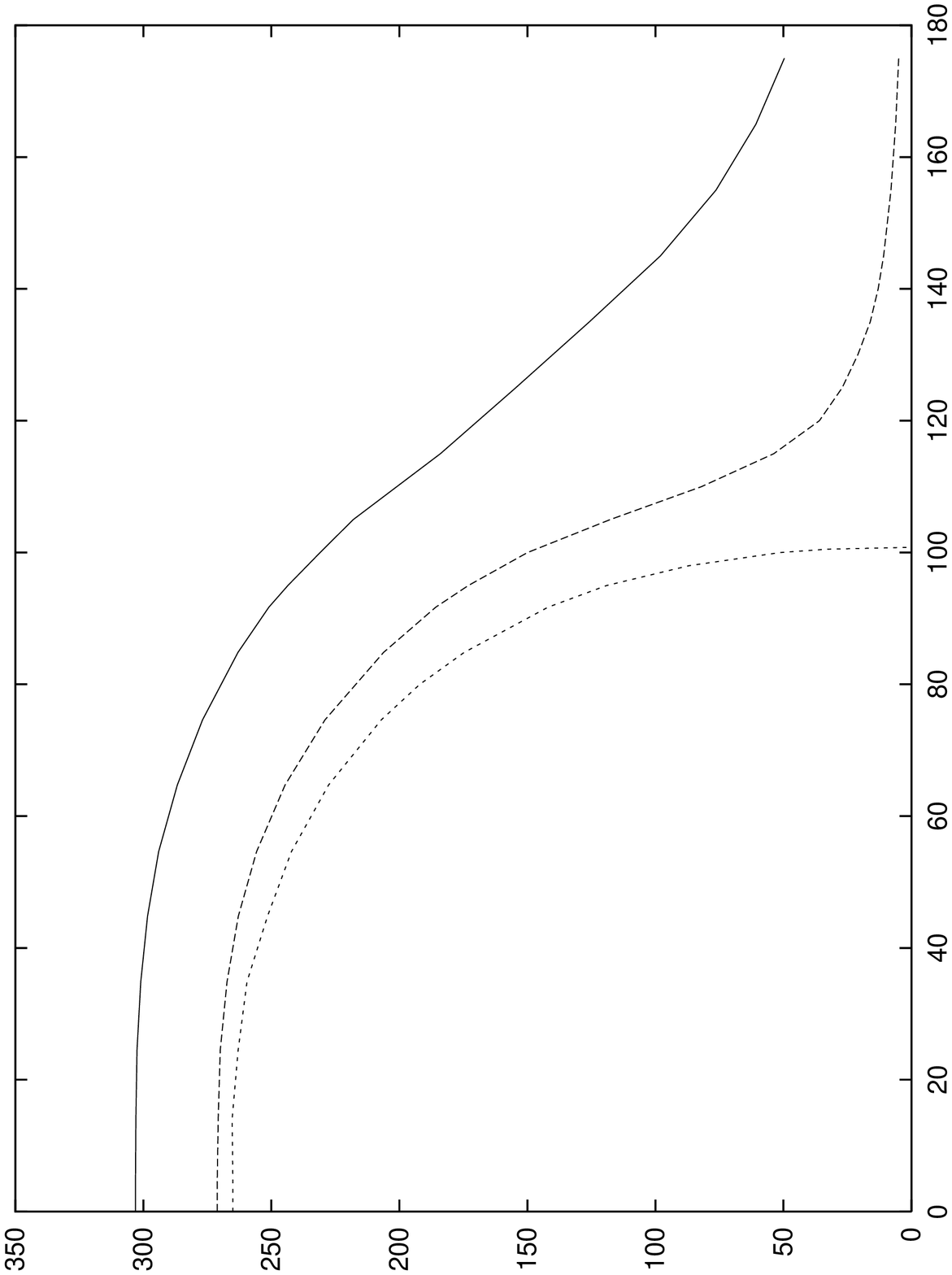}}
}
\put(-0.4,4.2){\bf $\ds{\frac{M_q}{\MeV}}$}
\put(5.8,-0.5){\bf $\ds{T/\MeV}$}
\put(5.2,1.0){\footnotesize\bf $m_\pi=0$}
\put(8.1,1.0){\footnotesize\bf $m_\pi=45\MeV$}
\put(5.8,5.9){\footnotesize\bf $m_\pi=135\MeV$}

\end{picture}
\end{center}
\caption{\footnotesize The plot shows the constituent quark mass $M_q$ 
  as a function of $T$ for three different values of the average light
  current quark mass $\hat{m}$. The solid line corresponds to the
  realistic value $\hat{m}=\hat{m}_{\rm phys}$ whereas the dotted line
  represents the situation without explicit chiral symmetry breaking,
  i.e., $\hat{m}=0$. The intermediate, dashed line assumes
  $\hat{m}=\hat{m}_{\rm phys}/10$.}
\label{mq_T}
\end{figure}

\begin{figure}
\unitlength1.0cm
\begin{center}
\begin{picture}(13.,7.0)
\put(0.0,0.0){
\epsfysize=11.cm
\rotate[r]{\epsffile{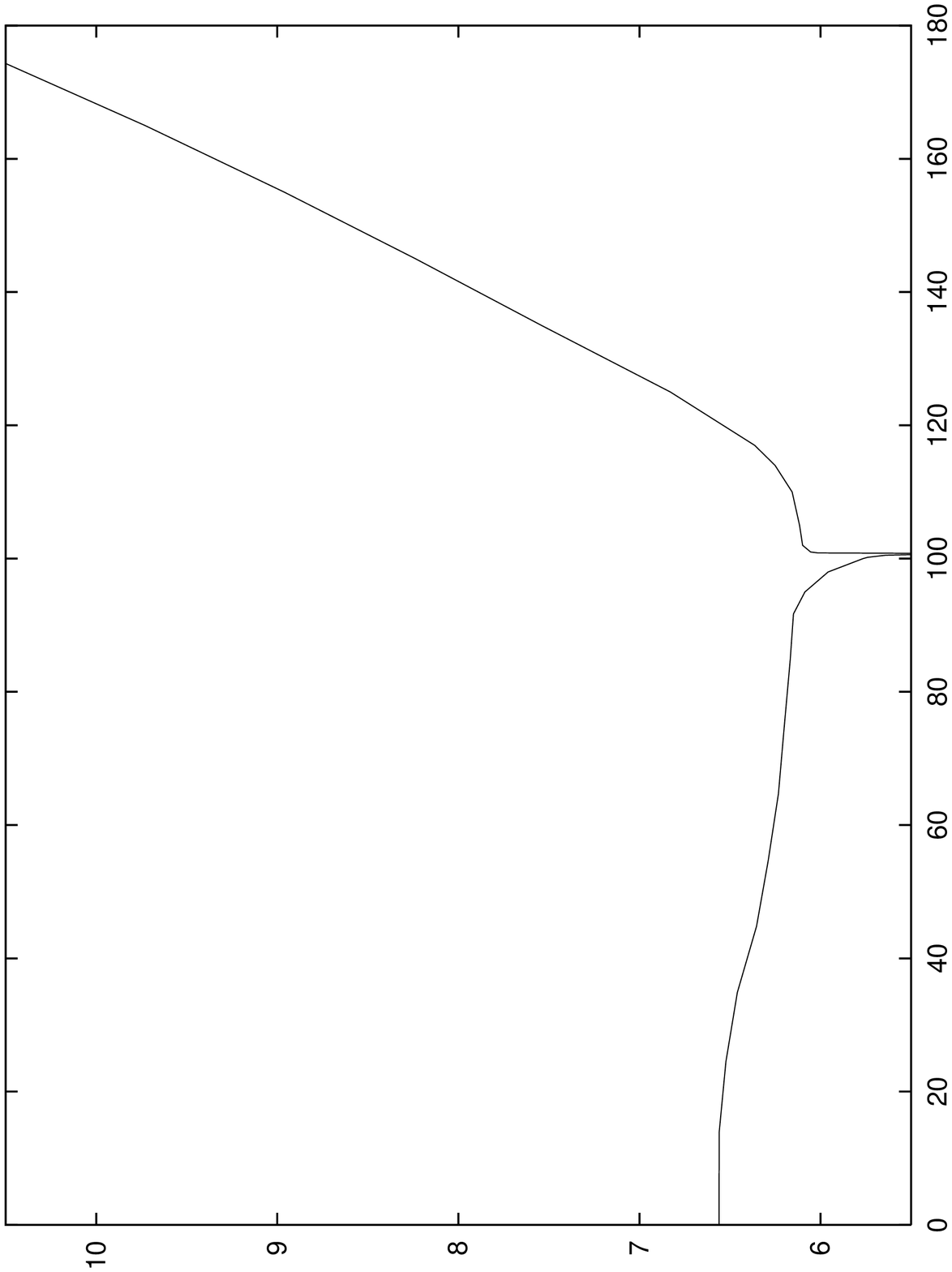}}
}
\put(0.2,4.2){\bf $\ds{h}$}
\put(5.8,-0.5){\bf $\ds{T/\MeV}$}
\end{picture}
\end{center}
\caption{\footnotesize The plot shows the Yukawa coupling, $h$, as a
  function of temperature $T$ in the chiral limit.}
\label{h2_T}
\end{figure}

\begin{figure}
\unitlength1.0cm
\begin{center}
\begin{picture}(13.,7.0)

\put(0.0,0.0){
\epsfysize=11.cm
\rotate[r]{\epsffile{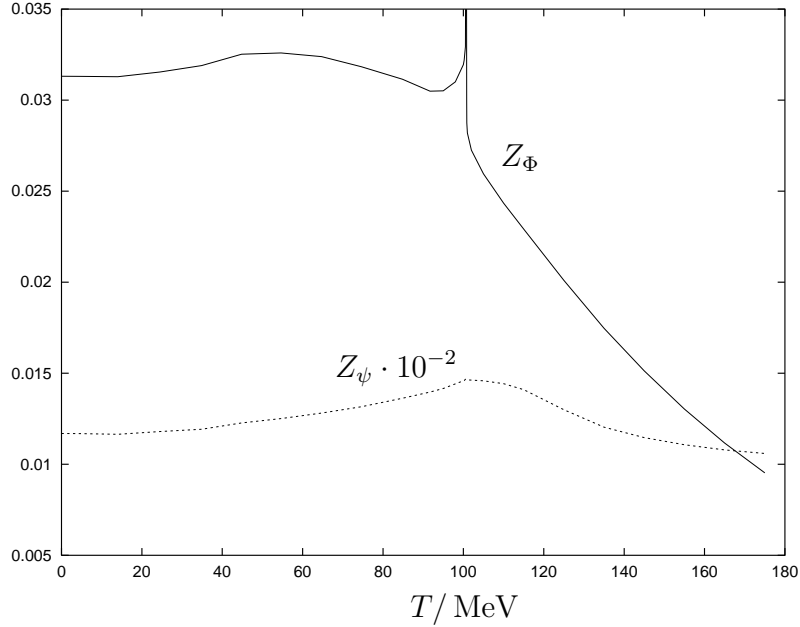}}
}
\put(5.8,-0.5){\bf $\ds{T/\MeV}$}
\put(4.8,2.7){\bf $\ds{Z_\psi\cdot10^{-2}}$}
\put(7.0,5.5){\bf $\ds{Z_\Phi}$}
\end{picture}
\end{center}
\caption{\footnotesize The plot shows the scalar (solid line) and
  quark (dashed line) wave function renormalization constants,
  $Z_\Phi(T)$ and $Z_\psi(T)\cdot10^{-2}$, respectively, as functions
  of temperature $T$ for $\hat{m}=0$.}
\label{Z_T}
\end{figure}

Our results for the chiral condensate $\VEV{\ol{\psi}\psi}$ as a
function of temperature for different values of the average current
quark mass are presented in figure~\ref{ccc_T} in the introduction. We
will compare $\VEV{\ol{\psi}\psi}(T,\hat{m})$ with its universal
scaling form for the $O(4)$ Heisenberg model in
section~\ref{CriticalBehavior}.  Another interesting quantity is the
temperature dependence of the constituent quark mass. Figure
\ref{mq_T} shows $M_q(T)$ for $\hat{m}=0$, $\hat{m}=\hat{m}_{\rm
  phys}/10$ and $\hat{m}=\hat{m}_{\rm phys}$, respectively.
Its behavior is related to the temperature dependence of the
renormalized order parameter $\sigma_{0}(T)\equiv\sigma_{0,k=0}(T)$
and the renormalized Yukawa coupling $h(T)\equiv h_{k=0}(T)$.  The
temperature dependence of $h$ in the chiral limit can be found in
fig.~\ref{h2_T}.
Near the critical temperature one notices a characteristic dip.  This
results from the long wavelength pion fluctuations through a
non--analytic behavior of the mesonic wave function renormalization
$Z_\Phi(T)\equiv Z_{\Phi,k=0}(T)$ which is displayed in
figure~\ref{Z_T}.
There we also present the temperature dependence of the fermionic wave
function renormalization $Z_\psi(T)\equiv Z_{\psi,k=0}(T)$.  Away from
the chiral limit we take the effective quark mass dependence of
$h_k(T)$, $Z_{\Phi,k}(T)$ and $Z_{\psi,k}(T)$ into account by stopping
their evolution when $k$ reaches the temperature dependent pion mass.
In this way we observe no substantial quark mass dependence of these
quantities except for $Z_\Phi(T)$, and consequently for $h(T)$, in the
vicinity of the critical temperature.  A more complete truncation
would incorporate field dependent wave function renormalization
constants and a field dependent Yukawa coupling.

\begin{figure}
\unitlength1.0cm
\begin{center}
\begin{picture}(13.,7.0)
\put(0.0,0.0){
\epsfysize=11.cm
\rotate[r]{\epsffile{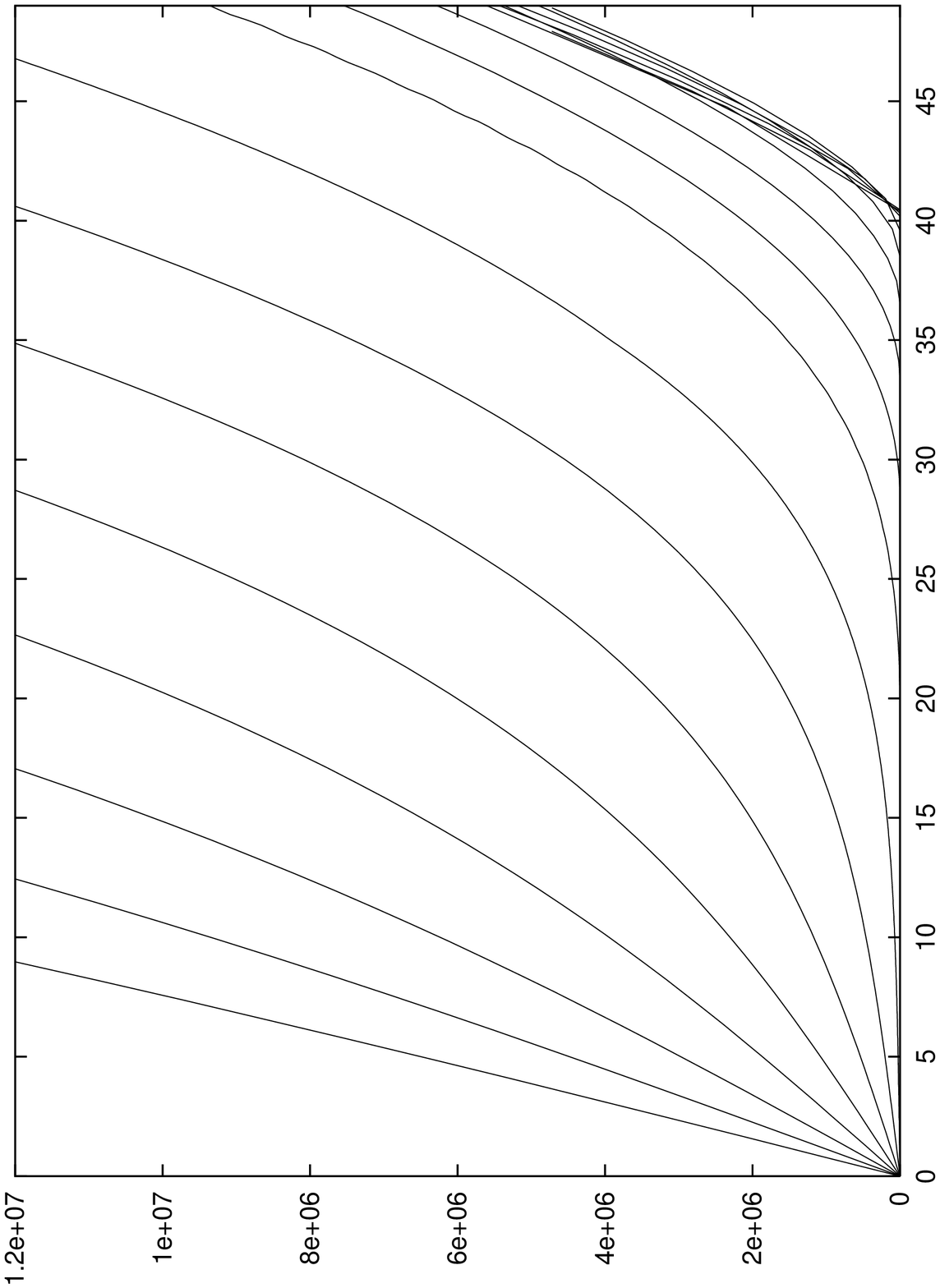}}
}
\put(-1.3,4.2){\bf $\ds{\frac{\partial U(T)/\partial \phi_R}
    {\MeV^{3}}}$}
\put(5.8,-0.5){\bf $\ds{\phi_R/\MeV}$}
\end{picture}
\end{center}
\caption{\footnotesize The plot shows the derivative of the
  meson potential $U(T)$ with respect to the renormalized field
  $\phi_R=(Z_\Phi\rho/2)^{1/2}$ for different values of $T$.  The
  first curve on the left corresponds to $T=175 \MeV$. The successive
  curves to the right differ in temperature by $\Delta T=10 \MeV$ down
  to $T=5 \MeV$. }
\label{Usig}
\end{figure}

Our ability to compute the complete temperature dependent effective
meson potential $U$ is demonstrated in fig.~\ref{Usig} where we
display the derivative of the potential with respect to the
renormalized field $\phi_R=(Z_\Phi\rho/2)^{1/2}$, for different values
of $T$.
The curves cover a temperature range $T = (5 - 175) \MeV$.  The first
one to the left corresponds to $T=175 \MeV$ and neighboring curves
differ in temperature by $\Delta T = 10 \MeV$. One observes only a
weak dependence of $\partial U(T)/\partial\phi_R$ on the temperature
for $T\lta60\MeV$.  Evaluated for $\phi_R=\sigma_{0}$ this
function connects the renormalized field expectation value with
$m_{\pi}(T)$, the source $\jmath$ and the mesonic wave function
renormalization $Z_{\Phi}(T)$ according to
\begin{equation}
  \label{Usigeq}
  \ds{\frac{\partial U(T)}{\partial\phi_R}}
  (\phi_R=\sigma_{0})=
  \ds{\frac{2\jmath}{Z_{\Phi}^{1/2}(T)}}=4 \sigma_{0}(T) 
  m_{\pi}^2(T) \; .
\end{equation} 

We close this section with a short assessment of the validity of our
effective quark meson model as an effective description of two flavor
QCD at non--vanishing temperature.  The identification of
qualitatively different scale intervals which appear in the context of
chiral symmetry breaking, as presented in section
\ref{ASemiQuantitativePicture} for the zero temperature case, can be
generalized to $T \neq 0$: For scales below $k_{\Phi}$ there exists a
hybrid description in terms of quarks and mesons. For $k_{\chi SB}
\leq k \lta 600 \MeV$ chiral symmetry remains unbroken where the
symmetry breaking scale $k_{\chi SB}(T)$ decreases with increasing
temperature. Also the constituent quark mass decreases with $T$
(cf.~figure~\ref{mq_T}). The running Yukawa coupling depends only
mildly on temperature for $T\lta120\MeV$ (see fig.~\ref{h2_T}).  (Only
near the critical temperature and for $\hat{m}=0$ the running is
extended because of massless pion fluctuations.) On the other hand,
for $k\lta4T$ the effective three--dimensional gauge coupling
increases faster than at $T=0$ leading to an increase of $\Lambda_{\rm
  QCD}(T)$ with $T$~\cite{RW1}. As $k$ gets closer to the scale
$\Lambda_{\rm QCD}(T)$ it is no longer justified to neglect in the
quark sector confinement effects which go beyond the dynamics of our
present quark meson model.  Here it is important to note that the
quarks remain quantitatively relevant for the evolution of the meson
degrees of freedom only for scales $k \gta T/0.6$
(cf.~fig.~\ref{Thresh}, section~\ref{FiniteTemperatureFormalism}).  In
the limit $k \ll T/0.6$ all fermionic Matsubara modes decouple from
the evolution of the meson potential according to the temperature
dependent version of eq.\ (\ref{AAA68}). Possible sizeable confinement
corrections to the meson physics may occur if $\Lambda_{\rm QCD}(T)$
becomes larger than the maximum of $M_q(T)$ and $T/0.6$. From
fig.~\ref{mq_T} we infer that this is particularly dangerous for small
$\hat{m}$ in a temperature interval around $T_c$. Nevertheless, the
situation is not dramatically different from the zero temperature case
since only a relatively small range of $k$ is concerned. We do not
expect that the neglected QCD non--localities lead to qualitative
changes.  Quantitative modifications, especially for small $\hat{m}$
and $\abs{T-T_c}$ remain possible. This would only effect the
non--universal amplitudes (see sect.~\ref{CriticalBehavior}). The size
of these corrections depends on the strength of (non--local)
deviations of the quark propagator and the Yukawa coupling from the
values computed in the quark meson model.

\subsection{Universal critical equation of state}
\label{CriticalBehavior}

In this section we study the linear quark meson model in the vicinity
of the critical temperature $T_c$ close to the chiral limit
$\hat{m}=0$. In this region we find that the sigma mass
$m_\sigma^{-1}$ is much larger than the inverse temperature $T^{-1}$,
and one observes an effectively three--dimensional behavior of the
high temperature quantum field theory.  We also note that the fermions
are no longer present in the dimensionally reduced system as has been
discussed in section \ref{FiniteTemperatureFormalism}. We therefore
have to deal with a purely bosonic $O(4)$--symmetric linear sigma
model.  At the phase transition the correlation length becomes
infinite and the effective three--dimensional theory is dominated by
classical statistical fluctuations. In particular, the critical
exponents which describe the singular behavior of various quantities
near the second order phase transition are those of the corresponding
classical system.

Many properties of this system are universal, i.e.~they only depend
on its symmetry ($O(4)$), the dimensionality of space (three) and its
degrees of freedom (four real scalar components). Universality means
that the long--range properties of the system do not depend on the
details of the specific model like its short distance
interactions. Nevertheless, important properties as the value of the
critical temperature are non--universal. We emphasize that although we
have to deal with an effectively three--dimensional bosonic theory,
the non--universal properties of the system crucially depend on the
details of the four--dimensional theory and, in particular, on the
fermions. 

Our aim is a computation of the critical equation of state which
relates for arbitrary $T$ near $T_c$ 
the derivative of the free energy or effective potential $U$
to the average current quark mass $\hat{m}$. The equation of state
then permits to study the temperature and quark mass dependence of
properties of the chiral phase transition.

At the critical temperature and in the chiral limit there is no scale
present in the theory. In the vicinity of $T_c$ and for small enough
$\hat{m}$ one therefore expects a scaling behavior of the effective
average potential $u_k$~\cite{TW94-1} and accordingly a universal
scaling form of the equation of state (cf.\ section \ref{uceos}). 
There are only two independent
scales close to the transition point which can be related to the
deviation from the critical temperature, $T-T_c$, and to the explicit
symmetry breaking through the quark mass $\hat{m}$.  As a consequence,
the properly rescaled potential can only depend on one scaling
variable.  A possible choice for the parameterization of the rescaled
``unrenormalized'' potential is the use of the Widom scaling
variable~\cite{Widom}
\begin{equation}
  \label{XXX20}
  x=\frac{\left( T-T_c\right)/T_c}
  {\left(2\ol{\sigma}_0/T_c\right)^{1/\beta}}\; .
\end{equation}
Here $\beta$ is the critical exponent of the order parameter
$\ol{\sigma}_0$ in the chiral limit $\hat{m}=0$ (see equation
(\ref{NNN21})).  With
$U^\prime(\rho=2\ol{\sigma}_0^2)=\jmath/(2\ol{\sigma}_0)$ the Widom
scaling form of the equation of state reads~\cite{Widom}
\begin{equation}
  \label{XXX21}
  \frac{\jmath}{T_c^3}=
  \left(\frac{2\ol{\sigma}_0}{T_c}\right)^\delta f(x)
\end{equation}
where the exponent $\delta$ is related to the behavior of the order
parameter according to (\ref{NNN21b}).  The equation of state
(\ref{XXX21}) is written for convenience directly in terms of
four--dimensional quantities.  They are related to the corresponding
effective variables of the three--dimensional theory by appropriate
powers of $T_c$.  The source $\jmath$ is determined by the average
current quark mass $\hat{m}$ as $\jmath=2\ol{m}^2_{k_\Phi}\hat{m}$.
The mass term at the compositeness scale, $\ol{m}^2_{k_\Phi}$, also
relates the chiral condensate to the order parameter according to
$\VEV{\ol{\psi}\psi}=-2\ol{m}^2_{k_\Phi}(\ol{\sigma}_0-\hat{m})$.  The
critical temperature of the linear quark meson model was found in
section \ref{TheQuarkMesonModelAtTNeq0} to be $T_c=100.7\MeV$.

\begin{figure}
\unitlength1.0cm
\begin{center}
\begin{picture}(17.,12.)
\put(0.3,5.5){$\ds{\frac{2\ol{\sigma}_0/T_c}
{(\jmath/T_c^3 D)^{1/\delta}}}$}
\put(8.5,-0.2){$\ds{\frac{(T-T_c)/T_c}
{(\jmath/T_c^3 B^{\delta} D)^{1/\beta \delta}}}$}
\put(8.,2.5){\footnotesize $\mbox{average action}$}
\put(4.19,11.2){\footnotesize $\mbox{average action}$}
\put(13.8,2.2){\footnotesize $\epsilon$}
\put(3.53,11.19){\footnotesize $\epsilon$}
\put(11.3,2.05){\footnotesize $\mbox{MC}$}
\put(3.9,10.){\footnotesize $\mbox{MC}$}
\put(13.2,1.8){\footnotesize $\mbox{mf}$}
\put(6.5,9.2){\footnotesize $\mbox{mf}$}
\put(-1.,-5.9){
\epsfysize=21.cm
\epsfxsize=18.cm
\epsffile{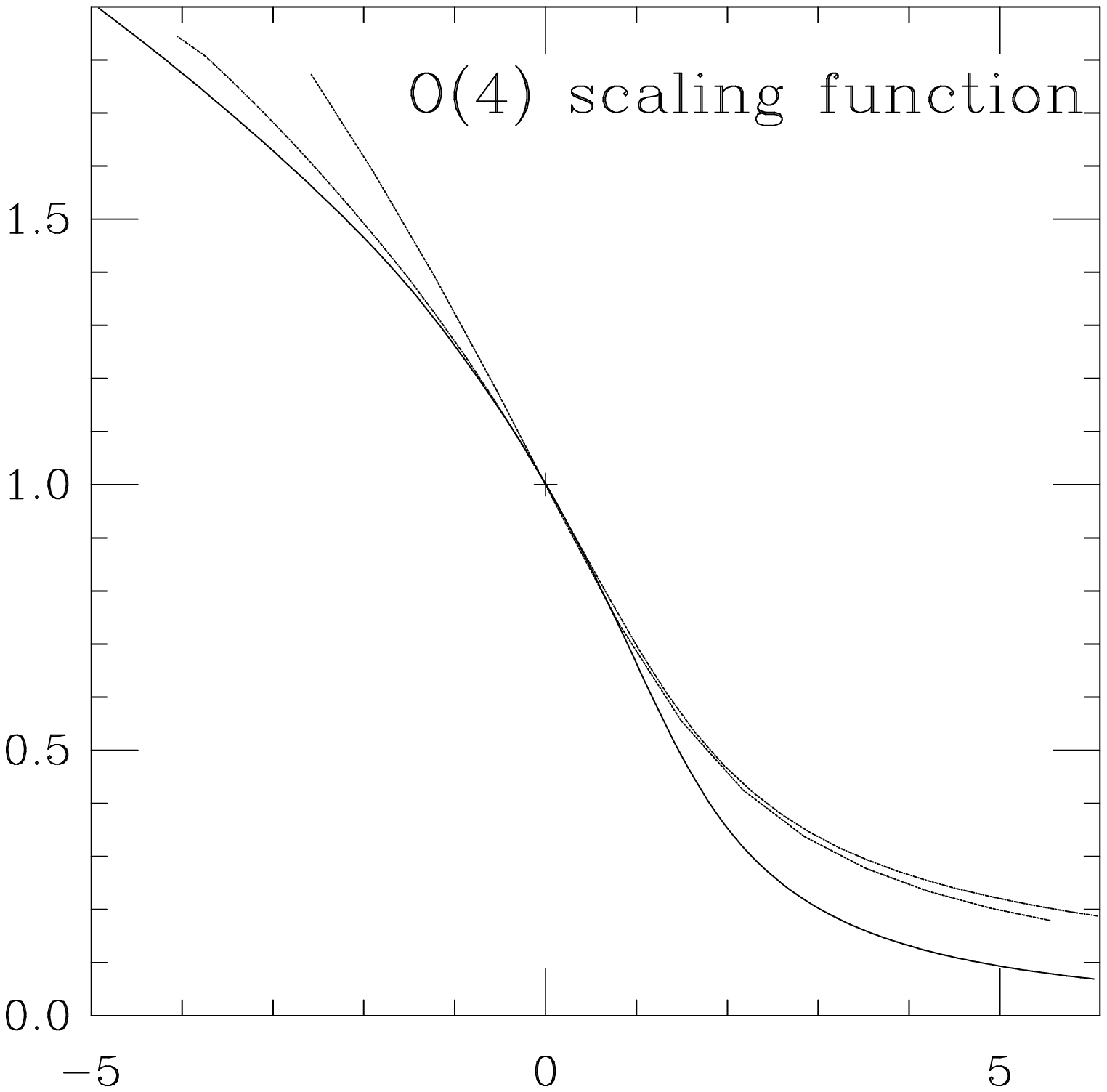}}
\put(1.25,0.483){
\epsfysize=13.45cm
\epsfxsize=11.22cm
\rotate[r]{\epsffile{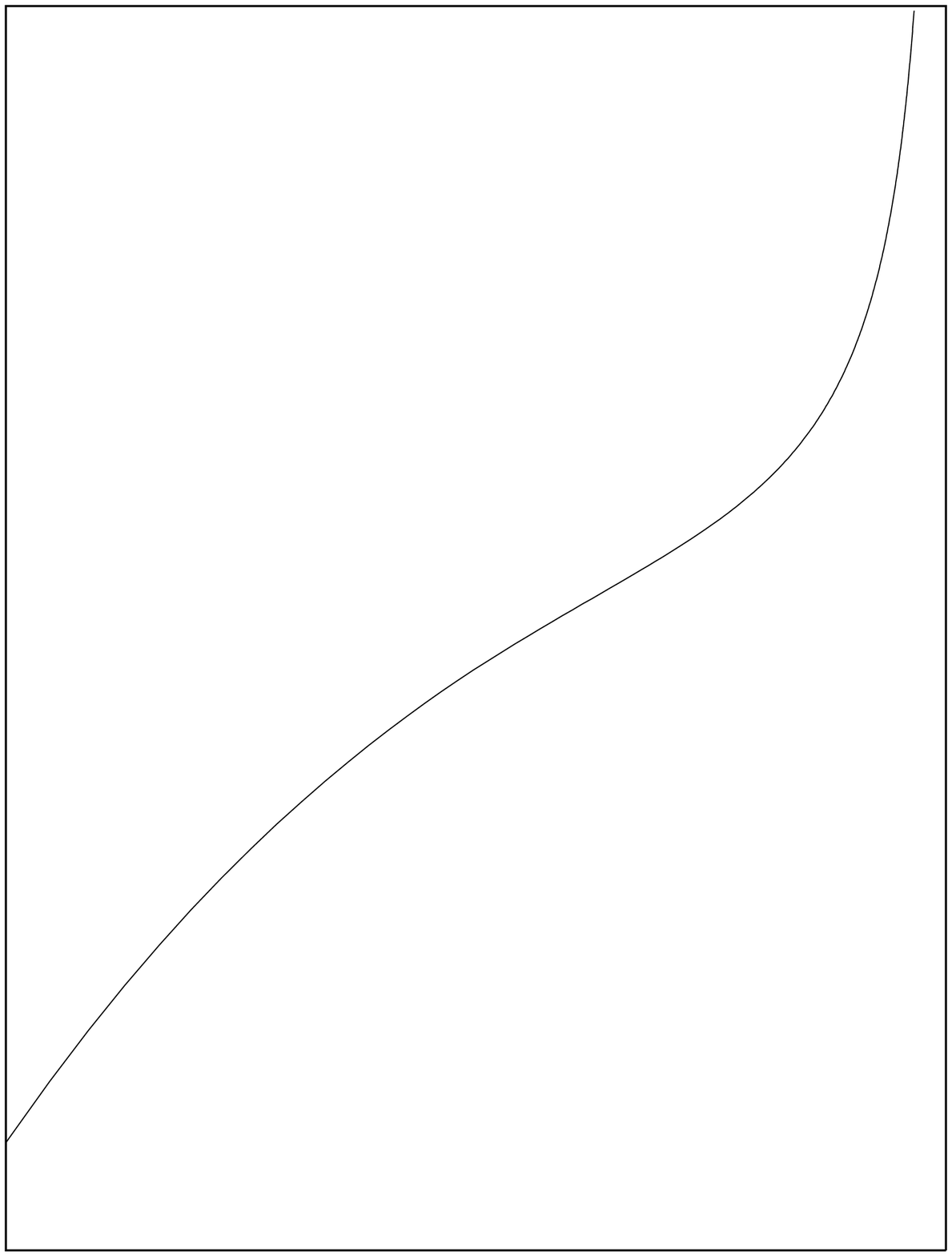}}}
\end{picture}
\end{center}
\caption[]{\footnotesize
  The figure shows a comparison of our results, denoted by ``average
  action'', with results of other methods for the scaling function of
  the three--dimensional $O(4)$ Heisenberg model. We have labeled the
  axes for convenience in terms of the expectation value
  $\ol{\sigma}_0$ and the source $\jmath$ of the corresponding
  four--dimensional theory.  The constants $B$ and $D$ specify the
  non--universal amplitudes of the model (cf.~eq.~\ref{norm}).  The
  curve labeled by ``MC'' represents a fit to lattice Monte Carlo
  data. The second order epsilon expansion \cite{BWW73-1} and mean
  field results are denoted by ``$\epsilon$'' and ``mf'',
  respectively.  Apart from our results the curves are taken
  from ref.~\cite{Tou}.
  \label{scalfunc}
  }
\end{figure}

The scaling function $f$ is universal up to the model specific
normalization of $x$ and itself. Accordingly, all models in the same
universality class can be related by a rescaling of $\ol{\sigma}_0$
and $T-T_c$. The non--universal normalizations for the quark meson
model discussed here are defined according to
\begin{equation}
  \label{norm}
  f(0)=D\quad, \qquad f(-B^{-1/\beta})=0\; .
\end{equation}
We find $D=1.82\cdot10^{-4}$, $B=7.41$ and our result for $\beta$ is
given in table~\ref{tab2}. Apart from the immediate vicinity of the
zero of $f(x)$ we find the following two parameter fit
(cf.\ section \ref{uceos}) for the scaling function,
\begin{equation}
  \label{ffit}
  \begin{array}{rcl}
    \ds{f_{\rm fit}(x)}&=&\ds{1.816 \cdot 10^{-4} (1+136.1\, x)^2 \,
      (1+160.9\, \theta\,
      x)^{\Delta}}\nnn
    &&
    \ds{(1+160.9\, (0.9446\, \theta^{\Delta})^{-1/(\gamma-2-\Delta)} 
      \, x)^{\gamma-2-\Delta}}
  \end{array}
\end{equation}
to reproduce the numerical results for $f$ and $df/dx$ at the $1-2\%$
level with $\theta=0.625$ $(0.656)$, $\Delta=-0.490$ $(-0.550)$ for $x
> 0$ $(x < 0)$ and $\gamma$ as given in table \ref{tab2}.  The
universal properties of the scaling function can be compared with
results obtained by other methods for the three--dimensional $O(4)$
Heisenberg model.  In figure \ref{scalfunc} we display our results
along with those obtained from lattice Monte Carlo simulation
\cite{Tou}, second order epsilon expansion \cite{BWW73-1} and mean
field theory.
We observe a good agreement of average action, lattice and epsilon
expansion results within a few per cent for $T < T_c$. Above $T_c$ the
average action and the lattice curve go quite close to each other with
a substantial deviation from the epsilon expansion and mean field
scaling function\footnote{ We note that the question of a better
  agreement of the curves for $T < T_c$ or $T > T_c$ depends on the
  chosen non--universal normalization conditions for $x$ and $f$ (cf.
  eq.\ (\ref{norm})).}.

Before we use the scaling function $f(x)$ to discuss the general
temperature and quark mass dependent case, we consider the limits
$T=T_c$ and $\hat{m}=0$, respectively.  In these limits the behavior
of the various quantities is determined solely by critical amplitudes
and exponents. In the spontaneously broken phase ($T<T_c$) and in the
chiral limit we observe that the renormalized and unrenormalized order
parameters scale according to
\begin{equation}
  \label{NNN21}
  \begin{array}{rcl}
    \ds{\frac{2\sigma_0(T)}{T_c}} &=& \ds{
      \left(2E\right)^{1/2}
        \left(\frac{T_c-T}{T_c}\right)^{\nu/2}
      }\; ,\nnn
    \ds{\frac{2\ol{\sigma}_0(T)}{T_c}} &=& \ds{
      B \left(\frac{T_c-T}{T_c}\right)^{\beta}
      }\; ,
  \end{array}
\end{equation}
respectively, with $E=0.814$ and the value of $B$ given above.  In the
symmetric phase the renormalized mass $m=m_\pi=m_\si$ and the
unrenormalized mass $\ol{m}=Z_\Phi^{1/2}m$ behave as
\begin{equation}
  \label{NNN21a}
  \begin{array}{rcl}
    \ds{\frac{m(T)}{T_c}} &=& \ds{
      \left(\xi^+\right)^{-1}
      \left(\frac{T-T_c}{T_c}\right)^\nu
      }\; ,\nnn
    \ds{\frac{\ol{m}(T)}{T_c}} &=& \ds{
      \left( C^+\right)^{-1/2}
      \left(\frac{T-T_c}{T_c}\right)^{\gamma/2}
       \; , }
  \end{array}
\end{equation}
where $\xi^+=0.270$, $C^+=2.79$. For $T=T_c$ and
non--vanishing current quark mass we have
\begin{equation}
  \label{NNN21b}
  \begin{array}{rcl}
    \ds{\frac{2\ol{\sigma}_0}{T_c}} &=& \ds{
      D^{-1/\delta}
        \left(\frac{\jmath}{T_c^3}\right)^{1/\delta}
      }
  \end{array}
\end{equation}
with the value of $D$ given above. 

Though the five amplitudes $E$, $B$, $\xi^+$, $C^+$ and $D$ are not
universal there are ratios of amplitudes which are invariant under a
rescaling of $\ol{\sigma}_0$ and $T-T_c$. Our results for the
universal amplitude ratios are
\begin{equation}
  \label{ABC01}
  \begin{array}{rcl}
    \ds{R_\chi} &=& \ds{C^+ D B^{\delta-1}=1.02}\; ,\nnn
    \ds{\tilde{R}_\xi} &=& \ds{
      (\xi^+)^{\beta/\nu}D^{1/(\delta+1)}B=0.852}\; ,\nnn
    \ds{\xi^+ E} &=& \ds{0.220}\; .
  \end{array}
\end{equation}
Those for the critical exponents are given in table \ref{tab2}.
\begin{table}
\begin{center}
\begin{tabular}{|c||l|l|l|l|l|} \hline
   &
  $\nu$ &
  $\gamma$ &
  $\delta$ &
  $\beta$ &
  $\eta$
  \\[0.5mm] \hline\hline
  average action &
  $0.787$ &
  $1.548$ &
  $4.80$ &
  $0.407$ &
  $0.0344$
  \\ \hline
  FD &
  $0.73(2)$ &
  $1.44(4)$ &
  $4.82(5)$ &
  $0.38(1)$ &
  $0.03(1)$
  \\ \hline
  MC &
  $0.7479(90)$ &
  $1.477(18)$ &
  $4.851(22)$ &
  $0.3836(46)$ &
  $0.0254(38)$
  \\ \hline
\end{tabular}
\caption[]{\footnotesize The table shows the critical exponents 
  corresponding to the three--dimensional $O(4)$--Heisenberg model.
  Our results are denoted by ``average action'' whereas ``FD''
  labels the exponents obtained from perturbation series at fixed
  dimension to seven loops \cite{BMN78-1}. 
  The bottom line contains lattice Monte Carlo
  results \cite{KK95-1}.
  \label{tab2}}
\end{center}
\end{table}
Here the value of $\eta$ is obtained from the temperature dependent
version of eq.~(\ref{AAA69}) (cf.~appendix~\ref{AnomalousDimensions})
at the critical temperature. For comparison table~\ref{tab2} also
gives the results from perturbation series at fixed dimension to 
seven--loop order~\cite{BMN78-1,ZJ} as well as lattice Monte Carlo
results~\cite{KK95-1} which have been used for the lattice form of the
scaling function in fig.~\ref{scalfunc}.~\footnote{See also
  ref.~\cite{MT97-1} and references therein for a recent calculation
  of critical exponents using similar methods as in this work. For
  high precision estimates of the critical exponents see also
  ref.~\cite{BC95-1,Rei95-1}.} There are only two independent
amplitudes and critical exponents, respectively. They are related by
the usual scaling relations of the three--dimensional scalar
$O(N)$--model~\cite{ZJ} which we have explicitly verified by the
independent calculation of our exponents.

We turn to the discussion of the scaling behavior of the chiral
condensate $\VEV{\ol{\psi}\psi}$ for the general case of a temperature
and quark mass dependence.  In figure~\ref{ccc_T} in the introduction
we have displayed our results for the scaling equation of state in
terms of the chiral condensate\footnote{In the literature also a
  different definition of the chiral condensate is used, corresponding
  to $\VEV{\ol{\psi}\psi}= -\ol{m}^2_{k_\Phi}T_c[\jmath/(T_c^3
  f(x))]^{1/\delta}$.}
\begin{equation}
  \label{XXX30}
  \VEV{\ol{\psi}\psi}=
  -\ol{m}^2_{k_\Phi}T_c
  \left(\frac{\jmath/T_c^3}{f(x)}\right)^{1/\delta}+
    \jmath
\end{equation}
as a function of $T/T_c=1+x(\jmath/T_c^3 f(x))^{1/\beta\delta}$ for
different quark masses or, equivalently, different values of $\jmath$.
The curves shown in figure \ref{ccc_T} correspond to quark masses
$\hat{m}=0$, $\hat{m}=\hat{m}_{\rm phys}/10$, $\hat{m}=\hat{m}_{\rm
  phys}$ and $\hat{m}=3.5\hat{m}_{\rm phys}$ or, equivalently, to zero
temperature pion masses $m_\pi=0$, $m_\pi=45\MeV$, $m_\pi=135\MeV$ and
$m_\pi=230\MeV$, respectively (cf.~figure~\ref{mm}). One observes that
the second order phase transition with a vanishing order parameter at
$T_c$ for $\hat{m}=0$ is turned into a smooth crossover in the
presence of non--zero quark masses.

The scaling form (\ref{XXX30}) for the chiral condensate is exact only
in the limit $T\to T_c$, $\jmath\ra0$.  It is interesting to find the
range of temperatures and quark masses for which $\VEV{\ol{\psi}\psi}$
approximately shows the scaling behavior (\ref{XXX30}).  This can be
infered from a comparison (see fig.\ \ref{ccc_T}) with our full
non--universal solution for the $T$ and $\jmath$ dependence of
$\VEV{\ol{\psi}\psi}$ as described in
section~\ref{TheQuarkMesonModelAtTNeq0}. For $m_\pi=0$ one observes
approximate scaling behavior for temperatures $T\gta90\MeV$. This
situation persists up to a pion mass of $m_\pi=45\MeV$. Even for the
realistic case, $m_\pi=135\MeV$, and to a somewhat lesser extent for
$m_\pi=230\MeV$ the scaling curve reasonably reflects the physical
behavior for $T\gta T_c$. For temperatures below $T_c$, however, the
zero temperature mass scales become important and the scaling
arguments leading to universality break down.

The above comparison may help to shed some light on the use of
universality arguments away from the critical temperature and the
chiral limit. One observes that for temperatures above $T_c$ the
scaling assumption leads to quantitatively reasonable results even for
a pion mass almost twice as large as the physical value. This in turn
has been used for two flavor lattice QCD as theoretical input to guide
extrapolation of results to light current quark masses.  From
simulations based on a range of pion masses $0.3\lta
m_\pi/m_\rho\lta0.7$ and temperatures $0<T\lta250\MeV$ a
``pseudocritical temperature'' of approximately $140\MeV$ with a weak
quark mass dependence is reported~\cite{MILC97-1}. Here the
``pseudocritical temperature'' $T_{pc}$ is defined as the inflection
point of $\VEV{\ol{\psi}\psi}$ as a function of temperature.  The
values of the lattice action parameters used in~\cite{MILC97-1} with
$N_t=6$ were $a\hat{m}=0.0125$, $6/g^2=5.415$ and $a\hat{m}=0.025$,
$6/g^2=5.445$. For comparison with lattice data we have displayed in
figure \ref{ccc_T} the temperature dependence of the chiral condensate
for a pion mass $m_\pi=230\MeV$.  From the free energy of the linear
quark meson model we obtain in this case a pseudocritical temperature
of about $150\MeV$ in reasonable agreement with the results of
ref.~\cite{MILC97-1}.  In contrast, for the critical temperature in
the chiral limit we obtain $T_c=100.7\MeV$.  This value is
considerably smaller than the lattice results of about $(140 - 150)
\MeV$ obtained by extrapolating to zero quark mass in
ref.~\cite{MILC97-1}.  We point out that for pion masses as large as
$230\MeV$ the condensate $\VEV{\ol{\psi}\psi}(T)$ is almost linear
around the inflection point for quite a large range of temperature.
This makes a precise determination of $T_c$ somewhat difficult.
Furthermore, figure \ref{ccc_T} shows that the scaling form of
$\VEV{\ol{\psi}\psi}(T)$ underestimates the slope of the physical
curve. Used as a fit with $T_c$ as a parameter this can lead to an
overestimate of the pseudocritical temperature in the chiral limit.
We also mention here the results of ref.~\cite{Got97-1}.  There two
values of the pseudocritical temperature, $T_{pc}=150(9)\MeV$ and
$T_{pc}=140(8)$, corresponding to $a\hat{m}=0.0125$, $6/g^2=5.54(2)$
and $a\hat{m}=0.00625$, $6/g^2=5.49(2)$, respectively, (both for
$N_t=8$) were computed.  These values show a somewhat stronger quark
mass dependence of $T_{pc}$ and were used for a linear extrapolation
to the chiral limit yielding $T_c=128(9)\MeV$.

The linear quark meson model exhibits a second order phase transition
for two quark flavors in the chiral limit. As a consequence the model
predicts a scaling behavior near the critical temperature and the
chiral limit which can, in principle, be tested in lattice
simulations. For the quark masses used in the present lattice studies
the order and universality class of the transition in two flavor QCD
remain a partially open question. Though there are results from the
lattice giving support for critical scaling~\cite{Kar94-1,IKKY97-1}
there are also recent simulations with two flavors that reveal
significant finite size effects and problems with
$O(4)$ scaling~\cite{BKLO96-1,Uka97-1}.

\subsection{Conclusions}
\label{AdditionalDegreesOfFreedom}

So far we have investigated the chiral phase transition of QCD as
described by the linear $O(4)$--model containing the three pions and
the sigma resonance as well as the up and down quarks as degrees of
freedom. Of course, it is clear that the spectrum of QCD is much
richer than the states incorporated in our model. It is therefore
important to ask to what extent the neglected degrees of freedom like
the strange quark, strange (pseudo)scalar mesons, (axial)vector
mesons, baryons, etc., might be important for the chiral dynamics of
QCD.  Before doing so it is perhaps instructive to first look into the
opposite direction and investigate the difference between the linear
quark meson model described here and chiral perturbation theory based
on the non--linear sigma model~\cite{GL82-1}.  In some sense, chiral
perturbation theory is the minimal model of chiral symmetry breaking
containing only the Goldstone degrees of freedom. By construction it
is therefore only valid in the spontaneously broken phase and can not
be expected to yield realistic results for temperatures close to $T_c$
or for the symmetric phase.  However, for small temperatures (and
momentum scales) the non--linear model is expected to describe the
low--energy and low--temperature limit of QCD reliably as an expansion
in powers of the light quark masses. For vanishing temperature it has
been demonstrated recently~\cite{QuMa,JW97-1} that the results of
chiral perturbation theory can be reproduced within the linear meson
model once certain higher dimensional operators in its effective
action are taken into account for the three flavor case.  Moreover,
some of the parameters of chiral perturbation theory
($L_4,\ldots,L_8$) can be expressed and therefore also numerically
computed in terms of those of the linear model. For non--vanishing
temperature one expects agreement only for low $T$ whereas deviations
from chiral perturbation theory should become large close to $T_c$.
Yet, even for $T\ll T_c$ small quantitative deviations should exist
because of the contributions of (constituent) quark and sigma meson
fluctuations in the linear model which are not taken into account in
chiral perturbation theory.

\begin{figure}
\unitlength1.0cm
\begin{center}
\begin{picture}(13.,7.0)

\put(0.0,0.0){
\epsfysize=11.cm
\rotate[r]{\epsffile{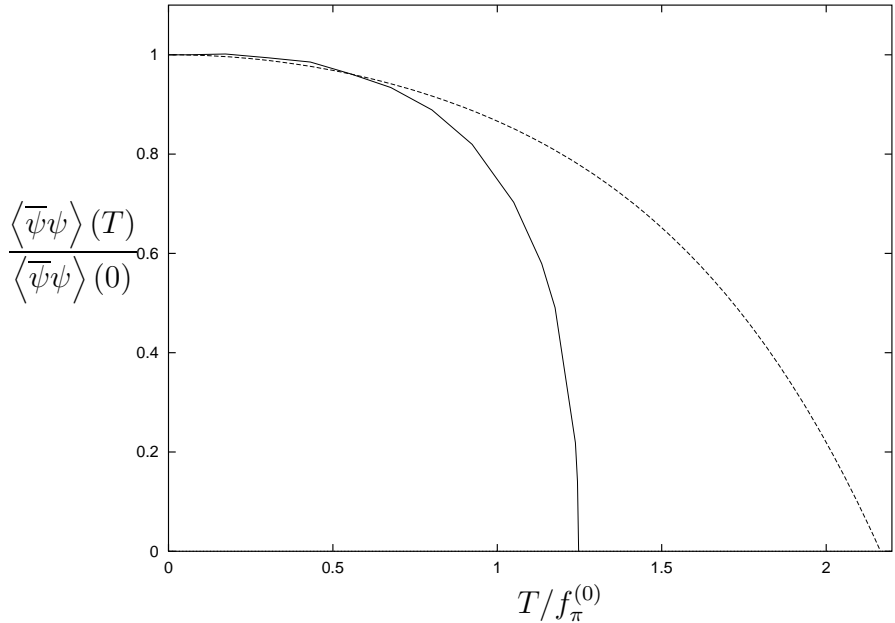}}
}
\put(-1.0,4.2){\bf 
  $\ds{\frac{\VEV{\ol{\psi}\psi}(T)}{\VEV{\ol{\psi}\psi}(0)} }$}
\put(5.8,-0.5){\bf $\ds{T/f_\pi^{(0)}}$}

\end{picture}
\end{center}
\caption{\footnotesize The plot displays the chiral condensate
  $\VEV{\ol{\psi}\psi}$ as a function of $T/f_\pi^{(0)}$. The solid
  line corresponds to our results for vanishing average current quark
  mass $\hat{m}=0$ whereas the dashed line shows the corresponding
  three--loop chiral perturbation theory result for
  $\Gamma_1=470\MeV$.}
\label{cc_T}
\end{figure}

From~\cite{GL87-1} we infer the three--loop result for the
temperature dependence of the chiral condensate in the chiral limit
for $N$ light flavors
\begin{equation}
  \label{BBB100}
  \begin{array}{rcl}
    \ds{\VEV{\ol{\psi}\psi}(T)_{\chi PT}} &=& \ds{
      \VEV{\ol{\psi}\psi}_{\chi PT}(0)
      \Bigg\{1-\frac{N^2-1}{N}\frac{T^2}{12F_0^2}-
        \frac{N^2-1}{2N^2}
        \left(\frac{T^2}{12F_0^2}\right)^2}\nnn
    &+& \ds{
      N(N^2-1)\left(\frac{T^2}{12F_0^2}\right)^3
      \ln\frac{T}{\Gamma_1}
      \Bigg\} +\Oc(T^8)}\; .
  \end{array}
\end{equation}
The scale $\Gamma_1$ can be determined from the $D$--wave isospin zero
$\pi\pi$ scattering length and is given by $\Gamma_1=(470\pm100)\MeV$.
The constant $F_0$ is (in the chiral limit) identical to the pion
decay constant $F_0=f_\pi^{(0)}=80.8\MeV$. In figure \ref{cc_T} we
have plotted the chiral condensate as a function of $T/F_0$ for both,
chiral perturbation theory according to (\ref{BBB100}) and for the
linear quark meson model.
As expected the agreement for small $T$ is very good. Nevertheless,
the anticipated small numerical deviations present even for $T\ll T_c$
due to quark and sigma meson loop contributions are manifest.  For
larger values of $T$, say for $T\gta0.8f_\pi^{(0)}$ the deviations
become significant because of the intrinsic inability of chiral
perturbation theory to correctly reproduce the critical behavior of
the system near its second order phase transition.

Within the language of chiral perturbation theory the neglected
effects of thermal quark fluctuations may be described by an effective
temperature dependence of the parameter $F_0(T)$. We notice that the
temperature at which these corrections become important equals
approximately one third of the constituent quark mass $M_q(T)$ or the
sigma mass $m_\sigma(T)$, respectively, in perfect agreement with
fig.~\ref{Thresh}. As suggested by this figure the onset of the
effects from thermal fluctuations of heavy particles with a
$T$--dependent mass $m_H(T)$ is rather sudden for $T\gta m_H(T)/3$.
These considerations also apply to our two flavor quark meson model.
Within full QCD we expect temperature dependent initial values at
$k_\Phi$.

The dominant contribution to the temperature dependence of the initial
values presumably arises from the influence of the mesons containing
strange quarks as well as the strange quark itself.  Here the quantity
$\ol{m}^2_{k_\Phi}$ seems to be the most important one.  (The
temperature dependence of higher couplings like $\lambda(T)$ is not
very relevant if the IR attractive behavior remains valid, i.e.~if
$Z_{\Phi,k_\Phi}$ remains small for the range of temperatures
considered. We neglect a possible $T$--dependence of the current quark
mass $\hat{m}$.) In particular, for three flavors the potential
$U_{k_\Phi}$ contains a term
\begin{equation}
  \label{LLL12}
  -\frac{1}{2}\ol{\nu}_{k_\Phi}
  \left(\det\Phi+\det\Phi^\dagger\right)=
  -\ol{\nu}_{k_\Phi}\vph_s\Phi_{uu}\Phi_{dd}+\ldots
\end{equation}
which reflects the axial $U_A(1)$ anomaly. It yields a contribution to
the effective mass term proportional to the expectation value
$\VEV{\Phi_{ss}}\equiv\vph_s$, i.e.
\begin{equation}
  \label{LLL13}
  \Delta\ol{m}^2_{k_\Phi}=
  -\frac{1}{2}\ol{\nu}_{k_\Phi}\vph_s\; .
\end{equation}
Both, $\ol{\nu}_{k_\Phi}$ and $\vph_s$, depend on $T$.  We expect
these corrections to become relevant only for temperatures exceeding
$m_K(T)/3$ or $M_s(T)/3$. We note that the temperature dependent kaon
and strange quark masses, $m_K(T)$ and $M_s(T)$, respectively, may be
somewhat different from their zero temperature values but we do not
expect them to be much smaller. A typical value for these scales is
around $500\MeV$. Correspondingly, the thermal fluctuations neglected
in our model should become important for $T\gta170\MeV$. It is even
conceivable that a discontinuity appears in $\vph_s(T)$ for
sufficiently high $T$ (say $T\simeq170\MeV$). This would be reflected
by a discontinuity in the initial values of the $O(4)$--model leading
to a first order transition within this model.  Obviously, these
questions should be addressed in the framework of the three flavor
$SU_L(3)\times SU_R(3)$ quark meson model. Work in this direction is
in progress.

We note that the temperature dependence of $\ol{\nu}(T)\vph_s(T)$ is
closely related to the question of an effective high temperature
restoration of the axial $U_A(1)$ symmetry~\cite{PW84-1,Shu94-1}.  The
$\eta^\prime$ mass term is directly proportional to this combination,
$m_{\eta^\prime}^2(T)-m_\pi^2(T)\simeq\frac{3}{2}\ol{\nu}(T)
\vph_s(T)$ \cite{JuWet96}. Approximate $U_A(1)$ restoration would occur
if $\vph_s(T)$ or $\ol{\nu}(T)$ would decrease sizeable for large $T$.
For realistic QCD this question should be addressed by a three flavor
study. Within two flavor QCD the combination $\ol{\nu}_k\vph_s$ is
replaced by an effective anomalous mass term $\ol{\nu}_k^{(2)}$. The
temperature dependence of $\ol{\nu}^{(2)}(T)$ could be studied by
introducing quarks and the axial anomaly in the two flavor matrix
model of ref.~\cite{BW97-1}.  We add that this question has also been
studied within full two flavor QCD in lattice
simulations~\cite{BKLO96-1,MILC97-2,KLS97-1}. So far there does not
seem to be much evidence for a restoration of the $U_A(1)$ symmetry
near $T_c$ but no final conclusion can be drawn yet.

To summarize, we have found that the effective two flavor quark meson
model presumably gives a good description of the temperature effects
in two flavor QCD for a temperature range $T\lta170\MeV$. Its
reliability should be best for low temperature where our results agree
with chiral perturbation theory.  However, the range of validity is
considerably extended as compared to chiral perturbation theory and
includes, in particular, the critical temperature of the second order
phase transition in the chiral limit.  We have explicitly connected
the universal critical behavior for small $\abs{T-T_c}$ and small
current quark masses with the renormalized couplings at $T=0$ and
realistic quark masses. The main quantitative uncertainties from
neglected fluctuations presumably concern the values of $f_\pi^{(0)}$
and $T_c$ which, in turn, influence the non--universal amplitudes $B$
and $D$ in the critical region. We believe that our overall picture is
rather solid. Where applicable our results compare well with numerical
simulations of full two flavor QCD.

%% file: outlook.tex
\sect{Outlook \label{outlook}}

The use of non--perturbative flow equations within the framework of
the effective average action provides a practical means for 
non--perturbative studies within a field theoretic context.
In this work its application culminated in the computation
of the equation of state for two flavor QCD within the 
effective quark meson model.
In particular, the method is free of infrared divergences and allows
a precise description of critical phenomena in high temperature 
quantum field theory or in statistical physics.

It is a valuable merit of the approach presented in this work
to yield directly the whole equation of state of the theory
under investigation. We have given a detailed quantitative picture
of the non--analytic properties of the effective potential
or the free energy for second order phase transitions. We
have demonstrated that along the lines presented in section 
\ref{secfirst} also the universal equation of state
for weak first order phase transitions can be reliably
computed. Furthermore, for $k > 0$ the effective average
action realizes the concept of a coarse grained free energy.
We, therefore, were able to extract quantities like the ``classical''
interface tension used in the standard approach to phase conversion
by spontaneous bubble nucleation. 

Apart from their phenomenological relevance and their intrinsic
appeal, critical phenomena also provide a severe test for the 
reliability of a non--perturbative method.
The present level of truncation used in this work is based on
a lowest order derivative expansion and the most general potential
term\footnote{For the scalar matrix models discussed in section
\ref{secfirst} we kept the most general form of the potential only for
the relevant $\rho$ dependence.}. 
Where available we observe agreement
with recent lattice results for the critical equation of state 
within a few per cent accuracy. So far the main drawback of the method
is a precise error estimate which is a natural difficulty in a 
non--perturbative calculation with couplings that cannot be
considered as small. However, the structure of the exact flow equation
(\ref{ERGE}) is of considerable help in this context. For instance,
for the $O(N)$ symmetric $N$-component model one easily verifies
that the ansatz (cf.\ section \ref{secsec} for notation) 
\be
\Gamma_k = \int d^dx \bigl\{
U_k(\rho) + \frac{1}{2} \partial^{\mu} \phi_a
Z_k(\rho,-\partial^{\nu}\partial_{\nu})
\partial_{\mu} \phi^a +\frac{1}{4}
\partial^{\mu} \rho
Y_k(\rho,-\partial^{\nu}\partial_{\nu})
\partial_{\mu} \rho \bigr\}.
\label{exa} \ee
contains the most general terms which contribute to
$\Gamma^{(2)}_k$ when evaluated for a constant background field 
\cite{TW94-1}. With the above ansatz one therefore obtains an exact 
flow equation 
for the effective average potential $U_k$ containing the 
unknown functions $Z_k(\rho,q^2)$ and
$Y_k(\rho,q^2)$, which play the role of field and momentum
dependent wave function renormalization constants. 
Knowledge of these functions
would yield an exact critical equation of state from the solution
of the flow equation for $U_k$.
These functions involve, however, more complicated momentum
dependences of $1 PI$ vertices with three or four legs with 
non--vanishing momenta. Along the lines presented in section
\ref{secsec} recent calculations for the critical equation of state
$(N=1)$ have been successively extended to include a wave function
renormalization with a most general field dependence\footnote{
For $N=1$ the $Y_k$ term is absent.} \cite{BSW}. These results seem
to improve the exponents and the equation of state within
the expected accuracy of a few per cent. In conjunction with the 
smallness of the anomalous dimension ($\eta \simeq 0.04$) one is lead 
to expect only
small corrections from the momentum dependence of the
wave function renormalization. A crucial test 
for the stability of the results would be to change a given
parameterization of the momentum dependence of the wave function 
renormalization in a reasonable range\footnote{One expects a very mild
momentum dependence of $Z_k(\rho=\rho_0,q^2)$ for $q^2$ between
zero and $k^2$ due to the presence of the infrared cutoff.
For $q^2 \gg k^2$ the anomalous dimension should determine
the behavior near the transition, 
$Z_k(\rho=\rho_0,q^2)\sim (q^2)^{-\eta /2}$.}. In this way 
we hope to obtain reliable error estimates for the critical
equation of state, critical exponents etc. 

We also note that the approach is valid for arbitrary dimensions
$d$. All evolution equations presented in this work are formulated 
for general $d$.
In particular, two dimensional field theories can be treated
with this method. This allows, e.g., the comparison with known exact 
results. We have calculated the critical
properties for the two dimensional $O(N)$ model in the ``replica
limit'' \cite{EA75} $N \to 0$, which describes the critical swelling 
of long polymer chains in two space dimensions \cite{deG2}. 
For instance, we find the exponent $\nu$ 
(cf. section \ref{secsec}) to be given by
$\nu=0.782$ \cite{BG95} in the two dimensional model which 
compares rather well 
to the exact result $\nu_{(\rm{exact})}=0.75$ \cite{ZJ}. 
We also note, e.g., that  
Gr{\"a}ter and Wetterich \cite{GW94} 
have used the effective average action method
in two dimensions 
to describe the
Kosterlitz--Thouless phase transition by employing an expansion 
around the minimum of the effective average potential. 
The validity of the method 
for arbitrary dimensions $d$ may also be used to get a deeper insight
into the $\epsilon$-expansion approach \cite{WilFis72}.         

The investigation of three dimensional scalar models
for complex $2 \times 2$ matrices
in section \ref{secfirst} can be extended in several directions.
The generalization to complex
$N \times N$ matrices for arbitrary $N$ is straightforward.
This opens the possibility to study the large--$N$ behavior
and to compare with $1/N$--expansions\footnote{See refs.\ 
\cite{Fer,Nis} for a renormalization group study within the large--$N$ 
limit of hermitean matrix models.}. 
The most interesting generalizations 
in the context of chiral symmetry breaking are the systems with
reduced $SU(N) \times SU(N)$ symmetry (cf.\ section \ref{2qcd}). 
Another interesting application
concerns Anderson localization. The problem of an electron in
a random potential can be mapped onto a field theoretic matrix model
with e.g.\ $U(2N)$ symmetry \cite{Weg}. To get insight into the involved
replica limit in the context of flow equations, we have computed
the free energy for the three dimensional $O(N)$ model in the 
limit $N \to 0$ \cite{BG95} (cf.\ also section \ref{secsec}). 
Further studies are necessary to ensure
the validity of this limit for matrix models within the approach
presented in this work.

Of course the approach is not restricted to the study of the universal
behavior. The ability of the method to cover both the non--universal
properties of the four dimensional theory at low temperature and the 
universal properties near $T_c$ and the chiral limit has been crucial
for the investigation of the QCD chiral phase transition in section
\ref{2qcd}. We expect our results within the effective quark meson
model to give a reliable description of the equation of
state for two flavor QCD.
The employed method can be generalized to the (more) realistic
description including the strange quark. The relevant generalizations
have been outlined in section \ref{2qcd} and the investigation 
is work in progress \cite{BJW97-2}. 
We expect our approach to be very well suited to the non--perturbative
investigation of the three flavor case. At non--zero temperature
we should be able to reliably answer a number of urging questions,
like the order of the phase transition or if one has to deal with
a crossover phenomenon. Connected to this, we are well prepared
to answer the question of a large
scale that could lead to a disoriented chiral condensate
(cf.\ section \ref{mainin}). The three flavor results will
also give details about the speculative ``effective restoration''
of the axial $U_A(1)$ around $T_c$. If the approximate partial
infrared fixed point, which we have observed for the two flavor  
case, is realized for $N_f=3$ our equation of state will
develop a considerable degree of predictive power. Of course,
this concerns also the zero temperature low--energy properties
of QCD. In particular, these results will be very interesting for
the scalar spectrum which is experimentally
not very well understood. Further
investigations, extending the description in terms of quarks
and the low--lying scalar and pseudoscalar mesons, have
to include fields for the lightest vector and pseudovector mesons.

We finally note that a computation of the equation of state
within QCD, i.e.\ as a function of $\alpha_s(1 \GeV)$ and the 
quark masses, depends on a reliable computation of the scalar
mass term $\overline{m}^2(k_{\Phi})/k_{\Phi}^2$ and the
Yukawa coupling $h(k_{\Phi})$ at the compositeness scale.
Integrating out the gluons completely for the determination
of the quark meson model at $k_{\Phi}$ is, however, a difficult
task. Present investigations
in this direction within the QCD framework for the effective average 
action proposed in ref.\ \cite{Wet95-2} confirm the
relatively simple ansatz at $k_{\Phi}$ for the resulting quark
propagator we have used in section \ref{2qcd} \cite{BWet97}.

\section{Acknowledgments \label{ack}}

I first want to thank C.~Wetterich. I gained a lot from discussing 
and working with him. Our collaboration on the topics presented
in this work has been always fruitful and fun. 
I also want to thank F.~Wegner for his continuous interest and for his
substantial influence on the statistical physics aspects of this work.
I have considerably profited from the joined Graduate College of the
high energy physics and the statistical physics department of the ITP.
I also want to thank D.~Jungnickel for many valuable discussions
and for collaborating on the QCD chiral phase transition. I am grateful
to N.~Tetradis for his collaboration on the scaling equation of state
and on the coarse grained free energy. In addition I thank J.~Adams,
S.~Bornholdt and F.~Freire for their co-operation on numerics.
I also thank M.\ Gr{\"a}ter for many discussions on matrix models.
I would like to express my gratitude to all people at the ITP for
providing a most stimulating environment.

%% file: append.tex
\appendix{Threshold functions}
\label{ThresholdFunctions}

In this appendix we list the various definitions of dimensionless
threshold functions appearing in the flow equations and the
expressions for the anomalous dimensions for $T=0$. They involve the
inverse scalar average propagator\footnote{We note that 
the definition of $P_k(q)$ in section \ref{scale} obtains 
from the definition used here by a rescaling
with the factor $Z_k$.} $P(q)$ 
\begin{equation}
  \label{BBB01}
  P(q)=
  \frac{q^2}{1-\exp\left\{-\frac{q^2}{k^2}\right\}}
\end{equation}
and the corresponding fermionic function $P_F$ which can be chosen
as~\cite{Ju95-7} 
\begin{equation}
  \label{JJJ000}
  P_F(q)=P(q)\equiv q^2\left(1+r_F(q)\right)^2\; .
\end{equation}
We abbreviate
\begin{equation}
  \label{LLL23}
  x= q^2\; ,\;\; P(x)\equiv P(q)\; ,\;\;
  \dot{P}(x)\equiv\frac{\prl}{\prl x}P(x)\; ,\;\;
  \widehat{\frac{\prl}{\prl t}}\dot{P}\equiv
  \frac{\prl}{\prl x}\widehat{\frac{\prl}{\prl t}}P\; ,
\end{equation}
etc., and use the formal definition
\begin{equation}
  \label{LLL24}
  \ds{\widehat{\frac{\prl}{\prl t}}}
  \equiv \ds{
    \frac{1}{Z_{\Phi,k}}\frac{\prl R_k}{\prl t}
    \frac{\prl}{\prl P} }
  + \ds{
    \frac{2}{Z_{\psi,k}} \frac{P_F}{1+r_F}
    \frac{\prl \left[Z_{\psi,k} r_F\right]}{\prl t}
    \frac{\prl}{\prl P_F} }\; .
\end{equation}
The bosonic threshold functions read
\begin{equation}
  \label{LLL20}
  \begin{array}{rcl}
    \ds{l_n^d (w;\eta_\Phi)} &=& \ds{
      l_n^d (w) - \eta_\Phi\hat{l}_n^d (w)
      }\nnn
    &=& \ds{
      \frac{n+\delta_{n,0}}{2} 
      k^{2n-d} \int_0^\infty d x\, x^{\frac{d}{2}-1}
      \left(\frac{1}{Z_{\Phi,k}} \frac{\prl R_k}{\prl t}\right)
        \left( P+w k^2\right)^{-(n+1)} }\nnn
    \ds{ l_{n_1,n_2}^d(w_1,w_2;\eta_\Phi)} &=& \ds{
      \l_{n_1,n_2}^d(w_1,w_2)
      -\eta_\Phi \hat{l}_{n_1,n_2}^d(w_1,w_2)}\nnn
    &=& \ds{
      -\hal k^{2(n_1+n_2)-d}
      \int_0^\infty dx\, x^{\frac{d}{2}-1}
      \widehat{\frac{\prl}{\prl t}} \left\{\left(
      P+w_1k^2\right)^{-n_1}
      \left( P+w_2k^2\right)^{-n_2} \right\}}
  \end{array}
\end{equation}
where $n,n_1,n_2\ge0$ is assumed. For $n\neq0$ the
functions $l_n^d$ may also be written as
\begin{equation}
  \label{LLL21}
  l_n^d (w;\eta_\Phi) =
  -\hal k^{2n-d} \int_0^{\infty} dx x^{\frac{d}{2}-1}
  \widehat{\frac{\prl}{\prl t}} \left( P+w k^2\right)^{-n}\; .
\end{equation}
The fermionic integrals $l_n^{(F)d} (w;\eta_\psi)=l_n^{(F)d} (w)-
\eta_\psi\check{l}_n^{(F)d} (w)$ are defined analogously as
\begin{equation}
  \label{JJJ001}
  \begin{array}{rcl}
    \ds{l_n^{(F)d} (w;\eta_\psi)} &=& \ds{
      \left(n+\delta_{n,0}\right)
      k^{2n-d} \int_0^\infty d x\, x^{\frac{d}{2}-1}
      \frac{1}{Z_{\psi,k}}\frac{P_F}{1+r_F}
      \frac{\prl\left[ Z_{\psi,k}r_F\right]}{\prl t}
        \left(P+w k^2\right)^{-(n+1)} }\; .
  \end{array}
\end{equation}
Furthermore one has
\begin{equation}
  \label{LLL22}
  \begin{array}{rcl}
    \ds{l_{n_1,n_2}^{(FB)d}(w_1,w_2;\eta_\psi,\eta_\Phi)} &=& \ds{
      l_{n_1,n_2}^{(FB)d}(w_1,w_2)
      -\eta_\psi \check{l}_{n_1,n_2}^{(FB)d}(w_1,w_2)
      -\eta_\Phi \hat{l}_{n_1,n_2}^{(FB)d}(w_1,w_2) 
      }\nnn
    && \ds{ \hspace{-2cm}
      = -\hal k^{2(n_1+n_2)-d}
      \int_0^\infty dx\, x^{\frac{d}{2}-1}
      \widehat{\frac{\prl}{\prl t}}\left\{
      \frac{1}{[P_F(x)+k^2w_1]^{n_1} [P(x)+k^2w_2]^{n_2} } \right\}
      }\nnn
    \ds{m_{n_1,n_2}^d (w_1,w_2;\eta_\Phi)} &\equiv& \ds{
      m_{n_1,n_2}^d (w_1,w_2) - \eta_\Phi 
      \hat{m}_{n_1,n_2}^d (w_1,w_2) 
      }\nnn
    && \ds{\hspace{-2cm} 
      = -\hal k^{2(n_1+n_2-1)-d}
      \int_0^\infty dx\, x^{\frac{d}{2}}
      \widehat{\frac{\prl}{\prl t}} \left\{
      \frac{\dot{P} (x)}
      {[P(x)+k^2 w_1]^{n_1} }
      \frac{\dot{P} (x)}
      {[P(x)+k^2 w_2]^{n_2}} \right\} 
      }\nnn
    \ds{m_4^{(F)d} (w;\eta_\psi)} &=& \ds{
      m_4^{(F)d} (w)-\eta_\psi \check{m}_4^{(F)d} (w) 
      }\nnn
    &=& \ds{ 
      -\hal k^{4-d}
      \int_0^\infty dx\, x^{\frac{d}{2}+1}
      \widehat{\frac{\prl}{\prl t}} \left(
      \frac{\prl}{\prl x}
      \frac{1+r_F(x)}{P_F(x)+k^2w}\right)^2
      \label{m4Fd} 
      }\nnn
    \ds{m_{n_1,n_2}^{(FB)d}(w_1,w_2;\eta_\psi,\eta_\Phi)} &=& \ds{
      m_{n_1,n_2}^{(FB)d}(w_1,w_2)
      -\eta_\psi \check{m}_{n_1,n_2}^{(FB)d}(w_1,w_2)
      -\eta_\Phi \hat{m}_{n_1,n_2}^{(FB)d}(w_1,w_2) 
      }\nnn
    && \ds{ \hspace{-2cm}
      = -\hal k^{2(n_1+n_2-1)-d}
      \int_0^\infty dx\, x^{\frac{d}{2}}
      \widehat{\frac{\prl}{\prl t}}\left\{
      \frac{1+r_F(x)}{[P_F(x)+k^2w_1]^{n_1}}
      \frac{\dot{P}(x)}{[P(x)+k^2w_2]^{n_2}} \right\} \; . 
      }
  \end{array}
\end{equation}
The dependence of the threshold functions on the anomalous dimensions
arises from the $t$--derivative acting on $Z_{\Phi,k}$ and
$Z_{\psi,k}$ within $R_k$ and $Z_{\psi,k}r_{F}$, respectively.  We
furthermore use the abbreviations
\begin{equation}
  \label{LLL25}
  \begin{array}{rcl}
  \ds{l_n^d(\eta_\Phi)\equiv l_n^d(0;\eta_\Phi)} &,& \ds{
    l_{n}^{(F)d}(\eta_\psi)\equiv l_{n}^{(F)d}(0;\eta_\psi)}\nnn
  \ds{l_n^d(w)\equiv l_n^d(w;0)} &,& \ds{
  l_n^d\equiv l_n^d(0;0)}
  \end{array}
\end{equation}
etc.~and note that in four dimensions the integrals
\begin{equation}
  \label{LLL26}
  l_2^4(0,0)=l_2^{(F)4}(0,0)=
  l_{1,1}^{(FB)4}(0,0)=m_4^{(F)4}(0)=m_{1,2}^{(FB)4}(0,0)=1
\end{equation}
are independent of the particular choice of the infrared cutoff.

\appendix{Temperature dependent threshold functions}
\label{AnomalousDimensions}

Non--vanishing temperature manifests itself in the flow equations
(\ref{AAA68}), (\ref{AAA91})---(\ref{AAA69}) only through a change to
$T$--dependent threshold functions. In this appendix we will define
these functions and discuss some subtleties regarding the definition
of the anomalous dimensions and the Yukawa coupling for $T\neq0$. The
corresponding $T=0$ threshold functions can be found in
appendix~\ref{ThresholdFunctions} where also some of our notation is
fixed.

The flow equation (\ref{AAA68}) for the effective average potential
involves a bosonic and a fermionic threshold function whose
generalization to finite temperature is straightforward
\begin{equation}
  \label{JJJ00}
  \begin{array}{rcl}
    \ds{l_n^d(w,\tilde{T};\eta_\Phi)} &=& \ds{
      \frac{(n+\delta_{n,0})}{2}
      \frac{v_{d-1}}{v_d}k^{2n-d+1}\tilde{T}
      \sum_{l\in\ZZZ}\int_0^\infty
      d x\,x^{\frac{d-3}{2}}
      Z_{\Phi,k}^{-1}
      \frac{\prl_t R_k(y)}
      {\left[ P(y)+k^2 w\right]^{n+1}}\; ,
      }\nnn
    \ds{l_n^{(F)d}(w,\tilde{T};\eta_\Phi)} &=& \ds{
      \left(n+\delta_{n,0}\right)
      \frac{v_{d-1}}{v_d}k^{2n-d+1}}\nnn
    &\times& \ds{
      \tilde{T}\sum_{l\in\ZZZ}\int_0^\infty
      d x\,x^{\frac{d-3}{2}}
      Z_{\psi,k}^{-1}\frac{P_F(y_F)}{1+r_F(y_F)}
      \frac{\prl_t\left[ Z_{\psi,k}r_F(y_F)\right]}
      {\left[P_F(y_F)+k^2 w\right]^{n+1}}
      }
  \end{array}
\end{equation}
where $\tilde{T}=T/k$ and
\begin{equation}
  \label{JJJ01}
  \begin{array}{rcl}
    \ds{y} &=& \ds{x+(2 l\pi T)^2}\; ,\nnn
    \ds{y_F} &=& \ds{x+(2 l+1)^2\pi^2 T^2}\; ,
  \end{array}
\end{equation}

The computation of the anomalous dimensions $\eta_\Phi$, $\eta_\psi$
and the flow equation for the Yukawa coupling $h$ at non--vanishing
temperature requires some care. The anomalous dimensions determine the
IR cutoff scale dependence of $Z_{\Phi,k}$ and $Z_{\psi,k}$ according
to $\eta_\Phi=-\prl_t\ln Z_{\Phi,k}$, $\eta_\psi=-\prl_t\ln
Z_{\psi,k}$ with $t=\ln k/k_\Phi$.  It is important to realize that
for a computation of the scale dependence of the effective
three--dimensional $Z_{\Phi,k}$ and $Z_{\psi,k}$ for $T\neq0$ momentum
dependent wave function renormalization constants of the
four--dimensional theory are required.  This is a consequence of the
fact that in the three--dimensional model each of the infinite number
of different Matsubara modes of a four--dimensional bosonic or
fermionic field $\phi(Q)$ corresponds to a different value of
$Q_0=2\pi lT$ or $Q_0=(2l+1)\pi T$, respectively, with
$Q^2=Q_0^2+\vec{Q}^{\,2}$ and $l\in\ZZ$.  We will therefore allow for
momentum dependent wave function renormalizations, i.e.~for a kinetic
part of $\Gamma_k$ of the form
\begin{equation}
  \label{PPP00}
  \Gamma^{\rm kin}_k=
  \int\frac{d^d q}{(2\pi)^d}\left\{
  Z_{\Phi,k}(q^2)q^2\Tr\left(\Phi^\dagger(q)\Phi(q)\right)+
  Z_{\psi,k}(q^2)\ol{\psi}(q)\gamma^\mu q_\mu\psi(q)\right\}
\end{equation}
in momentum space.

In the $O(4)$--model the evolution equation for $Z_{\Phi,k}(Q)$ may
then be obtained by considering a background field configuration with
a small momentum dependence,
\begin{equation}
  \label{PPP01}
  \Phi_j(x)=\vph\delta_{j 1}+
  \left(\delta\vph e^{-i Q x}+
    \delta\vph^* e^{i Q x}\right)\delta_{j 2}\; ;\;\;\;
    j=1,\ldots,4
\end{equation}
supplemented by
\begin{equation}
  \label{PPP02}
  \psi_a=\ol{\psi}_a=0\; ;\;\;\; a=1,2\; .
\end{equation}
Expanding around this configuration at the minimum of the effective
average potential $U_k$ we observe that $\delta\vph$ corresponds to an
excitation in the Goldstone boson direction. The exact inverse
two--point function $\Gamma_k^{(2)}$ turns out to be block--diagonal
with respect to scalar and fermion indices for this configuration. It
therefore decays into corresponding matrices $\Gamma_{S k}^{(2)}$ and
$\Gamma_{F k}^{(2)}$ acting in the scalar and fermion subspaces,
respectively.  The scale dependence of the scalar wave function
renormalization for non--vanishing $T$ is obtained from (\ref{ERGE})
and (\ref{PPP00}) for the configuration (\ref{PPP01}) as
\begin{equation}
  \label{PPP03}
  \begin{array}{rcl}
    \ds{\frac{\prl}{\prl t}Z_{\Phi,k}(Q)} &=& \ds{
      \frac{1}{\vec{Q}^2}\Bigg(\lim_{\delta\vph\delta\vph^*\ra0}
      \frac{\delta}{\delta(\delta\vph\delta\vph^*)}
      \Bigg\{\frac{1}{2}\Tr\left[\left(
      \Gamma_{S k}^{(2)}+R_k\right)^{-1}
      \frac{\prl}{\prl t}R_k\right]}\nnn
    &-& \ds{
      \Tr\left[\left(\Gamma_{F k}^{(2)}+R_{F k}\right)^{-1}
      \frac{\prl}{\prl t}R_{F k}\right]\Bigg\}-
      (\vec{Q}\ra0)\Bigg) }\; .
  \end{array}
\end{equation}
In the three--dimensional theory there is now a different scalar wave
function renormalization $Z_{\Phi,k}(Q_0,\vec{Q})$ for each Matsubara
mode $Q_0$. As in the four--dimensional model for $T=0$ we neglect the
momentum dependence of the wave function renormalization constants and
evaluate $Z_{\Phi,k}$ for $\vec{Q}=0$ for each Matsubara mode. We will
furthermore simplify the truncation of the effective average action by
choosing the Matsubara zero--mode wave function renormalization
constant for all Matsubara modes, i.e., approximate
\begin{equation}
  \label{PPP03a}
  Z_{\Phi,k}(\tilde{T})=Z_{\Phi,k}(Q_0^2=0,\vec{Q}^2=0)\; .
\end{equation}
This is justified by the rapid decoupling of all massive Matsubara
modes within a small range of $k$ for fixed $T$ as discussed in
section~\ref{FiniteTemperatureFormalism}. This results in the
expression (\ref{AAA69}) for $\eta_\Phi$ but now with temperature
dependent threshold functions ($\tilde{T}=T/k$)
\begin{equation}
  \label{PPP04a}
  \begin{array}{rcl}
    \ds{m_{n_1,n_2}^{d}(w_1,w_2,\tilde{T};\eta_\Phi)}
    &=& \ds{
      m_{n_1,n_2}^{d}(w_1,w_2,\tilde{T})-\eta_\Phi 
      \hat{m}_{n_1,n_2}^{d}(w_1,w_2,\tilde{T}) 
      }\nnn
    && \ds{\hspace{-1cm}=
      -\hal k^{2(n_1+n_2-1)-d+1}
      \frac{d v_{d-1}}{(d-1)v_d}
      }\nnn
    && \ds{\hspace{-1cm}\times
      \tilde{T}\sum_{l\in\ZZZ}
      \int_0^\infty dx\, x^{\frac{d-1}{2}} 
      \widehat{\frac{\prl}{\prl t}} \left\{
      \frac{\dot{P}(y)}
      {[P(y)+k^2 w_1]^{n_1} }
      \frac{\dot{P} (y)}
      {[P(y)+k^2 w_2]^{n_2}} \right\} }\nnn
    \ds{m_{4}^{(F)d}(w,\tilde{T};\eta_\psi)}
    &=& \ds{
      m_4^{(F)d}(w,\tilde{T})-
      \eta_\psi\check{m}_4^{(F)d}(w,\tilde{T}) 
      }\nnn
    && \ds{ \hspace{-1cm}=
      -\frac{1}{2} k^{5-d}\frac{d v_{d-1}}{(d-1)v_d}
      \tilde{T}\sum_{l\in\ZZZ}
      \int_0^\infty dx\, x^{\frac{d-1}{2}}y_F
      \widehat{\frac{\prl}{\prl t}} \left(
      \frac{\prl}{\prl x}
      \frac{1+r_F(y_F)}{P_F(y_F)+k^2w}\right)^2}\; .
  \end{array}
\end{equation}
For further technical details we refer the reader to
ref.~\cite{}.

The fermion anomalous dimension and the flow equation for the Yukawa
coupling can be obtained by considering a
field configuration
\begin{equation}
  \label{PPP05}
  \begin{array}{rcl}
    \ds{\Phi_j(x)} &=& \ds{\vph\delta_{j 1}\; ;\;\;\;
      j=1,\ldots,4}\; ,\nnn
    \ds{\psi_a(x)} &=& \ds{
      \psi_a e^{-i Q x}}\; ,\nnn
    \ds{\ol{\psi}_a(x)} &=& \ds{
      \ol{\psi}_a e^{i Q x}\; ;\;\;\;
      a=1,2}\; .
  \end{array}
\end{equation}
The derivation follows similar lines as for the scalar anomalous
dimension discussed above.  For computational details we refer the
reader to ref.~\cite{Ju95-7}. An important difference as compared to
$Z_{\Phi,k}$ relates to the fact that there are no fermionic zero
modes. It would therefore be inconsistent to define
$Z_{\psi,k}(\tilde{T})$ or $h_k(\tilde{T})$ at $Q_0=0$ if $Q$ denotes
the external fermion momentum. Yet, we will again resort to the
approximation of using the same wave function renormalization constant
and Yukawa coupling for all fermionic Matsubara modes. For the same
reason as for $Z_{\Phi,k}(\tilde{T})$ we will use for this purpose the
mode with the lowest $T$--dependent mass, i.e.~define
\begin{equation}
  \label{PPP06}
  \begin{array}{rcl}
  \ds{Z_{\psi,k}(\tilde{T})} &=& \ds{
    Z_{\psi,k}(Q_0^2=\pi^2 T^2,\vec{Q}^{\,2}=0)}\; ,\nnn
  \ds{h_k(\tilde{T})} &=& \ds{
    h_k(Q_0^2=\pi^2 T^2,\vec{Q}^{\,2}=0)}\; ,
  \end{array}
\end{equation}
where we have neglected a possible dependence of $h_k$ on the external
scalar momentum of the Yukawa vertex.  This yields the expressions
(\ref{AAA70}) and (\ref{AAA69}) for the flow of $h^2$ and $\eta_\psi$,
respectively, but now with the $T$--dependent threshold functions
\begin{equation}
  \label{PPP08}
  \begin{array}{rcl}
    \ds{m_{1,2}^{(F B)d}(w_1,w_2,\tilde{T};\eta_\psi,\eta_\Phi)}
    &=& \ds{
      m_{1,2}^{(F B)d}(w_1,w_2,\tilde{T})
      }\nnn
    &-& \ds{
      \eta_\Phi\hat{m}_{1,2}^{(F B)d}(w_1,w_2,\tilde{T})-
      \eta_\psi\check{m}_{1,2}^{(F B)d}(w_1,w_2,\tilde{T})
      }\nnn
    && \ds{ \hspace{-4cm}
      = -\frac{1}{2} k^{2(n_1+n_2)-d-1}
      \frac{d v_{d-1}}{(d-1)v_d}
      \tilde{T}\sum_{l\in\ZZZ}
      \int_0^\infty dx\, x^{\frac{d-1}{2}}
      }\nnn
    && \ds{\hspace{-4cm}\times
      \widehat{\frac{\prl}{\prl t}}\left\{
      \frac{1+r_F(y_F)}{[P_F(y_F)+k^2w_1]^{n_1}}
      \frac{\dot{P}(y)}{[P(y)+k^2w_2]^{n_2}} \right\}
      }\; , \nnn
    \ds{l_{n_1,n_2}^{(FB)d}(w_1,w_2,\tilde{T};\eta_\psi,\eta_\Phi)} 
    &=& \ds{
      l_{n_1,n_2}^{(FB)d}(w_1,w_2,\tilde{T})}\nnn
    &-& \ds{
      \eta_\psi \check{l}_{n_1,n_2}^{(FB)d}(w_1,w_2,\tilde{T})
      -\eta_\Phi \hat{l}_{n_1,n_2}^{(FB)d}(w_1,w_2,\tilde{T}) 
      }\nnn
    && \ds{ \hspace{-4cm}
      = -\hal k^{2(n_1+n_2)-d+1}
      \frac{v_{d-1}}{v_d}\tilde{T} 
      \times\sum_{l\in\ZZZ}
      \int_0^\infty dx\, x^{\frac{d-3}{2}}
      }\nnn
    && \ds{ \hspace{-4cm}
      \widehat{\frac{\prl}{\prl t}}\left\{
      \frac{1}{[P_F(y_F)+k^2w_1]^{n_1} [P(y)+k^2w_2]^{n_2} } \right\}
      }\; .
  \end{array}
\end{equation}

\appendix{Pole structure of the $l^d_n$ integrals \label{poles}}

The integrals
\bea
\ds{l^d_n(\om)} &=&
\ds{- n \int^{\infty}_0 dy y^{\frac{d}{2}+1} 
\frac{\partial r(y)}{\partial y}
\left[ y(1+r(y)) + \om \right]^{-(n+1)}}, 
\nnn 
\ds{\hat{l}^d_n(\om)}
&= &\ds{ \frac{n}{2} \int^{\infty}_0 dy y^{\frac{d}{2}} }
r(y)
\left[ y(1+r(y)) + \om \right]^{-(n+1)}
\label{threshldn} 
\eea
with
\be
r(y)=\ds{\frac{e^{-y}}{1-e^{-y}}}\quad , \quad
\ds{\frac{\partial r(y)}{\partial y}=
-\frac{e^{-y}}{(1-e^{-y})^2}}
\ee
exhibit for $d \le 2(n+1)$ a singularity at $\om=-1$.
The massless dimensionless average propagator
$y(1+r(y))$ is a monotonic function of $y$ that takes
on its minimum at $y=0$ with $\lim_{y \to 0}
y(1+r(y))=1$. We define new variables $\dt$ and $z$,
\be
\dt=\om + 1\quad,\quad z=y(1+r(y))-1
\label{deltaz}
\ee
and substitute in (\ref{threshldn}),
\bea
\ds{l^d_n(\dt)} &=&
\ds{\int^{\infty}_0 dz G^d_n(z) (z+\dt)^{-(n+1)}}, 
\nnn 
\ds{\hat{l}^d_n(\dt)}
&= &\ds{ \int^{\infty}_0 dz \hat{G}^d_n(z) (z+\dt)^{-(n+1)}}
\label{subst} 
\eea
with
\bea
G^d_n(z)&=&\ds{-\frac{n y^{\frac{d}{2}+1} 
\ds{\frac{\partial r(y)}{\partial y}}}{1+r(y)+
y\ds{\frac{\partial r(y)}{\partial y}}} },\nnn
\hat{G}^d_n(z)&=&\ds{\frac{n y^{\frac{d}{2}} 
r(y)}{2 \left(1+r(y)+
y\ds{\frac{\partial r(y)}{\partial y}}\right)} }
\eea
and $y=y(z)$. For $d < 2(n+1)$ the integrals (\ref{subst})
have a pole at $\dt = 0$. (The singularity becomes
logarithmic in $\dt$ for $d = 2(n+1)$). In this case
for $\dt \to 0$ the dominant contribution to the 
integral comes from the region $y \simeq 0$ or equivalently
$z \simeq 0$. To find an approximate expression for
$l^d_n$ and $\hat{l}^d_n$ near the pole we expand
the regular part of $G^d_n$ and $\hat{G}^d_n$ around
$z=0$. With
\bea
r(y)&=&\ds{\frac{1}{y}\left(1-\hal y+\frac{1}{12}y^2+O(y^4)
\right)},\nnn
\ds{\frac{\partial r(y)}{\partial y}}&=&\ds{-\frac{1}{y^2}
\left(1-\frac{1}{12} y^2 +O(y^4)\right)}
\eea
one obtains
\bea
G^d_n(z)&=&\ds{n y^{\frac{d}{2}-1}\left(
2-\frac{2}{3}y+\frac{1}{18}y^2+O(y^3)\right)},\nnn
\hat{G}^d_n(z)&=&\ds{\frac{n}{2} y^{\frac{d}{2}-1}
\left(2-\frac{5}{3}y+\frac{13}{18}y^2+O(y^3)\right)}.
\label{gy}
\eea
The inversion of $z(y)$ given by (\ref{deltaz})
can be done by expanding
\be
z=\ds{\hal y + \frac{1}{12} y^2 + O(y^4)}
\ee
and we find
\be
y=2 z - \frac{2}{3} z^2 + \frac{4}{9} z^3 + O(z^4).
\label{inv}
\ee
Insertion of (\ref{inv}) in (\ref{gy}) yields
\bea
G^d_n(z)&=&\ds{ 2 n (2 z)^{\frac{d}{2}-1}
\left(1-\frac{1}{3}\left(\frac{d}{2}+1\right)z
+ \left[\frac{1}{3}+\frac{1}{18}\left(\frac{d}{2}-1\right)
\left(\frac{d}{2}+6\right)  \right] z^2 +O(z^3)\right)},\nnn
\hat{G}^d_n(z)&=&\ds{ n (2 z)^{\frac{d}{2}-1}
\left(1-\frac{1}{3}\left(\frac{d}{2}+4\right)z
+ \left[2+\frac{1}{18}\left(\frac{d}{2}-1\right)
\left(\frac{d}{2}+12\right)  \right] z^2 +O(z^3)\right)}.
\label{gz}
\eea
We consider for $d=3$ and $n \ge 1$ the zeroth order
expression for $l^d_n$ and $\hat{l}^d_n$ that obtains from
the first term in (\ref{gz}) and the exchange of 
summation and integration in (\ref{subst}). Near the
pole one finds
\bea
l^3_n(\dt) &\simeq& \ds{2^{3/2} n \int^{\infty}_0 dz
z^{1/2}(z+\dt)^{-(n+1)}}\nnn
\hat{l}^3_n(\dt) &=& \ds{\hal l^3_n(\dt)}.
\eea
The leading contributions to $l^3_1$, $l^3_2$ and
$l^3_3$ are therefore given by
\bea
l^3_1(\dt)&=&2^{1/2} \pi \dt^{-1/2},\nnn
l^3_2(\dt)&=&2^{-1/2} \pi \dt^{-3/2},\nnn
l^3_3(\dt)&=&2^{-5/2} 3\pi \dt^{-5/2}.\nnn
\eea
We have verified this numerically.

\appendix{The quark mass term}
\label{Source}
In this appendix we determine the source $\jmath={\rm
  diag}(j_u,j_d,\ldots)$ as a function of the average current quark
mass $\hat{m}$. In this context it is important to note that the
source $\jmath$ does not depend on the IR cutoff scale $k$.  Since
$\jmath$ is determined by the properties of the quark meson model at
the compositeness scale $k_{\Phi}$ and also enters directly the value
of the pion mass, which is determined at $k=0$, this relation provides
a bridge between the short and long distance properties of the quark
meson model.  This will allow us to compute the chiral condensate
$\VEV{\ol{\psi}\psi}$ or the parameter $B_0$ of chiral perturbation
theory \cite{GL82-1}. (We expect, however, sizeable corrections when
going from two to three flavors.  They arise because of the relevance
of strange quark physics at scales near $k_{\Phi}$.). In a more
general context we need the proportionality coefficient $a_q$ between
the source $\jmath_q$ and the current quark mass $m_q$,
$q=u,d,\ldots$, taken at the renormalization scale\footnote{We will
  occasionally use the notation $\hat{m}(\mu)$, $m_q(\mu)$ or
  $\VEV{\ol{\psi}\psi}(\mu)\equiv\VEV{\ol{\psi}\psi}_{k=0}(\mu)$ (not
  to be confused with
  $\VEV{\ol{\psi}\psi}_k\equiv\VEV{\ol{\psi}\psi}_k(\mu=k_\Phi)$) in
  order to indicate the renormalization scale $\mu$. If no argument is
  given $\mu=k_\Phi$ is assumed.} $\mu=k_{\Phi}$,
\begin{equation}
  \label{PPF00} 
  \jmath_q=\frac{Z_{\psi,k_{\Phi}}}{\ol{h}_{k_{\Phi}}} a_q m_q \; .
\end{equation}  
For a computation of the coefficient $a_q$ we need to look into the
details of the introduction of composite meson fields in QCD
\cite{EW94-1,JW96-4}. Let us assume that at the scale $k_{\Phi}$ a
part of the QCD average action for quarks $\Gamma_{k_\Phi}[\psi]$
factorizes in the quark bilinear
\begin{equation}
  \label{PPF01}
  \chi_{ab}(q)=-\int\frac{d^4p}{(2\pi)^4}\tilde{g}(p,q)
  \ol{\psi}_{Lb}(p) \psi_{Ra}(p+q)
\end{equation}
such that
\begin{equation}
  \label{Fk}
  \Gamma_{k_\Phi}[\psi]=-F_{k_\Phi}[\chi]+
  \Gamma_{k_\Phi}^{\prime}[\psi]\; .
\end{equation}
We can then introduce meson fields by inserting the identity
\begin{equation}
  \label{PPF03}
  N \int D\Phi \exp\left(-F_{k_\Phi}[\chi+\Phi]\right)=1
\end{equation}
into the path integral which formally defines $\Gamma_{k_\Phi}[\psi]$.
(Here $N$ is a field independent normalization factor.) This
effectively replaces in (\ref{Fk}) the term $-F_{k_\Phi}[\chi]$
by\footnote{The summation over internal indices as well as the
  integration over momenta has been suppressed. For complex
  $\chi_{ab}(q)$ similar terms have to be supplemented in the
  expansion. See ref.~\cite{EW94-1,JW96-4} for a more detailed
  description. In our Euclidean conventions one has
  $\chi^{\dagger}_{ab}\sim +\tilde{g}^* \ol{\psi}_{Rb} \psi_{La}$.}
\begin{equation}
  \label{PPF04}
  -F_{k_\Phi}[\chi]+F_{k_\Phi}[\chi+\Phi]=\ds{
    \frac{\partial{F_{k_\Phi}[\chi]}}
    {\partial\chi_{ab}(q)}\Phi_{ab}(q)
    +\frac{1}{2}\frac{\partial^2F_{k_\Phi}[\chi]}
    {\partial\chi_{ab}(q)\partial\chi_{cd}(q^\prime)}
  \Phi_{ab}(q)\Phi_{cd}(q^{\prime})+\ldots\; .}
\end{equation}
The original multi--quark interaction $-F_{k_\Phi}[\chi]$ is canceled
by the lowest order term in the Taylor expansion in $\Phi$.  Instead,
we have substituted mesonic self--interactions $F_{k_\Phi}[\Phi]$ and
interactions between mesons and quarks corresponding to the terms in
the expansion which contain powers of $\chi$ and $\Phi$. In
particular, we may specialize to the case where the derivative terms
in $F_{k_\Phi}$ are small and consider a local form $F_{k_\Phi}=\int
d^4x f_{k_\Phi}(\chi)$. A quark mass term is linear in $\chi$ and
translates into a source term for $\Phi$
\begin{equation}
  \label{source} 
  \ds{-\frac{Z_{\psi,k_\Phi}}{\tilde{g}}\mbox{Tr}
    (\chi^{\dagger}m+m^{\dagger}\chi)}\longrightarrow
  \ds{-\frac{Z_{\psi,k_\Phi}}{\tilde{g}}\mbox{Tr}
    (\Phi^{\dagger}m+m^{\dagger}\Phi)}
  =-\ds{\frac{1}{2}\mbox{Tr}(\Phi^{\dagger}\jmath+\jmath^{\dagger}\Phi)
    }
\end{equation}
where $m={\rm diag}(m_u,m_d,\ldots)$.  A factorizing four fermion
interaction yields
\begin{equation}
  \label{PPF06}
  \ol{m}_{k_\Phi}^2\mbox{Tr}\chi^{\dagger}\chi \longrightarrow
  \ol{m}_{k_\Phi}^2\mbox{Tr}\Phi^{\dagger}\Phi
  +\ol{m}_{k_\Phi}^2\mbox{Tr}(\Phi^{\dagger}\chi+\chi^{\dagger}\Phi)\; .
\end{equation}
The second term corresponds to the Yukawa interaction with
$\ol{h}_{k_\Phi}=\ol{m}_{k_\Phi}^2\tilde{g}$. We can therefore extract
$a_q$ from eq.\ (\ref{source}) as
\begin{equation}
  \label{aq}
  a_q=2 \ol{m}_{k_\Phi}^2\; .
\end{equation}
We note that only the terms linear and quadratic in $\chi$ influence
the value of $a_q$. We could either restrict the composite fields from
the beginning to the ones contained in the $O(4)$ symmetric linear
$\sigma$-model or work with all the fields contained in a complex $2
\times 2$ matrix $\Phi$.  In the latter case the anomaly term would
contribute to both the masses and the Yukawa coupling. The net result
is the same with $\ol{m}_{k_\Phi}^2$ denoting the relevant mass term
for the $O(4)$ vector. For our conventions with $\ol{h}_{k_\Phi}=1$ we
have to normalize with $\tilde{g}=\ol{m}_{k_\Phi}^{\ -2}$.  Finally an
eight fermion interaction becomes
\begin{equation}
  \label{eightfer}
  \ds{\frac{1}{2}\ol{\lambda}_{k_\Phi}\, (\mbox{Tr} \chi^{\dagger}\chi)^2}
  \longrightarrow \ds{\frac{1}{2}\, \ol{\lambda}_{k_\Phi}(\mbox{Tr}
    \Phi^{\dagger}\Phi)^2+\, \ol{\lambda}_{k_\Phi} 
    \mbox{Tr}\Phi^{\dagger}\Phi
    \,  \mbox{Tr}(\chi^{\dagger}\Phi+\Phi^{\dagger}\chi)+\ldots
    }\; .
\end{equation}
We see here the appearance of terms quadratic in the quarks involving
higher powers of $\Phi$.

There is an alternative, equivalent way of understanding the relation
between $\jmath$ and $m_q$: The quark masses in the picture with
mesons must be equal at the scale $k_{\Phi}$ to the current quark mass
$m_q(k_{\Phi})$. Let us consider an $O(4)$ symmetric fermionic
interaction $\ol{m}_{k_\Phi}^2 \mbox{Tr}
\chi^{\dagger}\chi+\frac{1}{2}\ol{\lambda}_{k_\Phi}(\mbox{Tr}
\chi^{\dagger}\chi)^2$ which leads to a meson potential
\begin{equation}
  \label{PPF07}
  U_{k_\Phi}=\ol{m}_{k_\Phi}^2\Tr\Phi^\dagger\Phi+
  \frac{1}{2}\ol{\lambda}_{k_\Phi}
  \left(\Tr\Phi^\dagger\Phi\right)^2\; .
\end{equation}
In the mesonic picture the quarks acquire masses through the Yukawa
coupling to $\Phi$
\begin{equation}
  \label{source2}
  M_{k}=\ds{\frac{\ol{h}_{k}}{Z_{\psi,k}}
    \left(1+\frac{\ol{\lambda}_{k}}
      {\ol{m}_{k}^2}\mbox{Tr}\langle\Phi^{\dagger}\rangle_k
      \VEV{\Phi}_k\right) \VEV{\Phi}_k }
\end{equation}
where the second term arises from the higher order coupling in
(\ref{eightfer}). Here $\VEV{\Phi}_k=\mbox{diag}
(\varphi_u,\varphi_d,\ldots)$ is the expectation value at the coarse
graining scale $k$ in the presence of the source term and
$M_k=\mbox{diag}(M_u,M_d,\ldots)$.  It is sufficient to specify the
dependence of $U_{k_\Phi}$ on real diagonal fields $\Phi_{qq}$. Then
the $\varphi_q$ are determined from the condition
\begin{equation}
  \label{PPF08}
  \ds{\frac{\partial U_{k_\Phi}}{\partial \Phi_{qq}}(\varphi_q)}
  =2\left(\ol{m}_{k_\Phi}^2+
  \ol{\lambda}_{k_\Phi}\sum_{q^{\prime}}\varphi_{q^{\prime}}^2
  \right)
  \varphi_q=\jmath_q\; .
\end{equation}
Identifying $M_{k=k_{\Phi}}$ in (\ref{source2}) with $m(k_\Phi)$ one
has
\begin{equation}
  \label{PPF09}
  a_q\left(1+\ds{\frac{\ol{\lambda}_{k_\Phi}}{\ol{m}_{k_\Phi}^2}
    \sum_{q^{\prime}}
    \varphi_{q^{\prime}}^2}\right)=\ds{\frac{\jmath_q}{\varphi_q}}=
  2\ol{m}_{k_\Phi}^2
  +2\ol{\lambda}_{k_\Phi}
  \sum_{q^{\prime}}\varphi_{q^{\prime}}^2
\end{equation}
and we recover (\ref{aq}) or, in our normalization with
$Z_{\psi,k_{\Phi}}=1$, $\ol{h}_{k_{\Phi}}=1$,
\begin{equation}
  \label{PPF10}
  \jmath=2 \ol{m}_{k_\Phi}^2 \hat{m}\; .
\end{equation}
It is remarkable that higher order terms (e.g. $\sim
\ol{\lambda}_{k_\Phi}$) do not influence the relation between $\jmath$
and $\hat{m}$.  Only the quadratic term $\ol{m}_{k_\Phi}^2$ enters
which is in our scenario the only relevant coupling. This feature is
an important ingredient for the predictive power of the model as far
as the absolute size of the current quark mass is concerned. An
appearance of higher order couplings in $a_q$ would make it very hard
to compute this quantity.  We emphasize that within our formalism
there is no difference of principle between the current quark mass and
the constituent quark mass.  Whereas the current quark mass
$m_q(k_\Phi)$ at the normalization scale $\mu=k_\Phi$ corresponds to
$M_{q,k}$ at the compositeness scale $k_{\Phi}$ the constituent quark
mass is $M_{q,k=0}$. As $k$ is lowered from $k_{\Phi}$ to zero one
observes that the quark mass increases, similarly to the running
current quark mass.  Once chiral symmetry breaking sets in at the
scale $k_{\chi SB}$ there is a large increase in the quark masses,
especially for $M_u$ and $M_d$.

The formalism of composite fields also provides the link
\cite{EW94-1} to the chiral condensate $\VEV{\ol{\psi}\psi}$ since
the expectation value $\VEV{\Phi}$ is related to the expectation
value of a composite quark-antiquark operator. For 
$\ol{\lambda}=0$ one has \cite{JW96-4}
\begin{equation}
  \label{PPF11}
  \VEV{\Phi}_k+\langle\Phi^{\dagger}\rangle_k=
  -\ds{\frac{1}{\ol{m}_{k_\Phi}^2}}
  \VEV{\ol{\psi}\psi}_k(k_{\Phi})+m_q(k_{\Phi})+m_q^{\dagger}(k_{\Phi})
\end{equation}
with $\VEV{\ol{\psi}\psi}_k(k_\Phi)$ a suitably regularized operator
normalized at $\mu=k_{\Phi}$.